\documentclass[letterpaper,11pt,reqno]{article}
\usepackage[portrait,margin=1in]{geometry}
\usepackage{graphicx,color,fancyhdr}
\usepackage{xcolor}
\usepackage{amsmath,amssymb,amsthm,amsfonts,amsbsy,latexsym,dsfont}
\usepackage{mathrsfs}
\usepackage[colorlinks=true,linkcolor=black,citecolor=black,urlcolor=black]{hyperref} 
\usepackage{multirow, tabularx}
\usepackage{amsmath,bm}
\usepackage{subcaption, caption}
\usepackage{enumerate}
\usepackage{authblk}
\usepackage[shortlabels]{enumitem}
\setlist[enumerate]{topsep=0pt}
\usepackage{setspace}

\usepackage[utf8]{inputenc}
\usepackage{cite}
\usepackage{comment}
\usepackage{placeins}

\newtheorem{lemma}{Lemma}
\newtheorem*{remark}{Remark}

\newcommand{\bbm}{\begin{bmatrix}}
\newcommand{\bpm}{\begin{pmatrix}}
\newcommand{\ebm}{\end{bmatrix}}
\newcommand{\epm}{\end{pmatrix}}

 \newcommand{\del}[2]{\frac{\partial #1}{\partial #2}}
 \newcommand{\dsdel}[2]{\displaystyle\frac{\partial #1}{\partial #2}}

  \newcommand{\doubledelsame}[2]{\displaystyle\frac{\partial^2 #1}{\partial #2^2}}
   \newcommand{\doubledelsamesmall}[2]{\frac{\partial^2 #1}{\partial #2^2}}

\newcommand{\dsddt}[1]{\displaystyle\frac{d #1}{dt}}

\numberwithin{equation}{section}

\title{Pattern Formation in a Spatial Public Goods Dilemma due to Diffusive or Directed Motion}

\author[,1]{Yuxuan Zhao\thanks{\href{mailto:yuxuan44@illinois.edu}{yuxuan44@illinois.edu}}}
\author[1]{Kaisheng Zhu}
\author[1]{Yefei Zhang}
\author[,1,2]{Daniel B. Cooney\thanks{\href{mailto:dbcoone2@illinois.edu}{dbcoone2@illinois.edu}}}

\affil[1]{\small{Department of Mathematics, University of Illinois Urbana-Champaign, Urbana, IL 61801, USA}}
\affil[2]{\small{Carl R. Woese Institute for Genomic Biology, University of Illinois Urbana-Champaign, Urbana, IL, 61801, USA}}

\begin{document}

\maketitle

\begin{abstract}
The costly provision of public goods serves as a model problem for the evolution of cooperative behavior, presenting a social dilemma between the collective benefits of shared resources and the individual incentive to free-ride in resource production. The spatial structure of populations can also impact cooperation over public goods, as diffusion of public goods and intentional motion of individuals towards regions with greater resources can interact with population and public goods dynamics to produce heterogeneous patterns in the spatial distribution of strategies and resources. In this paper, we build off a model introduced by Young and Belmonte for the reaction dynamics of interacting individuals and explicit public good, deriving a system of PDEs that describes the spatial profiles of strategies and the public good in the presence of both diffusive motion of individuals and resources and chemotaxis-like directed motion of individuals in response to gradients in the concentration of public goods. Through linear stability analysis, we show that spatial patterns in strategic and public goods profiles can emerge due to either Turing instability with high defector diffusivity or a directed-motion instability through strong sensitivity of cooperators towards increasing resource concentration. We further explore the emergent spatial patterns with a mix of weakly nonlinear stability analysis and numerical simulation, showing that, for a wide range of reaction parameters, diffusion-driven instability appears to increase cooperation and public goods across the spatial domain, while directed motion of cooperators towards public goods tends to decrease cooperation and environmental quality across the environment.
\end{abstract}

\begin{spacing}{0.01}
    \renewcommand{\baselinestretch}{0.1}\normalsize
    \tableofcontents
    \addtocontents{toc}{\protect\setcounter{tocdepth}{2}}
\end{spacing}

\section{Introduction} \label{sec:intro}

The production of public goods presents a social dilemma in the evolution of cooperative behavior, as the costly production of shared resources provides a tug-of-war between the collective benefit imparted by the resource and the individual incentives to cheat and rely on resources provided by peers. Such public goods dilemmas arise in many biological and social systems, ranging across scales from the production of beneficial enzymes in microbial populations \cite{helbing2009outbreak,craig2008stable,maclean2010mixture,Vulic200106}, collective action problems in animal groups \cite{Dobata201102,levin2014public}, and spillover effects of economic activity and international trade \cite{conybeare1984public}. Many mathematical models have used evolutionary game theory or related eco-evolutionary models to describe the evolution of cooperative behavior in public goods dilemmas \cite{frank2010general,ARCHETTI20129,Fehr200001,Hauert200804,hauert2008ecological,YOUNG201812,santos2008social}, and this modeling work has proposed many mechanisms for the emergence and maintenance for the cooperative maintenance of public goods. Such models of public goods dilemmas have particularly interesting features when considering the effects of space on public goods provision, as the spatial diffusion of public goods can benefit free-riding individuals \cite{driscoll2010theory,allen2013spatial,yao2025spatial} and individuals can also perform directed or diffusive motion towards regions of greater concentrations of public goods \cite{youk2014secreting,YOUNG201812,funk2019directed,yao2026pattern}.

Such spatial models of public goods dilemmas have often considered systems of partial differential equations (PDEs) to describe the spatial dynamics of the strategic composition of populations, using the payoffs from public goods games to describe local reproduction or imitation dynamics and considering spatial motion through either purely diffusive motion \cite{wakano2009spatial,wakano2011pattern,park2019ecological,gokhale2020eco,aydogmus2017preservation,aydogmus2018discovering} or through directed motion towards cooperators or away from defectors \cite{funk2019directed}. These papers have demonstrated that the strategic dynamics can interact with the rules for spatial motion to produce spatial patterns in the profiles of strategies, with spatial clustering of individuals often resulting in an increased overall level of cooperation in the population \cite{wakano2009spatial,wakano2011pattern}. Similar spatial models have also been considered for two-player snowdrift games that feature coexistence of cooperators and defectors in the absence of space, showing that spatial patterns can tend to emerge due to Turing instability under purely diffusive motion \cite{yao2026pattern,cheng2024evolution,cheng2026spatio,fahimipour2022sharp,lawton2024interspecific} or due to instabilities induced through directed motion through mechanisms of chemotaxis-like environmental-driven motion with environmental feedback \cite{yao2025spatial} or payoff-driven motion in which individuals move towards regions featuring higher payoff for their strategy \cite{helbing2009pattern,helbing2009outbreak,deforest2013spatial,deforest2018game,yao2026pattern}. The case of payoff-driven motion is particularly notable as it can tend to feature negative diffusive defects due to cooperators directing their motion towards regions with increasing densities of cooperators, often leading to a physically unrealistic scenario of short-wave / infinite-wavenumber instabilities \cite{helbing2009pattern,fahimipour2022sharp,yao2026pattern}. 

While public goods games and other evolutionary game models can capture many features of the spatial dynamics of cooperation, models that explicitly incorporate the spatial profile of a public good can further permit exploration of key features of strategic and movement dynamics specific to problem of public goods provision. Recent work has explored spatial models of public goods that incorporate the costly production of public goods, the diffusion of the public good across space, and the directed and diffusive motion of individuals across continuous space or metapopulation networks \cite{YOUNG201812,gerlee2019persistence,driscoll2010theory}. In particular, Young and Belmonte formulated a metapopulation model in which cooperators pay a cost to produce a public good, both cooperators and defectors consume the good, and individuals with each strategy can perform directed motion towards patches with greater concentrations of public goods using a piecewise movement rule. Young and Belmonte considered a public good that did not diffuse between patches, and, under the assumption of fast public good dynamics, they were able to use a quasi-steady state approximation to reduce their model to a system of ODEs for the densities of cooperators and defectors in each patch of their metapopulation, showing that sufficiently fast motion of defectors could help to establish long-time survival of cooperation in their population \cite{YOUNG201812}.

In this paper, we look to further explore the spatial dynamics for public goods models inspired by questions about the local ecological dynamics of cooperators, defectors, and public goods and the rules for motion of individuals and public goods in space. We consider a modified version of the metapopulation model introduced by Young and Belmonte that allows us to explore the spatial patterns generated by diffusion of the public good and both diffusive and directed motion for cooperators and defectors, formulating the rules for spatial motion as biased random walks of individuals to neighboring patches. We derive a PDE model for the spatial profiles of cooperators, defectors, and the public good in the continuum limit of infinitely many patches and infinitesimal distance between patches, capturing undirected spatial motion through diffusion terms and having both cooperators and defectors perform a chemotaxis-like directed motion up spatial gradients of the concentration of the public good. Our full model takes the form of a system of three reaction-diffusion-taxis PDEs which bear resemblance to prior models exploring collective behavior and directed motion in multi-species systems \cite{keller1970initiation,hillen2009user,amorim2015modeling,alonso2016analysis,painter2019mathematical,deng2026reaction,wang2015global,owolabi2026multi}, and we can also explore a special case of the model that reduces to a three-type reaction-diffusion system that have often been used to study Turing pattern formation with three interacting species \cite{turing1952chemical,murray2003mathematical,satnoianu2000turing,maini2019turing,piskovsky2025turing}. We perform linear stability analysis to derive conditions for spatial pattern formation in both our full model and the special case of the reaction-diffusion system, showing that there can be Turing patterns in the reaction diffusion model due to increased diffusivity of defectors and that we can see a chemotaxis-like instability in the full model when cooperators have a sufficiently strong sensitivity to moving towards increasing concentration of the public good. 

The remainder of the paper is organized in the following manner. In Section \ref{section 2}, we recall the baseline Young-Belmonte model, we describe our modified version of this model to describe biased random walks along the metapopulation graph, and we present the PDE model we obtain in the continuum limit. We then present the stability analysis of our PDE models for both the case of Turing instability and a taxis-driven instability in Section \ref{sec:linear-stability-analysis}, and we provide a weakly nonlinear stability analysis of both instabilities in Section \ref{sec:weakly-nonlinear-analysis}. We provide numerical simulations of the reaction-diffusion model and our full reaction-diffusion-taxis model in Section \ref{section:simulations}, along with comparisons between our simulation results and the predictions of weakly nonlinear analysis. We discuss our results in Section \ref{sec:Discussion} and we provide additional derivations in the Appendix, highlighting the derivation of our PDE model (Section \ref{sec:PDEderivation}), comparing possible conditions for instability in our reaction-diffusion model (Section \ref{sec:linear-Routh-Hurwitz}), providing the asymptotic expansions of our PDE model for weakly nonlinear analysis (Section \ref{sec:weakly-nonlinear-expansions}), and analyzing the correlation between concentrations of cooperators, defectors, and the public good near the onset of the pattern-forming instability (Section \ref{sec:eigenvectors}). 

\section{Model Construction}\label{section 2}

In this section, we will develop a spatial model for the public goods dilemma. We first introduce the ODE model proposed by Young and Belmonte \cite{YOUNG201812}, which considers the dynamics of cooperators, defectors, and a public good on a discrete system of patches. We incorporate the reaction dynamics considered in the Young-Belmonte model, and we build on their framework by incorporating (potentially biased) random walks of individuals and the public good between patches, using these assumptions of movement between patches to derive a PDE model in the continuum limit.

We start our discussion in Section \ref{sec:YoungBelmontebackground} by formulating the reaction dynamics of the Young-Belmonte model and highlighting the assumptions used by Young and Belmonte for directed motion between patches. We then proceed to formulate our new assumptions on motion between patches in Section \ref{section 2.1} and present our PDE model that we obtain in the continuum limit in Section \ref{section 2.2}. Finally, we summarize some of the key properties of the reaction dynamics for cooperators, defectors, and the public good in Section \ref{sec:ODEreaction}, which will prepare us for the linear stability analysis of our ODE reaction dynamics and spatial PDE models that is presented in Section \ref{sec:linear-stability-analysis}.

\subsection{Background on Young-Belmonte Model for Public Goods Dilemma on Metapopulation}
\label{sec:YoungBelmontebackground}

Following the approach introduced by Young and Belmonte, we consider a population structure consisting of multiple spatial sites that we refer to as patches. We assume that there are $N$ patches, which we index by $i\in\{1,\dots,N\}$. Within each patch, there are two types of individuals. The first type, called cooperators, are those who contribute to the production of public goods; their population size in patch $i$ is denoted by $u_i$. The second type, called defectors, are those who consume public goods without contributing; their population size in patch $i$ is denoted by $v_i$. The amount of public goods in patch $i$ is represented by $\phi_i$.

We assume that the growth of cooperators and defectors is proportional to the amount of available public goods, with growth rates $r_u$ and $r_v$, respectively. Both types also experience natural death rates $\mu_u$ and $\mu_v$. Crowding effects are taken into account, with both types sharing the same loss rate $\gamma(u_i+v_i)$. In addition, cooperators suffer an extra cost for producing public goods, modeled as an additional loss rate $c$.

The dynamics of the public goods are described as follows. Cooperators produce public goods at a rate equal to the additional cost $c$. Public goods are consumed by both types at rate $\kappa$, and they also decay naturally at rate $\delta$.

Now, putting these assumptions together, we can formulate an ODE system for patch $i$ as follows
\begin{equation}\label{eq:Young-Belmonte-ODE}
    \begin{aligned}
        \frac{du_i}{dt}&=u_i\left[r_u\phi_i-c-\gamma(u_i+v_i)-\mu_{u}\right],\\
        \frac{dv_i}{dt}&=v_i\left[r_v\phi_i-\gamma(u_i+v_i)-\mu_{v}\right],\\
        \frac{d\phi_i}{dt}&=cu_i-\left[\kappa(u_i+v_i)+\delta\right]\phi_i.
    \end{aligned}
\end{equation}
The crowding term makes it possible for each patch to reach an internal coexistence equilibrium, where the size of each population is inversely related to $\gamma$. Our Table~\ref{table 1} shows default values for all parameters in Equation \eqref{eq:Young-Belmonte-ODE} that we adopt from their use in the paper by Young and Belmonte \cite{YOUNG201812}. In particular, the choices of the resource utilization rates $r_u$ and $r_v$ and the background death rates $\mu_u$ and $\mu_v$ for cooperators and defectors were chosen in the paper of \cite{YOUNG201812} to produce a stable coexistence equilibrium state featuring positive densities of both cooperators and defectors, and we will continue to follow this assumption due to our interest in exploring how rules for spatial motion of individuals and resources can destabilize a spatially uniform coexistence equilibrium. 

\begin{table}[!ht]
    \centering
    \begin{tabular}{lll}
        \hline
        Parameters &  & Default values \\
        \hline
        $r_u$ & Growth rate of cooperators & $5.0$ \\
        $r_v$ & Growth rate of defectors & $6.0$ \\
        $c$ & Producing rate & $1.0$ \\
        $\gamma$ & Crowding effect rate & $1.0$\\
        $\mu_u$ & Natural death rate of cooperators & $2.0$ \\
        $\mu_v$ & Natural death rate of defectors & $3.7$ \\
        $\kappa$ & Consuming rate of goods & $1.0$ \\
        $\delta$ & Decay rate of public goods & $10^{-3}$ \\
        \hline
    \end{tabular}
    \caption{Default values for parameters for the reaction dynamics from Equation \eqref{eq:Young-Belmonte-ODE}, based on the values assumed in the model of Young and Belmonte \cite{YOUNG201812}.}\label{table 1}
\end{table}

Young and Belmonte consider the case of fast dynamics of the public good to make a quasi-steady state that the level of public good in patch $i$ takes the value
\begin{equation}
    \phi_i = \frac{c u_i}{\kappa (u_i + v_i) + \delta},
\end{equation}
allowing them to rewrite the per-capita growth rates for cooperators and defectors on patch $i$ as
\begin{subequations}
    \begin{align}
        F_u\left(u_i,v_i\right) &= \frac{r_u c u_i}{\kappa (u_i + v_i) + \delta} - c - \gamma \left(u_i + v_i\right) - \mu_u \\
        F_v\left(u_i,v_i\right) &= \frac{r_v c u_i}{\kappa (u_i + v_i) + \delta} - \gamma \left(u_i + v_i\right) - \mu_v.
    \end{align}
\end{subequations}
Young and Belmonte then use these per-capita growth rates as a way to describe fitness-driven directed motion, introducing a system of ODEs describing both the reaction dynamics and directed motion across patches of their metapopulation network:
\begin{subequations}
    \begin{align}
        \dsddt{u_i} &= u_i F_u\left(u_i,v_i\right) + \sum_{j=1}^N \alpha_{ij} H_u \left( u_i,v_i,u_j,v_j\right) \\
        \dsddt{v_i} &= u_i F_u\left(u_i,v_i\right) + \sum_{j=1}^N \alpha_{ij} H_u \left( u_i,v_i,u_j,v_j\right),
    \end{align}
\end{subequations}
where $\alpha_{ij}$ captures the connectivity or relative movement rates between patches $i$ and $j$ and the functions $H_u(\cdot)$ and $H_v(\cdot)$ describe how the fitnesses of individuals on each patch impact the rate of directed motion. The particular choice of directed motion model considered by Young and Belmonte involved taking $H_u$ and $H_v$ to depend on the per-capita fitness functions $F_u$ and $F_v$ in the following manner
\begin{subequations}
    \begin{align}
        H_u\left( u_i,v_i,u_j,v_j \right) &= \left[ F_u \left( u_i,v_i\right) - F_u\left( u_j,v_j \right)  \right]_{+} u_j -  \left[ F_u \left( u_i,v_i\right) - F_u\left( u_j,v_j \right)  \right]_{-} u_i \\
        H_v\left( u_i,v_i,u_j,v_j \right) &= \left[ F_v \left( u_i,v_i\right) - F_v\left( u_j,v_j \right)  \right]_{+} v_j -  \left[ F_v \left( u_i,v_i\right) - F_v\left( u_j,v_j \right)  \right]_{-} v_i,
    \end{align}
\end{subequations}
where $[g(\cdot)]_+$ and $[g(\cdot)]_-$ denote the positive and negative part of the function $g(\cdot)$, respectively. The key assumption in their model of movement between edges is that individuals will move from patch $j$ to patch $i$ only if the per-capita fitness of their strategy is greater in patch $i$ than in patch $j$, and, if that is the case, the rate of motion will be proportional to the difference in per-capita fitness of their strategy between the two patches. Young and Belmonte are able to analyze this piecewise model on their metapopulation network to characterize the long-time behavior under these movement rules, but the piecewise characterization of directed motion provides a potential obstacle to the derivation of continuum limits of the model and the use of PDE techniques to further study the resulting dynamics combining the public goods dilemma with directed spatial motion. 

In this paper, we explore a modified version of the Young-Belmonte model, in which we assume the same reaction terms within patches but consider different roles for spatial motion of individuals and public goods between patches. In particular, we will allow the public good to diffuse along edges of the metapopulation network, and we will formulate a model for biased random walks of cooperators and defectors between patches based on the profile of public goods across the metapopulation, allowing us to deduce the effects of diffusive motion of individuals and intentional motion towards more beneficial environments. We also restrict attention to the case of metapopulation networks consisting of spatial lattices, providing us the opportunity to derive PDE limits of the model in the case of infinitely many patches and infinitesimal distance between the patches.  

\subsection{Formulation of Public Goods Model Based on Biased Random Walks Between Patches}\label{section 2.1}

We assume that all colonies form an $n$-dimensional lattice of nodes, with $n=1$ or $2$. To incorporate spatial distribution, we assume that cooperators and defectors in each patch can migrate to neighboring colonies. To clarify migration behavior, we assign to each patch three quantities, $A_i$, $B_i$, and $C_i$, which are functions of $u_i$, $v_i$, and $\phi_i$, respectively. These quantities represent the attractiveness of patch $i$ to cooperators, defectors, and public goods, and we refer to them as the attractions of patch $i$.

Let $N(i)$ denote the set of all neighboring colonies of patch $i$, excluding $i$ itself. We assume that cooperators, defectors, and the public good can consider to move to a neighboring patch between times $t$ and $t + \Delta t$ with probabilities $\chi_u$, $\chi_v$, and $\chi_{\phi}$, respectively. We assume that an individual can either choose to move to one of their neighboring patches (a patch in the neighborhood $N(i)$ of patch $i$) or to choose to stay at the same patch due to attractiveness of neighboring patches and the current patch. Similarly, for the delivery of goods, we consider the possibility that goods either stay in the same patch or are transported to one of the neighboring colonies. Given that individuals or public good considers to move in a given time-step, we can define the conditional probability that an individual or a unit of public good moves to a patch $j \in N(i) \cup \{i\}$ by the following expressions: 
\begin{equation}\label{eq:moving-probabilities}
    \begin{aligned}
        \mathds{P}_{u}(i\to j)&=\frac{f(w_u, A_j)}{f(w_u, A_i)+\sum_{k\in N(i)}f(w_u, A_k)},\\
        \mathds{P}_{v}(i\to j)&=\frac{f(w_v, B_j)}{f(w_v, B_i)+\sum_{k\in N(i)}f(w_v, B_k)},\\
        \mathds{P}_{\phi}(i\to j)&=\frac{f(w_\phi, C_j)}{f(w_\phi, C_i)+\sum_{k\in N(i)}f(w_\phi, C_k)},
    \end{aligned}
\end{equation}
where $f$ is a positive weighting function, $w_u$, $w_v$ and $w_\phi$ are parameters that controls the influence of attractiveness, with $$f(0, A_i) = f(0, B_i) = f(0, C_i) = 1.$$
In particular, individuals or public good can remain in the same location with probability $\mathds{P}_u(i\to i)$, $\mathds{P}_v(i\to i)$, and $\mathds{P}_{\phi}(i \to i)$ with probability that is weighted by the attractiveness of the current location. 

Based on our assumptions, we consider the a modified version of the Belmonte-Young model to describe public goods dynamics with spatial movement on a lattice metapopulation. We start with a discrete-time model with time-steps of length $\Delta t$, and we consider how the population and public goods concentrations at patch $i$ change due to the following events that occur between a time $t$ and time $t + \Delta t$. 
\begin{enumerate}
    \item First, the cooperators, defectors, and public good undergo their reaction dynamics based on the rates described in Equation \eqref{eq:Young-Belmonte-ODE} and the current values of $u_i(t)$, $v_i(t)$, and $\phi_t(t)$ at patch $i$ at the beginning of the time-step.
    \item Next, individuals and public goods present after the reaction events can perform one step of a biased random walk based on the movement probabilities $\chi_u$, $\chi_v$, and $\chi_{\phi}$ and the conditional probabilities between neighboring patches defined in Equation \eqref{eq:moving-probabilities}.
\end{enumerate}
In particular, we choose to separate the reaction and movement dynamics into this sequence of events so that individuals cannot both die and be chosen to move within the same discrete time-step, which will help guarantee non-negativity of the quantities $u_i(t+\Delta t)$, $v_i(t+\Delta t)$, and $\phi_t(t+\Delta t)$ across the spatial domain. We will show that this choice of movement rules will lead to higher-order terms in the time-step $\Delta t$ and the spatial grid step $\ell$, and therefore our resulting PDE model will feature separate terms describing the continuous-time dynamics of reactions and movement. 

At time $t + \Delta t$, the values of $u$, $v$, and $\phi$ first change according the reaction dynamics motivated by Equation \eqref{eq:Young-Belmonte-ODE}. After this local update, they are redistributed across the grid due to migration or transport. Therefore, we can write the expected concentrations $u_i(t+\Delta t)$, $v_i(t+\Delta t)$, $\phi_t(t+\Delta t)$ of the cooperators, defectors, and public good at patch $i$ and time $t+\Delta t$ as 
\begin{equation}\label{eq:discrete-main-equations}
    \begin{aligned}
        u_i(t+\Delta t)=&\,\,\displaystyle \left\{ u_i(t)\ + u_i(t)\left[r_u\phi_i(t)-c-\gamma(u_i(t)+v_i(t))-\mu_u\right]\Delta t \right\} \left( 1 - \chi_u + \chi_u \mathds{P}_u(i \to i) \right) \\
        &\,\,\displaystyle+\chi_u \sum_{j\in N(i)}\mathds{P}_{u}(j\to i) \left\{ u_j(t) +  u_j(t)\left[r_u\phi_j(t)-c-\gamma(u_j(t)+v_j(t))-\mu_u\right]\Delta t\right\}\\
        v_i(t+\Delta t)=&\,\,\displaystyle \left\{ v_i(t) + v_i(t)\left[r_v\phi_i(t)-\gamma(u_i(t)+v_i(t))-\mu_v\right]\Delta t \right\}\left( 1 - \chi_v + \chi_v \chi_u \mathds{P}_v(i \to i) \right) \\
        &\,\,\displaystyle+\chi_v \sum_{j\in N(i)}\mathds{P}_{v}(j\to i) \left\{ v_j(t) + v_j(t)\left[r_v\phi_j(t)-\gamma(u_j(t)+v_j(t))-\mu_v\right]\Delta t \right\} \\
        \phi_i(t+\Delta t)=&\,\,\displaystyle  \left\{ \phi_i(t) + \left(cu_i(t)-\left[\kappa(u_i(t)+v_i(t))+\delta\right]\phi_i(t)\right)\Delta t \right\} \left( 1 - \chi_{\phi} + \chi_{\phi}  \mathds{P}_{\phi}(i \to i) \right) \\
        &\,\,\displaystyle+ \chi_{\phi} \sum_{j\in N(i)}\mathds{P}_{\phi}(j\to i) \left\{ \phi_j(t) + \left(cu_j(t)-\left[\kappa(u_j(t)+v_j(t))+\delta\right]\phi_j(t)\right)\Delta t \right\} \\
    \end{aligned}
\end{equation}

\subsection{Continuum Limit}\label{section 2.2}

We can then derive our PDE model for the spatial dynamics of cooperators, defectors, and the public good by considering the limit as the time-step $\Delta t \to 0$ and the spacing between grid points $\ell \to 0$, assuming the parabolic scaling relation that $\frac{\ell^2}{\Delta t} \to D$ for some positive constant $D$ in the joint limit as $\ell, \Delta t \to 0$. We present the full derivation of this continuum limit in Section \ref{sec:PDEderivation} of the appendix, and the argument makes substantial use of writing our rules for spatial movement in terms of the discrete Laplacian operator on our spatial lattice, which is given by
\begin{equation}
    \Delta_{\ell} := \ell^{-2} \left[ \sum_{j \in N(i)}u_j(t)-\sum_{j\in N(i)} u_i(t) \right].
\end{equation}
Our derivation follows an approach that has often been used to derive chemotaxis-like models from biased random walk models for social systems \cite{short2010nonlinear,alsenafi2018convection,tse2015localized}, and that has also been used recently to derive a model for spatial evolutionary games with environmental feedback and environment-driven motion \cite{yao2025spatial}.

With our assumptions that $\Delta t, \ell \to 0$ and $\frac{\ell^2}{\Delta t} \to D$, we show in Section \ref{sec:PDEderivation} that we obtain following the continuum limit of Equation \eqref{eq:discrete-main-equations}, which describes the spatial profiles of cooperators $u(t,\mathbf{x})$, defectors $v(t,\mathbf{x})$, and the public good $\phi(t,\mathbf{x})$ through a system of three nonlinear PDEs:
\begin{equation}\label{eq:general-continuum-PDE}
    \begin{aligned}
        \frac{\partial u(t,\mathbf{x})}{\partial t}&=D_u\nabla\cdot\left[\nabla u-2u\nabla\ln f(w_u,A)\right]+u\left[r_u\phi-c-\gamma(u+v)-\mu_u\right],\\
        \frac{\partial v(t,\mathbf{x})}{\partial t}&=D_v\nabla\cdot\left[\nabla v-2v\nabla\ln f(w_v,B)\right]+v\left[r_v\phi-\gamma(u+v)-\mu_v\right],\\
        \frac{\partial \phi(t,\mathbf{x})}{\partial t}&=D_\phi\nabla\cdot\left[\nabla \phi-2\phi\nabla\ln f(w_\phi,C)\right]+cu-\left[\kappa(u+v)+\delta\right]\phi,
    \end{aligned}
\end{equation}
where $D_u = \chi_u D/3$, $D_v = \chi_v D/3$ and $D_\phi = \chi_\phi D/3$ (in one-dimensional space) are positive diffusion coefficients that respectively govern the overall rate of motion for cooperators, defectors, and the public good. The system of PDEs provided by Equation \eqref{eq:general-continuum-PDE} constitutes the continuum limit of our discrete model, and will serve as the baseline model we explore for the spatial public goods dilemma in the presence of diffusion of the public good and the possibility of both diffusive and directed motion for cooperators and defectors. 

We see that the PDE for the change in the density of cooperators $u(t,\mathbf{x})$ has terms describing spatial movement that can be decomposed into purely diffusive spatial movement of the form $D_u \nabla^2 u = D_u \Delta u$ and a term describing directed motion $-2 D_u \nabla \cdot \left(u \nabla \ln f(w_u,A) \right)$ of cooperators towards nearby regions with increases attractiveness $f(w_u,A)$ for their strategy. An analogous decomposition holds for the densities of defectors and the public good, and the directed motion terms will be nonlinear provided that the attractiveness functions $A(t,\mathbf{x})$, $B(t,\mathbf{x})$, or $C(t,\mathbf{x})$ are functions of any of the profiles of $u(t,\mathbf{x})$, $v(t,\mathbf{x})$, and $\phi(t,\mathbf{x})$. More generally, the choice of the sensitivity functions $f$ sensitivity parameter $w_{\cdot}$, and attractiveness functions will shape the possibility of pattern-forming spatial dynamics. We will focus on the case of exponential mappings $f(w_u,A) = e^{w_u A}$, $f(w_v,B) = e^{w_v B}$, and $f(w_{\phi}, C) = e^{w_{\phi} C}$ for convenience, as this choice of attractiveness-mapping function allows us to eliminate the natural logarithm that appear in each of our three PDEs. This choice is common in the derivation of the Keller-Segel model for chemotaxis based on biased random walks \cite{hillen2009user,bubba2020discrete} as well as related models of directed motion in social systems \cite{alsenafi2018convection}, while linear attractiveness mappings have been used to describe logarithmic chemotaxis models \cite{rodriguez2010local} and related models in social systems \cite{SHORT200808,short2010nonlinear}.

\subsection{Properties of ODE Reaction System}
\label{sec:ODEreaction}

Before analyzing the dynamical behavior of our PDE model for spatial public goods, we can explore the expected behavior of the reaction dynamics of cooperators, defectors, and public goods in the absence of space. In particular, we can look to characterize the equilibrium points of the ODE model, which we can use as a basis for exploring the stability of spatially uniform equilibrium states due to various rules for spatial movement in the full PDE model.  

The equilibrium states of the ODE system given in Equation \eqref{eq:Young-Belmonte-ODE} can be obtained by setting the right-hand side to zero, which gives us the following system of algebraic equations satisfied by equilibrium states $(u_0,v_0,\phi_0)$:
\begin{equation} \label{eq:equilibrium-relationship-ODE}
    \begin{aligned}
        u\left[r_{u}\phi-c-\gamma(u+v)-\mu_{u}\right]&=0,\\
        v\left[r_{v}\phi-\gamma(u+v)-\mu_{v}\right]&=0,\\
        cu-\left[\kappa(u+v)+\delta\right]\phi&=0.
    \end{aligned}
\end{equation}
There are four non-negative equilibrium solutions to this system of ODEs. The first is $(u_0, v_0, \phi_0) = (0, 0, 0)$, which we denote by $E_0$. This equilibrium features no population and no public good, and we refer to this equilibrium as the extinction state. We also see that there is a pair of equilibrium described by the following expression
\begin{equation}\label{eq:second-third-equilibrium}
    \begin{aligned}
        u_0&=\frac{\left(cr_u-\gamma \delta-\kappa c-\kappa\mu_u\right)\pm\sqrt{\left(cr_u-\gamma\delta-\kappa c-\kappa\mu_u\right)^2-4\kappa\delta\gamma(c+\mu_u)}}{2\gamma\kappa},\\
        v_0&=0,\\
        \phi_0&=\frac{cu_0}{\kappa u_0+\delta},
    \end{aligned}
\end{equation}
which we denote by $E_1$ (for the positive root) and by $E_2$ for the negative root. Both of these equilibria feature zero defectors, with only cooperators and public good present in this state. The final equilibrium features coexistence of cooperators, defectors, and the public good, with the level of each quantity given by 
\begin{equation}\label{eq:coexistence-equilibrium}
    \begin{aligned}
        u_0&=\frac{cr_v}{\gamma\kappa s^2}+\frac{\delta}{\kappa s}-\frac{\mu_v}{\gamma s}\\
        v_0&=\frac{cr_v}{\gamma\kappa s}-\frac{cr_v}{\gamma\kappa s^2}-\frac{\mu_v}{\gamma}+\frac{\mu_v}{\gamma s}-\frac{\delta}{\kappa s}\\
        \phi_0&=\frac{c+\mu_u-\mu_v}{r_u - r_v},
    \end{aligned}
\end{equation}
where $s=c\left(r_u-r_v\right)/\left[\kappa\left(c+\mu_u-\mu_v\right)\right]=c/(\kappa\phi_0)$. This equilibrium is denoted by $E_3$, and we will focus on this coexistence equilibrium as a jumping off point for possible spatial patterns emerging in the profiles of our three quantities. This equilibrium is of particular interest for the purpose of analyzing the qualitative effects of spatial pattern formation, as we can explore how the densities of both cooperators and defectors will change due to the spatial movement rules that each follow under different schemes of directed and undirected motion. In particular, we can explore how competition between these two strategies is shaped by the interaction between spatial movement and the reaction dynamics of public good production by cooperators and the consumption of public goods by both types.

It is possible to check that, for each equilibrium point described above, a spatial state of the form $(u(x),v(x),\phi(x)) = (u_0,v_0,\phi_0)$ for each point $x$ in our spatial domain will constitute a homogeneous equilibrium state for our PDE model in Equation \eqref{eq:general-continuum-PDE}. For this reason, we will study the stability of these ODE equilibrium points under the reaction dynamics, and then explore how the corresponding spatially uniform states can be destabilized due to diffusive or directed motion in our PDE model. 

\section{Linear Stability Analysis of the ODE and PDE Models}\label{sec:linear-stability-analysis}

In this section, we will study the stability of the equilibrium points under the reaction dynamics (Section \ref{sec:ODE-stability}), as well as the stability of the uniform coexistence state for our PDE model (Section \ref{sec:PDElinear}), focusing on both the special case of the reaction-diffusion system (Section \ref{RD:Linearstability}) and the case of the full PDE model with directed motion (Section \ref{sec:linear-directed}). Before proceeding to the stability calculations, we will first set the state of the one-dimensional spatial setting in which we explore pattern-forming behavior. 

In the following sections, we consider a specific form of the attraction function under which our general PDE system simplifies to a more tractable model. We set
\begin{equation}
    f(w_u,A)=e^{w_uA},\,\,f(w_v,B)=e^{w_vB},\,\,f(w_\phi,C)=e^{w_\phi C}.
\end{equation}
To begin our discussion on the choice of attraction functions, we first make the following assumptions
\begin{enumerate}
    \item For both cooperators and defectors, the more goods there are in the patch $i$, the more individuals want to migrate to $i$. 
    \item The public good does not perform directed motion and only moves through purely diffusive motion in space.
\end{enumerate}
With our assumption, a suitable choice of attractiveness can be 
\begin{equation}
    A=\phi,\,\,B=\phi,\,\, C=\text{constant}.
\end{equation}
Then our general PDE model from Equation \eqref{eq:general-continuum-PDE} can be reduced to 
\begin{equation}\label{eq:DM-model}
    \begin{aligned}
        \frac{\partial u(t,x)}{\partial t}&=D_u\nabla\cdot\left(\nabla u-2w_uu\nabla\phi\right)+u\left[r_u\phi-c-\gamma(u+v)-\mu_u\right],\\
        \frac{\partial v(t,x)}{\partial t}&=D_v\nabla\cdot\left(\nabla v-2w_vv\nabla\phi\right)+v\left[r_v\phi-\gamma(u+v)-\mu_v\right],\\
        \frac{\partial \phi(t,x)}{\partial t}&=D_\phi\nabla^2\phi+cu-\left[\kappa(u+v)+\delta\right]\phi,
    \end{aligned}
\end{equation}

For this PDE model, we will consider the analysis of the stability of spatially uniform states $(u(x),v(x),\phi(x)) = (u_0,v_0,\phi_0)$ with concentrations of cooperators, defectors, and the public good at levels given by the coexistence equilibrium from the reaction dynamics. To do this, we will consider solutions of the form
\begin{subequations}
    \begin{align}
        u(t,x) &= u_0 + \varepsilon \tilde{u}(t,x) \\
        v(t,x) &= v_0 + \varepsilon \tilde{v}(t,x) \\
        \phi(t,x) &= \phi_0 + \varepsilon \tilde{\phi}(t,x)
    \end{align}
\end{subequations}
for a small parameter $\varepsilon$, allowing us to explore the behavior of solutions that are close to the uniform equilibrium coexistence state. Plugging in the ansatz from Equation \eqref{eq:DM-model} and collecting all $\mathcal{O}(\varepsilon)$ terms, we obtain the following system of PDEs as the linearization of our system around the uniform coexistence equilibrium: 
\begin{equation}\label{eq:linearized-pde}
    \begin{aligned}
        \frac{\partial\tilde{u}(t,x)}{\partial t}&=D_u\nabla^2\left(\tilde{u}-2w_uu_0\tilde{\phi}\right)+u_0\left[r_u\tilde{\phi}-\gamma(\tilde{u}+\tilde{v})\right],\\
        \frac{\partial\tilde{v}(t,x)}{\partial t}&=D_v\nabla^2\left(\tilde{v}-2w_vv_0\tilde{\phi}\right)+v_0\left[r_v\tilde{\phi}-\gamma(\tilde{u}+\tilde{v})\right],\\
        \frac{\partial\tilde{\phi}(t,x)}{\partial t}&=D_\phi\nabla^2\tilde{\phi}+c\tilde{u}-\left[\kappa(u_0+v_0)+\delta\right]\tilde{\phi}-\kappa\phi_0(\tilde{u}+\tilde{v}).
    \end{aligned}
\end{equation}
We will use this linearized system of PDEs to determine the conditions required for spatial pattern formation due to the effects of increased diffusivity of defectors or increased sensitivity of cooperators in moving towards increasing concentrations of the public good. 

Before studying the stability of the uniform coexistence equilibrium for the PDE system, we will first look to characterize the stability of the equilibrium points of the original ODE system of Equation \eqref{eq:Young-Belmonte-ODE}. We highlight conditions for stability of these equilibria in the next section, showing that the coexistence equilibrium will be stable under the default parameter values from Table \ref{table 1} and highlighting regions of potential bistability between equilibria in the ODE system.

\subsection{Equilibrium Stability of the ODE Model}\label{sec:ODE-stability}

We start our linear stability analysis for the ODE model by considering the stability of the extinction equilibrium and the cooperator-only equilibria in Section \ref{sec:edgeequilibria}. We then explore the stability of the coexistence equilibrium in Section \ref{sec:coexistenceequilibrium}, noting that this equilibrium will be stable for our default parameters and that this equilibrium will be bistable with the extinction state whenever it is stable.

\subsubsection{Stability of the Extinction and Cooperator-Only Equilibria}
\label{sec:edgeequilibria}

We first study the stability of the extinction equilibrium $E_0$. Linearizing the right-hand side of the ODE model \eqref{eq:Young-Belmonte-ODE} yields the following Jacobian matrix:
\begin{equation}
    J(E_0)=\begin{bmatrix}
    -c-\mu_u & 0 & 0\\
    0 & -\mu_v & 0\\
    c & 0 & -\delta
    \end{bmatrix}.
\end{equation}
Since $J(E_0)$ is a lower triangular matrix with all negative diagonal entries, all of its eigenvalues are negative whenever we consider positive reaction parameters. Therefore, the extinction equilibrium is always locally asymptotically stable for all biologically feasible parameters, so any stable equilibria featuring positive population will coexist in bistability with the extinction state. This extinction equilibrium is always stable because the per-capita rate of population growth of both cooperators and defectors is proportional to the concentration of public goods, so it is not possible for a small population of individuals to grow when there is only a small concentration of public goods present.

Next, we consider the stability of the two cooperator-only equilibria $E_1$ and $E_2$. We again linearize the right-hand side of the ODE model \eqref{eq:Young-Belmonte-ODE} and obtain the following Jacobian matrix:
\begin{equation}
    J(E_i)=\begin{bmatrix}
        -\gamma u_0 & -\gamma u_0 & r_u u_0\\
        0 & r_v\phi_0 - \gamma u_0 - \mu_v & 0\\
        c-\kappa\phi_0 & -\kappa \phi_0 & -\kappa u_0 - \delta
    \end{bmatrix}
\end{equation}
for $i \in \{1,2\}$. Because there is only one nonzero entry in the middle row, we see that one eigenvalue of this Jacobian matrix is given by
\begin{equation}
    r_v\phi_0 - \gamma u_0 - \mu_v = (r_v-r_u)\phi_0+c+\mu_u-\mu_v,
\end{equation}
which corresponds to the eigenvalue determining whether a small population of defectors will be able to successfully invade the equilibrium coexistence of cooperators and public good. Therefore we see that a sufficient condition for the instability of $E_1$ or $E_2$ is
\begin{equation}
\label{eq:defectorinvade}
    (r_v-r_u)\phi_0+c+\mu_u-\mu_v>0.
\end{equation}
If this inequality does not hold, then we can determine stability of our cooperator-only equilibria by considering the eigenvalues of the minor matrix
\begin{equation}
J_{22}(E_i) = \begin{bmatrix}
- \gamma u_0 & r_u u_0 \\
c - \kappa \phi_0 & - \kappa u_0 - \delta
\end{bmatrix}.
\end{equation}
As the trace of this matrix is always negative for positive reaction parameters, we see that conditions for stability of $E_1$ and $E_2$ of the form $(u_0,0,\phi_0)$ are that $(r_v-r_u)\phi_0+c+\mu_u-\mu_v < 0$ and $\det\left(J_{22}(E_i) \right) > 0$.

Under the default parameter values from Table~\ref{table 1}, the ODE model \eqref{eq:Young-Belmonte-ODE}, the two cooperator-only equilibria take the approximate values
\begin{equation}
    E_1 \approx (1.9975,0,0.9995),\quad E_2 \approx (0.0015,0,0.6003),
\end{equation}
and we can check numerically that both equilibria are unstable. In particular, we see that, for the equilibrium $E_1$, the inequality from Equation \eqref{eq:defectorinvade} holds, and therefore this cooperator-only equilibrium will be unstable to invasion by a small group of defectors due to the abundance of public goods. For the equilibrium $E_2$, instability occurs even in the subsystem consisting only of cooperators and the public good, as additional cooperators will be able to invade this equilibrium to a relatively high abundance of public goods proportional to the low density of cooperators present in the state $E_2$.

\subsubsection{Stability of the Coexistence Equilibrium}
\label{sec:coexistenceequilibrium}

For our system from Equation \eqref{eq:Young-Belmonte-ODE}, we see that the linearization of the reaction dynamics near the coexistence equilibrium $E_3=(u_0, v_0, \phi_0)$ is given by the following Jacobian matrix 
\begin{equation}\label{eq:Jacobian}
    J=\begin{bmatrix}
        -\gamma u_0 & -\gamma u_0 & r_u u_0\\
        -\gamma v_0 & -\gamma v_0 & r_v v_0\\
        c-\kappa\phi_0 & -\kappa\phi_0 & -\kappa (u_0+v_0)-\delta
    \end{bmatrix}.
\end{equation}
We can calculate the trace, the sum of the principal minors of order two, and the determinant of $J$ as 
\begin{equation}\label{eq:R-H-ODE}
    \begin{aligned}
        \operatorname{tr}(J)&=-(\gamma+\kappa)(u_0+v_0)-\delta<0,\\
        S_2(J)&=\gamma\kappa(u_0+v_0)^2+\gamma\delta(u_0+v_0)+\kappa\phi_0(r_uu_0+r_vv_0)-cr_uu_0,\\
        \det(J)&=\gamma cu_0v_0(r_u-r_v).
    \end{aligned}
\end{equation}

We can then use these quantities to analyze the stability of the equilibrium, noting from the Routh-Hurwitz criteria that this equilibrium will be stable provided that the following three conditions hold:

\begin{equation}\label{eq:reaction-stability-condition}
\begin{aligned}
(i)&: \mathrm{tr}(J) < 0 \\
(ii)&: \det(J)  < 0 \\
(iii)&: \mathrm{tr}(J) S_2(J) < \det(J)
\end{aligned}
\end{equation}

\begin{remark}
    In order to have that $E_3$ is stable under the dynamics of the ODE system, we require that $\det(J)<0$ and $S_2>0$.  These inequalities imply that $r_u<r_v$ and
    \begin{equation}
        \gamma\kappa(u_0+v_0)^2+\gamma\delta(u_0+v_0)+\kappa\phi_0(r_uu_0+r_vv_0)>cr_uu_0
    \end{equation}
are necessary conditions for the stability of the coexistence equilibrium. In addition, we can rearrange the equilibrium relationship for the ODE for the change in $\phi(t)$ in Equation \eqref{eq:equilibrium-relationship-ODE} to see that
    \begin{equation}
        \big(c-\kappa\phi_0)u_0= \left( \kappa v_0 + \delta \right) \phi_0,
    \end{equation}
    so we can deduce that $c>\kappa\phi_0$ for the case of the coexistence equilibrium under the assumption that all of the reaction parameters in our ODE model are positive. We will use these observations about necessary conditions for existence and stability of the coexistence equilibrium to further understand conditions for the instability of uniform coexistence states under the dynamics of our PDE model. 
\end{remark}

To calculate the coexistence equilibrium and evaluate its stability, we can look to the parameter values we assume in Table~\ref{table 1}. We can use parameter values and the equilibrium relation 
\begin{equation}
    u_0\approx 0.3507,\,\,\, v_0\approx 0.1493,\,\,\,\phi_0=0.7.
    \nonumber
\end{equation}
We can show numerically that the Jacobian $J$ at $E_3$ has eigenvalues with negative real parts, and hence the ODE equilibrium $E_3$ is stable for the parameters given in Table \ref{table 1}. Therefore, in the following sections, we focus on $E_3$, which will describe the population and public good densities in the spatially homogeneous coexistence equilibrium in our PDE model \eqref{eq:DM-model}. We want to examine whether spatial movement can induce pattern formation starting close to a coexistence state of the population, allowing us to explore how cooperators and defectors are helped or harmed by the emergence of spatial heterogeneity.

\begin{remark}
Because the extinction equilibrium is stable for all non-negative reaction parameters, we see that the coexistence equilibrium can only coexist in bistability with the extinction equilibrium. This raises the potential concern that diffusive or directed motion may destabilize a spatially uniform coexistence equilibrium, resulting in the initial emergence of spatial patterns that later yield to a long-time extinction state. Such transient pattern formation that destabilizes one uniform equilibrium but results in the achievement of another uniform equilibrium has been observed in several examples of reaction-diffusion systems with bistable reaction dynamics \cite{krause2024turing,oliver2024effects}, and spatial instability of a positive uniform equilibrium has shown to give rise to an extinction equilibrium in a reaction-diffusion model for an economic-demographic system \cite{zincenko2018economic,zincenko2021turing}. While we numerically observe the achievement of heterogeneous patterned states starting from small perturbations of the uniform coexistence state in this paper, it is still important to understand the possible long-time outcomes of spatial models with multiple uniform states that correspond to stable equilibrium values of the reaction dynamics. 
\end{remark}

\subsection{Linear Stability Analysis of Our PDE Model}\label{sec:PDElinear}

In this subsection, we investigate the linear stability spatially uniform steady states of our PDE model under a regular spatial domain $\Omega$ such as $[0,L]$. We will present our analysis of diffusion-driven instabilities in the special case of reaction-diffusion dynamics in Section \ref{RD:Linearstability}, and we study the case of spatial pattern formation due to differences in the strength of directed motion in Section \ref{sec:linear-directed}. Finally, we provide additional commentary on the possibility of additional pattern-forming instabilities for different cases of the movement parameters.

For the two cases we examine, we will impose zero-flux boundary conditions on the interval $[0,L]$:
\begin{equation}
    \left.\frac{\partial u}{\partial x}\right|_{x=0,\,L}=\left.\frac{\partial v}{\partial x}\right|_{x=0,\,L}=\left.\frac{\partial \phi}{\partial x}\right|_{x=0,\,L}=0.
\end{equation}
This assumption ensures that no net flow of individuals or resources occurs across the domain boundaries, and allows us to consider a family of sinusoidal perturbations for our ansatz for the small perturbations $(\tilde{u}(t,x), \tilde{v}(t,x), \tilde{\phi}(t,x))$ from the uniform equilibrium state to complete our linear stability analysis.
We therefore consider perturbations of the form 
\begin{equation}\label{eq:LSA-perturbation}
    \begin{aligned}
        \tilde{u}(t,x)&= \delta_u e^{\lambda t}\cos\left(\frac{\pi k}{L}x\right)\\
        \tilde{v}(t,x)&=  \delta_v e^{\lambda t}\cos\left(\frac{\pi k}{L}x\right)\\
        \tilde{\phi}(t,x)&=  \delta_\phi e^{\lambda t}\cos\left(\frac{\pi k}{L}x\right)
    \end{aligned}
\end{equation}
for any wavenumber $k\in\mathbb{Z}$, and we substitute the ansatz from Equation \eqref{eq:LSA-perturbation} into the linearized system Equation \eqref{eq:linearized-pde}. Plugging in this ansatz then allows us to see that $(\tilde{u}(t,x), \tilde{v}(t,x), \tilde{\phi}(t,x))$  will solve our linearized system provided that the temporal growth rate $\lambda$ is a solution to the following eigenvalue problem:
\begin{equation}\label{eq:eigenvalue-problem}
    \lambda\begin{bmatrix}
        \delta_u\\
        \delta_v\\
        \delta_\phi
    \end{bmatrix}=M(k)\begin{bmatrix}
        \delta_u\\
        \delta_v\\
        \delta_\phi,
    \end{bmatrix}
\end{equation}
where the linearization matrix $M(k)$ takes the form 
\begin{equation} \label{eq:Mkgeneral}
M(k)=\begin{bmatrix}
        -\gamma u_0 - D_u \left( \frac{\pi k}{L} \right)^2 & -\gamma u_0 & r_u u_0 + 2 w_u D_w u_0 \left( \frac{\pi k}{L} \right)^2\\
        -\gamma v_0 & -\gamma v_0 - D_v \left( \frac{\pi k}{L} \right)^2& r_v v_0 + 2 w_v D_v v_0 \left( \frac{\pi k}{L} \right)^2\\
        c-\kappa\phi_0 & -\kappa\phi_0 & -\kappa (u_0+v_0)-\delta - D_{\phi} \left( \frac{\pi k}{L} \right)^2
    \end{bmatrix}.
\end{equation}
In particular, we can understand the components of the linerization matrix can be written in the form
\begin{equation}
M(k) = J + \left( \frac{\pi k}{L} \right)^2 R,
\end{equation}
where the reaction matrix $J$ is given by Equation \eqref{eq:Jacobian} and the movement matrix $R$ is given by
\begin{equation}\label{eq:diffusion-matrix}
    R=\begin{bmatrix}
        -D_u & 0 & 2w_uD_uu_0\\
        0 & -D_v & 2w_vD_vv_0\\
        0 & 0 & -D_\phi
    \end{bmatrix}.
\end{equation}
We note that the matrix $J$ has eigenvalues only with negative real part due to the stability of the coexistence equilibrium under the reaction dynamics, and we can see that the matrix $R$ will have all negative eigenvalues provided that the diffusivities $D_u$, $D_v$, and $D_{\phi}$ are all positive. We will therefore look to see whether there are spatial movement parameters and a finite wavenumber $k$ for which it is possible that the matrix $M = J + \left( \frac{\pi k}{L} \right)^2 R$ can have a positive real part. 

\begin{remark}
    As $k\to\infty$, the term $\left(\frac{\pi k}{L}\right)^2 R$ dominates $M$. Hence the eigenvalues of $M$ converge to $\left(\frac{\pi k}{L}\right)^2$ times the eigenvalues of $R$, which are all negative. Therefore, there exists some $K\in\mathbb{N}$ such that all eigenvalues of $M$ have negative real parts for all $|k|>K$. In other words, at most finitely many integer wavenumbers can destabilize $M$. Therefore we see that short-wave instabilities are not possible in either our reaction-diffusion model or in our model of directed motion towards increasing levels of public goods, which stands in contrast to the unphysical wavenumber instabilities that can arise in evolutionary game models with nonlinear diffusion or payoff-driven directed motion \cite{helbing2009pattern,xu2017strong,funk2019directed,fahimipour2022sharp,yao2025spatial}.
\end{remark}

\subsubsection{Linear Stability Analysis of the Unbiased Model}\label{RD:Linearstability}

We start by analyzing the stability of the uniform coexistence state in the case of the reaction-diffusion dynamics for the unbiased model. We assume that $w_u = w_v = 0$, and we look to determine how the change in the diffusivity of defectors $D_v$ can induce pattern formation. Our choice of $D_v$ as the bifurcation parameter of interest is based on the intuition gleaned from Turing pattern formation in the case of spatial evolutionary games \cite{yao2026pattern}, in which defectors play the role of inhibitors and cooperators play a role of activators in a system of two reaction-diffusion equations. 

In this case, our model \eqref{eq:DM-model} becomes
\begin{equation}\label{eq:RD-model}
    \begin{aligned}
        \frac{\partial u(t,x)}{\partial t}&=D_u\nabla^2u+u\left[r_u\phi-c-\gamma(u+v)-\mu_u\right],\\
        \frac{\partial v(t,x)}{\partial t}&=D_v\nabla^2v+v\left[r_v\phi-\gamma(u+v)-\mu_v\right],\\
        \frac{\partial \phi(t,x)}{\partial t}&=D_\phi\nabla^2\phi+cu-\left[\kappa(u+v)+\delta\right]\phi.
    \end{aligned}
\end{equation}
We are interested in how the diffusion coefficients determine the spatial instability. To see this, the linearized matrix $M$ from Equation \eqref{eq:Mkgeneral} reduces to
\begin{equation} \label{eq:MRDmodel}
    M=\begin{bmatrix}
        -\left(\frac{\pi k}{L}\right)^2 D_u - \gamma u_0 & -\gamma u_0 & r_u u_0\\
        -\gamma v_0 & -\left(\frac{\pi k}{L}\right)^2D_v - \gamma v_0 & r_v v_0\\
        c-\kappa\phi_0 & -\kappa \phi_0 & -\left(\frac{\pi k}{L}\right)^2D_\phi -\kappa(u_0 + v_0)-\delta
    \end{bmatrix}.
\end{equation}

To determine conditions for instability of our system, we will look to apply the Routh-Hurwitz criteria for the $3 \times 3$ matrix $M$. To do this, we must calculate the quantities 
\begin{subequations}
\begin{align}
Q_1(k) &= -\mathrm{tr}(M(k)) \\
Q_2(k) &= S_2(M(k)) \\
Q_3(k) &= -\det(M(k)),
\end{align}
\end{subequations}
and can establish instability of the uniform state to a perturbation with wavenumber $k$ by finding conditions under which one of the following three conditions holds
\begin{subequations}
\begin{align}
Q_1(k) &< 0 \\
Q_3(k) &< 0 \\
Q_3(k) &> Q_1(k) Q_2(k).
\end{align}
\end{subequations}
We can see from Equation \eqref{eq:MRDmodel} that the trace of our linearization $M(k)$ is given by
\begin{equation} \label{eq:Q1RD}
    \begin{aligned}
        Q_1(k)&=-\mathrm{tr}(M)=\left(\frac{\pi k}{L}\right)^2\left(D_u + D_v + D_\phi\right)+(\gamma+\kappa)(u_0+v_0)+\delta\\
        &=\underbrace{\left(\frac{\pi k}{L}\right)^2}_{:=Q_{1,v}(k)}D_v+\underbrace{\left(\frac{\pi k}{L}\right)^2\left(D_u + D_\phi\right)+(\gamma+\kappa)(u_0+v_0)+\delta}_{:=Q_{1,c}(k)} > 0,
    \end{aligned}
\end{equation}
and therefore the uniform state cannot be destabilized through the trace of $M(k)$ becoming positive. We may only expect to see the emergence of patterns when the determinant of $M(k)$ becomes positive or because the inequality $Q_3(k) > Q_1(k) Q_2(k)$ is satisfied, so we calculate the sum of the principal minors and determinant of the linearization matrix. We first see that the sum of the principal minors is given by
\begin{equation}
    Q_2(k) =Q_{2,v}(k) D_v+Q_{2,c}(k),
\end{equation}
where 
\begin{equation} \label{eq:Q2RD}
    \begin{aligned}
        Q_{2,v}(k)=&\,\,\left(\frac{\pi k}{L}\right)^4\left(D_u+D_\phi\right)+\left(\frac{\pi k}{L}\right)^2\left[\kappa\left(u_0+v_0\right)+\gamma u_0+\delta\right]>0,\\
        Q_{2,c}(k)=&\,\,\left[\left(\frac{\pi k}{L}\right)^2D_\phi +\delta\right]\left[\left(\frac{\pi k}{L}\right)^2D_u+\gamma\left(u_0+v_0\right)\right]\\
        &\,\,+\left(\frac{\pi k}{L}\right)^2D_u\left[\kappa (u_0+v_0)+\gamma v_0\right]+\gamma\kappa (u_0+v_0)^2+\kappa\phi_0\left(r_uu_0+r_vv_0\right)-cr_uu_0\\
        =&\,\,\left(\frac{\pi k}{L}\right)^4D_uD_\phi + \left(\frac{\pi k}{L}\right)^2\left[D_u\delta+D_\phi\gamma
        (u_0+v_0)\right]+\left(\frac{\pi k}{L}\right)^2D_u\left[\kappa (u_0+v_0)+\gamma v_0\right]\\
&\,\,+\underbrace{\gamma\kappa(u_0+v_0)^2+\gamma\delta(u_0+v_0)+\kappa\phi_0(r_uu_0+r_vv_0)-cr_uu_0}_{S_2>0\text{ in \eqref{eq:R-H-ODE}}}>0,
    \end{aligned}
\end{equation}
and therefore we can deduce that $Q_2(k) = S_2(k) > 0$. Next, we look to derive an expression for the quantity $Q_3(k) = - \det(M(k))$. By developing along the first column, we can calculate the determinant of linearization matrix to see that
\begin{equation}
    \begin{aligned}
        Q_3(k)=&\,\, -Q_{3,v}(k) D_v + Q_{3,c}(k),
    \end{aligned}
\end{equation}
where 
\begin{equation} \label{eq:Q3RD}
    \begin{aligned}
        Q_{3,v}(k)=&\,\,\left(\frac{\pi k}{L}\right)^2\Biggl\{-\left[\left(\frac{\pi k}{L}\right)^2D_u +\gamma u_0\right]\left[\left(\frac{\pi k}{L}\right)^2 D_\phi  + \kappa(u_0+v_0)+\delta \right]\\
        &\,\,\ \hspace{25mm} + r_u u_0(c-\kappa\phi_0)\Biggr\},\\
        Q_{3,c}(k)=&\,\,\left(\frac{\pi k}{L}\right)^2v_0D_u \left\{\kappa\phi_0 r_v+\gamma\left[\left(\frac{\pi k}{L}\right)^2D_\phi+\kappa(u_0+v_0) + \delta\right]\right\}+\gamma cu_0v_0 (r_v-r_u)>0.
    \end{aligned}
\end{equation}
Therefore we see that $Q_{3,c}(k) > 0$ for the parameter regime in which the coexistence equilibrium $E_3$ is stable under the reaction dynamics, but the sign of $Q_{3,v}(k)$ does not have a conclusive sign just based on the stability conditions of the coexistence state. If $Q_{3,v}(k) > 0$, we can see that it is possible to achieve the condition $Q_3(k) < 0$ for a given wavenumber $k$ by choosing the defector diffusivity to satisfy the following condition:
\begin{equation}\label{eq:D_v-threshold}
    \begin{aligned}
        D_v&>D_v^\ast(k):=\frac{Q_{3,c}}{Q_{3,v}}\\
        &=\frac{\left(\frac{\pi k}{L}\right)^2v_0D_u \left\{\kappa\phi_0 r_v+\gamma\left[\left(\frac{\pi k}{L}\right)^2D_\phi+\kappa(u_0+v_0) + \delta\right]\right\}+\gamma cu_0v_0 (r_v-r_u)}{\left(\frac{\pi k}{L}\right)^2\biggl\{r_u u_0(c - \kappa\phi_0)-\left[\left(\frac{\pi k}{L}\right)^2D_u +\gamma u_0\right]\left[\left(\frac{\pi k}{L}\right)^2 D_\phi  + \kappa(u_0+v_0)+\delta \right] \biggr\}}.
    \end{aligned}
\end{equation}
In addition, we show in Section \ref{sec:linear-Routh-Hurwitz} of the appendix that the condition $Q_{3}(k) > Q_1(k) Q_2(k)$ cannot hold when $Q_{3,v} > 0$ for sufficiently small $D_u$ and $D_{\phi}$, so we are able to focus on studying the case of the formation of Turing patterns for the parameter regime in which $Q_{3,v}(k) > 0$ for some wavenumber and the spatially uniform equilibrium can be destabilized through the determinant of the linearization matrix becoming positive. In particular, this indicates that patterns will form in the unbiased model only through stationary Turing instabilities through an increase in the defector diffusivity $D_v$ for small cooperator and public good diffusivity. Therefore we expect the patterns that emerge from small perturbations of the uniform coexistence state to eventually settle upon a steady-state solution in the long-time limit, rather than resulting in the long-time oscillations characteristic of a Turing-Hopf instability.

Furthermore, we note that the term $-D_u D_{\phi} \left( \frac{\pi k}{L}\right)^6$ will dominate the behavior of $Q_{3,v}(k)$ for large $k$, so there will be a threshold wavenumber $k_c \in \mathbb{Z}_{\geq 0}$ such that $Q_{3,V}(k) < 0$ for all $k \geq k_c$. As Turing instability can occur for a wavenumber $k < k_c$ for which $Q_{3,v}(k) > 0$ when the defector diffusivity satisfies $D_v > D_{v}^*(k)$, we can introduce the following threshold on defector diffusivity 
\begin{equation}\label{eq:D_v-threshold-min}
D_v^* := \min_{\substack{k \in \mathbb{Z}_{> 0} \\ k < k_c}} D_v^*(k),
\end{equation}
and we can look to characterize the $D_v^*$ that governs the onset of pattern formation. Due to lengthy expressions in these threshold quantities, we will turn to a numerical approach to characterize the onset of Turing patterns in this model for the case of the parameters provided in Table \ref{table 1}.  

\begin{figure}[!ht]
    \centering
    \begin{subfigure}[t]{0.465\linewidth}
        \centering
        \includegraphics[width=\linewidth]{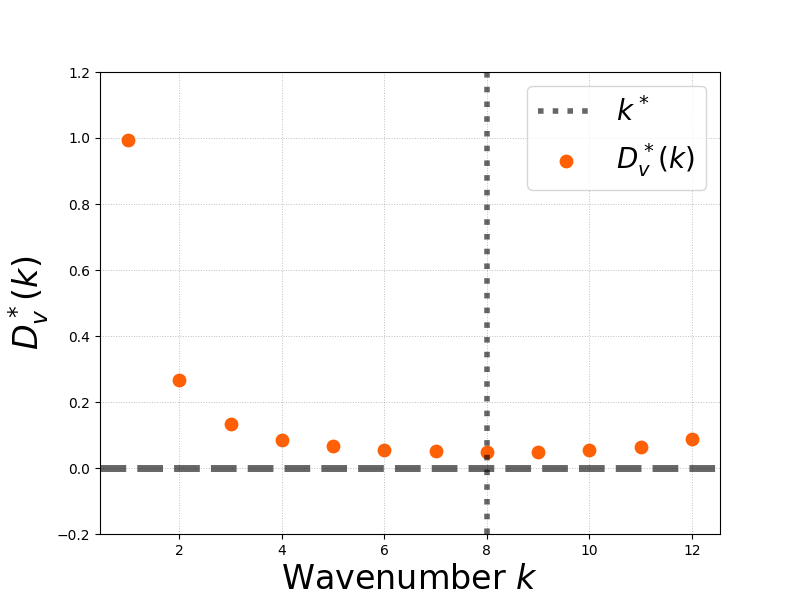}
    \end{subfigure}
    \hspace{0.05\linewidth}
    \begin{subfigure}[t]{0.465\linewidth}
        \centering
        \includegraphics[width=\linewidth]{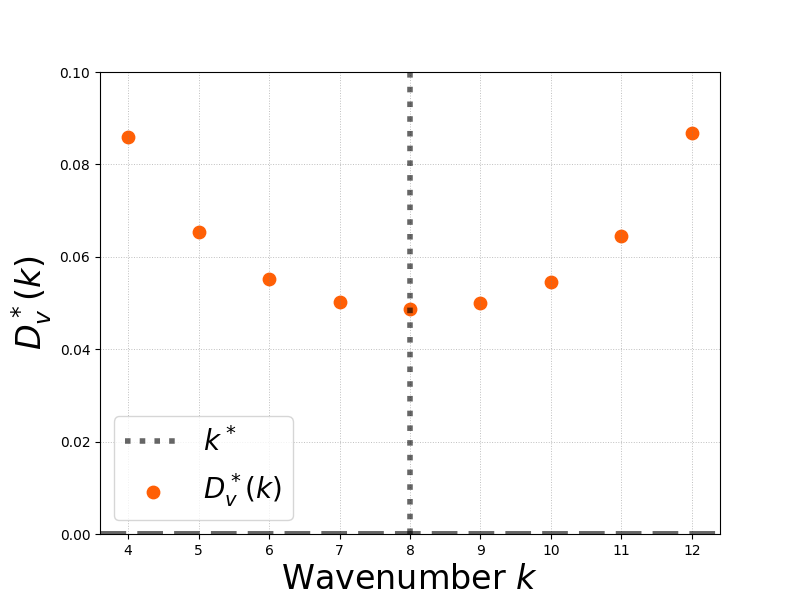}
    \end{subfigure}
    \caption{Plots of the threshold value $D_v^\ast(k)$ as a function of $k$ with $L=8$, both for the range of integer wavenumbers between $1$ and $12$. The right panel shows a zoomed-in view of the left panel near the minimal threshold $D_v^\ast$ is attained as $k=8$, with $D_v^\ast\approx 0.04861$. The orange circles represent the thresholds obtained from the condition $Q_3(k) = -\det(M(k))<0$, the horizontal dashed line in the left panel indicates the limit of zero diffusivity $D_v = 0$, and the vertical dotted lines in both panels highlights the threshold wavenumber for pattern formation given by $k^* = 8$. The parameters for the reaction dynamics are taken from Table~\ref{table 1}, and the PDE parameters are chosen as $D_u=D_\phi=0.01$.  }
    \label{fig:thresholds-unbiased}
\end{figure}

To explore this threshold for pattern formation numerically, we plot the $D_v^*(k)$ as a function of the wavenumber $k$ in Figure~\ref{fig:thresholds-unbiased} for the case of $D_u = D_{\phi} = 0.01$ and the parameters of the reaction dynamics in Table \ref{table 1}. We observe numerically that $Q_{3,v}(k) > 0$ only for integer wavenumbers less than $k_c = 13$, so we can determine the threshold for pattern formation $D_v^*$ by considering the minimum over the thresholds $D_v(k)$ among the integer wavenumbers $k\in\{1,2,\ldots,12\}$. By zooming in on the wavenumbers with the lowest thresholds in Figure \ref{fig:thresholds-unbiased}, we can see minimal threshold occurs at the wavenumber $k^*=8$. We can then calculate from Equation \eqref{eq:D_v-threshold} and the parameters in Table \ref{table 1} to determine that the threshold defector diffusivity for pattern formation is given by $D_v^* = D_v^*(8) \approx 0.04861$.

\begin{figure}[htbp]
    \centering
    \begin{subfigure}[t]{0.45\linewidth}
        \centering
        \includegraphics[width=\linewidth]{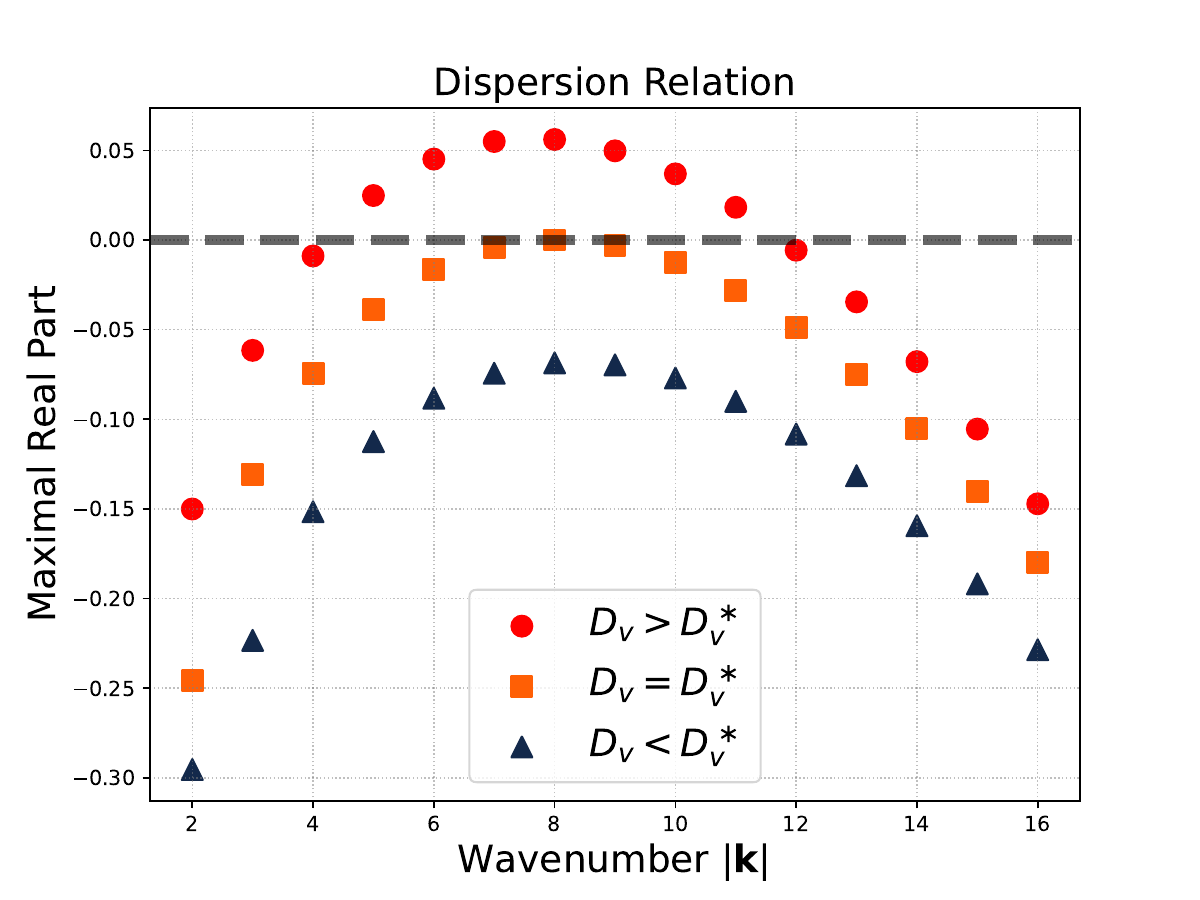}
        \caption{Dispersion relation}
    \end{subfigure}
    \hspace{0.05\linewidth}
    \begin{subfigure}[t]{0.45\linewidth}
        \centering
        \includegraphics[width=\linewidth]{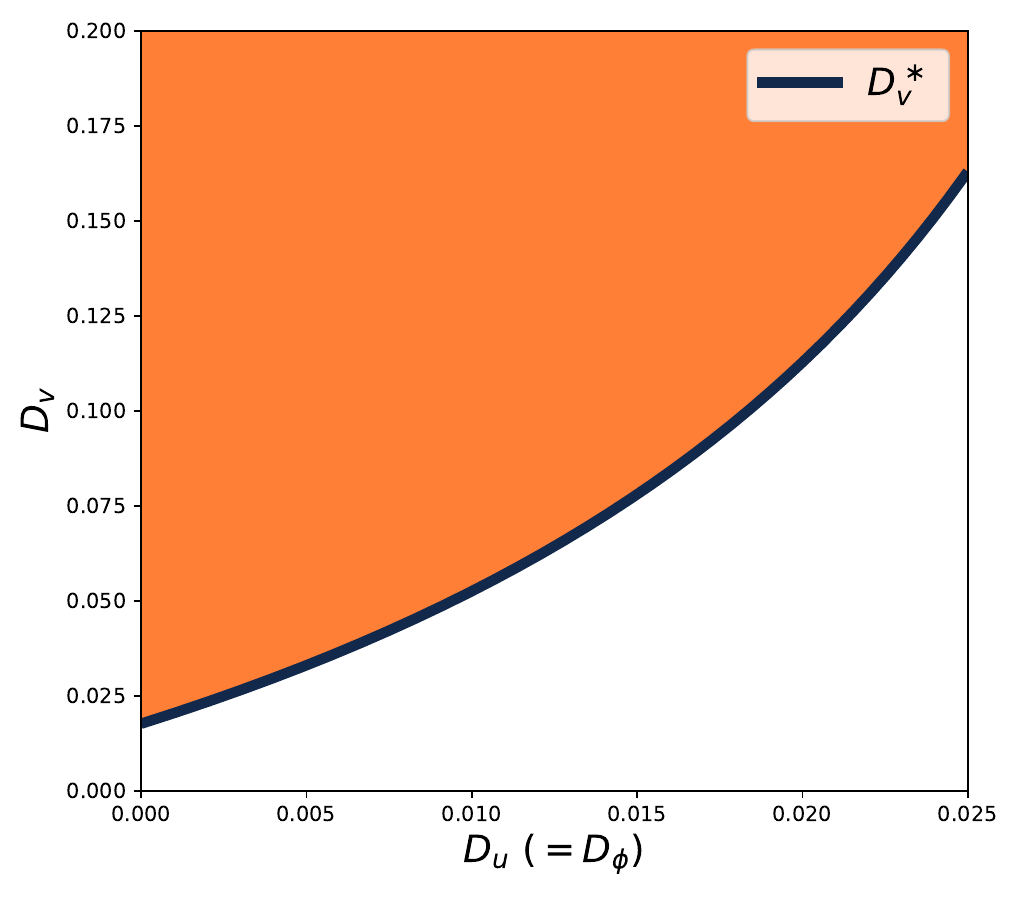}
        \caption{Instability condition}
    \end{subfigure}
    \caption{Dispersion relation and instability condition for the one-dimensional background domain $[0, L]$ with $L=8$ and wavenumber $k^\ast =8$. The parameters for the reaction dynamics are taken from Table~\ref{table 1}. In panel (a), we plot the dispersion relation describing the largest real part of the lineriazation matrix $M(k)$ as a function of the wavenumber $k$ for the choices of defector diffusivity of $D_v=0.03$, $0.04861$, and $0.08$. In panel (b), we plot the threshold $D_v^*$ for pattern-forming instability in the $(D_u, D_v)$-plane, with the orange region describing the pattern-formation regime and the white region describing the regime of a locally stable uniform coexistence equilibrium. The black curve represents the boundary determined by evaluating the maximum value of the threshold $D_v^*(k)$ from Equation \ref{eq:D_v-threshold} over all feasible integer wavenumbers. In particular, when $D_u=D_\phi=0.01$, the threshold value is $D_v^{\ast}\approx  0.04861$.}
    \label{figure 1}
\end{figure}

We can further illustrate the behavior of the pattern-forming threshold by plotting the real part of the dominant eigenvalue for different values of the defector diffusivity $D_v$ and wavenumber $k$ in Figure \ref{figure 1}(a) and by plotting the threshold defector diffusivity $D_v$ and the diffusivity of the cooperators and public good $D_u$ and $D_v$ in Figure \ref{figure 1}(b) under the assumption that $D_u = D_{\phi}$. The plot of the dispersion relation (featuring the maximum real part of the linearization matrix $M(k)$) allows us to see that that there is a region in parameter space for defector diffusivity $D_v$ sufficiently close to $D_v^*$ in which there is a single unstable wavenumber $k = k^*$, which will allow us to perform weakly nonlinear stability analysis to analyze the long-time behavior of our emergent patterns when we are sufficiently close to the threshold for pattern formation. By characterizing the threshold $D_v^*$ as a function of the equal diffusivities $D_u = D_{\phi}$, we will further be able to explore how the pattern-formation threshold can depend on changes in other parameters governing the movement of individuals or public goods in our reaction-diffusion model.
\FloatBarrier

\subsubsection{Linear Stability Analysis of the Biased Model}\label{sec:linear-directed}

We now consider the biased diffusion case, in which we assume $w_u > 0$ and $w_v > 0$. Since the parameters $w_u$ and $w_v$ control the sensitivity of public goods to cooperators and defectors, respectively, our aim is to investigate how these parameters influence the conditions for pattern-forming instability, with an emphasis on exploring how the sensitivity $w_u$ of cooperators in moving towards increasing concentrations of public good can help to destabilize the uniform coexistence state.

For our model with biased directed motion, the lineriazation matrix takes the following form
\begin{equation}
    M(k)=\begin{bmatrix}
        -\left(\frac{\pi k}{L}\right)^2 D_u - \gamma u_0 & -\gamma u_0 & 2\left(\frac{\pi k}{L}\right)^2w_uD_u u_0 + r_u u_0\\
        -\gamma v_0 & -\left(\frac{\pi k}{L}\right)^2D_v - \gamma v_0 & 2\left(\frac{\pi k}{L}\right)^2w_v D_v v_0 + r_v v_0\\
        c-\kappa\phi_0 & -\kappa \phi_0 & -\left(\frac{\pi k}{L}\right)^2D_\phi -\kappa(u_0 + v_0)-\delta
    \end{bmatrix},
\end{equation}
and we can look to use the Routh-Hurwitz criteria to determine conditions for instability. To apply the Routh-Hurwitz conditions, we introduce the following three quantities
\begin{equation}\label{biased-RH-condition}
    \begin{aligned}
        P_1(k)&=-\mathrm{tr}(M(k))=\left(\frac{\pi k}{L}\right)^2\left(D_u + D_v + D_\phi\right)+(\gamma+\kappa)(u_0+v_0)+\delta,\\
        P_2(k)&=-P_{2, u}(k)w_u + P_{2, v}(k)w_v + P_{2, c}(k),\\
        P_3(k)&= -\det(M(k)) =  -P_{3, u}(k)w_u + P_{3, v}(k)w_v + P_{3, c}(k),
    \end{aligned}
\end{equation}
where 
\begin{equation}
    \begin{aligned}
        P_{2, u}(k)=&\,\,2\left(\frac{\pi k}{L}\right)^2D_uu_0\left(c-\kappa\phi_0\right),\\
        P_{2, v}(k)=&\,\,2\left(\frac{\pi k}{L}\right)^2D_vv_0\kappa\phi_0,\\
        P_{2, c}(k)=&\,\,\left(\frac{\pi k}{L}\right)^4D_uD_v+\gamma\left(\frac{\pi k}{L}\right)^2\left(D_u v_0 + D_v u_0\right)\\
        &\,\,+\left[\left(\frac{\pi k}{L}\right)^2(D_u+D_v)+\gamma(u_0 + v_0)\right]\left[\left(\frac{\pi k}{L}\right)^2D_\phi +\kappa(u_0 + v_0)+\delta\right]\\
        &\,\,+\kappa\phi_0(r_u u_0 + r_vv_0)-r_uu_0 c,\\
        P_{3, u}(k)=&\,\,2\left(\frac{\pi k}{L}\right)^2D_uu_0\left[\left(\frac{\pi k}{L}\right)^2D_v(c-\kappa\phi_0)+\gamma cv_0\right],\\
        P_{3, v}(k)=&\,\,2\left(\frac{\pi k}{L}\right)^2D_vv_0\left[\left(\frac{\pi k}{L}\right)^2D_u\kappa\phi_0+\gamma cu_0\right],\\
        P_{3, c}(k)=&\,\,\left(\frac{\pi k}{L}\right)^4D_uD_v\left[\left(\frac{\pi k}{L}\right)^2D_\phi+\kappa(u_0+v_0)+\delta\right]\\
        &\,\,+\left(\frac{\pi k}{L}\right)^2\left(\kappa\phi_0(D_ur_vv_0+D_vr_uu_0)-cD_vr_uu_0\right)\\
        &\,\,+\gamma\left(\frac{\pi k}{L}\right)^2\left(D_uv_0+D_vu_0\right)\left[\left(\frac{\pi k}{L}\right)^2D_\phi+\kappa(u_0+v_0)+\delta\right]+\gamma cu_0v_0(r_v-r_u).
    \end{aligned}
\end{equation}

\begin{remark}
  The conditions for stability for the coexistence equilibrium $E_3$ under the ODE reaction dynamics as well as the equilibrium condition $\left[ c - \kappa \phi_0\right] u_0 = \left[ \kappa v_0 + \delta\right] \phi_0$ from the third line of Equation \eqref{eq:equilibrium-relationship-ODE} require us to assume that $r_v>r_u$ and $c>\kappa\phi_0$ if there is a spatially uniform coexistence equilibrium. Therefore we can deduce that the following coefficient functions satisfy $P_{2,u}(k)$, $P_{2,v}(k)$, $P_{3,u}(k)$, and $P_{3,v}(k) > 0$ for all relevant parameters. For the coefficient $P_{2,c}(k)$, we can find that
\begin{equation}
    \begin{aligned}
        P_{2, c}(k)=&\,\,\left(\frac{\pi k}{L}\right)^4D_uD_v+\gamma\left(\frac{\pi k}{L}\right)^2\left(D_uv_0 + D_vu_0\right)\\
        &\,\,+\left[\left(\frac{\pi k}{L}\right)^2(D_u+D_v)+\gamma(u_0 + v_0)\right]\left[\left(\frac{\pi k}{L}\right)^2D_\phi +\kappa(u_0 + v_0)+\delta\right]\\
        &\,\,+\kappa\phi_0(r_u u_0 + r_vv_0)-r_uu_0 c,\\
        =&\,\,\left(\frac{\pi k}{L}\right)^4(D_uD_v+D_vD_\phi+D_\phi D_u)+\gamma\left(\frac{\pi k}{L}\right)^2\left(D_uv_0 + D_vu_0\right)\\
        &\,\,+\gamma\left(\frac{\pi k}{L}\right)^2D_\phi(u_0+v_0)+\left(\frac{\pi k}{L}\right)^2(D_u+D_v)\left[\kappa(u_0+v_0)+\delta\right]\\
        &\,\,+\underbrace{\gamma\kappa(u_0+v_0)^2+\gamma\delta(u_0+v_0)+\kappa\phi_0(r_u u_0 + r_vv_0)-r_uu_0 c}_{S_2>0\text{ in \eqref{eq:R-H-ODE}}}>0,
    \end{aligned}
\end{equation}
as the sum of principal minors for the Jacobian matrix of the reaction dynamics will satisfy $S_2(J) > 0$ if the coexistence equilibrium is stable in the absence of spatial effects.

Finally, we can determine the sign of the coefficient $P_{3,c}(k)$ provided that we assume that the diffusivity of the cooperators, defectors, and the public good all take the same value (which is our case of interest in determining whether changes in the sensitivities $w_u$ or $w_v$ of directed motion alone are capable of generating spatial patterns). Assuming that the diffusivities of the three species $D_u=D_v=D_\phi=:\bar{D}$, then we can show that the coefficient $P_{3,c}(k)$ satisfies
\begin{equation}
    \begin{aligned}
        P_{3, c}(k)=&\,\,\left(\frac{\pi k}{L}\right)^4\bar{D}^2\left[\left(\frac{\pi k}{L}\right)^2\bar{D}+\kappa(u_0+v_0)+\delta\right]\\
        &\,\,+\left(\frac{\pi k}{L}\right)^2\bar{D}\left(\kappa\phi_0(r_vv_0+r_uu_0)-cr_uu_0\right)\\
        &\,\,+\gamma\left(\frac{\pi k}{L}\right)^2\bar{D}\left(v_0+u_0\right)\left[\left(\frac{\pi k}{L}\right)^2\bar{D}+\kappa(u_0+v_0)+\delta\right]+\gamma cu_0v_0(r_v-r_u)\\
        =&\,\,\left(\frac{\pi k}{L}\right)^4\bar{D}^2\left[\left(\frac{\pi k}{L}\right)^2\bar{D}+(\kappa+\gamma)(u_0+v_0)+\delta\right]\\
        &\,\,+\left(\frac{\pi k}{L}\right)^2\bar{D}\big[\underbrace{\gamma\kappa(u_0+v_0)^2+\gamma\delta(u_0+v_0)+\kappa\phi_0(r_uu_0+r_vv_0)-cr_uu_0}_{S_2>0\text{ in \eqref{eq:R-H-ODE}}}\big]\\
        &\,\,+\underbrace{\gamma cu_0v_0(r_v-r_u)}_{-\det(J)>0\text{ in \eqref{eq:R-H-ODE}}}>0.
    \end{aligned}
\end{equation}
\end{remark}

Therefore, all coefficients of $w_u$, $w_v$, and the constant term in Equation \eqref{biased-RH-condition} are positive when we have the same diffusivities for all three quantities. As we have shown that $Q_1(k) = -\mathrm{tr}(M(k))) > 0$ whenever the coexistence equilibrium is stable, we need to find values of $w_u$ that allow us to satisfy one of the remaining Routh-Hurwitz conditions for instability
\begin{subequations}
\begin{align}
P_3(k) &< 0 \\
P_3(k) &> P_1(k) P_2(k)
\end{align}
\end{subequations}
We can then use the expressions we derived for $Q_1(k)$, $Q_2(k)$, and $Q_3(k)$ to see that the uniform coexistence state will be unstable to perturbations with wavenumber $k$ provided that the cooperator's sensitivity $w_u$ towards climbing gradients in the concentration of public goods satisfies
\begin{equation}\label{eq:w_u-threshold}
    \begin{aligned}
        w_u>w_u^\ast(k):=\min\left\{\frac{P_{3, v}(k)}{P_{3, u}(k)}w_v +\frac{P_{3, c}(k)}{P_{3, u}(k)},\, \frac{\left(P_1(k) P_{2,v}(k) - P_{3,v}(k) \right) w_v + \left(P_1(k) P_{2,c}(k) - P_{3,c}(k) \right)}{P_1(k) P_{2,u}(k) - P_{3,u}(k)}\right\},
    \end{aligned}
\end{equation}
where the first term corresponds to the threshold on $w_u$ required to satisfy $Q_3(k) < 0$ and the second term corresponds to the threshold $w_v$ required to achieve $Q_{3}(k) > Q_{1}(k) Q_2(k)$. Furthermore, we can characterize the minimum cooperator movement sensitivity $w_u$ by considering the minimum threshold across wavenumbers $k$:
\begin{equation}\label{eq:w_u-threshold-min}
w_u > w_u^* = \min_{k \in \mathbb{Z}_{> 0}} w_u^*(k).
\end{equation}

\begin{figure}[htbp]
    \centering
    \begin{subfigure}[t]{0.465\linewidth}
        \centering
        \includegraphics[width=\linewidth]{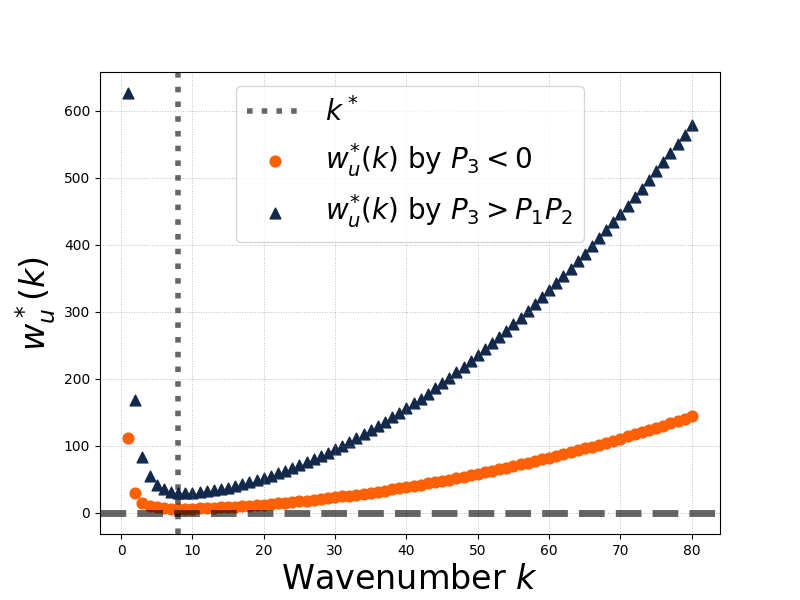}
    \end{subfigure}
    \hspace{0.05\linewidth}
    \begin{subfigure}[t]{0.465\linewidth}
        \centering
        \includegraphics[width=\linewidth]{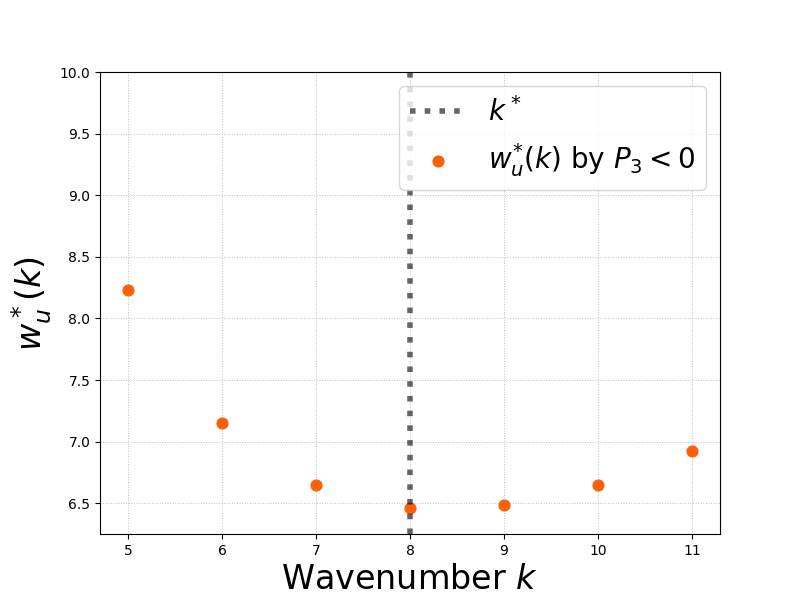}
    \end{subfigure}
    \caption{Threshold value $w_u^*(k)$ as a function of $k$ with $L=8$. The ODE parameters are taken from Table~\ref{table 1}, and the PDE parameters are chosen as $D_u=D_v=D_\phi=0.03$. The orange circles represent the thresholds obtained from the condition $P_3<0$, the dark blue triangles represent those obtained from $P_3>P_1P_2$, the horizontal black dashed line corresponds to a cooperator movement sensitivity of $w_u = 0$, and the vertical black dotted line corresponds to the critical wavenumber $k^* = 8$. The right panel shows a zoomed-in view of the left panel near the minimal $w_u^\ast(k)$. The minimal threshold is attained at $k=8$, with $w_u^\ast \approx 6.4603$.}
    \label{fig:thresholds-biased}
\end{figure}

\begin{remark}
It is somewhat easier to interpret the condition for instability of the uniform state due to the determinant becoming positive (corresponding to $Q_3(k) = -\det(M(k)) > 0$), as $\frac{P_{3, v}(k)}{P_{3, u}(k)}w_v +\frac{P_{3, c}(k)}{P_{3, u}(k)} > 0$ for all wavenumbers $k$. In particular, we see that this threshold quantity is increasing in $w_v$, so direct motion of defectors towards increasing public good abundance helps to suppress pattern formation. This calculation matches biological intuition, as if defectors are more capable of reaching regions with more abundant public goods, they will consume the public good more rapidly than cooperators and reduce the public good concentration back to the level expected in a uniform spatial equilibrium. It is especially notable that the threshold $w_u^*(k)$ is an affine function of the defector movement sensitivity $w_v$ with positive slope, so an increase in $w_v$ produces a corresponding proportional increase in the threshold cooperator movement sensitivity $w_u$ required for the onset of pattern formation.

By contrast, it is difficult to analytically characterize the sign of the the constant term and the coefficient of $w_v$ in the second condition for instability on the right-hand side of Equation \eqref{eq:w_u-threshold}. We will resort to the use of numerical calculations to establish the conditions under which $P_3(k) > P_1(k) P_2(k)$ for the parameters provided in Table \ref{table 1}. Notably, we will see in Figure \ref{fig:thresholds-biased} that the determinant condition is the condition achieving the minimum threshold for the parameters we consider, so our interpretation of the determinant condition may still be reflective of the underlying interactions between the directed motion of cooperators and defectors in the formation of spatial patterns in our PDE model. In particular, this suggests that we expect a stationary pattern-forming instability for the biased model in the case of increasing $w_u$, our choice of equal diffusivities for the three species, and our default set of parameters.
\end{remark}

In Figure \ref{fig:thresholds-biased}, we calculate the components of the thresholds from Equation \eqref{eq:w_u-threshold} on $w_u$ required to generate pattern formation due to achieving the conditions $P_3(k) < 0$ and $P_3(k) > P_1(k) P_2(k)$ for each wavenumber $k$. For the case of the reaction parameters from Table \ref{table 1} and equal diffusivities $D_u = D_v = D_{\phi}$ for cooperators, defectors, and the public good, we see that the determinant condition produces a lower threshold for the onset of pattern formation than the combined condition on $P_1(k)$, $P_2(k)$, and $P_3(k)$. We also further zoom in on the wavenumbers with low threshold movement sensitivities $w_u^*(k)$, which allows us to see that the critical wavenumber is given by $k^* = 8$ and allows us to apply Equation \eqref{eq:w_u-threshold} to see that the threshold cooperator movement sensitivity is given by $w_u^* \approx 6.4603$.

\begin{figure}[htbp]
    \centering
    \begin{subfigure}[t]{0.45\linewidth}
        \centering
        \includegraphics[width=\linewidth]{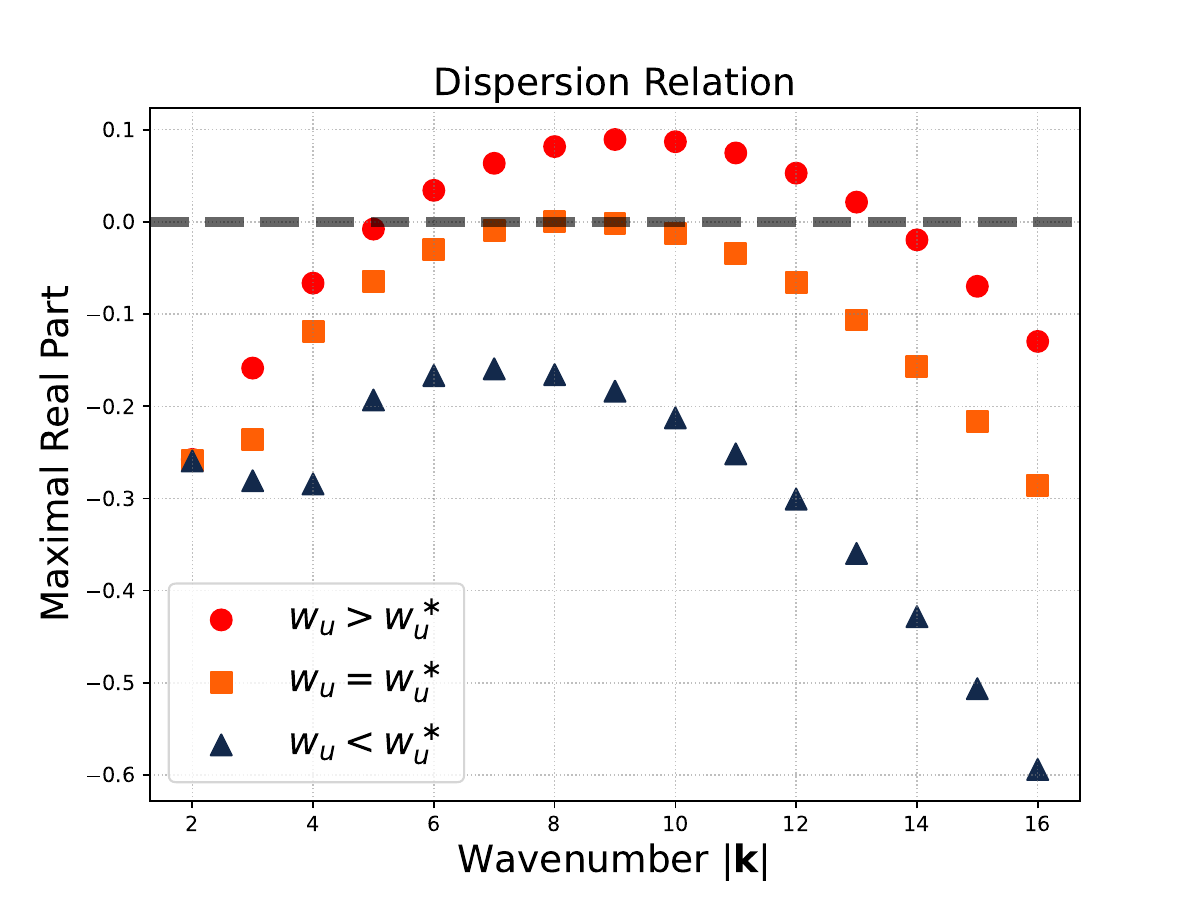}
        \caption{Dispersion relation}
    \end{subfigure}
    \hspace{0.05\linewidth}
    \begin{subfigure}[t]{0.45\linewidth}
        \centering
        \includegraphics[width=\linewidth]{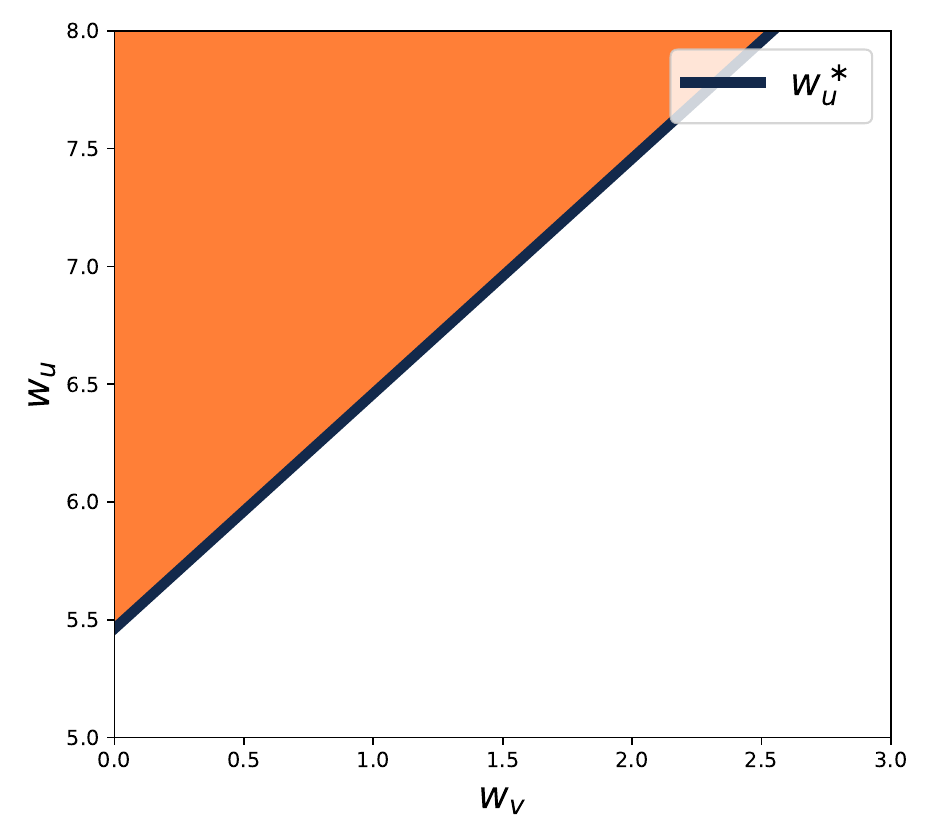}
        \caption{Stability condition}
    \end{subfigure}
    \caption{Dispersion relation and stability condition for the one-dimensional background domain $[0, L]$ with $L=8$ and wavenumber $k^\ast =8$. The ODE parameters are taken from Table~\ref{table 1}. The diffusion coefficients are set to be $D_u=D_v=D_\phi=0.03$.  In part (a), we choose $w_u=4$, $6.4603$, and $8$.  For the stability condition in the $(w_v, w_u)$-plane, the orange region corresponds to instability, while the white region corresponds to stability. The dark blue curve represents the boundary determined by Equation \eqref{eq:w_u-threshold}. In particular, when $w_v=1$, the threshold value is $w_u^{\ast}\approx 6.4603$.}
    \label{figure 2}
\end{figure}

We can further explore the conditions for instability by plotting the dispersion relation for our linearized system and displaying the threshold cooperator movement sensitivity $w_u^*$ for various defector movement sensitivities $w_v$ in Figure \ref{figure 2}. Since we want to understand how directed motion alone can produce linear instability, we set $D_u=D_v=D_\phi$ and explore how the uniform state can be destabilized by increasing the cooperator movement sensitivity $w_u^*$. In particular, when $L=8$, $w_v=1$, and we set $D_u=D_v=D_\phi=0.03$, we can compute the threshold of $w_u$ at which the uniform coexistence state becomes unstable as $w_u*\approx 6.4603$ and that the most unstable wavenumber is $k^* = 8$. We plot the dispersion relation for three values in $w_u$ in Figure \ref{figure 2}(left), showing that there are no unstable wavenumbers for $w_u = 4$, that there is a single unstable wavenumber $k^* = 8$ when $w_u$ is slightly above the threshold $w_u^*$, and that there is a range of unstable wavenumbers when $w_u = 8$. We then further illustrate the threshold condition $w_u^*$ by considering how this threshold depends on the defector movement sensitivity $w_v$, plotting in Figure \ref{figure 2}(right) the regions in the $(w_u,w_v)$ plane that produce instability in the biased model. Notably we observe a linear dependence of the threshold $w_u^*$ on the value of $w_v$, recapitulating the linear dependence of the analytical threshold condition on $w_u$ to achieve $\det(M(k)) > 0$ for the biased model.
\FloatBarrier

\subsubsection{Comments on Pattern Formation for More General Assumptions on Movement Parameters}
\label{sec:generalpatterncomment}

In the sections above, we have focused on the special cases of pattern formation for either the pure reaction-diffusion dynamics of the unbiased model and for the case of directed motion with equal diffusivities of the three species in the biased model. While we were able to derive conditions for the onset of instability in these two cases and used a mix of analytical and numerical approaches to demonstrate the formation of stationary patterns in these cases, it may be possible that other combinations of movement parameters can combine to produce spatial patterns, and that other choices of movement or reaction parameters could combine to produce Turing-Hopf bifurcations and corresponding oscillatory emerging patterns and the possibility of spatiotemporal chaos. A full characterization of the conditions for Turing and Turing-Hopf instabilities has been provided in a recent paper by Piskovsky \cite{piskovsky2025turing} for the case of reaction-diffusion systems with three species, so this approach could potentially be applied to understand the full range of stationary and oscillatory pattern-forming instabilities in the unbiased model for general values of $D_u$, $D_v$, and $D_{\phi}$. Due to the complexity of the analytical conditions from Piskovsky and the number of parameters present in the reaction dynamics of our public goods model would likely require a primarily numerical exploration of the instabilities possible across the set of possible reaction parameters. To perform a systematic exploration of possible Turing and Turing-Hopf patterns in the unbiased model, it may make sense to constrain a numerical sweep across reaction parameters that correspond to ecologically feasible coexistence equilibrium states to further understand how generic stationary Turing patterns within our parameter space \cite{gellner2023stable,gibbs2024can}.

For the case of the biased model, there is also room to explore the role played by directed motion in promoting spatial patterns for the case in which there are unequal diffusivities among species. In particular, one could look to characterize the minimal cooperator sensitivity $w_u^*$ to gradients in public good concentration required to generate spatial patterns when $D_v \ne D_u$ but $D_v$ is not sufficiently large to generate a Turing instability in the unbiased model. Studying general conditions for pattern formation in the unbiased model can show how the chemotactic mechanism can work in concert with the Turing mechanism to help produce spatial patterns when neither mechanism is sufficient on its own, and one can explore whether the qualitative properties of the resulting patterns match behavior seen due to instabilities resulting from each of the two mechanisms. A similar analysis of the role of differential diffusion and differential directed motion was explored in the case of evolutionary games with diffusive and payoff-driven motion Yao et al. \cite{yao2026pattern}, and further exploration on the mix of Turing and chemotactic instabilities could potentially generalize the approach of Piskovsky to understand general conditions for stationary and oscillatory pattern-forming instabilities in three-species models with both models of motion.

\section{Weakly Nonlinear Analysis of the PDE Model}\label{sec:weakly-nonlinear-analysis}

In this section, we apply weakly nonlinear analysis for our model near non-trivial equilibrium $E_3$ with one-dimensional background space $[0,L]$ with zero-flux boundary conditions. For both cases of the PDE model in which we found spatial instability, we will use perturbation expansions to study the approximate behavior of solutions to the PDE model when our bifurcation parameter is close to the threshold for the onset of pattern formation (exploring $D_v$ close to $D_v^*$ in the unbiased model and $w_u$ close to $w_u^*$ in the biased model). We focus on the case of the unbiased model in Section \ref{sec:weakly-nonlinear-unbiased} and the case of the biased model in Section \ref{sec:weakly-nonlinear-biased}, showing that the pattern-forming instabilities in both models correspond to supercritical pitchfork bifurcations under the default parameters from Table \ref{table 1}. We then explore the robustness of this prediction by changing two of the reaction parameters in Section \ref{sec:robustness-WNA}, showing that both subcritical and supercritical pitchfork bifurcations can be achieved as we vary the conversation rate $r_u$ of public good consumption into cooperator population growth or the baseline death rate $\mu_u$ for cooperators.

\subsection{Weakly Nonlinear Analysis of the Unbiased Model}\label{sec:weakly-nonlinear-unbiased}

We begin with the weakly nonlinear analysis of the unbiased model \eqref{eq:RD-model}, in this case $w_u=w_v=0$. Assume $\varepsilon>0$ small, use $D_v$ as the bifurcation parameter, then we use the following transformations:
\begin{equation}\label{eq:unbiased-transformations}
    T = \varepsilon^2 t,\quad D_v = D_v^\ast + \varepsilon^2 \tilde{D}.
\end{equation}
In these transformations, $\tilde{D}$ is the controlling parameter, $D_v^\ast$ is the threshold that can be obtained by Equation \eqref{eq:D_v-threshold}. We consider the following perturbation expansions for the profiles of cooperators, defectors, and the public good
\begin{equation}\label{eq:weakly-nonlinear-expansions}
    \begin{aligned}
        u(T,x)&=u_0+\varepsilon u_1(T,x)+\varepsilon^2 u_2(T,x)+\varepsilon^3 u_3(T,x)+\cdots,\\
        v(T,x)&=u_0+\varepsilon v_1(T,x)+\varepsilon^2 v_2(T,x)+\varepsilon^3 v_3(T,x)+\cdots,\\
        \phi(T,x)&=\phi_0+\varepsilon \phi_1(T,x)+\varepsilon^2 \phi_2(T,x)+\varepsilon^3 \phi_3(T,x)+\cdots,
    \end{aligned}
\end{equation}
where $(u_0,v_0,\phi_0)$ corresponds to the uniform equilibrium state and we consider solutions that depend only on the slow timescale $T = \varepsilon^2 t$ (and not the original timescale $t$) due to our assumption that $D_v$ is close to the threshold $D_v^*$ for the onset of pattern formation.

\begin{remark}
We chose the slow timescale $T = \varepsilon^2 t$ and the scaling of the perturbation of the defector diffusivity $D_v = D_v^* + \varepsilon^2 \tilde{D}$ due to the boundary condition associated with our one-dimensional domain. In particular, both Turing instabilities and chemotactic instabilities typically take the form of a pitchfork bifurcation for the case of a one-dimensional spatial domain with zero-flux boundary conditions \cite{woolley2022boundary,short2010nonlinear,cross1993pattern}, and the choice of $\mathcal{O}(\varepsilon^2)$ scaling of time and our bifurcation parameter will allow us to use the self-consistency of our perturbation expansion to derive an ODE for the amplitude of our emerging patterns for diffusivity $D_v$ close to the threshold for Turing instability. While we directly assume the scaling that will allow us to derive the appropriate amplitude equation for our two models, it was shown by Woolley \cite{woolley2022boundary} that it is also possible to study these bifurcations using a more systematic expansion for the bifurcation parameter of the form 
\begin{equation}
D_v = D_v^* + \epsilon \tilde{D}_1 + \epsilon^2 \tilde{D}_2 + \cdots
\end{equation}
and a hierarchy of time-scales $(T_1,T_2,\cdots)$ allowing for the following expansion of the $\mathcal{O}(1)$ time-derivative as
\begin{equation}
\dsdel{}{t} = \epsilon \dsdel{}{T_1} + \epsilon^2 \dsdel{}{T_2} + \cdots.
\end{equation}
With this approach, Woolley was able to demonstrate that the pattern-forming instability for two-species Turing patterns in one spatial dimension took the form of a pitchfork bifurcation for the case of zero-flux and periodic conditions and took the form of a transcritical bifurcation in the case of Dirichlet boundary conditions. 
\end{remark}

By plugging the perturbation ansatz of Equation \eqref{eq:weakly-nonlinear-expansions} into Equation \eqref{eq:DM-model}, we show in Section \ref{sec:weakly-nonlinear-expansions} of the appendix that the functions in our expansion will satisfy a hierarchy of partial differential equations for each order of $\varepsilon$. Up to cubic order in $\varepsilon$, this hierarchy of equations will be given by
\begin{subequations}
\begin{alignat}{2}
\mathcal{O}(1) &: \: \hspace{8.5mm} \bbm 0 \\ 0 \\ 0 \ebm =&&  \bbm  u_0 \left[ r_u \phi_0 - c - \gamma (u_0 + v_0) - \mu_u \right]  \\
 v_0 \left[ r_v \phi_0  - \gamma (u_0 + v_0) - \mu_v \right] \\
c u_0 -  \phi_0\left[ \kappa (u_0 + v_0) + \delta \right] \ebm \label{eq:RDorder1} \\
\mathcal{O}(\varepsilon) &: \:  -\mathcal{L}\begin{bmatrix}
        u_1\\
        v_1\\
        \phi_1
    \end{bmatrix} =&& \bbm 0 \\ 0 \\ 0 \ebm \label{eq:RDorderep} \\
\mathcal{O}(\varepsilon^2) &: \:     -\mathcal{L}\begin{bmatrix}
        u_2\\
        v_2\\
        \phi_2
    \end{bmatrix} =&&
    -\begin{bmatrix}
        -r_uu_1\phi_1 + \gamma u_1^2 + \gamma u_1v_1\\
        -r_vv_1\phi_1 + \gamma u_1v_1 +\gamma v_1^2\\
        \kappa\phi_1(u_1+v_1)
    \end{bmatrix} \label{eq:RDorderep2} \\
\mathcal{O}(\varepsilon^3) &: \:   -\mathcal{L}\begin{bmatrix}
            u_3\\
            v_3\\
            \phi_3
        \end{bmatrix} = && \,\,\begin{aligned}[t]
            &-\frac{\partial}{\partial T}\begin{bmatrix}
            u_1\\
            v_1\\
            \phi_1
        \end{bmatrix}+\begin{bmatrix}
            0\\
            \tilde{D}\frac{\partial^2 v_1}{\partial x^2}\\
            0
        \end{bmatrix}  \\
        &+\begin{bmatrix}
            u_2[r_u\phi_1-\gamma(u_1+v_1)]+u_1[r_u\phi_2-\gamma(u_2+v_2)]\\
            v_2[r_v\phi_1-\gamma(u_1+v_1)]+v_1[r_v\phi_2-\gamma(u_2+v_2)]\\
            -\kappa\phi_1(u_2+v_2)-\kappa\phi_2(u_1+v_1)
        \end{bmatrix}, 
        \end{aligned}
        \label{eq:RDorderep3}
\end{alignat}
\end{subequations}
where we consider the linearization $\mathcal{L}_{D_v^*}$ at the critical defector diffusivity $D_v^*$ given by
\begin{equation} \label{eq:linearization-reaction-diffusion}
\mathcal{L}_{D_v^*}= \begin{bmatrix}
       D_u \doubledelsamesmall{}{x} - \gamma u_0 & -\gamma u_0 & r_u u_0\\
        -\gamma v_0 & D_v^* \doubledelsamesmall{}{x} - \gamma v_0 & r_v v_0\\
        c-\kappa\phi_0 & -\kappa \phi_0 & D_\phi \doubledelsamesmall{}{x} -\kappa(u_0 + v_0)-\delta
    \end{bmatrix}. 
\end{equation}

For $D_v$ sufficiently close to $D_v^*$, we expect the solution of the linearized dynamics to be stationary on the original timescale $t$, and the growth of the amplitude of the sinusoidal pattern with the critical wavenumber $k^*$ to vary only on the slower timescale characterized by $T = \varepsilon^2 t$. In particular, we expect solutions to our equation at order $\mathcal{O}(\varepsilon)$ to take the form
\begin{equation}\label{eq:linearsolutionRD}
    \begin{bmatrix}
        u_1(T,x)\\
        v_1(T,x)\\
        \phi_1(T,x)
    \end{bmatrix}=A(T)\mathbf{q}\cos\left(\frac{\pi k^\ast}{L}x\right),
\end{equation}
where $A(T)$ is the amplitude of the sinusoidal profile of the $\mathcal{O}(\varepsilon)$-solution and the vector $\mathbf{q} = (q_u,q_v,q_{\phi})^T$ is an element of the kernel of the linearization matrix when applied to functions proportional to sinusoidal functions of the form $\cos\left(\frac{\pi k^\ast}{L}x\right)$, which is given by
\begin{equation} \label{eq:linearization-wavenumber-reaction-diffusion}
M_{D_v^*}(k) = \begin{bmatrix}
       - \left(\frac{\pi k^\ast}{L}\right)^2 D_u  - \gamma u_0 & -\gamma u_0 & r_u u_0\\
        -\gamma v_0 & - \left(\frac{\pi k^\ast}{L}\right)^2 D_v^*  - \gamma v_0 & r_v v_0\\
        c-\kappa\phi_0 & -\kappa \phi_0 & - \left(\frac{\pi k^\ast}{L}
        \right)^2 D_\phi  -\kappa(u_0 + v_0)-\delta
    \end{bmatrix}. 
\end{equation}

We can then plug our expression from Equation \eqref{eq:linearsolutionRD} for $u_1(T,x)$, $v_1(T,x)$, and $\phi_1(T,x)$ into the right-hand side of Equation \eqref{eq:RDorderep2} to see that our solutions at order $\mathcal{O}(\varepsilon^2)$ as an inhomogeneous system of ODEs for $u_2(T,x)$, $v_2(T,x)$, and $\phi_2(T,x)$. This allows us to rewrite Equation \eqref{eq:RDorderep2} as
\begin{equation}
   \begin{aligned}  - \mathcal{L}_{D_v^*} \bbm u_2 \\ v_2 \\ \phi_2 \ebm
=&\,\,-\begin{bmatrix}
        \gamma (q_u^2+q_uq_v)-r_uq_uq_\phi\\
        \gamma (q_uq_v+q_v^2)-r_vq_vq_\phi\\
        \kappa q_\phi(q_u+q_v)
    \end{bmatrix}A^2(T)\cos^2\left(\frac{\pi k^\ast}{L}x\right)\\
    \end{aligned}.
\end{equation}
We can then use the identity $\cos^2\left(\frac{\pi k^\ast}{L}x\right)=\frac{1}{2}\left(1+\cos\left(\frac{2\pi k^\ast}{L}x\right)\right)$  to rewrite this equation as 
\begin{equation}
    -\mathcal{L}\begin{bmatrix}
        u_2\\
        v_2\\
        \phi_2
    \end{bmatrix}=-\mathbf{s}_0A^2(T)-\mathbf{s}_2A^2(T)\cos\left(\frac{2\pi k^\ast}{L}x\right),
\end{equation}
where the coefficients $\mathbf{s}_0$ and $\mathbf{s}_2$ are given by
\begin{equation}
    %\begin{aligned}
        \mathbf{s}_0=\mathbf{s}_2:=\frac{1}{2}\begin{bmatrix}
        \gamma (q_u^2+q_uq_v)-r_uq_uq_\phi\\
        \gamma (q_uq_v+q_v^2)-r_vq_vq_\phi\\
        \kappa q_\phi(q_u+q_v)
    \end{bmatrix}.
\end{equation}
We guess a particular solution of $u_2, v_2, \phi_2$ in the form:
\begin{equation}\label{eq:unbiased-second-order}
    \begin{bmatrix}
        u_2(T,x)\\
        v_2(T,x)\\
        \phi_2(T,x)
    \end{bmatrix}=\mathbf{t}_0A^2(T)+\mathbf{t}_2A^2(T)\cos\left(\frac{2\pi k^\ast}{L}x\right).
\end{equation}

Because we are assuming that the spatially uniform equilibrium is stable under the reaction dynamics and that $k^*$ is the only unstable wavenumber for our choice of diffusivity $D_v$ close to $D_v^*$, we are able to deduce that the matrices $M_{D_v^*}(0) = J$ and $M_{D_v^*}(2k^*)$ are invertible, and we can therefore see that a particular solution to the $\mathcal{O}(\varepsilon^2)$ is given by the functions
\begin{subequations}\label{eq:unbiased-t}
\begin{align}
    \mathbf{t}_0 & =\left(M_{D_v^*}(0)\right)^{-1}\mathbf{s}_0=J^{-1}\mathbf{s}_0 \\
    \mathbf{t}_2 &= \left(M_{D_v^*}(2k^\ast)\right)^{-1}\mathbf{s}_2,
\end{align}
\end{subequations}
where we denote the entries of $\mathbf{t}_j$ by $\mathbf{t}_j=(t_{j,u}, t_{j, v}, t_{j, \phi})^T$ for $j=0, 2$. 

We can then use the solutions we have obtained for $$(u_1(T,x),v_1(T,x), \phi_1(T,x))^T\text{ and }(u_2(T,x),v_2(T,x), \phi_2(T,x))^T$$ to rewrite the right-hand side of our $\mathcal{O}(\varepsilon^3)$ equation. In particular, we plug these expressions into Equation \eqref{eq:RDorderep3} and use the identity $\cos\left( \frac{\pi k^*}{L} x \right)^3 = \frac{3}{4} \cos\left( \frac{\pi k^*}{L} x \right) + \frac{1}{4} \cos\left( \frac{3 \pi k^*}{L} x \right)$ to see that it is possible to write this equation in the form
\begin{equation} \label{eq:RDep3righthandexplicit}
- \mathcal{L}_{D_v^*} \bbm u_3 \\ v_3 \\ \phi_3 \ebm = \mathbf{H}_1 \cos\left( \frac{\pi k^*}{L} x \right)  + \mathbf{H}_3 \cos\left( \frac{3 \pi k^*}{L} x \right),
\end{equation}
where the entries vector $\mathbf{H}_1=(H_{1,u}, H_{1, v}, H_{1, \phi})^T$ of coefficients for the term proportional to $\cos\left( \frac{\pi k^*}{L} x \right)$ are given by
\begin{equation}\label{eq:H1RD}
    \begin{aligned}
        H_{1, u}=&\,\,-q_u A'(T)\\
        &\,\,+\left[ t_{0,u}\left(r_uq_\phi-\gamma(q_u+q_v)\right)+q_u\left(r_ut_{0, \phi}-\gamma(t_{0, u}+t_{0, v})\right)\right]A^3(T)\\
        &\,\,+\tfrac{1}{2}\left[t_{2, u}\left(r_uq_\phi-\gamma(q_u+q_v)\right) + q_u\left(r_ut_{2,\phi}-\gamma(t_{2, u}+t_{2, v})\right)\right]A^3(T),\\
        H_{1,v}=&\,\,-q_v A'(T)+\\
        &\,\,+\left[ t_{0,v}\left(r_vq_\phi-\gamma(q_u+q_v)\right)+q_v\left(r_vt_{0, \phi}-\gamma(t_{0, u}+t_{0, v})\right)\right]A^3(T)\\
        &\,\,+\tfrac{1}{2}\left[t_{2, v}\left(r_vq_\phi-\gamma(q_u+q_v)\right) + q_v\left(r_vt_{2,\phi}-\gamma(t_{2, u}+t_{2, v})\right)\right]A^3(T)\\
        &\,\,-\left(\tfrac{\pi k^\ast}{L}\right)^2 \tilde{D}q_v A(T),\\
        H_{1,\phi}=&\,\,-q_\phi A'(T)-\kappa\left[q_\phi \left(t_{0, u}+t_{0, v}\right)+t_{0,\phi}\left(q_u+q_v\right)\right]A^3(T)\\
        &\,\,-\tfrac{1}{2}\kappa\left[q_\phi\left(t_{2,u}+t_{2,v}\right)+t_{2,\phi}\left(q_u+q_v\right)\right]A^3(T).
    \end{aligned}
\end{equation}
To deduce the desired qualitative properties for our weakly nonlinear analysis, we do not need to use the coefficients of $\mathbf{H}_3$, the vector of coefficients for $\cos\left( \frac{3\pi k^*}{L} x \right)$. 

Because the right-hand side of Equation \eqref{eq:RDep3righthandexplicit} contains terms proportional to the function $\cos\left( \frac{\pi k^*}{L} x \right)$ that appears in the kernel of the linear operator $\mathcal{L}_{D_v^*}$, we need to ensure that this term does not produce a resonant solution. We can apply the Fredholm alternative to see that the term $\mathbf{H}_1 \cos\left( \frac{\pi k^*}{L} x \right)$ must be orthogonal to the kernel of the adjoint operator $\mathcal{L}_{D_v^*}^{\dag}$. Considering our solution space as $\left(\mathcal{L}^2\right)^3\left([0,L]\right)$, we then consider an element $\mathbf{p} \in \ker\left(\mathcal{L}_{D_v^*}^{\dag} \right)$, and we look to ensure that
\begin{equation}
\int_0^L \left[ \mathbf{H}_1 \cos\left( \frac{\pi k^*}{L} x \right)  \right] \cdot \left[ \mathbf{p}  \cos\left( \frac{\pi k^*}{L} x \right) \right] dx = 0 \Longrightarrow \frac{L}{2} \mathbf{H}_1 \cdot \mathbf{p} = 0,
\end{equation}
and we can therefore deduce that we need the coefficient function $\mathbf{H}_1$ to be orthogonal to the vector $\mathbf{p} = (p_u,p_v,p_{\phi})^T$. In particular, we can choose to scale the vector $\mathbf{p}$ so that  $\mathbf{p} \cdot \mathbf{q} = 1$, and we can use the condition $\mathbf{p} \cdot \mathbf{H}_1 = 0$ and the expression from Equation \eqref{eq:H1RD} to derive the following ODE for the amplitude function $A(T)$:
\begin{equation}\label{eq:unbiased-SL-equation}
    \alpha\frac{dA}{dT}=\eta\tilde{D} A+\beta A^3,
\end{equation}
where 
\begin{equation}\label{eq:unbiased-SL-coefficients}
    \begin{aligned}
        \alpha = &\,\, p_uq_u + p_vq_v + p_\phi q_\phi=1,\\
        \eta = &\,\,-\left(\tfrac{\pi k^\ast}{L}\right)^2 p_vq_v,\\
        \beta = &\,\,p_u\left[ t_{0,u}\left(r_uq_\phi-\gamma(q_u+q_v)\right)+q_u\left(r_ut_{0, \phi}-\gamma(t_{0, u}+t_{0, v})\right)\right]+\\
        &\,\,\tfrac{1}{2}p_u\left[t_{2, u}\left(r_uq_\phi-\gamma(q_u+q_v)\right) + q_u\left(r_ut_{2,\phi}-\gamma(t_{2, u}+t_{2, v})\right)\right]+\\
        &\,\,p_v\left[ t_{0,v}\left(r_vq_\phi-\gamma(q_u+q_v)\right)+q_v\left(r_vt_{0, \phi}-\gamma(t_{0, u}+t_{0, v})\right)\right]+\\
        &\,\,\tfrac{1}{2}p_v\left[t_{2, v}\left(r_vq_\phi-\gamma(q_u+q_v)\right) + q_v\left(r_vt_{2,\phi}-\gamma(t_{2, u}+t_{2, v})\right)\right]\\
        &\,\,-\kappa p_\phi\left[q_\phi \left(t_{0, u}+t_{0, v}\right)+t_{0,\phi}\left(q_u+q_v\right)\right]-\tfrac{1}{2}\kappa p_\phi\left[q_\phi\left(t_{2,u}+t_{2,v}\right)+t_{2,\phi}\left(q_u+q_v\right)\right].
    \end{aligned}
\end{equation}
Equation \eqref{eq:unbiased-SL-equation} is a cubic Stuart-Landau equation, and it describes how the amplitude of the pattern changes in the slow time variable $T = \varepsilon^2 t$. By characterizing the dynamics of $A(T)$ and determining the long-time behavior of the amplitude, we can help to characterize the qualitative behavior of our approximate solution from the perturbation expansion for parameters close to the pattern-forming threshold. We use this amplitude equation to classify the pattern-forming bifurcation, and will use the long-time value of the amplitude to help characterize the total population size for each strategy and the total concentration of the public good across the spatial domain. 

\begin{remark}
    In the derivation of the first and second-order perturbation solutions, we select a nonzero vector $\mathbf{q}$ in the kernel of $M_{D_v^\ast}(k^\ast)$. The choice of $\mathbf{q}$ is not unique, as it can be multiplied by any nonzero scalar. However, such a rescaling does not affect the resulting perturbation solutions, as shown in Appendix \ref{sec:eigenvectors}.
\end{remark}

Due to the complexity of the expressions, we determine the bifurcation type numerically. Using the reaction parameters from Table \ref{table 1} and holding fixed the values of $D_u$ and $D_\phi$ that we used for the linear stability analysis, we can compute the corresponding threshold values $D_v^\ast$ and use this to obtain the vectors $\mathbf{p}$ and $\mathbf{q}$ to derive the corresponding coefficients $\eta$ and $\beta$ for the Stuart-Landau equation. We then plot the values of $\eta$ and $\beta$ as a function of the cooperator diffusivity $D_u$, showing that $\eta > 0$ and $\beta < 0$ for the parameters in Table \ref{table 1} and the range of $D_u$ under consideration. These values of $\eta$ and $\beta$ suggest that the amplitude equation will undergo a supercritical pitchfork bifurcation as $\tilde{D}$ increases past $0$. In particular, this behavior of the amplitude equation rules out bistability of a pattern state and the uniform coexistence state for defector diffusivities $D_v$ slightly below the threshold required for Turing instability under the default parameters, with stable uniform states expected for $D_v < D_v^*$ and a stable patterned state expected for $D_v > D_v^*$. 

\begin{figure}[htbp]
    \centering
    
    \begin{subfigure}[t]{0.46\linewidth}
        \centering
        \includegraphics[width=\linewidth]{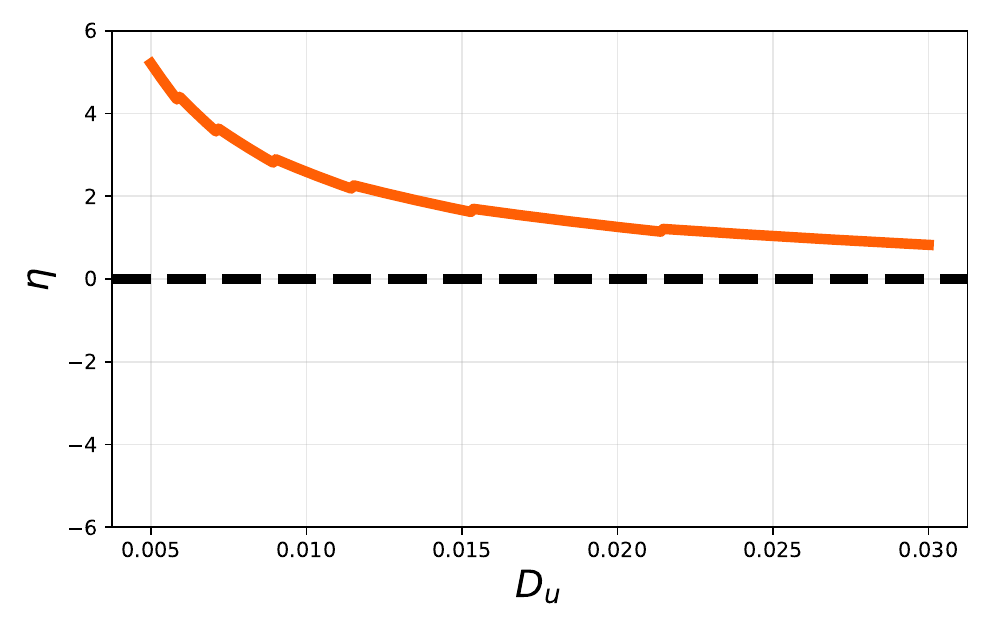}
        \caption{$\eta$ - $D_u$ graph}
    \end{subfigure}
    \hspace{0.05\linewidth}
    \begin{subfigure}[t]{0.46\linewidth}
        \centering
        \includegraphics[width=\linewidth]{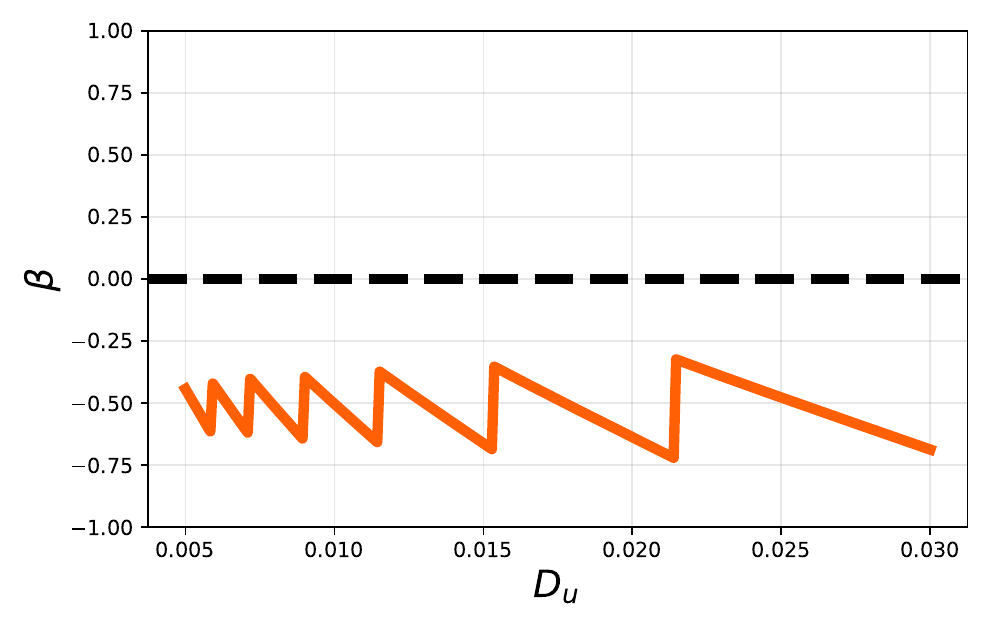}
        \caption{$\beta$ - $D_u$ graph}
    \end{subfigure}

    \vspace{1em}

    \caption{$\eta$ - $D_u$ graph and $\beta$ - $D_u$ graph on the one-dimensional domain $[0,L]$ with $L=8$. Wavenumbers $k^\ast$ are chosen based on Equations \eqref{eq:D_v-threshold} and \eqref{eq:D_v-threshold-min}. The parameters are $w_u=0$, $w_v=0$, and the ODE parameters are taken from Table~\ref{table 1}. The horizontal axis corresponds to the values of $D_u$. In particular, when $D_u = D_\phi = 0.01$, the critical wavenumber is $k^\ast=8$. The horizontal dashed lines represent $\eta = 0$ (left) and $\beta = 0$ (right), separating regimes in which we expect a supercritical pitchfork bifurcation ($\eta > 0$ and $\beta < 0$) and in which subcritical bifurcations may be possible.}
    \label{figure 3}
\end{figure}

In particular, for the special case $D_u = D_\phi = 0.01$, we compute the corresponding threshold value $D_v^\ast = 0.04861$, as in Figure~\ref{figure 1}. In this case, we can compute that 
\begin{equation}
    \mathbf{p}\approx (1.0852,-1.5536,0.8525)\quad \mathbf{q}\approx(0.9461,0.1688,0.2762),
\end{equation}
and the values of $\eta$ and $\beta$ are $\eta \approx 2.5887$ and $\beta \approx -0.4500$.

\begin{remark}
    We note that all components of the vector $\mathbf{q}$ have the same sign, which implies that the deviation of the spatial profiles of cooperators, defectors, and public goods from the uniform coexistence state will be positivity correlated. However, this correlation is not always guaranteed. In Appendix~\ref{sec:eigenvectors}, we calculate the ratios between the coefficients $q_u$, $q_v$, and $q_{\phi}$ for the solution of the densities cooperators, defectors, and the public good are always correlated for the $\mathcal{O}(\varepsilon)$ problem near the pattern-forming threshold for both the unbiased and biased models. Namely, we are able to show that, for both models, the ratios of $q_u$ between $q_{\phi}$ and between $q_v$ and $q_{\phi}$ take the form
    \begin{subequations}
    \begin{align}
    \label{eq:uphiratio}
    \frac{q_u}{q_{\phi}} &= \frac{\kappa\phi_0 v_0\Big[2\left(\frac{\pi k}{L}\right)^2w_vD_v+r_v\Big]+\Big[\left(\frac{\pi k}{L}\right)^2D_v+\gamma v_0\Big]\Big[\left(\frac{\pi k}{L}\right)^2D_\phi+\kappa\big(u_0+v_0\big)+\delta\Big]}{\gamma\kappa v_0\phi_0+\Big[\left(\frac{\pi k}{L}\right)^2D_v+\gamma v_0\Big]\big(c-\kappa \phi_0\big)} \\
    \label{eq:vphiratio} \frac{q_v}{q_{\phi}} &= \frac{\big(c-\kappa\phi_0\big)\Big[2\left(\frac{\pi k}{L}\right)^2w_vD_vv_0+r_vv_0\Big]-\gamma v_0\Big[\left(\frac{\pi k}{L}\right)^2D_\phi+\kappa\big(u_0+v_0\big)+\delta\Big]}{\gamma\kappa v_0\phi_0+\Big[\left(\frac{\pi k}{L}\right)^2D_v+\gamma v_0\Big]\big(c-\kappa \phi_0\big)}.
    \end{align}
    \end{subequations}
   Noting that we require that $c - \kappa \phi_0 > 0$ in order for the coexistence equilibrium to be biologically feasible, we can  deduce From Equation \eqref{eq:uphiratio} that the ratio $\frac{q_u}{q_{\phi}} > 0$. Therefore $q_u$ and $q_{\phi}$ will have the same sign, and we can therefore conclude that the spatial profiles of cooperators and defectors will be spatially correlated in both models when the bifurcation parameter is slightly above the threshold for the onset of pattern formation. 
   
    However, the lengthy expression for the ratio $\frac{q_v}{q_{\phi}}$ in Equation \eqref{eq:vphiratio} does not produce conclusive evidence for its sign and the corresponding correlation of the defector density with the density of cooperators and defectors. Instead, we can use numerical calculations to check the scenarios in which the profiles of cooperators and defectors are correlated or anti-correlated in space, and we will use our analytical expression primarily to deduce the impact that key parameters have on their relative effect in facilitating or impeding spatial correlation. Namely, we see that increasing the movement parameters for defectors $w_v$ and $D_v$ helps to facilitate spatial correlation in cooperators and defectors, while increasing diffusivity $D_{\phi}$ of the public good impedes the establishment of such a correlation.
\end{remark}

Because we have seen that $\eta>0$ and $\beta<0$ for the case of our default parameters from Table \ref{table 1}, we can see that the qualitative behavior of the amplitude equation has the following behaviors as $\tilde{D}$ passes through $0$.
\begin{enumerate}
    \item When $\tilde{D}=0$, Equation \eqref{eq:unbiased-SL-equation} only has the trivial equilibrium $A=0$, which is stable at the bifurcation point.

 \item When $\tilde{D}<0$, Equation \eqref{eq:unbiased-SL-equation} has only the trivial equilibrium $A=0$. In this case
    \begin{equation}
        F(A)=\eta\tilde{D}A+\beta A^3,\,\,\, 
        F'(0)=\eta\tilde{D}<0,
    \end{equation}
    so $A=0$ is stable.

    \item When $\tilde{D}>0$, Equation \eqref{eq:unbiased-SL-equation} has three equilibria:
    \begin{equation}
        A_0=0,\: \:  A_+=\sqrt{-\tfrac{\eta\tilde{D}}{\beta}},\: \: A_-=-\sqrt{-\tfrac{\eta\tilde{D}}{\beta}},
    \end{equation}
    with 
    \begin{equation}
        \begin{aligned}
            F'(A_0)&=\eta\tilde{D}>0,\\
            F'(A_+)&=F'(A_-)=-2\eta\tilde{D}<0.
        \end{aligned}
    \end{equation}
    Therefore, $A_0$ is unstable, while $A_+$ and $A_-$ are stable.
\end{enumerate}

\begin{figure}[!ht]
    \centering
    \includegraphics[width=0.5\textwidth]{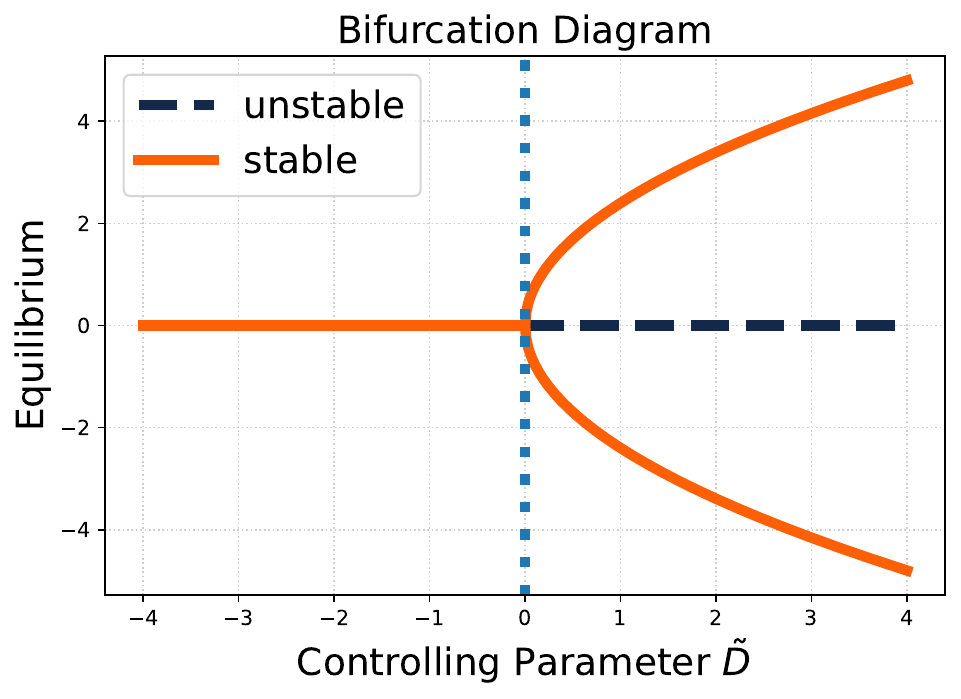}
    \caption{Pitchfork bifurcation diagram for the Stuart-Landau equation from Equation \eqref{eq:unbiased-SL-equation}. The horizontal axis represents the bifurcation parameter $\tilde{D}$, while the vertical axis shows the equilibrium amplitude $A$. The trivial branch $A=0$ is stable for $\tilde{D}<0$ but becomes unstable when $\tilde{D}>0$. For $\tilde{D}>0$, two nontrivial equilibria $A_\pm=\pm\sqrt{-\eta\tilde{D}/\beta}$ emerge, and both are stable. This corresponds to a supercritical pitchfork bifurcation.}
    \label{figure 4}
\end{figure}

We can illustrate the bifurcation structure by drawing the pitchfork bifurcation diagram, as shown in Figure~\ref{figure 4}. In particular, we see that, as $\tilde{D}$ increases through $0$, we see that the uniform state with $A(T) = 0$ becomes unstable, and two stable equilibria $A_{\pm}$ emerge continuously from the zero-amplitude state providing alternative stable states for the patterns. Due to our choice of zero-flux boundary conditions, the distinction between the positive amplitude solution $A_+$ and the negative amplitude solution $A_{-}$ depends on whether the pattern achieves local maxima or local minima at the endpoints of the domain $x = 0$ and $x = L$. 

\FloatBarrier

\subsection{Weakly Nonlinear Analysis of the Biased Model}\label{sec:weakly-nonlinear-biased}

\begin{figure}[htbp]
    \centering
    
    \begin{subfigure}[t]{0.46\linewidth}
        \centering
        \includegraphics[width=\linewidth]{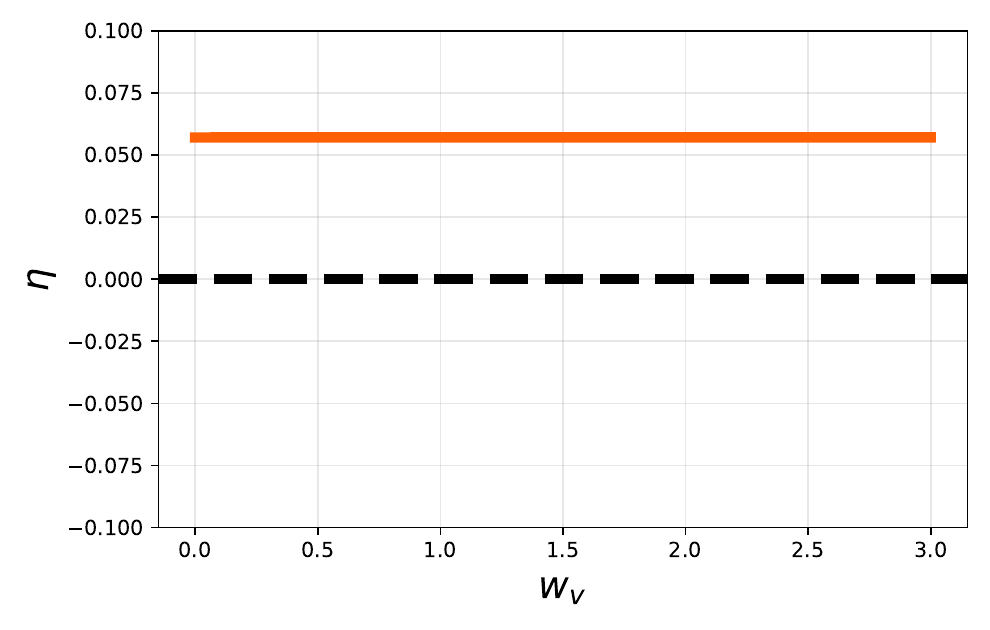}
        \caption{$\eta$ - $w_v$ graph}
    \end{subfigure}
    \hspace{0.05\linewidth}
    \begin{subfigure}[t]{0.46\linewidth}
        \centering
        \includegraphics[width=\linewidth]{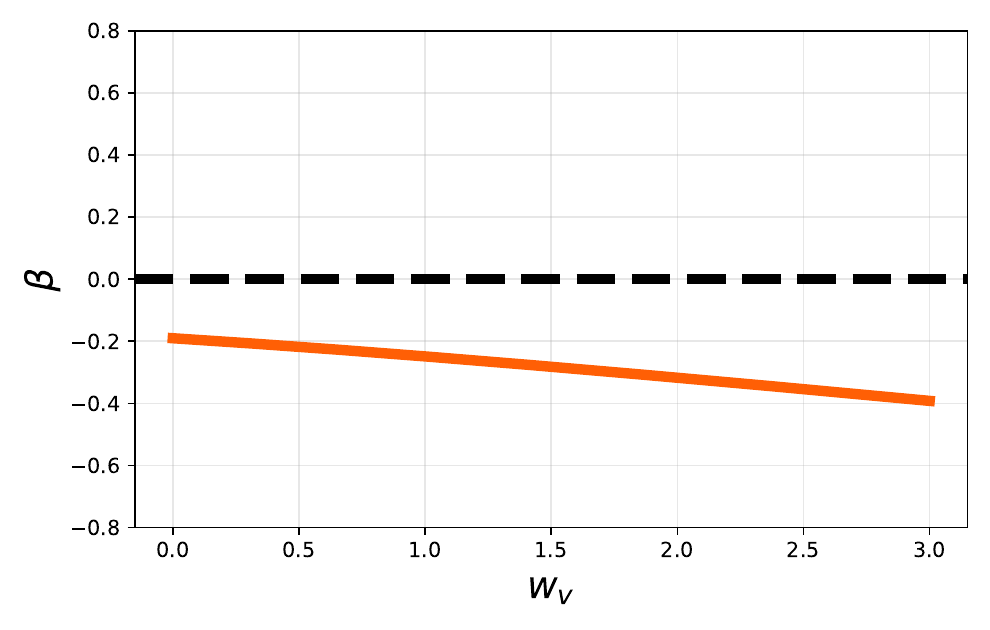}
        \caption{$\beta$ - $w_v$ graph}
    \end{subfigure}

    \vspace{1em}

    \caption{$\eta$ - $w_v$ graph, $\beta$ - $w_v$ graph on the one-dimensional domain $[0,L]$ with $L=8$.  Wavenumbers $k^\ast=8$ are chosen based on Equations \eqref{eq:w_u-threshold} and \eqref{eq:w_u-threshold-min}. The parameters are $D_u=D_v=D_\phi=0.03$, and the ODE parameters are taken from Table~\ref{table 1}.  The horizontal axis corresponds to the values of $w_v$.}
    \label{figure 5}
\end{figure}

We now consider the weakly nonlinear analysis of the biased model, in which $w_u,w_v\not=0$. In this case we use $w_u$ as the bifurcation parameter. And $w_u^\ast$ is the threshold that can be obtained by Equation \eqref{eq:w_u-threshold}. We apply the similar transformation as in Equation \eqref{eq:unbiased-transformations}:
\begin{equation}
    T = \varepsilon^2 t,\quad w_u = w_u^\ast + \varepsilon^2 \tilde{w},
\end{equation}
where $\tilde{w}$ is the controlling parameter. We also use the same perturbations as in Equation \eqref{eq:weakly-nonlinear-expansions}. Again, we show in Section \ref{sec:weakly-nonlinear-expansions} of the appendix that the functions in our expansion will satisfy a hierarchy of partial differential equations for each order of $\varepsilon$. Up to cubic order in $\varepsilon$, this hierarchy of equations will be given by
\begin{subequations}
\begin{alignat}{2}
\mathcal{O}(1) &: \: \hspace{8mm} \bbm 0 \\ 0 \\ 0 \ebm =&&  \bbm  u_0 \left[ r_u \phi_0 - c - \gamma (u_0 + v_0) - \mu_u \right]  \\
 v_0 \left[ r_v \phi_0  - \gamma (u_0 + v_0) - \mu_v \right] \\
c u_0 -  \phi_0\left[ \kappa (u_0 + v_0) + \delta \right] \ebm \label{eq:DMorder1} \\
\mathcal{O}(\varepsilon) &: \:  -\mathcal{L}\begin{bmatrix}
        u_1\\
        v_1\\
        \phi_1
    \end{bmatrix} =&& \bbm 0 \\ 0 \\ 0 \ebm \label{eq:DMorderep} \\
\mathcal{O}(\varepsilon^2) &: \:     -\mathcal{L}\begin{bmatrix}
        u_2\\
        v_2\\
        \phi_2
    \end{bmatrix} =&& %h(u_1,v_1,\phi_1)\colon\!\!\!\!
    -\begin{bmatrix}
        2w_u D_u\frac{\partial}{\partial x}\left(u_1\frac{\partial}{\partial x} \phi_1\right)\\
        2w_vD_v\frac{\partial}{\partial x}\left(v_1\frac{\partial}{\partial x} \phi_1\right)\\
        0
    \end{bmatrix}-\begin{bmatrix}
        -r_uu_1\phi_1 + \gamma u_1^2 + \gamma u_1v_1\\
        -r_vv_1\phi_1 + \gamma u_1v_1 +\gamma v_1^2\\
        \kappa\phi_1(u_1+v_1)
    \end{bmatrix} \label{eq:DMorderep2} \\
\mathcal{O}(\varepsilon^3) &: \:
-\mathcal{L}\begin{bmatrix}
u_3\\
v_3\\
\phi_3
\end{bmatrix}
=&&
\begin{aligned}[t]
&-\frac{\partial}{\partial T}
\begin{bmatrix}
u_1\\
v_1\\
\phi_1
\end{bmatrix}
-\begin{bmatrix}
    2\tilde{w}D_uu_0\frac{\partial^2\phi_1}{\partial x^2}\\
    0\\
    0
\end{bmatrix} \\[4pt]
&-
\begin{bmatrix}
2 w_u D_u \left[ \del{}{x} \left( u_1 \del{}{x} \phi_2 \right)
+ \del{}{x} \left( u_2 \del{}{x} \phi_1 \right) \right] \\
2 w_v D_v \left[ \del{}{x} \left( v_1 \del{}{x} \phi_2 \right)
+ \del{}{x} \left( v_2 \del{}{x} \phi_1 \right) \right] \\
0
\end{bmatrix}\\
&+\begin{bmatrix}
u_2[r_u\phi_1-\gamma(u_1+v_1)]
+u_1[r_u\phi_2-\gamma(u_2+v_2)]\\
v_2[r_v\phi_1-\gamma(u_1+v_1)]
+v_1[r_v\phi_2-\gamma(u_2+v_2)]\\
-\kappa\phi_1(u_2+v_2)-\kappa\phi_2(u_1+v_1)
\end{bmatrix},
\end{aligned}
\label{eq:DMorderep3}
\end{alignat}
\end{subequations}
where the linearization $\mathcal{L}_{w_u^*}$ at the critical sensitivity of public goods to the cooperators $w_u^\ast$ given by
\begin{equation}
    \mathcal{L}_{w_u^*}=\begin{bmatrix}
        D_u\frac{\partial^2}{\partial x^2} - \gamma u_0 & -\gamma u_0 & -2w_u^\ast D_u u_0\frac{\partial^2}{\partial x^2} + r_u u_0\\
        -\gamma v_0 & D_v\frac{\partial^2}{\partial x^2} - \gamma v_0 & -2w_v D_v v_0\frac{\partial^2}{\partial x^2} + r_v v_0\\
        c-\kappa\phi_0 & -\kappa \phi_0 & D_\phi\frac{\partial^2}{\partial x^2} -\kappa(u_0 + v_0)-\delta
    \end{bmatrix}.
    \nonumber
\end{equation}
We follow the same process as we did in Section \ref{sec:weakly-nonlinear-unbiased}. We expect solutions to our equation at order $\mathcal{O}(\varepsilon)$ to take the form
\begin{equation}\label{eq:linearsolutionDM}
    \begin{bmatrix}
        u_1(T,x)\\
        v_1(T,x)\\
        \phi_1(T,x)
    \end{bmatrix}=A(T)\mathbf{q}\cos\left(\frac{\pi k^\ast}{L}x\right),
\end{equation}
where $A(T)$ is the amplitude of the sinusoidal profile of the $\mathcal{O}(\varepsilon)$-solution and the vector $\mathbf{q} = (q_u,q_v,q_{\phi})^T$ is an element of the kernel of the linearization matrix when applied to functions proportional to sinusoidal functions of the form $\cos\left(\frac{\pi k^\ast}{L}x\right)$, which in the directed motion model is given by
\begin{equation} \label{eq:linearization-wavenumber-directed-motion}
M_{w_u^*}(k) = \begin{bmatrix}
        -\left(\frac{\pi k}{L}\right)^2 D_u - \gamma u_0 & -\gamma u_0 & 2\left(\frac{\pi k}{L}\right)^2w_uD_u u_0 + r_u u_0\\
        -\gamma v_0 & -\left(\frac{\pi k}{L}\right)^2D_v - \gamma v_0 & 2\left(\frac{\pi k}{L}\right)^2w_v D_v v_0 + r_v v_0\\
        c-\kappa\phi_0 & -\kappa \phi_0 & -\left(\frac{\pi k}{L}\right)^2D_\phi -\kappa(u_0 + v_0)-\delta
    \end{bmatrix}. 
\end{equation}
We can then plug our expression from Equation \eqref{eq:linearsolutionDM} for $u_1(T,x)$, $v_1(T,x)$, and $\phi_1(T,x)$ into the right-hand side of Equation \eqref{eq:DMorderep2} to see that our solutions at order $\mathcal{O}(\varepsilon^2)$ as an inhomogeneous system of ODEs for $u_2(T,x)$, $v_2(T,x)$, and $\phi_2(T,x)$. For the directed motion model, we see that the terms take the form 
\begin{equation}
    \begin{aligned}
        \frac{\partial}{\partial x}\left(u_1\frac{\partial}{\partial x} \phi_1\right)&=\frac{\partial u_1}{\partial x}\frac{\partial \phi_1}{\partial x}+u_1\frac{\partial^2\phi_1}{\partial x^2}\\
        &=q_uq_\phi A^2(T)\left(\frac{\pi k^\ast}{L}\right)^2\left[\sin^2\left(\frac{\pi k^\ast}{L}x\right)-\cos^2\left(\frac{\pi k^\ast}{L}x\right)\right]\\
        &=-q_uq_\phi A^2(T)\left(\frac{\pi k^\ast}{L}\right)^2\cos\left(\frac{2\pi k^\ast}{L}x\right),\\
        \frac{\partial}{\partial x}\left(v_1\frac{\partial}{\partial x} \phi_1\right)&=\frac{\partial v_1}{\partial x}\frac{\partial \phi_1}{\partial x}+v_1\frac{\partial^2\phi_1}{\partial x^2}\\
        &=q_vq_\phi A^2(T)\left(\frac{\pi k^\ast}{L}\right)^2\left[\sin^2\left(\frac{\pi k^\ast}{L}x\right)-\cos^2\left(\frac{\pi k^\ast}{L}x\right)\right]\\
        &=-q_vq_\phi A^2(T)\left(\frac{\pi k^\ast}{L}\right)^2\cos\left(\frac{2\pi k^\ast}{L}x\right).
    \end{aligned}
    \nonumber
\end{equation}
Since $\cos^2\left(\frac{\pi k^\ast}{L}x\right)=\frac{1}{2}\left(1+\cos\left(\frac{2\pi k^\ast}{L}x\right)\right)$, these allow us to rewrite Equation \eqref{eq:DMorderep2} similarly as
\begin{equation}\label{equation 3.32}
   - \mathcal{L}_{w^\ast_u} \bbm u_2 \\ v_2 \\ \phi_2 \ebm = -\mathbf{s}_0A^2(T)-\mathbf{s}_2A^2(T)\cos\left(\frac{2\pi k^\ast}{L}x\right),
\end{equation}
where in the biased model 
\begin{equation}
    \begin{aligned}
        \mathbf{s}_0&:=\frac{1}{2}\begin{bmatrix}
        \gamma (q_u^2+q_uq_v)-r_uq_uq_\phi\\
        \gamma (q_uq_v+q_v^2)-r_vq_vq_\phi\\
        \kappa q_\phi(q_u+q_v)
    \end{bmatrix},\\
    \mathbf{s}_2&:=\frac{1}{2}\begin{bmatrix}
        \gamma (q_u^2+q_uq_v)-r_uq_uq_\phi-4w_u D_uq_uq_\phi\left(\frac{\pi k^\ast}{L}\right)^2\\
        \gamma (q_uq_v+q_v^2)-r_vq_vq_\phi-4w_vD_vq_vq_\phi\left(\frac{\pi k^\ast}{L}\right)^2\\
        \kappa q_\phi(q_u+q_v)
    \end{bmatrix}.
    \end{aligned}
\end{equation}
Using these expressions for the inhomogeneous terms and a calculation that is analogous to the case of the unbiased motion model, we are able to see that
\begin{equation}\label{eq:secondsolutionDM}
    \begin{bmatrix}
        u_2(T,x)\\
        v_2(T,x)\\
        \phi_2(T,x)
    \end{bmatrix}=\mathbf{t}_0A^2(T)+\mathbf{t}_2A^2(T)\cos\left(\frac{2\pi k^\ast}{L}x\right),
\end{equation}
where 
\begin{subequations}\label{eq:biased-t}
\begin{align}
    \mathbf{t}_0 & =\left(M_{w_u^*}(0)\right)^{-1}\mathbf{s}_0=J^{-1}\mathbf{s}_0 \\
    \mathbf{t}_2 &= \left(M_{w_u^*}(2k^\ast)\right)^{-1}\mathbf{s}_2.
\end{align}
\end{subequations}
Again, we can plug expressions of $(u_1(T,x), v_1(T,x), \phi_1(T,x))^T$ and  $(u_2(T,x), v_2(T,x), \phi_2(T,x))^T$ into \eqref{eq:DMorderep3}, then we obtain 
\begin{equation}
    -\mathcal{L}_{w_u^\ast}\begin{bmatrix}
        u_3\\
        v_3\\
        \phi_3
    \end{bmatrix}=\mathbf{H}_1\cos\left(\frac{\pi k^\ast}{L}x\right)+\mathbf{H}_2\cos\left(\frac{3\pi k^\ast}{L}x\right),
\end{equation}
where $\mathbf{H}_1=(H_{1,u}, H_{1,v}, H_{1,\phi})^T$ with 
\begin{equation}
    \begin{aligned}
        H_{1, u}=&\,\,-q_u A'(T)+w_u^\ast D_u\left(\tfrac{\pi k^\ast}{L}\right)^2\left(2q_ut_{2,\phi}+q_\phi\left(2t_{0, u}-t_{2,u}\right)\right)A^3(T)\\
        &\,\,+\left[ t_{0,u}\left(r_uq_\phi-\gamma(q_u+q_v)\right)+q_u\left(r_ut_{0, \phi}-\gamma(t_{0, u}+t_{0, v})\right)\right]A^3(T)\\
        &\,\,+\tfrac{1}{2}\left[t_{2, u}\left(r_uq_\phi-\gamma(q_u+q_v)\right) + q_u\left(r_ut_{2,\phi}-\gamma(t_{2, u}+t_{2, v})\right)\right]A^3(T)\\
        &\,\,+2\left(\tfrac{\pi k^\ast}{L}\right)^2 D_u\tilde{w}u_0 q_\phi A(T),\\
        H_{1,v}=&\,\,-q_v A'(T)+w_vD_v\left(\tfrac{\pi k^\ast}{L}\right)^2\left(2q_vt_{2,\phi}+q_\phi\left(2t_{0, v}-t_{2,v}\right)\right)A^3(T)\\
        &\,\,+\left[ t_{0,v}\left(r_vq_\phi-\gamma(q_u+q_v)\right)+q_v\left(r_vt_{0, \phi}-\gamma(t_{0, u}+t_{0, v})\right)\right]A^3(T)\\
        &\,\,+\tfrac{1}{2}\left[t_{2, v}\left(r_vq_\phi-\gamma(q_u+q_v)\right) + q_v\left(r_vt_{2,\phi}-\gamma(t_{2, u}+t_{2, v})\right)\right]A^3(T),\\
        H_{1,\phi}=&\,\,-q_\phi A'(T)-\kappa\left[q_\phi \left(t_{0, u}+t_{0, v}\right)+t_{0,\phi}\left(q_u+q_v\right)\right]A^3(T)\\
        &\,\,-\tfrac{1}{2}\kappa\left[q_\phi\left(t_{2,u}+t_{2,v}\right)+t_{2,\phi}\left(q_u+q_v\right)\right]A^3(T).
    \end{aligned}
\end{equation}
After similar calculations with $\mathbf{p}=(p_u,p_v,p_\phi)\in\ker\big(\mathcal{L}_{w_u^\ast}^{\dagger}\big)$ and $\mathbf{p}\cdot\mathbf{q}=1$, we can obtain the Stuart-Landau equation as
\begin{equation}\label{eq:biased-SL-equation}
    \frac{dA}{dT}=\eta\tilde{w}A+\beta A^3,
\end{equation}
where the coefficients, however, are given by
\begin{equation}\label{eq:biased-SL-coefficients}
    \begin{aligned}
        \eta = &\,\,2p_u\left(\tfrac{\pi k^\ast}{L}\right)^2 D_uu_0 q_\phi,\\
        \beta = &\,\,w_u^\ast D_u\left(\tfrac{\pi k^\ast}{L}\right)^2\left(2p_uq_ut_{2,\phi}+p_uq_\phi\left(2t_{0, u}-t_{2,u}\right)\right)+\\
        &\,\,w_vD_v\left(\tfrac{\pi k^\ast}{L}\right)^2\left(2p_vq_vt_{2,\phi}+p_vq_\phi\left(2t_{0, v}-t_{2,v}\right)\right)+\\
        &\,\,p_u\left[ t_{0,u}\left(r_uq_\phi-\gamma(q_u+q_v)\right)+q_u\left(r_ut_{0, \phi}-\gamma(t_{0, u}+t_{0, v})\right)\right]+\\
        &\,\,\tfrac{1}{2}p_u\left[t_{2, u}\left(r_uq_\phi-\gamma(q_u+q_v)\right) + q_u\left(r_ut_{2,\phi}-\gamma(t_{2, u}+t_{2, v})\right)\right]+\\
        &\,\,p_v\left[ t_{0,v}\left(r_vq_\phi-\gamma(q_u+q_v)\right)+q_v\left(r_vt_{0, \phi}-\gamma(t_{0, u}+t_{0, v})\right)\right]+\\
        &\,\,\tfrac{1}{2}p_v\left[t_{2, v}\left(r_vq_\phi-\gamma(q_u+q_v)\right) + q_v\left(r_vt_{2,\phi}-\gamma(t_{2, u}+t_{2, v})\right)\right]\\
        &\,\,-\kappa p_\phi\left[q_\phi \left(t_{0, u}+t_{0, v}\right)+t_{0,\phi}\left(q_u+q_v\right)\right]-\tfrac{1}{2}\kappa p_\phi\left[q_\phi\left(t_{2,u}+t_{2,v}\right)+t_{2,\phi}\left(q_u+q_v\right)\right].
    \end{aligned}
\end{equation}
We use the same approach as in the unbiased model. However, we now fix $D_u$, $D_v$, and $D_\phi$ and vary $w_v$. We compute the corresponding threshold values of $w_u^\ast$, from which we obtain the values of $\eta$ and $\beta$, and determine the type of bifurcation, as shown in Figure~\ref{figure 5}. The plots in Figure \ref{figure 5} suggest that $\eta > 0$ and $\beta < 0$ for the range of values of defector's sensitivity $w_v$ to public goods we consider, indicating that we will expect a supercritical pitchfork bifurcation for the pattern-forming instability in this parameter regime. 

In particular, for the special case $w_v = 1.0$, we compute the corresponding threshold value $w_u^\ast = 6.4603$, as in Figure~\ref{figure 2}. In this case, we can compute that 
\begin{equation}
    \mathbf{p}\approx (1.2521,-2.9339,1.2394)\quad \mathbf{q}\approx([0.9620,0.1624,0.2194),
\end{equation}
and the values of $\eta$ and $\beta$ are $\eta \approx 0.05706$ and $\beta \approx -0.2302$. Since $\eta>0$ and $\beta<0$, allowing us to deduce that the amplitude equation undergoes a supercritical pitchfork bifurcation as the control parameter $\tilde{w}$ increases past $0$. In particular, we see that the unique stable equilibrium of the amplitude equation is $A_0 = 0$ for $\tilde{w} < 0$. This equilibrium is unstable when $\tilde{w} > 0$, and, in this parameter regime, we see two symmetric stable equilibria with amplitudes 
\begin{equation}
A_{\pm} = \pm \sqrt{- \frac{\eta \tilde{w}}{\beta}}
\end{equation}
that correspond to finite amplitude patterned states in the PDE model. 

\subsection{Robustness of the Weakly Nonlinear Analysis under Alternative Reaction Parameters}\label{sec:robustness-WNA}

So far, we have only explored the form of pitchfork bifurcations obtained for the default reaction parameters from Table \ref{table 1} and for various spatial movement parameters. We have seen that varying the cooperator diffusivity $D_u$ in the unbiased model and that varying the defector movement sensitivity $w_v$ does not impact the achievement of a supercritical pitchfork bifurcation as in the case of our default parameter set. However, we will now check how changing several reaction parameters could potentially impact the form of pattern-forming instabilities, exploring the role played in pattern-formation by the conversation rate $r_u$ of public good consumption into cooperator population growth and the background death rate $\mu_u$ of cooperators. We chose these parameters as representative of the growth and survival ability of cooperators, allowing us to explore how the qualitative properties of spatial patterns may depend on the fitness displayed by cooperators and defectors in the coexistence state and the relative risk of collapse to the locally stable extinction equilibrium for different reaction parameters.

Because we are interested in the instability of the coexistence state, we should consider parameter regimes for $r_u$ and $\mu_u$ for which the coexistence equilibrium exists and is stable under the reaction dynamics. As the coexistence equilibrium $E_3=(u_0, v_0, \phi_0)$ has a density of public goods given by
\[
    \phi_0 = \frac{c + \mu_u - \mu_v}{r_u - r_v},
\]
we can check the biological feasibility of the coexistence equilibrium by examining conditions under which the equilibrium public good concentration $\phi_0$ is stable. Furthermore, we also require from the stability conditions for the coexistence equilibrium $E_3$ in Equations \eqref{eq:R-H-ODE} and \eqref{eq:reaction-stability-condition} that $r_u < r_v$. We can therefore see that two necessary condition for the existence and stability of the coexistence equilibrium are given by
\begin{equation}
    r_u < r_v\quad \text{and} \quad \mu_u < \mu_u + c < \mu_v,
\end{equation}
and we can use numerical calculations to further characterize feasible ranges for varying either $r_u$ or $\mu_u$ under the assumption that all other parameters take the default values from Table \ref{table 1}.

\begin{figure}[htbp]
    \centering
    
    \begin{subfigure}[t]{0.46\linewidth}
        \centering
        \includegraphics[width=\linewidth]{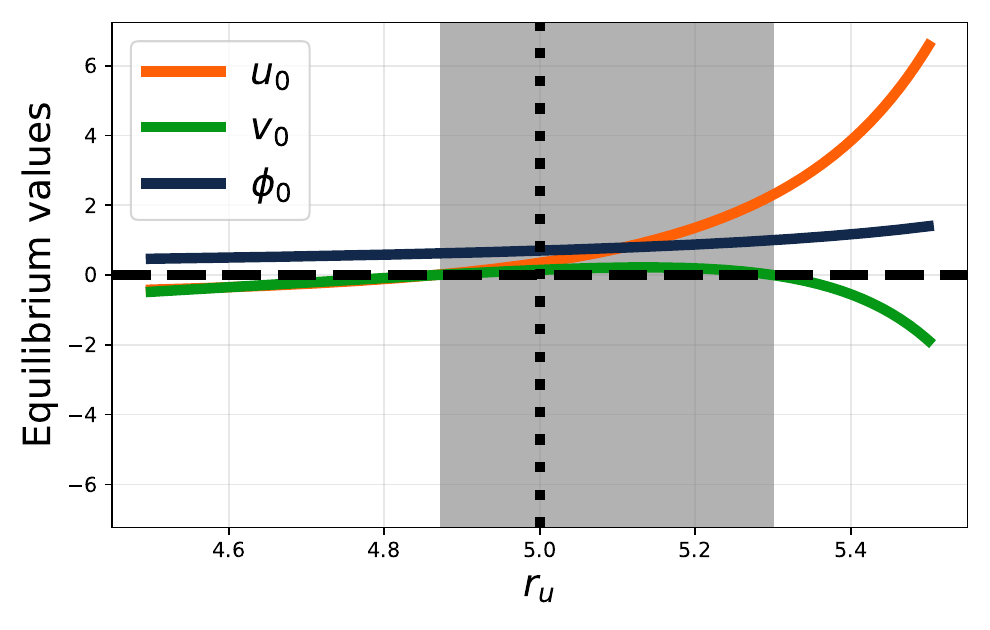}
    \end{subfigure}
    \hspace{0.05\linewidth}
    \begin{subfigure}[t]{0.46\linewidth}
        \centering
        \includegraphics[width=\linewidth]{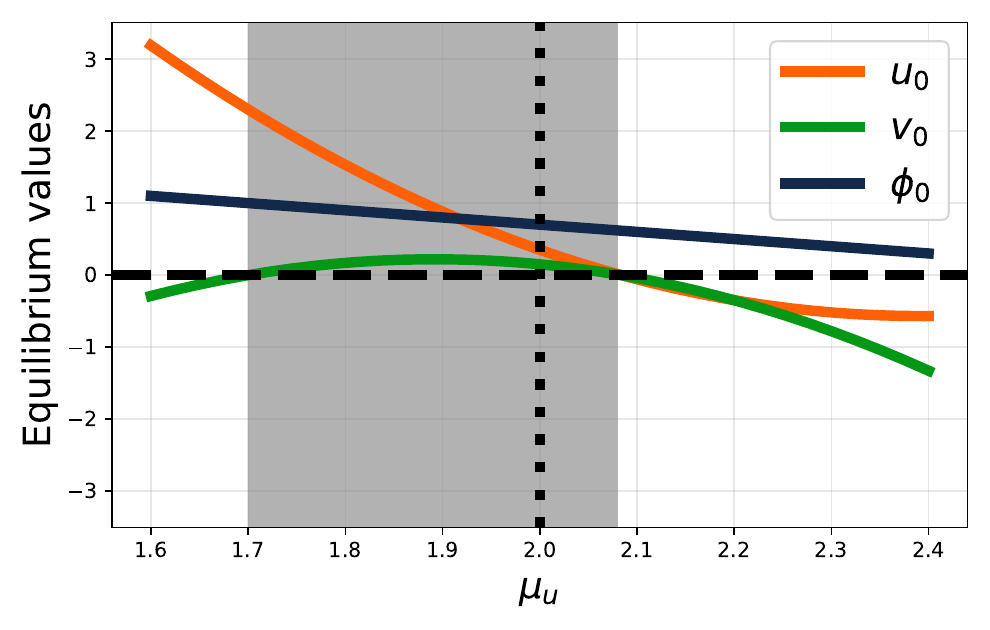}
    \end{subfigure}

    \vspace{1em}

    \caption{A plot of the cooperator population size $u_0$, the defector population size $v_0$, and the concentration of public goods $\phi_0$ for the coexistence equilibrium of the reaction dynamics, plotted as functions of the reaction parameters $r_u$ (left) and $\mu_u$ (right). The remaining reaction parameters are taken from Table~\ref{table 1}. The shaded regions indicate the parameter ranges in which a positive and stable coexistence equilibrium exists. The horizontal dashed lines represent an equilibrium population size or public concentration of $0$, indicating a boundary of biological feasibility for the coexitence equilibrium. The vertical dotted lines indicate the default parameter values of $r_u$ (left) or $\mu_u$ (right) listed in Table~\ref{table 1}.}
    \label{fig:equilibrium-vs-reaction-parameters}
\end{figure}

By considering a numerical sweep over the values of $r_u$ (and correspondingly $\mu_u$),  we calculate the coexistence equilibrium $(u_0, v_0, \phi_0)$ using Equation \eqref{eq:coexistence-equilibrium} and plot the resulting values of $u_0$, $v_0$, and $\phi_0$ in Figure \ref{fig:equilibrium-vs-reaction-parameters}. We then check whether this equilibrium corresponds to possible levels of population and public good density for each value of $r_u$ (or respectively $\mu_u$) and numerically determine the stability of these equilibria. The shaded regions in Figure \ref{fig:equilibrium-vs-reaction-parameters} indicate the parameter regions in which the reaction dynamics admit a positive and stable coexistence equilibrium. Notably, we find that the coexistence equilibrium intersects with the low-population all-cooperator equilibrium $E_2$ for sufficiently small cooperator consumption rate $r_u$ and intersects with the high-population all-cooperator equilibrium $E_1$ for sufficiently large $r_u$, while the coexistence equilibrium interpolates between the high-population all-cooperator equilibrium $E_1$ and the low-population all-cooperator equilibrium $E_2$ as we increase the value of background cooperator death rate $\mu_u$.

Now that we have established feasible regions for the parameters $r_u$ and $\mu_u$ for which the coexistence equilibrium will exist and be stable, we can look to perform weakly nonlinear analysis for the biased and unbiased models across these parameter ranges. We use a linear stability stability analysis for  for both models using Equations \eqref{eq:D_v-threshold-min} and \eqref{eq:w_u-threshold-min}, and we use the corresponding threshold movement parameters $w_u^*$ and $D_v^*$ and most unstable wavenumbers $k^*$ to perform our weakly nonlinear analysis for each $r_u$ and $\mu_u$ in our feasible range. In particular, we calculate the first-order coefficient $\eta$ and the cubic-order coefficient $\beta$ in the amplitude equation based on Equations \eqref{eq:unbiased-SL-coefficients} and \eqref{eq:biased-SL-coefficients}, and we use the sign of the cubic coefficient $\beta$ to determine whether the pattern-forming bifurcation constitutes a supercritical pitchfork bifurcation (if $\beta < 0$) or a subcritical pitchfork bifurcation (if $\beta > 0$). For both models, we see that the bifurcation remains supercritical for a wide range of $r_u$ and $\mu_u$ around the default parameter values from Table 1, and that several regions of subcritical bifurcations for high $r_u$ in the unbiased model, and for either low values of $r_u$ or high values of $\mu_u$ or $r_u$ in the biased model.

\begin{figure}[htbp]
    \centering
    
    \begin{subfigure}{0.45\linewidth}
        \centering
        \includegraphics[width=\linewidth]{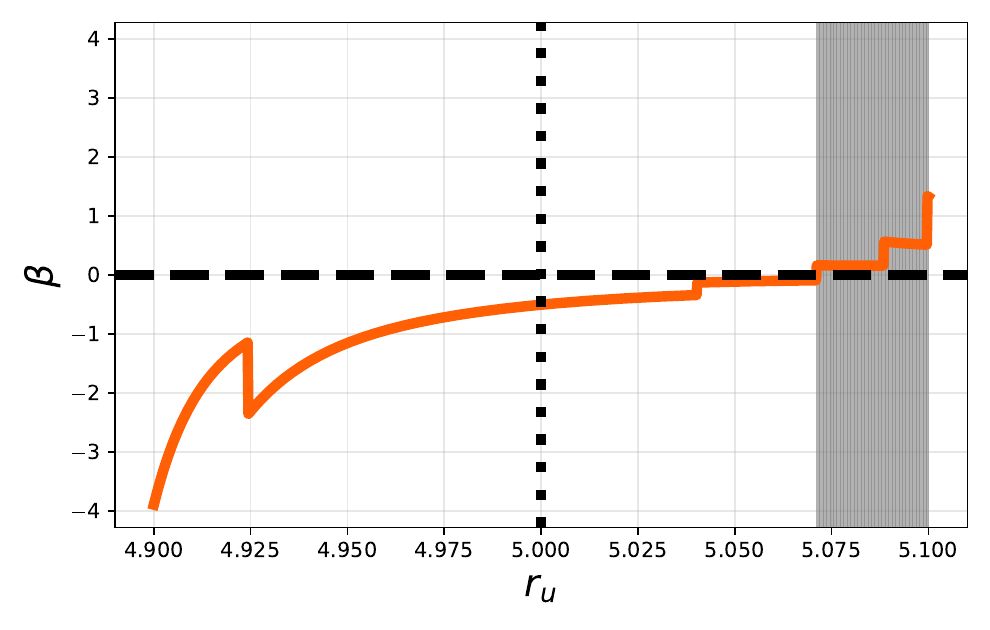}
        \caption{$r_u$ - $\beta$ diagram in the unbiased model}
    \end{subfigure}
    \begin{subfigure}{0.45\linewidth}
        \centering
        \includegraphics[width=\linewidth]{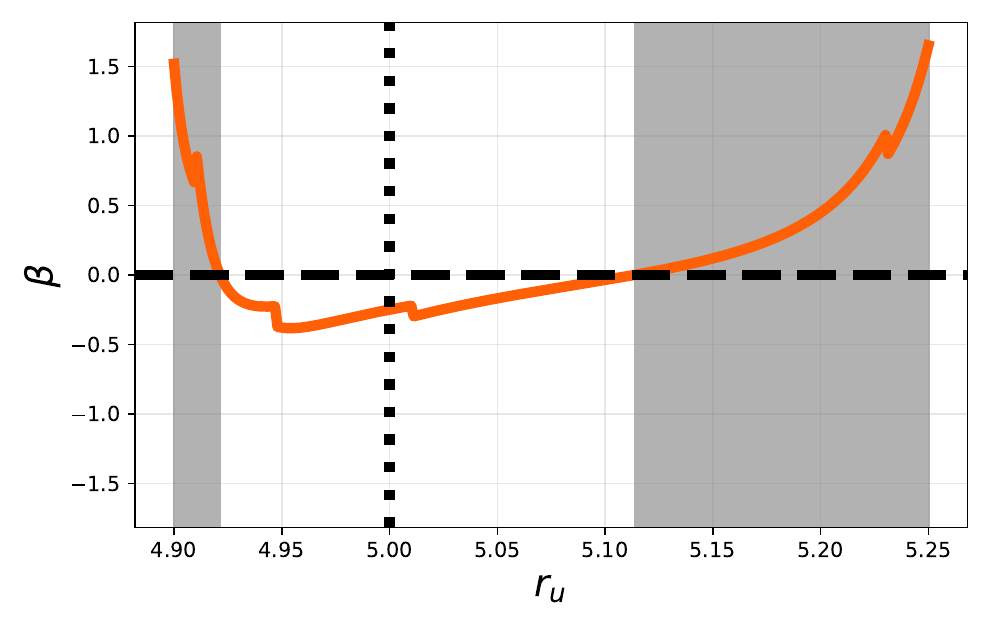}
        \caption{$r_u$ - $\beta$ diagram in the biased model}
    \end{subfigure}

    \vspace{1em}
    
    \begin{subfigure}{0.45\linewidth}
        \centering
        \includegraphics[width=\linewidth]{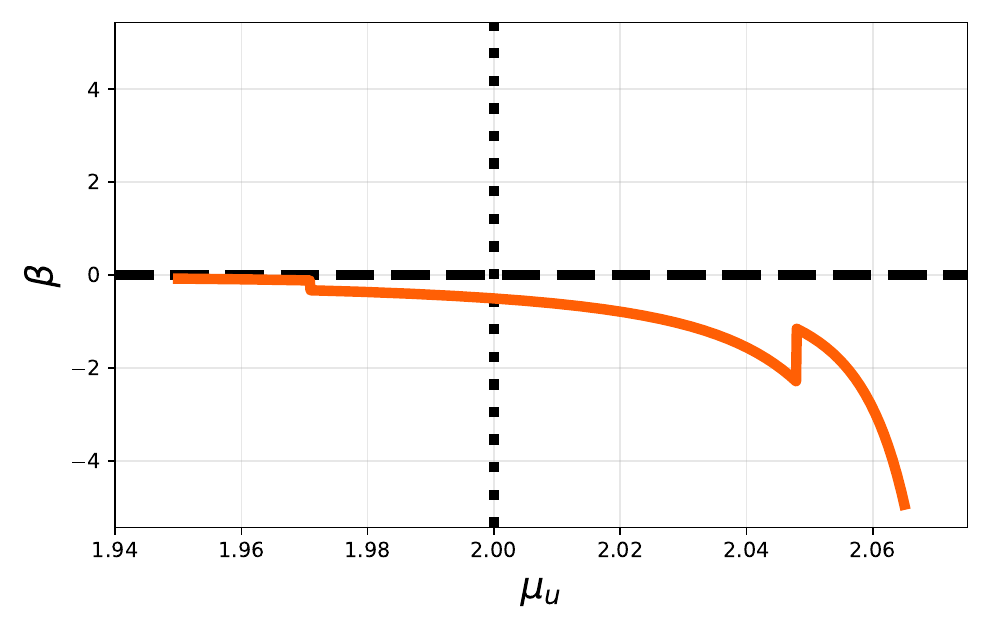}
        \caption{$\mu_u$ - $\beta$ diagram in the unbiased model}
    \end{subfigure}
       \begin{subfigure}{0.45\linewidth}
        \centering
        \includegraphics[width=\linewidth]{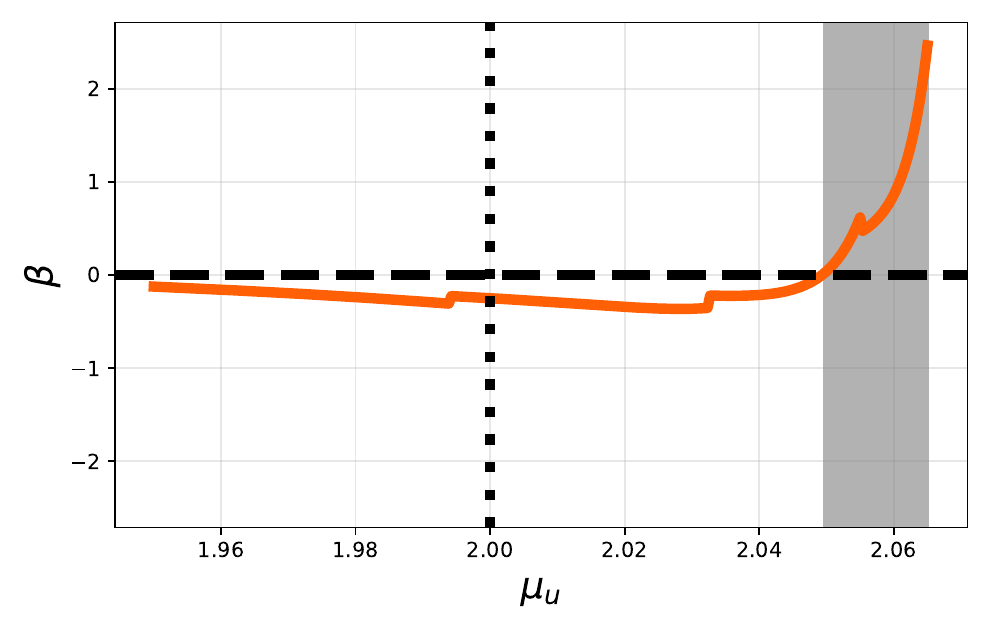}
        \caption{$\mu_u$ - $\beta$ diagram in the biased model}
    \end{subfigure}

    \caption{$r_u$-$\beta$ diagrams for the unbiased model (a) and the biased model (b), and $\mu_u$-$\beta$ diagrams for the unbiased model (c) and the biased model (d). The remaining reaction parameters are taken from Table~\ref{table 1}. For the PDE parameters, we choose $D_u=D_\phi=0.01$ and use $D_v$ as the bifurcation parameter in the unbiased model. For the biased model, we choose $D_u=D_v=D_\phi=0.03$, $w_v=1$, and use $w_u$ as the bifurcation parameter. The horizontal dashed lines indicate the value $\beta = 0$, which provides the boundary between a supercritical pitchfork bifurcation for $\beta < 0$ and a subcritical pitchfork bifurcation for $\beta > 0$. The regions featuring subcritical bifurcation are shaded in gray, and the vertical dotted lines indicate the default values of $r_u$ and $\mu_u$ given in Table~\ref{table 1}.}
    \label{fig:reaction-parameters-beta}
\end{figure}

Notably, we see that all regions of subcritical bifurcations occur in parameter intervals adjacent the parameter boundaries in which a further change of $r_u$ or $\mu_u$ will result in the disappearance of the coexistence equilibrium (and the corresponding extinction of defectors under the reaction dynamics. The presence of a subcritical pattern-forming bifurcation expands the region of the movement parameters $D_v$ or $w_u$ for which a stable patterned state can arise in the unbiased and biased models, respectively, allowing for a region of bistability between a patterned state and the uniform coexistence equilibrium when $D_v$ or $w_u$ are slightly below the respective critical levels of $D_v^*$ or $w_u^*$ that guarantees the local instability of the coexistence equilibrium. This presence of subthreshold patterns near the parameter regimes for elimination of the coexistence equilibrium state is reminiscent of behavior seen in other spatial models in which spatial pattern formation can serve as a resilience mechanism preventing extinction or competitive exclusion in a range of biological systems \cite{deangelis2012self,martinez2013vegetation,rietkerk2021evasion,brennan2025pattern,van2025vegetation,banerjee2026rethinking}.

\section{Numerical Simulation of Our Model}\label{section:simulations}

In this section, we use numerical simulations to solve our model~\eqref{eq:DM-model}. We consider the zero-flux boundary conditions on the interval $[0, L]$:
\begin{equation}
    \left.\frac{\partial u}{\partial x}\right|_{x=0,\,L}=\left.\frac{\partial v}{\partial x}\right|_{x=0,\,L}=\left.\frac{\partial \phi}{\partial x}\right|_{x=0,\,L}=0.
\end{equation}
To perform our numerical simulations of our model from Equation \eqref{eq:DM-model}, we use the built-in \texttt{pdepe} function in MATLAB that implements numerical solutions through the method of lines, a spatial Petrov-Galerkin discretization, and variable-step time integration \cite{skeel1990method,PDEPEdocumentation}. For our spatial grid, we discretize the interval $[0,L]$ into 128 equal subintervals. We consider initial conditions by generating a population and public goods densities at each grid point $x_i$ that are drawn from a random perturbation from the coexistence equilibrium point $E_{3}$ according to the following formula
\begin{equation}\label{eq:initial-condition}
    \begin{aligned}
        u(0,x)&=u_0+\zeta\,\mathrm{Ran}_u(x)\\
        v(0,x)&=v_0+\zeta\,\mathrm{Ran}_v(x)\\
        \phi(0,x)&=\phi_0+\zeta\,\mathrm{Ran}_\phi(x),
    \end{aligned}
\end{equation}
where $\zeta$ is a small parameter and $\mathrm{Ran}_{u}$, $\mathrm{Ran}_{v}$,and $\mathrm{Ran}_{\phi}$ are drawn independently from the uniform distribution on $[-1,1]$.  To ensure reproducibility across simulations, we fix the random-number generator seed in MATLAB so that the random initial conditions are the same each time.

We first represent the results from our numerical simulation of the reaction-diffusion dynamics of the unbiased model in Section \ref{section 4.1}, and then we present numerical dynamics of directed motion in the biased model in Section \ref{section 4.2}. In both of these sections, we provide comparisons between the results of numerical simulations with the analytical predictions provided from the weakly nonlinear stability analysis from Section \ref{sec:weakly-nonlinear-analysis}. We further provided additional comparison between the weakly nonlinear predictions and numerical simulations in Section \ref{section 4.3}, highlighting how the rules for diffusive and directed motion impact qualitative properties of the spatial profiles of cooperators, defectors, and the public good. 

\subsection{Simulations of the Unbiased Model}\label{section 4.1}

\begin{figure}[!htbp]
    \centering
    
    \begin{subfigure}{0.455\linewidth}
        \centering
        \includegraphics[width=\linewidth]{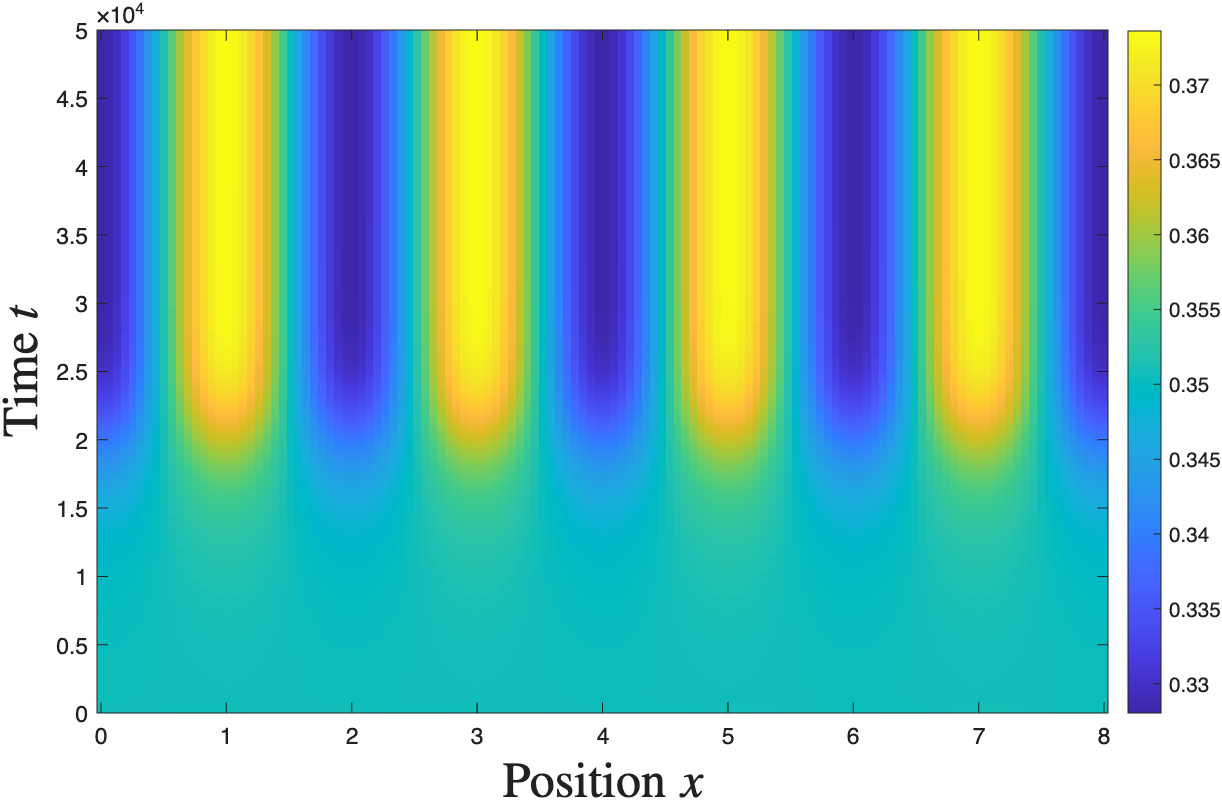}
        \caption{Heat map of $u(t,x)$}
    \end{subfigure}
   \begin{subfigure}{0.45\linewidth}
        \centering
        \includegraphics[width=\linewidth]{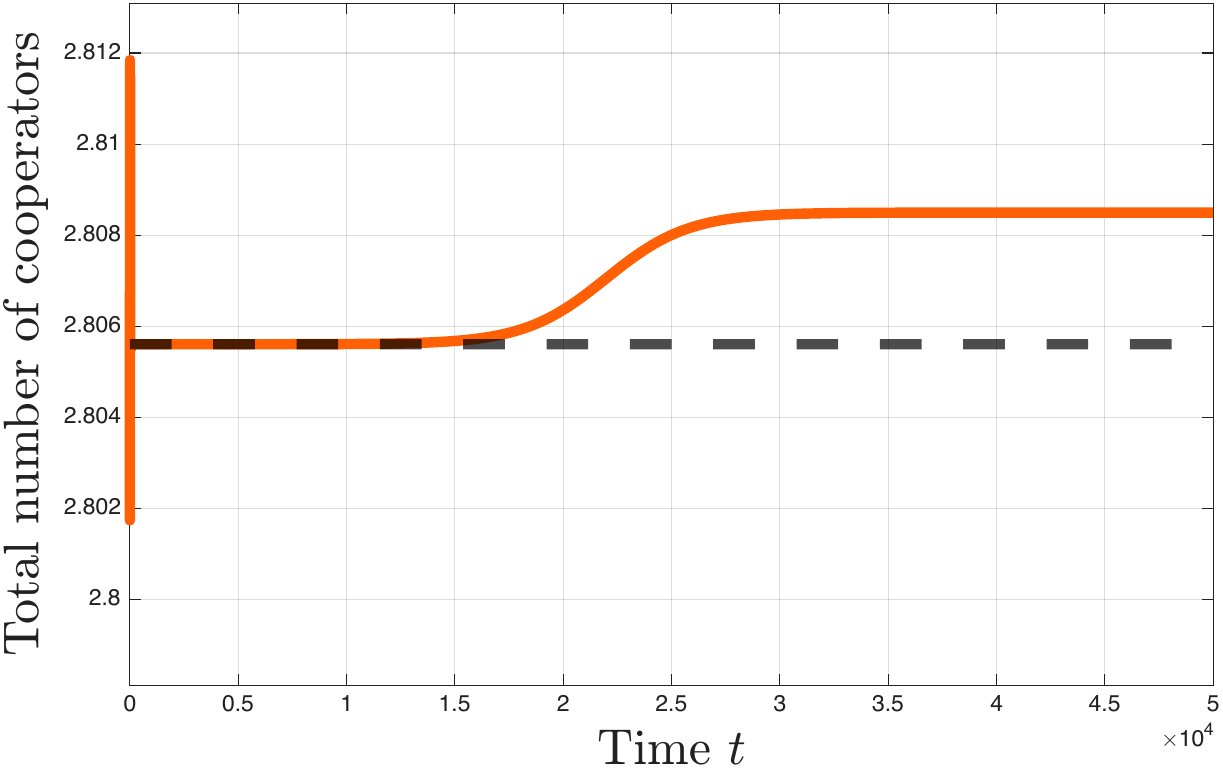}
        \caption{Total cooperators $\int_0^L u(t,x)\,\mathrm{d}x$}
    \end{subfigure}

    \vspace{0.5em}
    
    \begin{subfigure}{0.455\linewidth}
        \centering
        \includegraphics[width=\linewidth]{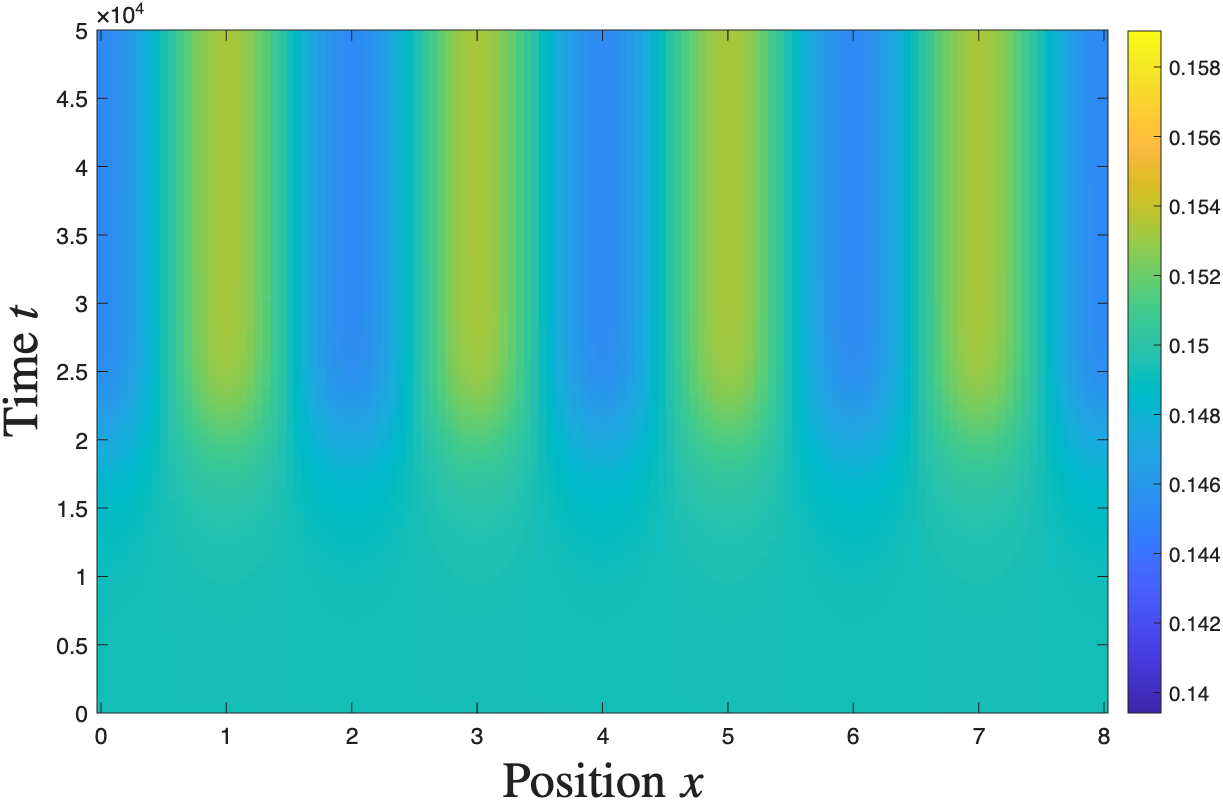}
        \caption{Heat map of $v(t,x)$}
    \end{subfigure}
       \begin{subfigure}{0.45\linewidth}
        \centering
        \includegraphics[width=\linewidth]{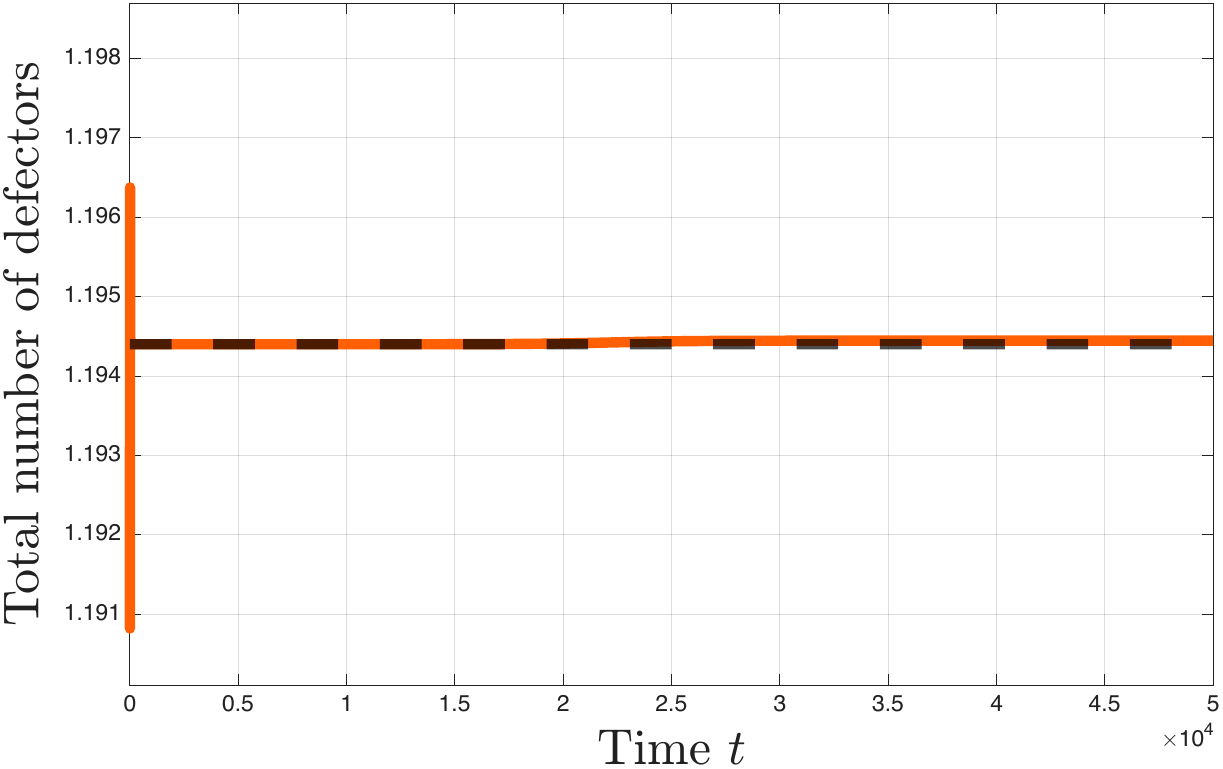}
        \caption{Total defectors $\int_0^L v(t,x)\,\mathrm{d}x$}
    \end{subfigure}

    \vspace{0.5em}

    \begin{subfigure}{0.455\linewidth}
        \centering
        \includegraphics[width=\linewidth]{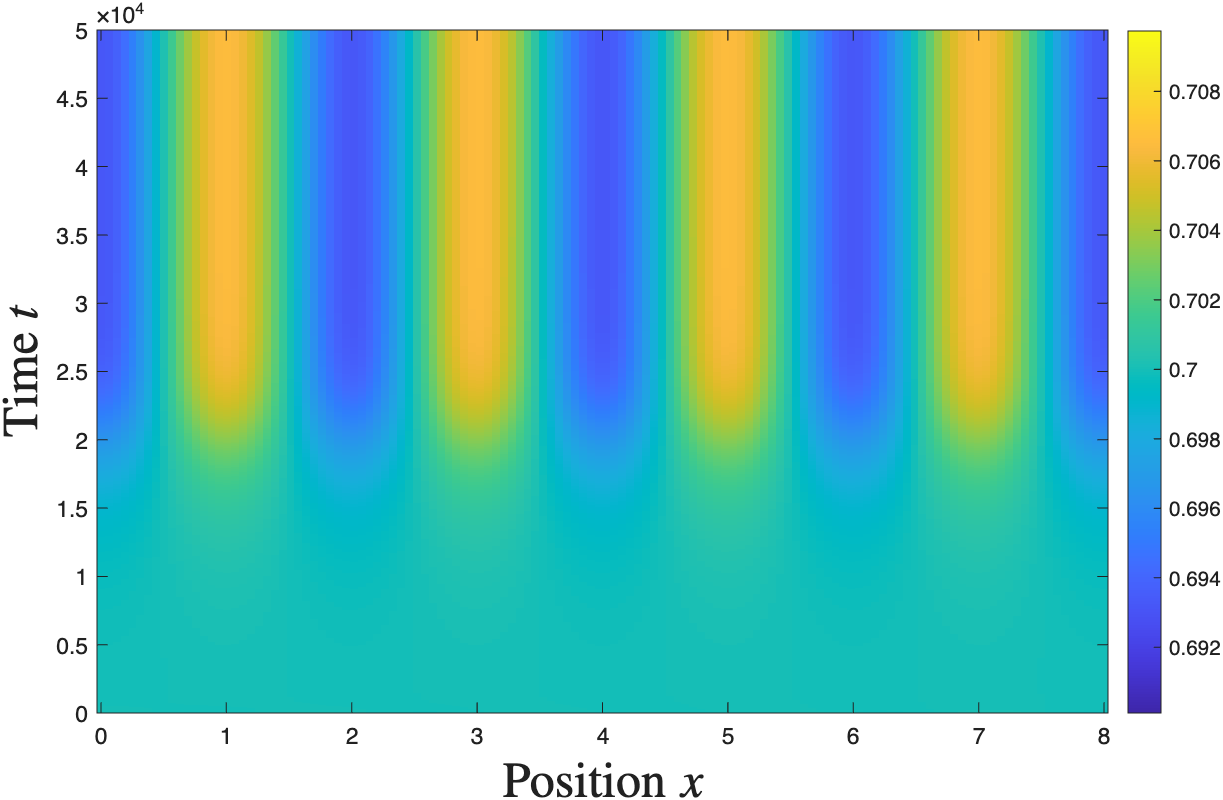}
        \caption{Heat map of $\phi(t,x)$}
        \end{subfigure}
         \begin{subfigure}{0.45\linewidth}
        \centering
        \includegraphics[width=\linewidth]{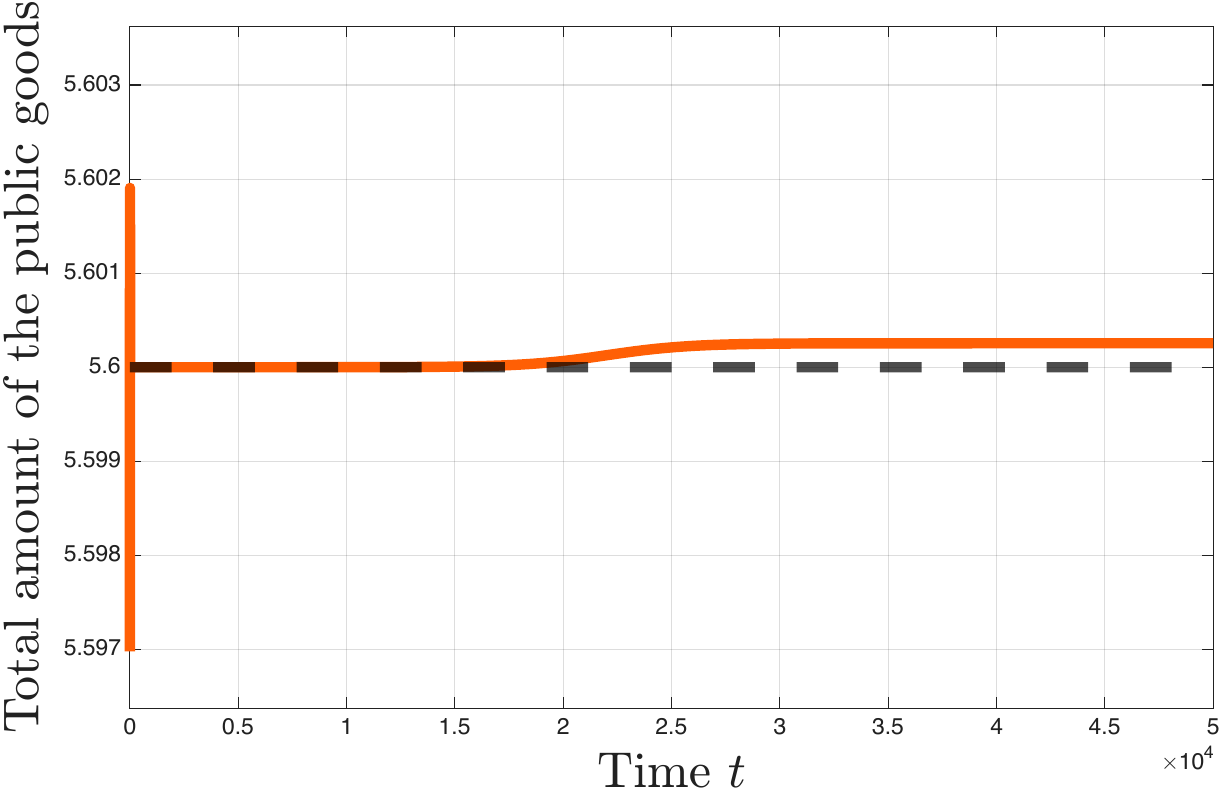}
        \caption{Total public good $\int_0^L \phi(t,x)\,\mathrm{d}x$}
    \end{subfigure}

    \caption{Numerical simulation results of the unbiased model on the one-dimensional domain $[0,L]$ with $L=8$. The parameters are $\zeta=0.01$, $w_u=0$, $w_v=0$, $D_u=D_v=0.01$, $D_v=D_v^\ast(k^\ast)+\varepsilon^2$ with critical wavenumber $k^\ast=8$ and $\varepsilon=0.01$, and the ODE parameters are taken from Table~\ref{table 1}. In the left panels, we present heatmaps of the spatial profiles of cooperators $u(t,x)$, defectors $v(t,x)$, and the public good $\phi(t,x)$, with the horizontal axes describing the spatial interval $[0,L]$ and the vertical axes indicating time increasing from bottom to top. The right panels show the total mass of cooperators $\int_0^L u(t,x) dx$, defectors $\int_0^L v(t,x) dx$, and the public good $\int_0^L \phi(t,x) dx$ across the spatial domain, each plotted as a function of time $t$. The horizontal dashed lines in the right-hand panels represent the total population sizes of cooperators and defectors, and the total concentration of public goods achieved at the spatially uniform coexistence state, which are respectively given by $Lu_0$, $Lv_0$, and $L\phi_0$.}
    \label{fig:simulation-unbiased}
\end{figure}

We start by presenting the emergence of spatial patterns in numerical simulations for the reaction-diffusion model starting with a small initial perturbation from the uniform coexistence equilibrium. In the left panels of Figure \ref{fig:simulation-unbiased}, we plot heat maps of $u$, $v$ and $\phi$ over time, showing how four clusters of cooperators, defectors, and public good emerge over time and approach a spatially heterogeneous steady state. Thus, each horizontal line represents the spatial distribution of $u$, $v$, and $\phi$ over $[0, L]$ at a given time. We used the same parameters as used for the linear stability analysis displayed in Figure \ref{figure 1}, for which instability was achieved by the critical wavenumber $k^* = 8$, corresponding to a prediction of a spatial pattern with four peaks over the interval $[0,L]$. We therefore see good agreement with the emergent pattern and the qualitative properties suggested by linear stability analysis for a simulation with parameters close to the onset of Turing instability.

In addition, we calculate the total mass of cooperators, defectors, and the public good across the domain as a function of time, represented by the integrals $\int_0^L u(t,x)\,dx$, $\int_0^L v(t,x)\,dx$, and $\int_0^L \phi(t,x)\,dx$ on the interval $[0,L]$. We plot these quantities in the right panels of Figure \ref{fig:simulation-unbiased},  finding that the total concentration of cooperators, defectors, and public good initially increase in time before settling into a steady-state value greater than the corresponding concentrations achieved in the spatially uniform steady state. This numerical result suggests that the Turing patterning mechanism can help to promote the population size for each strategy and promote greater provision of public goods across the spatial domain.  

\begin{figure}[!htbp]
    \centering
    \begin{subfigure}[b]{0.48\textwidth}
       \includegraphics[width=\linewidth]{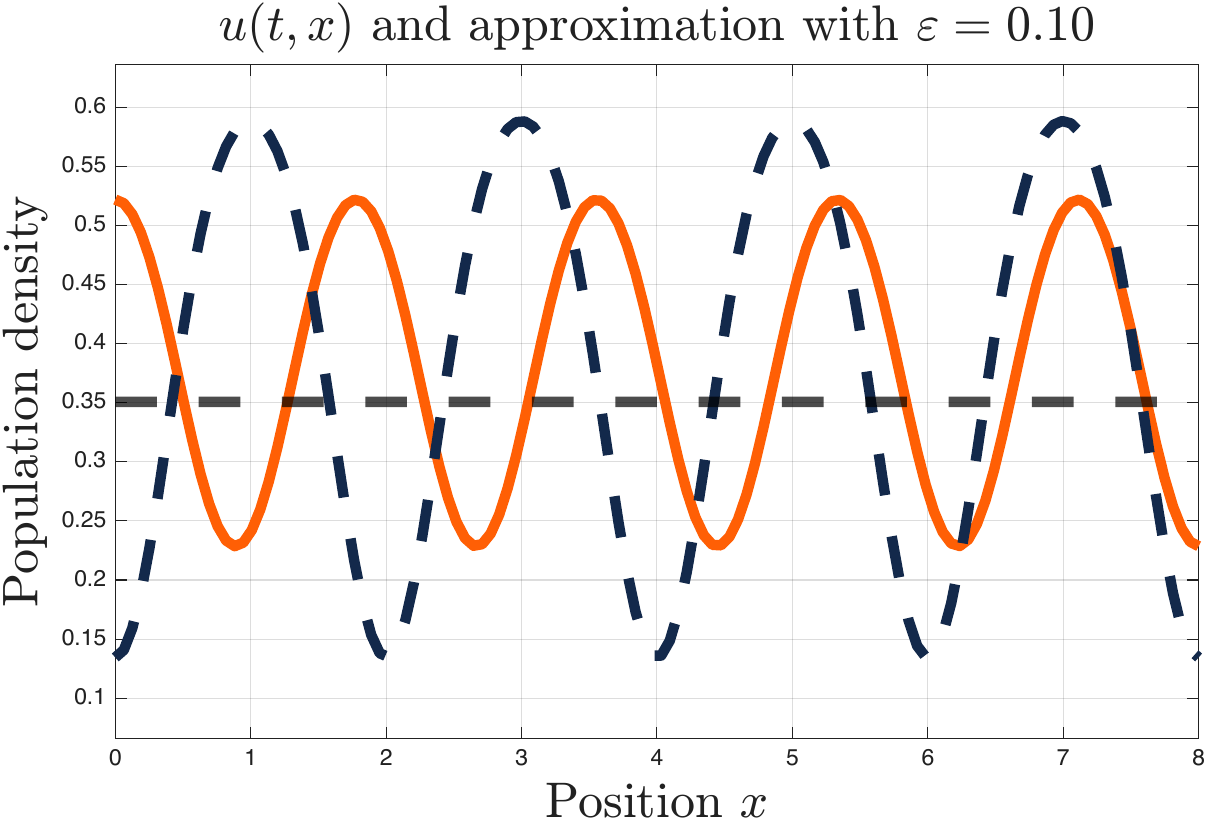}
    \end{subfigure}
    \hfill
    \begin{subfigure}[b]{0.48\textwidth}
        \includegraphics[width=\linewidth]{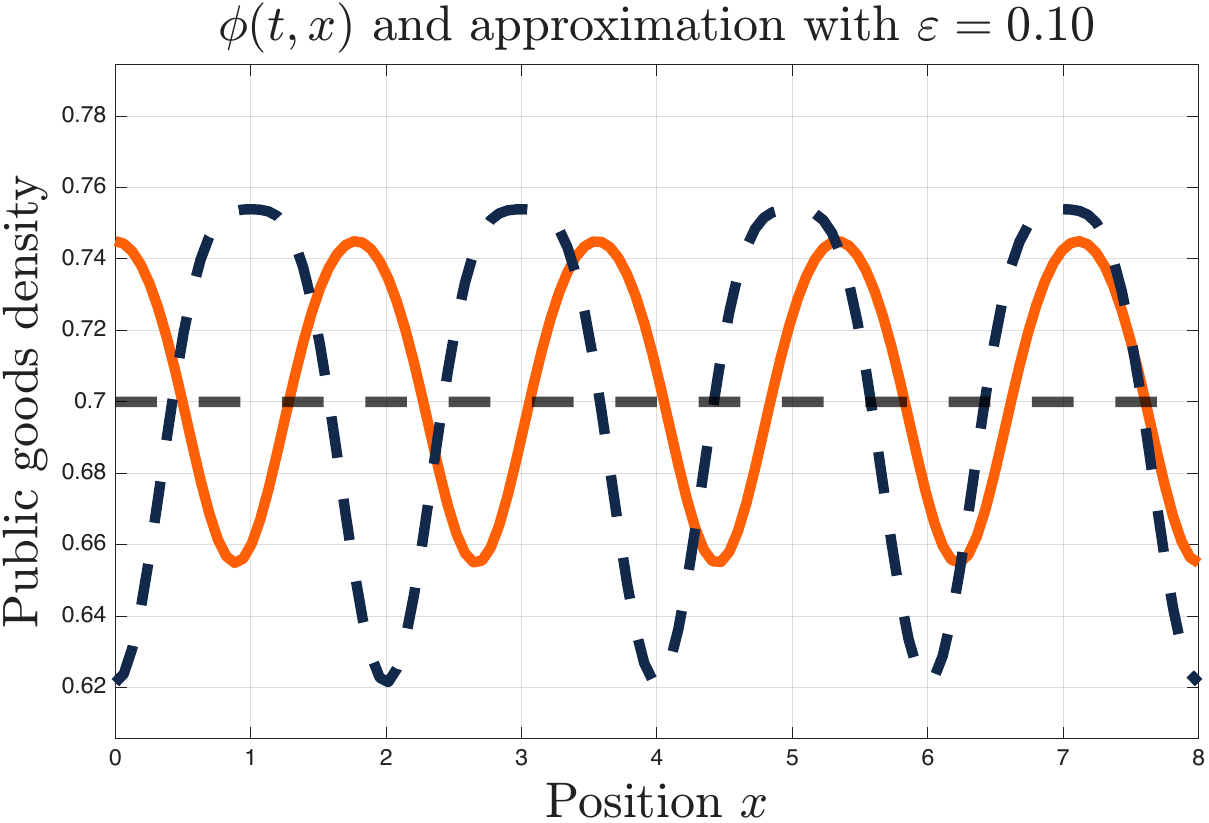}
    \end{subfigure}\vspace{-24pt}
    \begin{subfigure}[b]{\textwidth}
        \centering
        \vspace{2em}
        \caption*{(a) The results when $\zeta=\varepsilon = 0.1$ at $t=50{,}000$}
    \end{subfigure}

    \vspace{1em}

    \begin{subfigure}[b]{0.48\textwidth}
       \includegraphics[width=\linewidth]{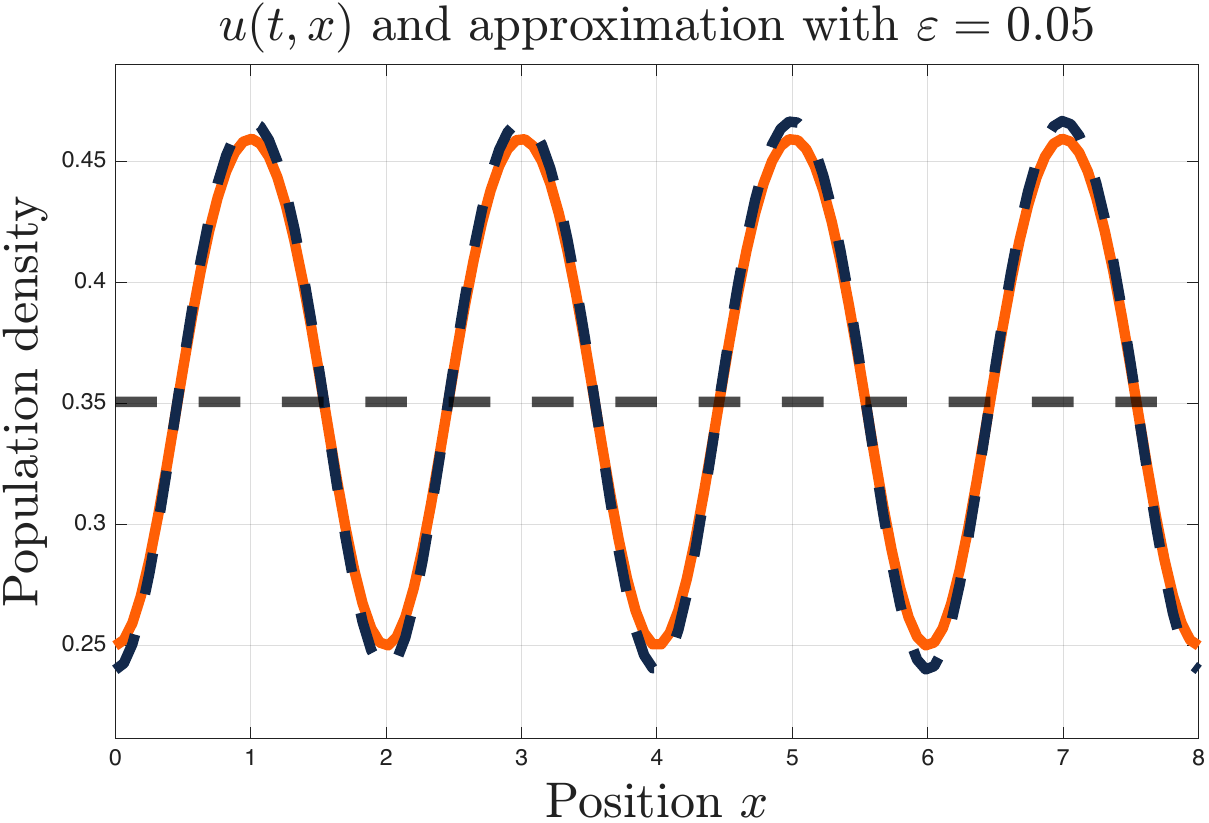}
    \end{subfigure}
    \hfill
    \begin{subfigure}[b]{0.48\textwidth}
        \includegraphics[width=\linewidth]{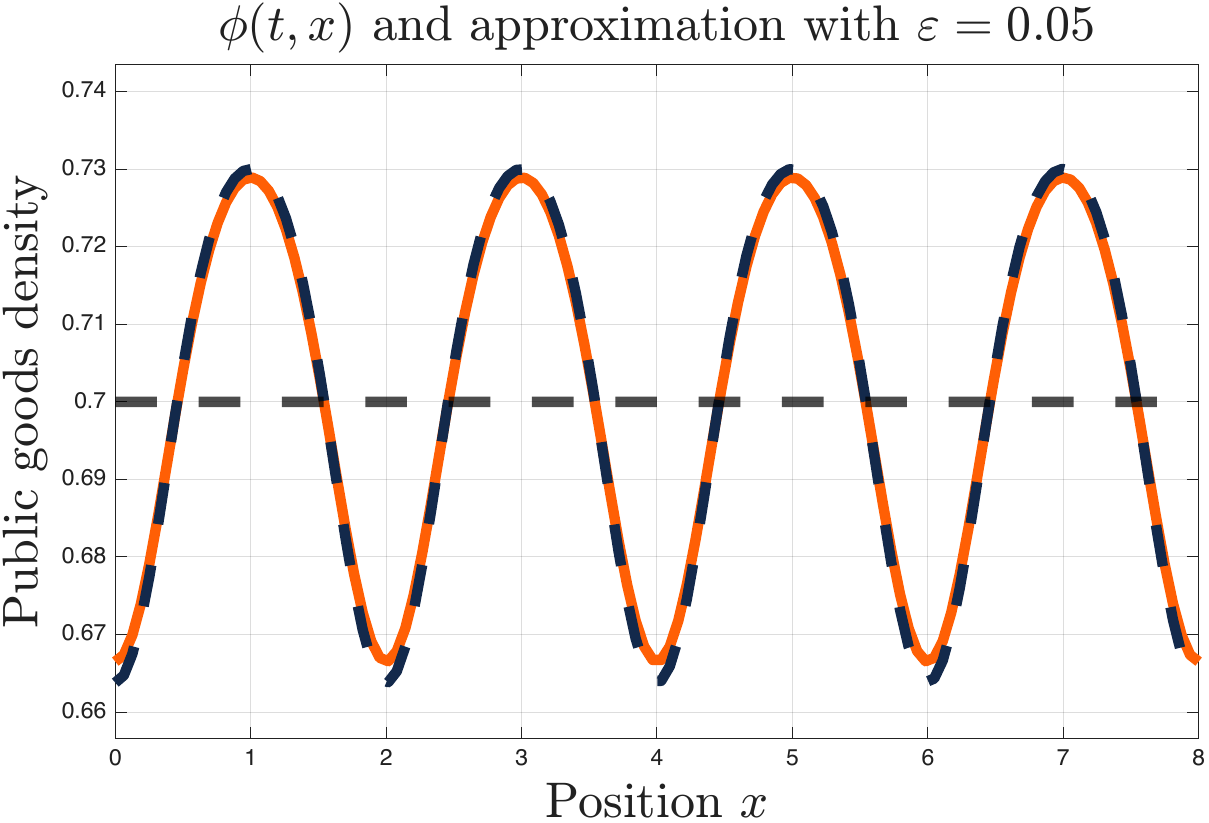}
    \end{subfigure}\vspace{-24pt}
    \begin{subfigure}[b]{\textwidth}
        \centering
        \vspace{2em}
        \caption*{(b) The results when $\zeta=\varepsilon = 0.05$ at $t=50{,}000$}
    \end{subfigure}

    \vspace{1em}

   \begin{subfigure}[b]{0.48\textwidth}
        \includegraphics[width=\linewidth]{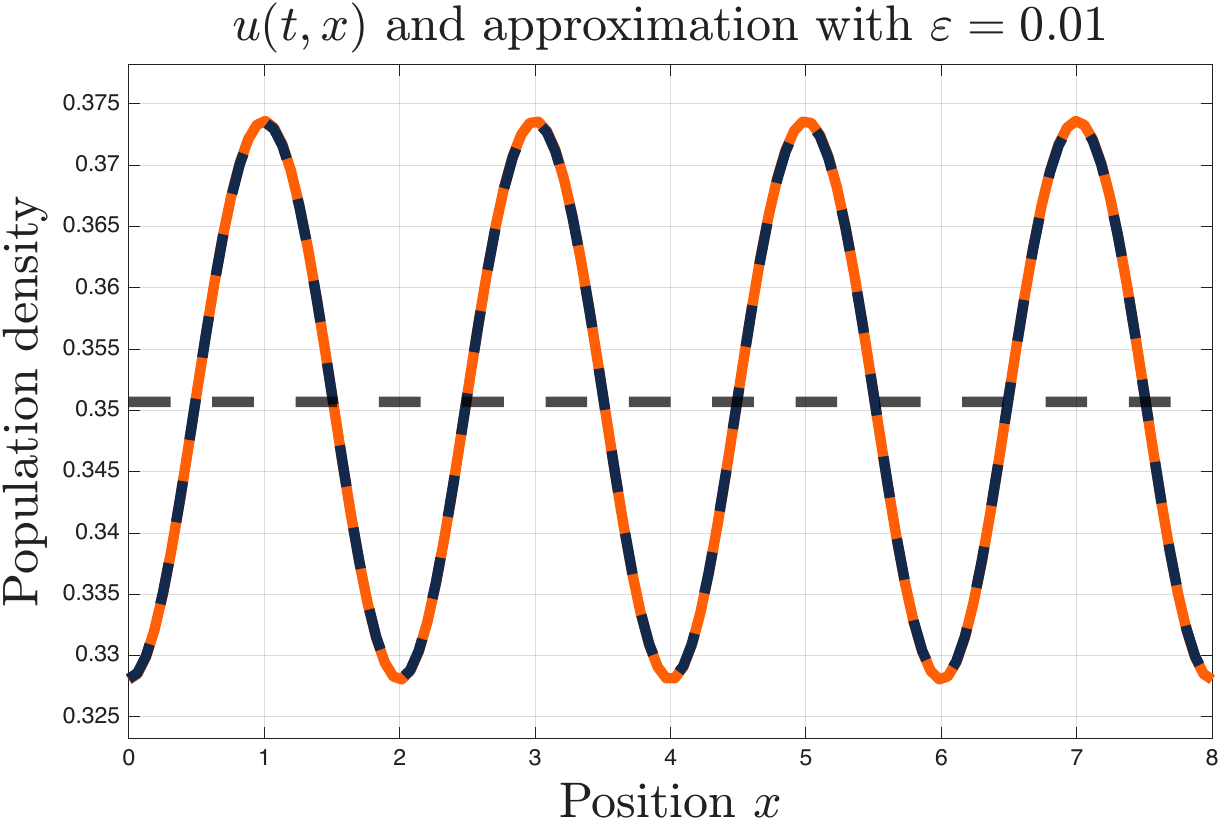}
    \end{subfigure}
    \hfill
    \begin{subfigure}[b]{0.48\textwidth}
        \includegraphics[width=\linewidth]{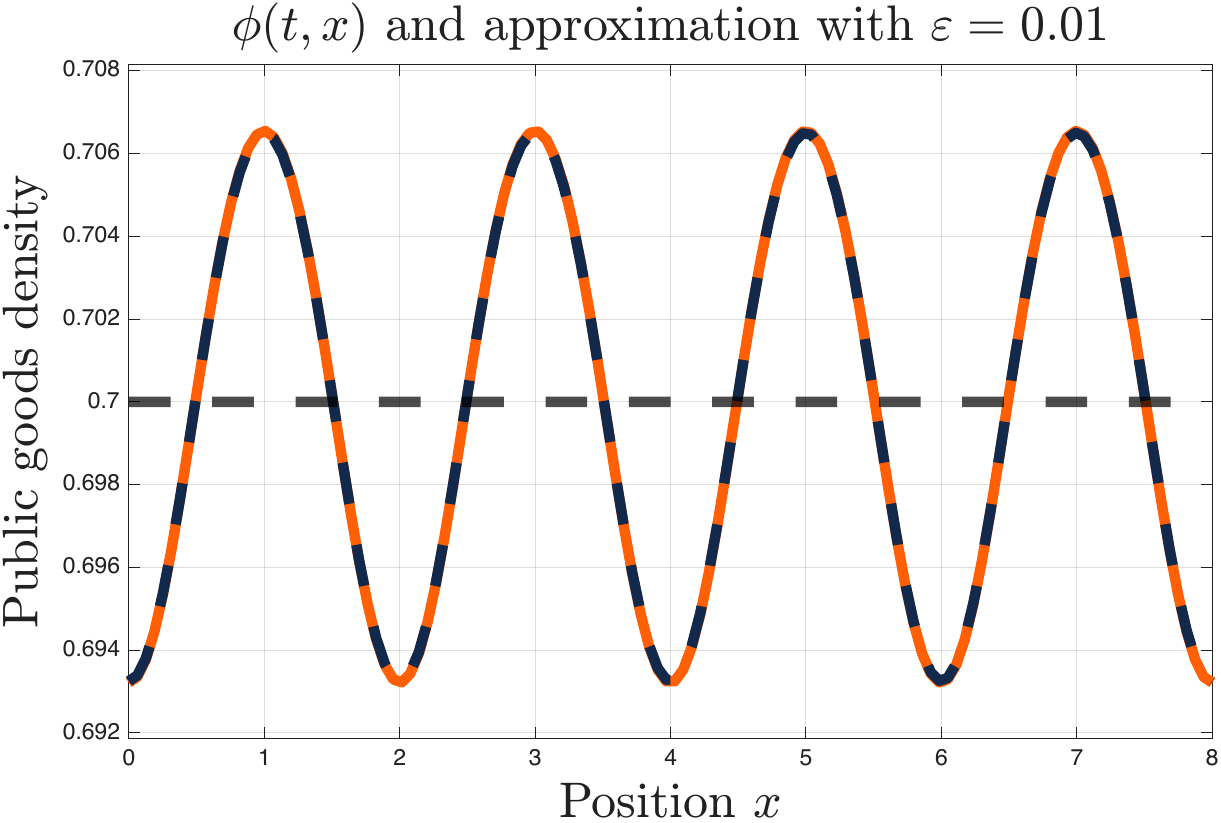}
    \end{subfigure}\vspace{-24pt}
    \begin{subfigure}[b]{\textwidth}
        \centering
        \vspace{2em}
        \caption*{(c) The results when $\zeta= \varepsilon = 0.01$ at $t=50{,}000$}
    \end{subfigure}

    \vspace{1em}

    \caption{Snapshots of $u$, $\phi$ and their approximations at $t=50,000$. Here the orange lines represent the numerical solutions of our unbiased model from Equation \eqref{eq:DM-model} at $t = 50,000$, while the dark blue lines represent the approximations of the solutions obtained by the weakly nonlinear analysis. We set $D_v=D_v^\ast(k^\ast)+\varepsilon^2$ with critical wavenumber $k^\ast=8$ for different values of $\varepsilon$, while using the same values of parameters and the same random initial condition as in Figure~\ref{fig:simulation-unbiased}.}\label{fig:weakly-nonlinear-unbiased}
\end{figure}

We can also provide a comparison between our numerical simulations and the analytical predictions for our patterned steady states that can be deduced from weakly nonlinear analysis. From the weakly nonlinear an analysis, we can consider the approximation provided by the asymptotic expansion of Equation \eqref{eq:weakly-nonlinear-expansions} and the expressions for the solutions at orders $\mathcal{O}(\varepsilon)$ and $\mathcal{O}(\varepsilon^2)$ given by Equations \eqref{eq:linearsolutionRD} and \eqref{eq:unbiased-second-order}, which tells us that an approximate solution for the unbiased model for small $\varepsilon$ is given by
\begin{equation}\label{eq:weakly-nonlinear-approximations}
    \begin{aligned}
        u(T, x)&=u_0 + \varepsilon q_u A(T)\cos\left(\frac{\pi k^\ast}{L}x\right) + \varepsilon^2A^2(T)\left[t_{0, u} + t_{2, u}\cos\left(\frac{2\pi k^\ast}{L}x\right)\right]+\mathcal{O}(\varepsilon^3),\\
        v(T, x)&=v_0 + \varepsilon q_v A(T)\cos\left(\frac{\pi k^\ast}{L}x\right) + \varepsilon^2A^2(T)\left[t_{0, v} + t_{2, v}\cos\left(\frac{2\pi k^\ast}{L}x\right)\right]+\mathcal{O}(\varepsilon^3),\\
        \phi(T, x)&=\phi_0 + \varepsilon q_\phi A(T)\cos\left(\frac{\pi k^\ast}{L}x\right) + \varepsilon^2A^2(T)\left[t_{0, \phi} + t_{2, \phi}\cos\left(\frac{2\pi k^\ast}{L}x\right)\right]+\mathcal{O}(\varepsilon^3).
    \end{aligned}
\end{equation}
We then look to compare these approximate steady-state profiles with the profiles obtained from numerical simulation at large time $t=50{,}000$ and for several values of the perturbation parameter $\varepsilon$, as shown in Figure~\ref{fig:weakly-nonlinear-unbiased}. We find that, when $\varepsilon$ is relatively large, the prediction from Equation \eqref{eq:weakly-nonlinear-approximations} does not provide a good approximation for the numerical steady states, as even the location and number of peaks in the patterns do not agree. For two smaller values of $\varepsilon$, we see that there is good agreement between the analytical approximation and the numerical solutions, indicating that we can look to the properties of the approximate states to glean qualitative information about the emerging patterns for parameters close to the onset of instability. 

\begin{remark}
In the comparison we provide between the weakly nonlinear prediction, we need to account for the fact that there are two stable equilibria $A_{\pm}$ for the Stuart-Landau equation when the diffusivity $D_v$ of defectors is sufficiently large. In particular, we need to choose which of our two equilibrium values $A_{\pm}$ we should use to compare the approximate solutions from Equation \eqref{eq:weakly-nonlinear-approximations} with the long-time outcome achieved under the numerical simulations. To generate Figure \ref{fig:weakly-nonlinear-unbiased}, we perform this approach manually by checking whether $A_{+}$ or $A_{-}$ appears to more closely match the numerically computed spatial profiles, and this choice between the positive and negative amplitude cases allows us to distinguish between the possibility of a spatial pattern with increased or decreased population and public goods near the spatial boundaries relative to the level expected in the spatially uniform state. We can also look to explore whether the steady-state amplitude selected by the numerical dynamics can be predicted from the initial condition provided through a small random perturbation of the uniform state, examining whether the long-time steady state can be determined from a numerical measurement of the amplitude of the initial profiles of strategies and public goods. 
\end{remark}

\begin{figure}[!ht]
    \centering
    \includegraphics[width=0.55\textwidth]{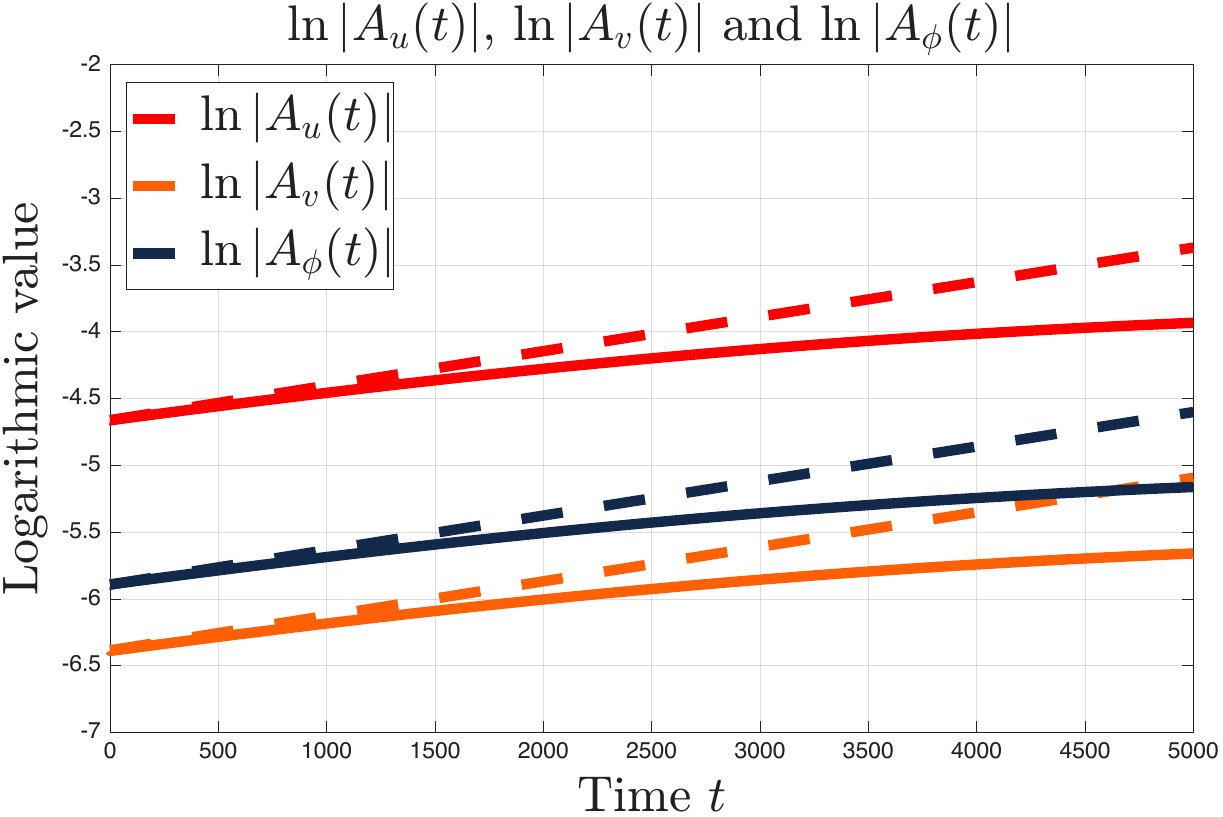}
    \caption{Logarithmic projections of the three variables $u$, $v$, and $\phi$ onto the critical cosine mode with wavenumber $k^\ast$ as in \eqref{eq:projections-on-cos}. The solid colored curves represent the numerical amplitudes obtained from simulations, while the dashed lines indicate exponential growth with slope $\lambda(k^\ast)$, as predicted by linear stability analysis. The close agreement in the early-time regime confirms that the numerical dynamics follow the theoretical growth rate before nonlinear saturation sets in.}
    \label{fig:projection-unbiased}
\end{figure}

We can also explore the time-dependent behavior of solutions, examining how solutions start from the perturbations of the uniform state up, start to grow exponentially with a spatial profile following the sinusoidal shape predicted by the most unstable wavenumber, and then settle upon a spatially heterogeneous equilibrium as described by the weakly nonlinear approximation. To quantify this initial exponential growth phase that occurs in proportion to the sinusoidal profile, we project each component of the solution onto the cosine shape with the critical wavenumber $k^*$ required for the onset of Turing instability. Specifically, we numerically compute the amplitudes $A_u(t)$, $A_v(t)$, and $A_{\phi}(t)$ for the spatial profiles of cooperators, defectors, and the public good by calculating the following deviations of the numerical solutions for $u(t,x)$, $v(t,x)$, and $\phi_t(x)$ from the spatially uniform state $u_0$, $v_0$, and $\phi_0$:
\begin{equation}\label{eq:projections-on-cos}
    \begin{aligned}
        A_u(t)&=\frac{2}{L}\int_{0}^{L}\left(u(t,x)-u_0\right)\cos\left(\frac{\pi k^\ast}{L}x\right)\,dx\\
        A_v(t)&=\frac{2}{L}\int_{0}^{L}\left(v(t,x)-v_0\right)\cos\left(\frac{\pi k^\ast}{L}x\right)\,dx\\
        A_\phi(t)&=\frac{2}{L}\int_{0}^{L}\left(\phi(t,x)-\phi_0\right)\cos\left(\frac{\pi k^\ast}{L}x\right)\,dx.
    \end{aligned}
\end{equation}
To illustrate the initial exponential growth of these cosine profiles, we plot the natural logarithms of these numerical amplitudes $\ln|A_u(t)|$, $\ln|A_v(t)|$, and $\ln|A_\phi(t)|$ as a function of time and compare them to the exponential growth rate predicted from the linear stability analysis from Section \ref{RD:Linearstability}. We see that the numerical curves initially match closely with the predicted exponential growth close to the uniform steady state, but that the growth of the numerical amplitudes eventually saturates as the profiles of individuals and the public good reach the steady states observed in Figure \ref{fig:projection-unbiased}. 

\subsection{Simulations of the Biased Model}\label{section 4.2}

\begin{figure}[htbp]
    \centering
    
    \begin{subfigure}{0.45\linewidth}
        \centering
        \includegraphics[width=\linewidth]{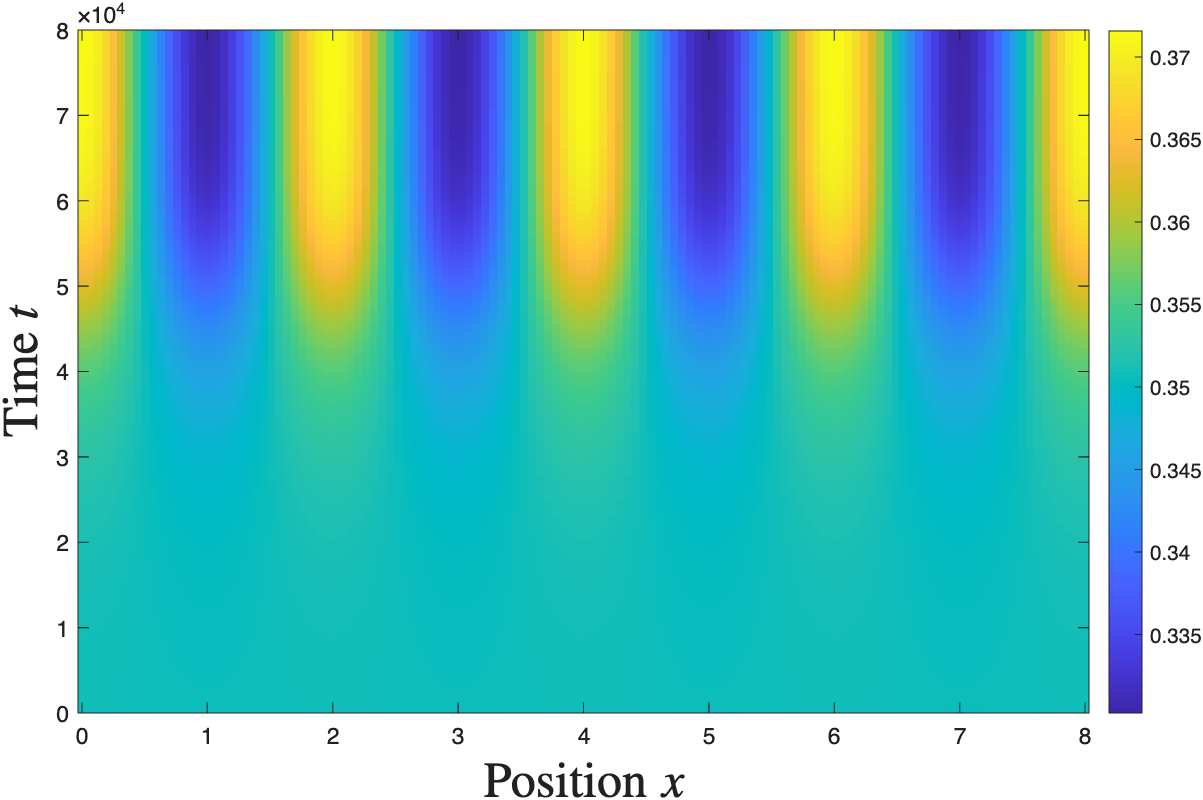}
        \caption{Heat map of $u(t,x)$}
    \end{subfigure}
    \begin{subfigure}{0.45\linewidth}
        \centering
        \includegraphics[width=\linewidth]{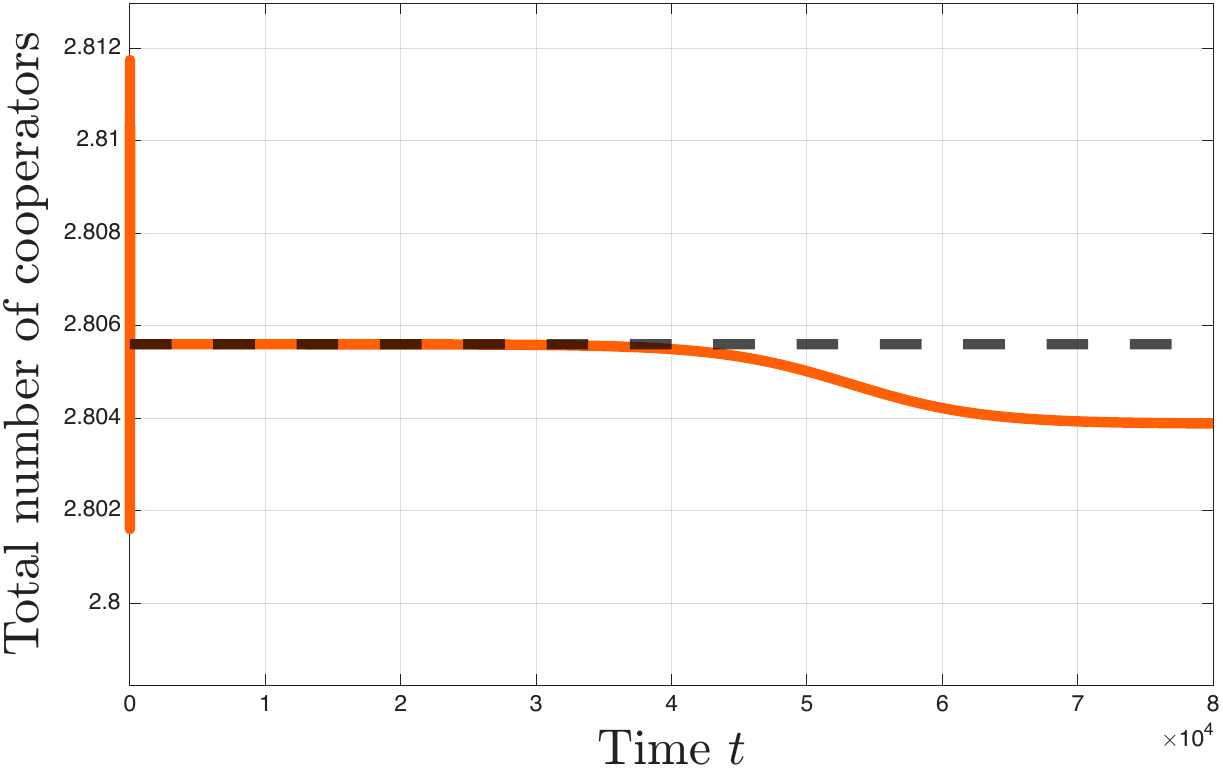}
        \caption{Total cooperators $\int_0^L u(t,x)\,\mathrm{d}x$}
    \end{subfigure}

    \vspace{0.5em}
    
    \begin{subfigure}{0.45\linewidth}
        \centering
        \includegraphics[width=\linewidth]{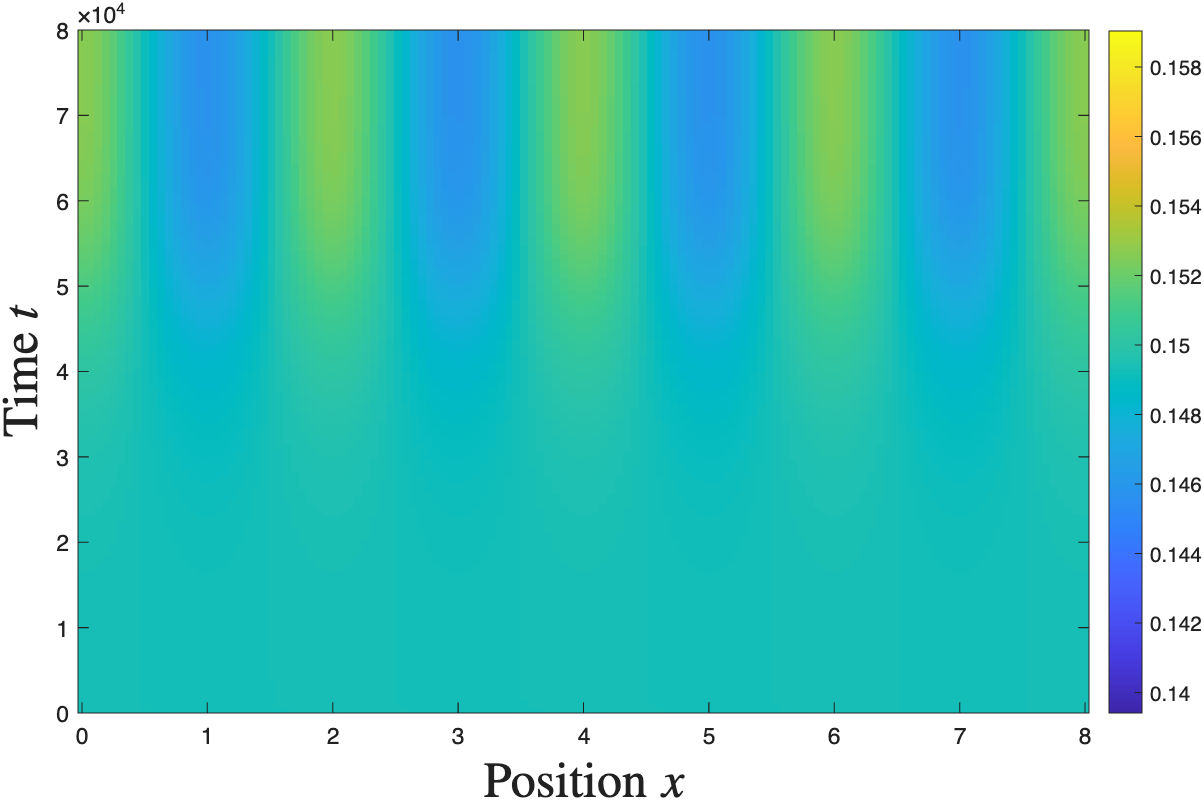}
        \caption{Heat map of $v(t,x)$}
    \end{subfigure}
       \begin{subfigure}{0.45\linewidth}
        \centering
        \includegraphics[width=\linewidth]{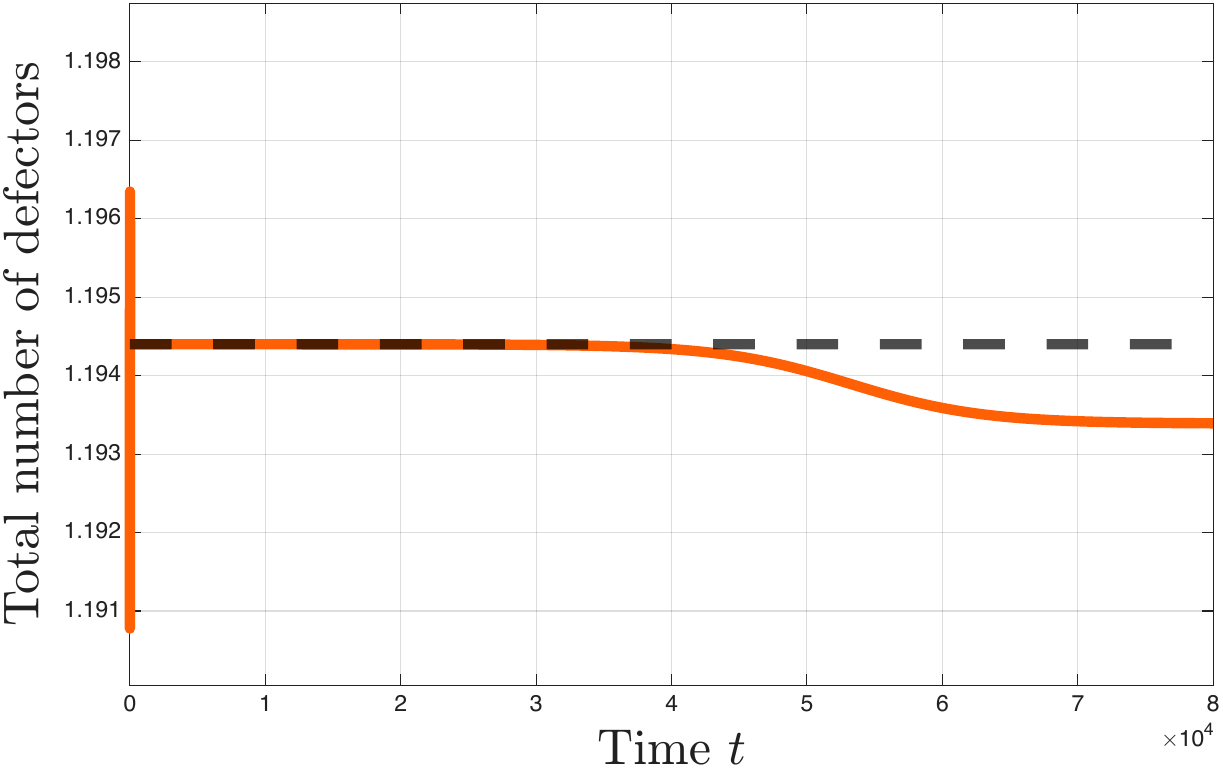}
        \caption{Total defectors $\int_0^L v(t,x)\,\mathrm{d}x$}
    \end{subfigure}
    
    \vspace{0.5em}

    \begin{subfigure}{0.45\linewidth}
        \centering
        \includegraphics[width=\linewidth]{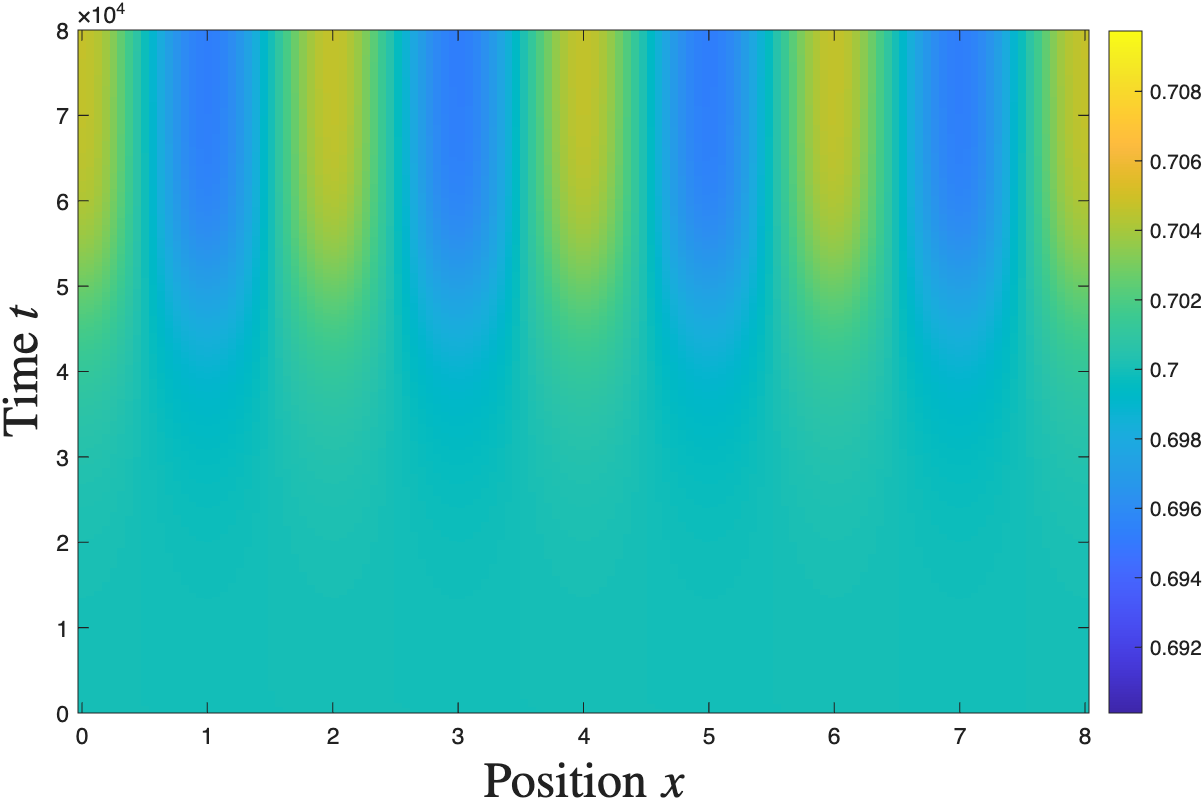}
        \caption{Heat map of $\phi(t,x)$}
        \end{subfigure}
         \begin{subfigure}{0.45\linewidth}
        \centering
        \includegraphics[width=\linewidth]{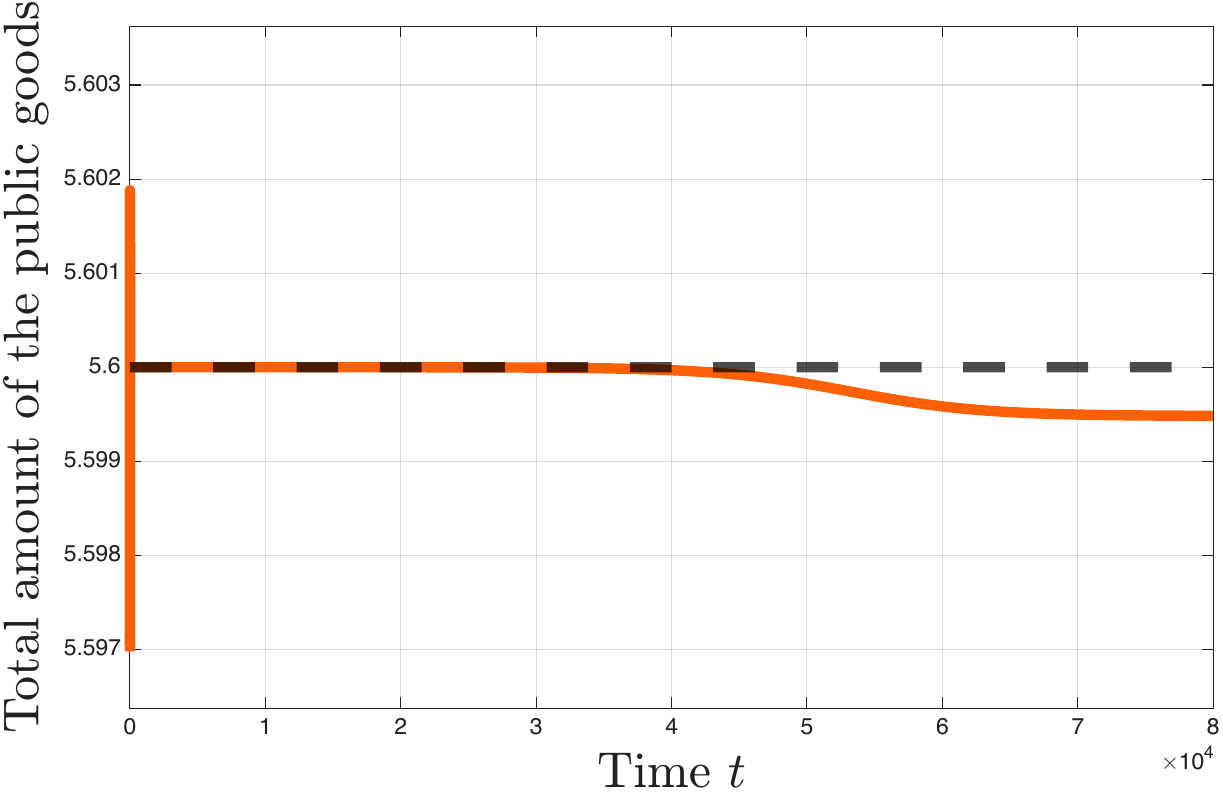}
        \caption{Total public good $\int_0^L \phi(t,x)\,\mathrm{d}x$}
    \end{subfigure}

    \caption{Numerical simulation results of the biased model on the one-dimensional domain $[0,L]$ with $L=8$. The parameters are $w_u=w_u^\ast(k^\ast)+\varepsilon^2$ with critical wavenumber $k^\ast=8$, $\varepsilon=0.01$, $\zeta=0.01$, $w_v=1$, $D_u=D_v=D_\phi=0.03$, and the ODE parameters are taken from Table~\ref{table 1}. Initial conditions are random. In the left panels, we present heatmaps of the spatial profiles of cooperators $u(t,x)$, defectors $v(t,x)$, and the public good $\phi(t,x)$, with the horizontal axes describing the spatial interval $[0,L]$ and the vertical axes indicating time increasing from bottom to top. The right panels show the total mass of cooperators $\int_0^L u(t,x) dx$, defectors $\int_0^L v(t,x) dx$, and the public good $\int_0^L \phi(t,x) dx$ across the spatial domain, each plotted as a function of time $t$.The horizontal dashed lines in the right-hand panels represent the total populations of cooperators and defectors and the total public-good concentration in the spatially uniform coexistence state, which are given by $Lu_0$, $Lv_0$, and $L\phi_0$, respectively.}
    \label{fig:simulation-biased}
\end{figure}

We now present the results for similar numerical simulations for our biased model, exploring the spatial patterns that emerge when cooperators have sufficiently strong sensitivity $w_u$ to gradients in the concentration of public goods. We present heatmaps of the densities of cooperators $u(t,x)$, defectors $v(t,x)$, and $\phi(t,x)$ change in time in the left panels of Figure \ref{fig:simulation-biased}, showing how four clusters of cooperators, defectors, and the public good emerge from an initial condition featuring a near uniform spread of individuals. These simulations were run for the parameters used for our linear stability analysis introduced in Figure \ref{fig:thresholds-biased}, with the four emerging clusters across the domain $[0,L]$ agreeing with the analytical prediction of a dominant wavenumber of $k^* = 8$. In the right panels of Figure \ref{fig:simulation-biased}, we also present the temporal dynamics for the total concentrations of the cooperators and defectors, and public good across the spatial domain as represented by the integrals $\int_0^L u(t,x)\,dx$, $\int_0^L v(t,x)\,dx$, and $\int_0^L \phi(t,x)\,dx$. We see that all three total concentrations decrease in time and settle to steady state values below the values that are achieved in the spatially uniform coexistence state, indicating that increased directed motion of cooperators towards regions with more abundant public goods can actually diminish overall population size and the overall public goods provision across the spatial domain. 

\begin{figure}[htbp]
    \centering

    \begin{subfigure}[b]{0.48\textwidth}
        \includegraphics[width=\linewidth]{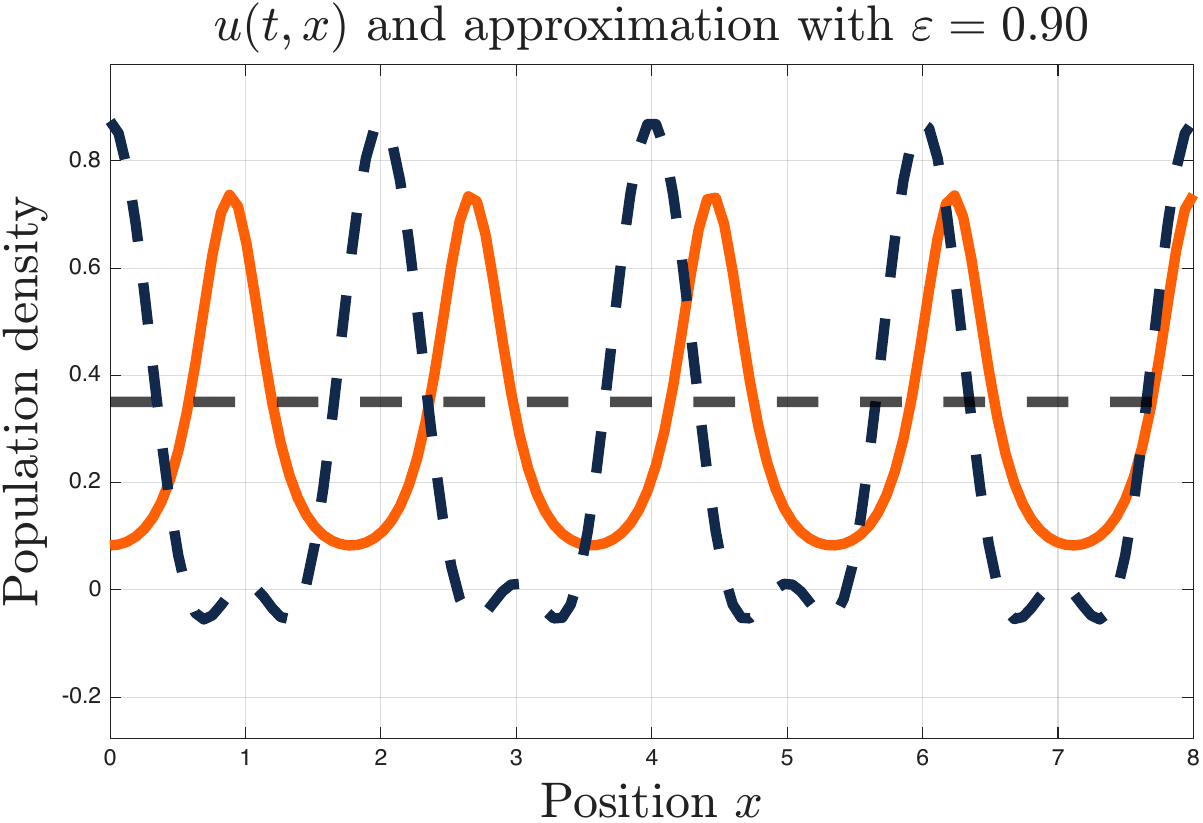}
    \end{subfigure}
    \hfill
    \begin{subfigure}[b]{0.48\textwidth}
        \includegraphics[width=\linewidth]{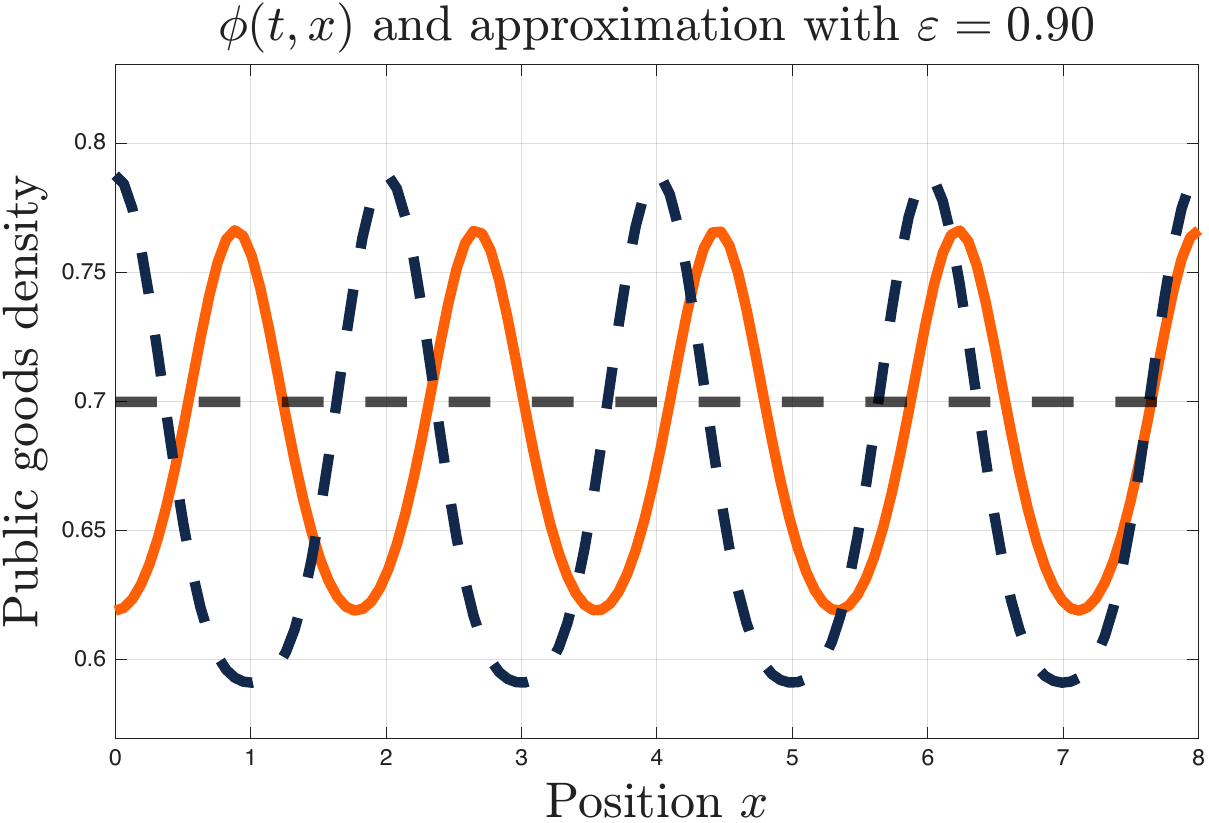}
    \end{subfigure}\vspace{-24pt}
    \begin{subfigure}[b]{\textwidth}
        \centering
        \vspace{2em}
        \caption*{(a) The result when $\zeta= \varepsilon=0.9$ at $t=80{,}000$}
    \end{subfigure}

    \vspace{0.2em}
    
    \begin{subfigure}[b]{0.48\textwidth}
        \includegraphics[width=\linewidth]{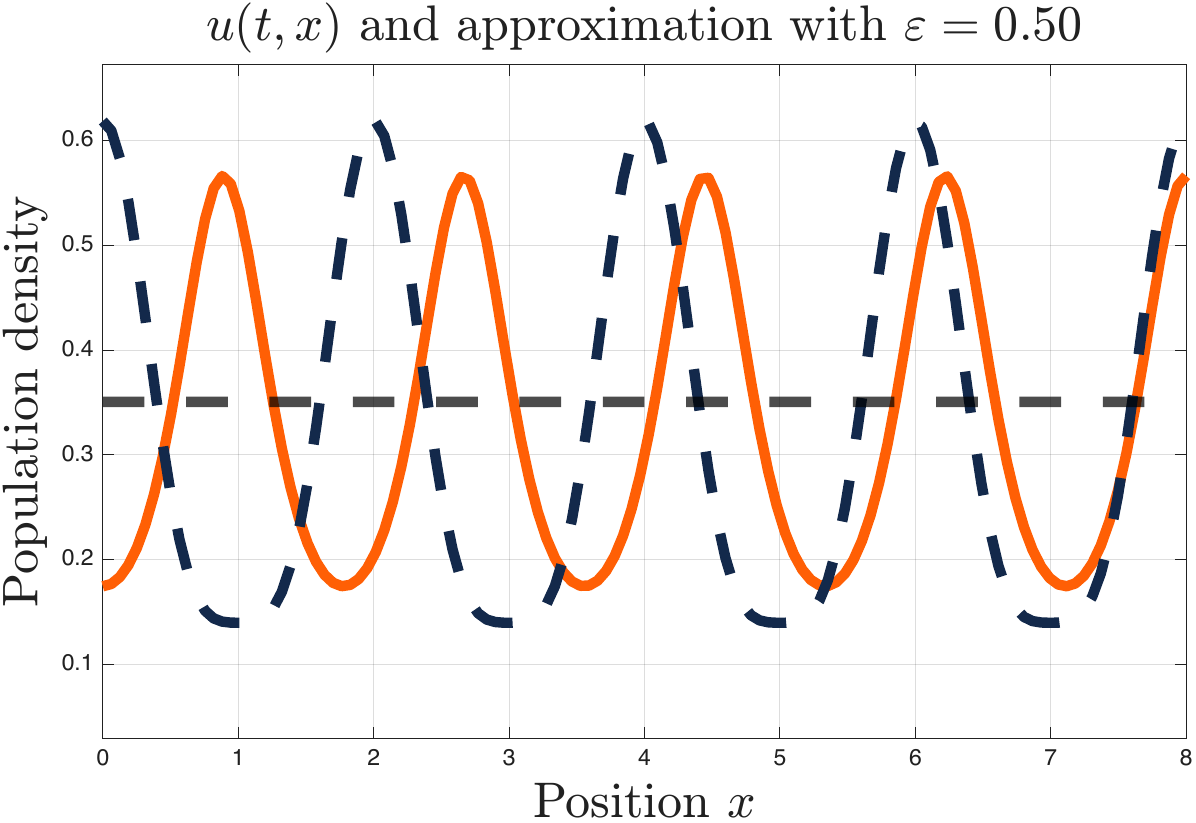}
    \end{subfigure}
    \hfill
    \begin{subfigure}[b]{0.48\textwidth}
        \includegraphics[width=\linewidth]{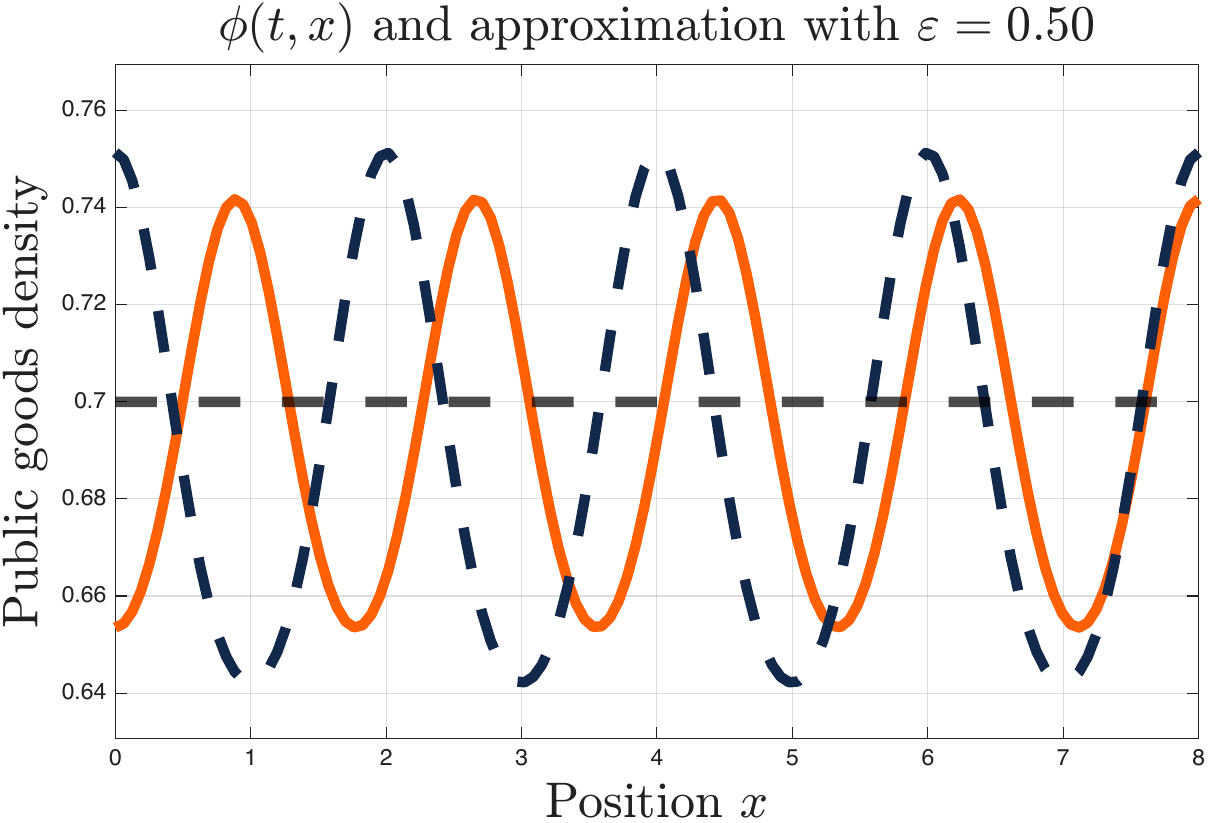}
    \end{subfigure}\vspace{-24pt}
    \begin{subfigure}[b]{\textwidth}
        \centering
        \vspace{1.8em}
        \caption*{(b) The result when $\zeta= \varepsilon=0.5$ at $t=80{,}000$}
    \end{subfigure}

    \vspace{0.2em}

    \begin{subfigure}[b]{0.48\textwidth}
        \includegraphics[width=\linewidth]{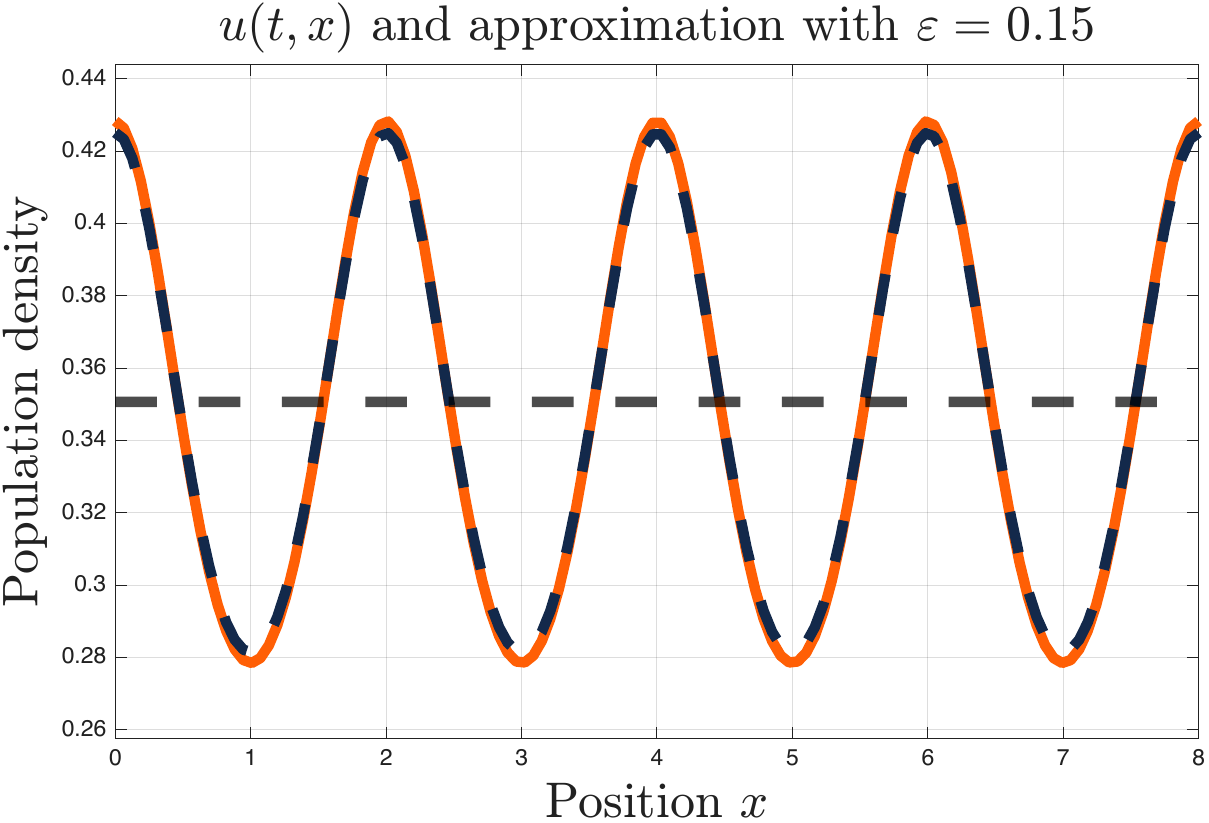}
    \end{subfigure}
    \hfill
    \begin{subfigure}[b]{0.48\textwidth}
        \includegraphics[width=\linewidth]{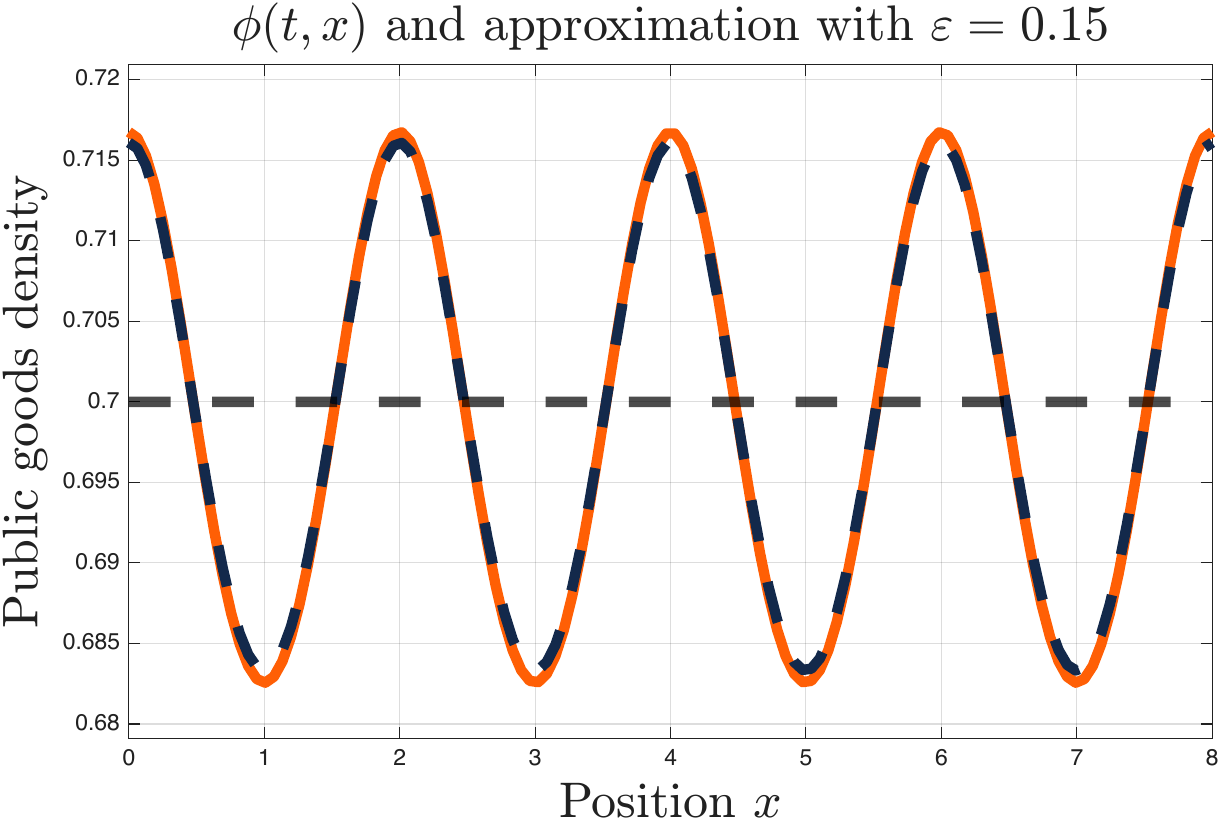}
    \end{subfigure}\vspace{-24pt}
    \begin{subfigure}[b]{\textwidth}
        \centering
        \vspace{1.8em}
        \caption*{(c) The result when $\zeta= \varepsilon=0.15$ at $t=80{,}000$}
    \end{subfigure}

    \vspace{0em}

    \caption{Snapshots of the numerically calculated spatial profiles in the biased model of cooperators $u(t,x)$ and the the public good concentration $\phi(t,x)$ and their approximations at $t=80{,}000$. Here the solid orange curves represent solutions of our model \eqref{eq:DM-model}, the dashed dark blue curves represent the approximation for the steady-state solution obtained by the weakly nonlinear analysis up to order $\mathcal{O}(\varepsilon^2)$, and the horizontal black dashed lines represent the values of the size of the cooperator population $u_0$ (left) and the concentration of public goods $\phi_0$ (right) at the spatially uniform coexistence equilibrium. We set $w_u=w_u^\ast(k^\ast)+\varepsilon^2$ with critical wavenumber $k^\ast=8$ for different values of $\varepsilon$, while use the same values of parameters and the same random initial condition as in Figure~\ref{fig:simulation-biased}.}\label{fig:weakly-nonlinear-biased}
\end{figure}
%\FloatBarrier

\subsection{Effects of Pattern-Forming Mechanisms on Cooperation and Public Goods Provision}\label{section 4.3}

We have seen so far that the simulations of the two models suggest that individuals performing biased or unbiased directed motion may lead to different qualitative behaviors in the long-time support for cooperation and public good provision. We now look to explore these differences in a bit more detail, making use of the weakly nonlinear expansions and numerical simulations to determine how rules for spatial motion can impact the long-time success of the population and help determine spatial patterns that emerge in parameter regimes further from the onset of pattern formation. 

While weakly nonlinear stability analysis is typically used to determine the form of pattern-forming bifurcation and to validate numerical simulations of models close to the instability threshold, it is also possible to glean additional qualitative information about emergent patterns based on the approximate equilibrium solutions achieved through the weakly nonlinear expansion. In particular, we can integrate our expressions from Equation \eqref{eq:weakly-nonlinear-approximations} for the weakly nonlinear expansions $u(T,x)$, $v(T,x)$, and $\phi(T,x)$  across the interval $[0,L]$ to see that the total mass of cooperators, defectors, and public good is given by
\begin{subequations} \label{eq:solution-masses}
\begin{align}
\int_0^L u(T,x) dx &= L u_0 + \varepsilon^2 L A(T)^2 t_{0,u} + \mathcal{O}\left( \varepsilon^3\right) \\
\int_0^L v(T,x) dx &= L v_0 + \varepsilon^2 L A(T)^2 t_{0,v} + \mathcal{O}\left( \varepsilon^3\right) \\
\int_0^L \phi(T,x) dx &= L \phi_0 + \varepsilon^2 L A(T)^2 t_{0,\phi} + \mathcal{O}\left( \varepsilon^3\right),
\end{align}
\end{subequations}
where we observe that the integrals of the sinusoidal terms in Equation \eqref{eq:weakly-nonlinear-approximations} all vanish. We therefore see that the signs of the terms $t_{0,u}$, $t_{0,v}$, and $t_{0,\phi}$ will determine whether the size of the cooperator and defector populations and the total concentration of public goods exceeds the masses $Lu_0$, $Lv_0$, and $L \phi_0$ achieved in the uniform state.

\begin{figure}[!htbp]
    \centering
    \begin{subfigure}[b]{0.48\textwidth}
        \includegraphics[width=\linewidth]{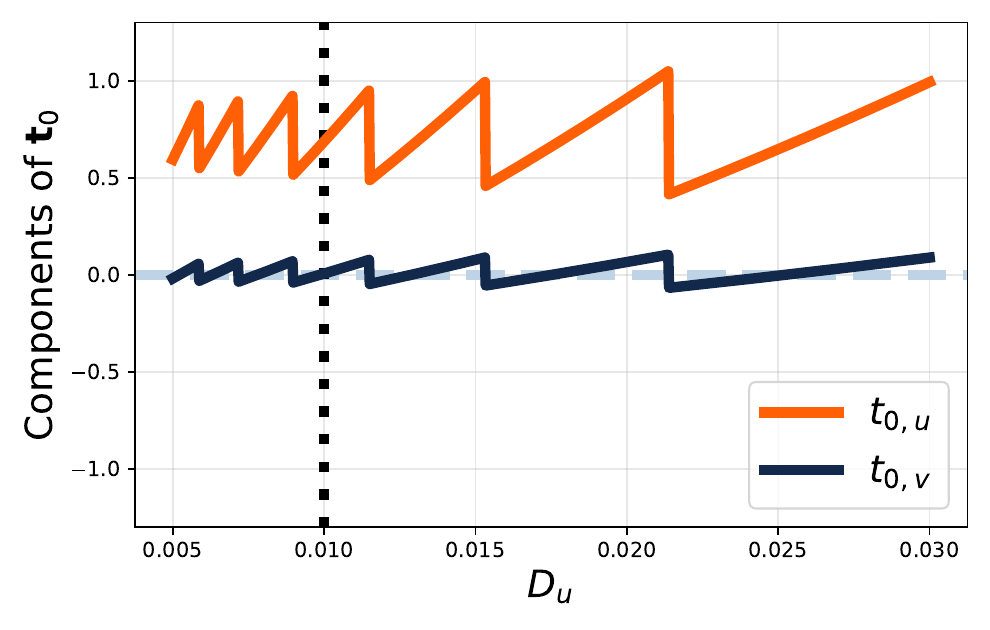}
        \caption*{(a1) $t_{0,u}$ and $t_{0,v}$ in the unbiased model}
    \end{subfigure}
    \hfill
    \begin{subfigure}[b]{0.48\textwidth}
        \includegraphics[width=\linewidth]{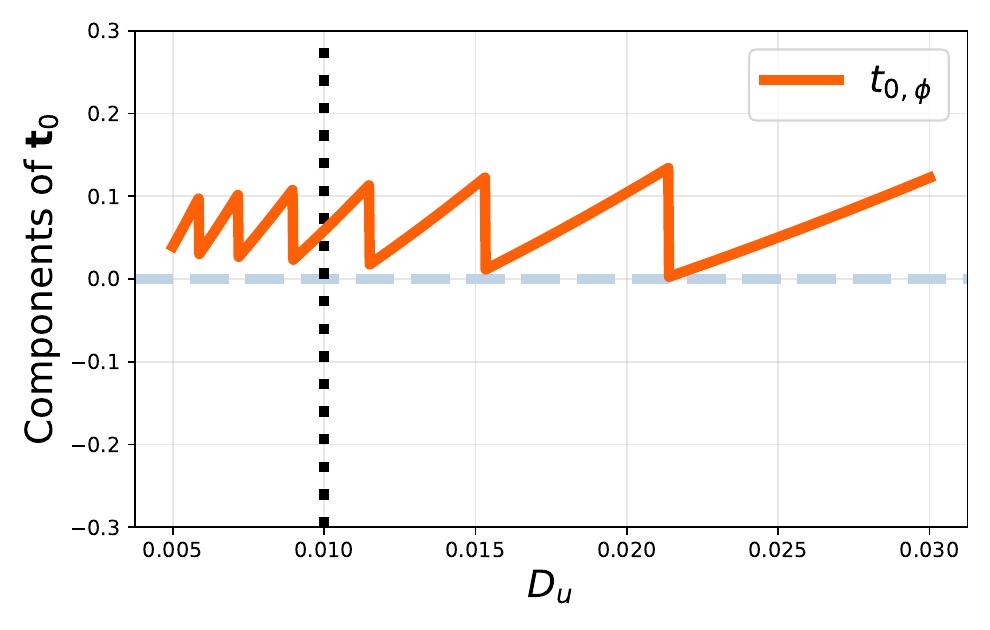}
        \caption*{(a2) $t_{0,\phi}$ in the unbiased model}
    \end{subfigure}\vspace{-12pt}
    \begin{subfigure}[b]{\textwidth}
        \centering
        \vspace{1.6em}
        \caption*{(a) The values of $t_{0,u}$, $t_{0,v}$, and $t_{0,\phi}$ in the unbiased model}
    \end{subfigure}

    \vspace{1em}

    \begin{subfigure}[b]{0.48\textwidth}
        \includegraphics[width=\linewidth]{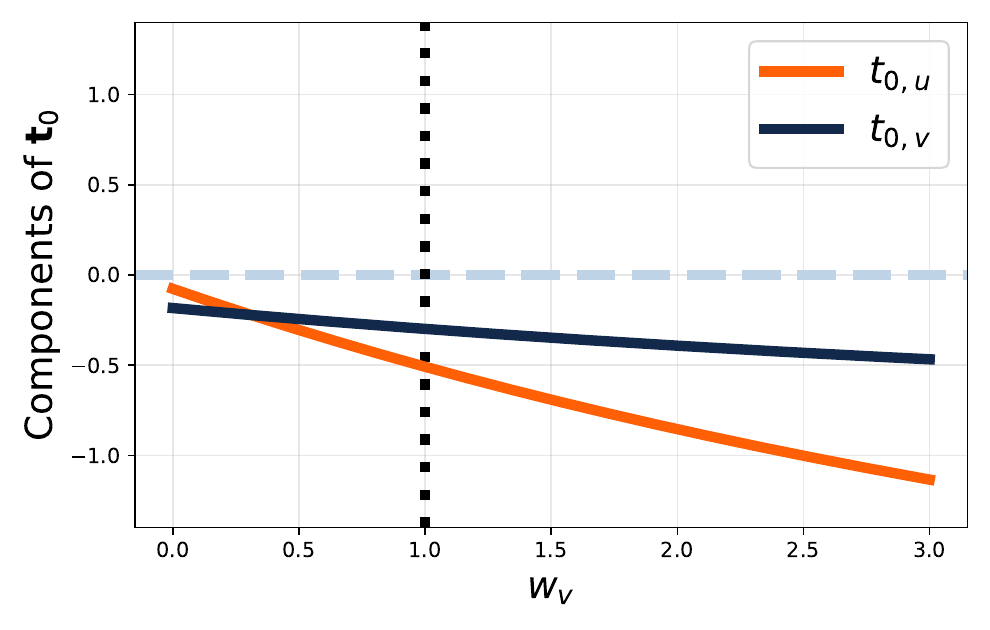}
        \caption*{(b1) $t_{0,u}$ and $t_{0,v}$ in the biased model}
    \end{subfigure}
    \hfill
    \begin{subfigure}[b]{0.48\textwidth}
        \includegraphics[width=\linewidth]{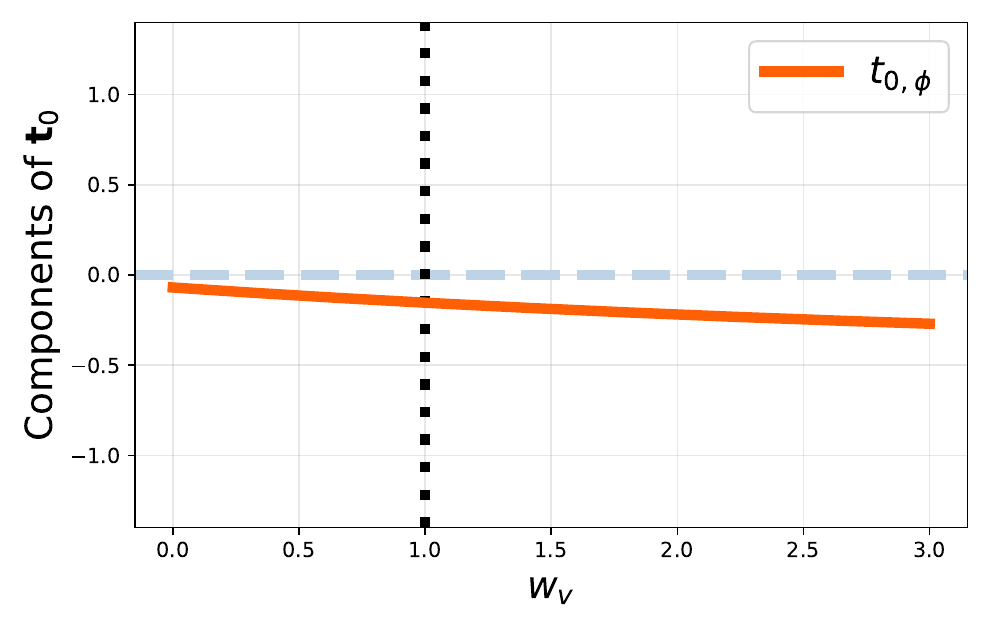}
        \caption*{(b2) $t_{0,\phi}$ in the biased model}
    \end{subfigure}\vspace{-12pt}
    \begin{subfigure}[b]{\textwidth}
        \centering
        \vspace{2em}
        \caption*{(b) The values of $t_{0,u}$, $t_{0,v}$, and $t_{0,\phi}$ in the biased model}
    \end{subfigure}

    \caption{Calculation of the components of the vector $\mathbf{t}_0$ in the $\mathcal{O}(\varepsilon^2)$ term of the solutions of the weakly nonlinear analysis for both the unbiased and biased models. We plot these quantities as functions of the bifurcation parameter ($D_v$ for the unbiased model and $w_u$ for the biased model). For each value of $D_u$ or $w_v$, we compute the linear instability threshold $D_v^\ast$ (based on \eqref{eq:D_v-threshold-min}) and $w_u^\ast$ (based on \eqref{eq:w_u-threshold-min}), and then evaluate the components of $\mathbf{t}_0$ using \eqref{eq:unbiased-t} for the unbiased model and \eqref{eq:biased-t} for the biased model. The horizontal dashed lines represent $0$, while the vertical dotted lines indicate the values of $D_u$ or $w_v$ used in the numerical simulations shown in Figures~\ref{fig:simulation-unbiased} and~\ref{fig:simulation-biased} for the unbiased and biased models, respectively. The horizontal axes correspond to $D_u$ in panel (a) and $w_v$ in panel (b), while the vertical axes show the values of the components of $\mathbf{t}_0$. The parameter values are the same as those used in Figure~\ref{figure 3} for panel (a) and Figure~\ref{figure 5} for panel (b).}\label{fig:t-0-components}
\end{figure}

We can then use the expressions for $t_{0,u}$, $t_{0,v}$, and $t_{0,\phi}$ we found in Section \ref{sec:weakly-nonlinear-analysis} to explore when cooperation and public goods provision can improve in the presence of diffusive or directed motion rules. We provide a comparison of these values in Figure~\ref{fig:t-0-components} for our two models, considering how these constants depend on the diffusivity of cooperators $D_u$ in unbiased model and on the sensitivity $w_v$ of defectors towards climbing gradients in public good concentration in the biased model. For the unbiased model, we see that the weakly nonlinear analysis predicts an increase in the total mass of cooperators, defectors, and the public good across the spatial domain relative to the spatially uniform state for the parameters used in Figure~\ref{fig:simulation-unbiased}, while we see that the mass of defectors can drop below that of the uniform equilibrium for different values of cooperator diffusivity. For the biased model, we see that the masses of cooperators, defectors, and the public good are all lower than the mass predicted for the spatial state for the case of the parameters in Figure \ref{fig:simulation-biased} and for all defector movement sensitivities $w_v$ considered in Figure \ref{fig:t-0-components}. 

\begin{figure}[htbp]
    \centering
    \begin{subfigure}{0.48\linewidth}
        \centering
        \includegraphics[width=\linewidth]{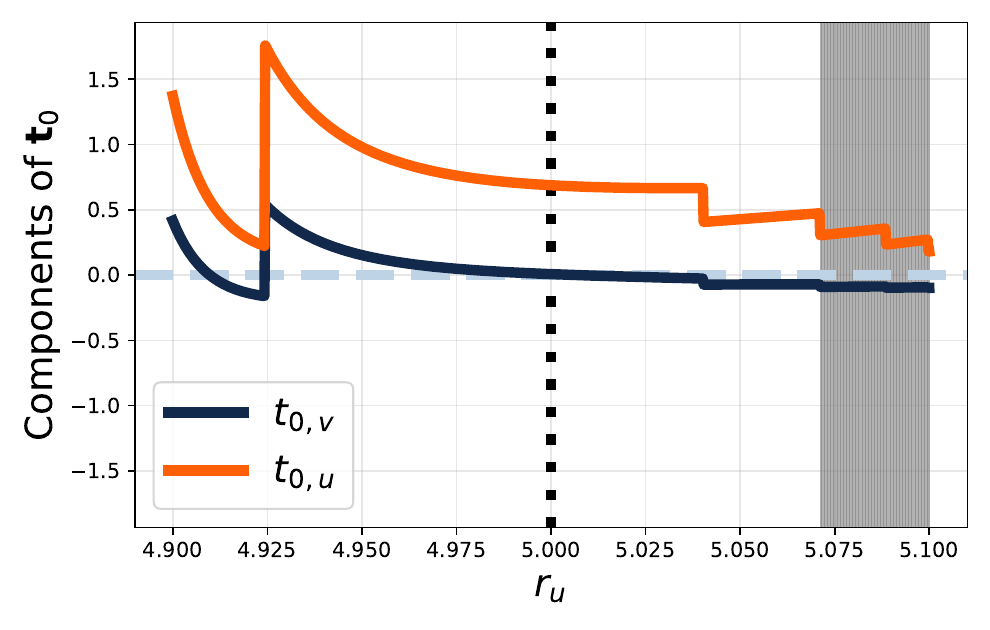}
        \caption{Unbiased model: $t_{0,u}$ and $t_{0,v}$.}
    \end{subfigure}
    \hfill
    \begin{subfigure}{0.48\linewidth}
        \centering
        \includegraphics[width=\linewidth]{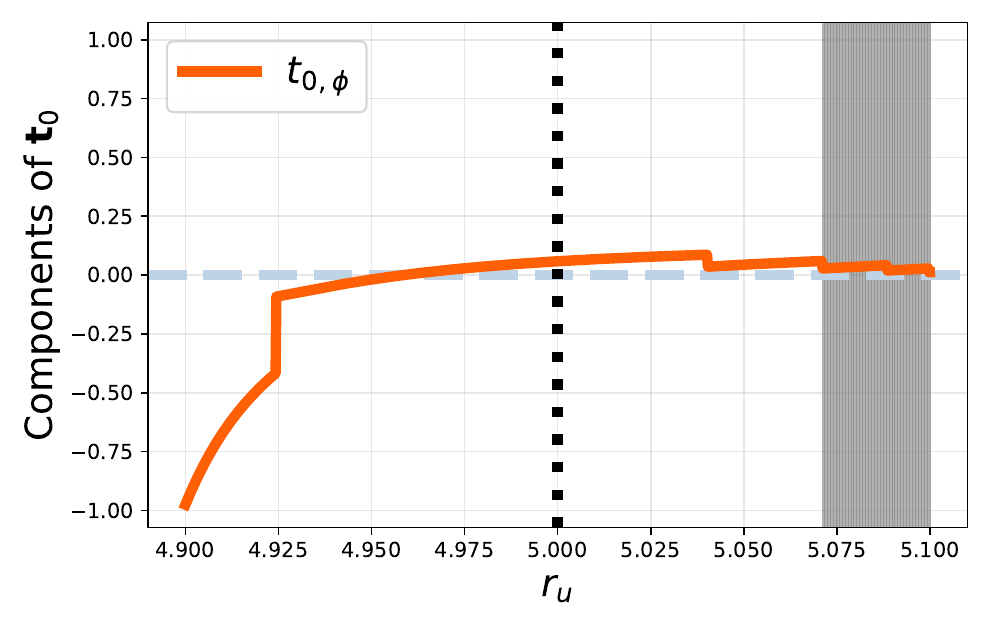}
        \caption{Unbiased model: $t_{0,\phi}$.}
    \end{subfigure}

    \vspace{0.3em}

    \begin{subfigure}{0.48\linewidth}
        \centering
        \includegraphics[width=\linewidth]{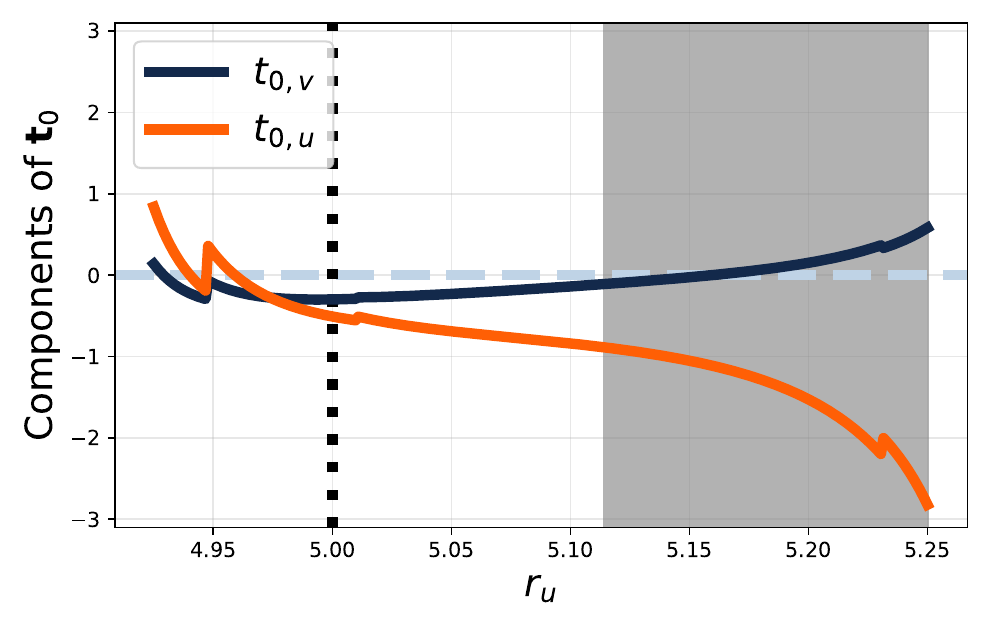}
        \caption{Biased model: $t_{0,u}$ and $t_{0,v}$.}
    \end{subfigure}
    \hfill
    \begin{subfigure}{0.48\linewidth}
        \centering
        \includegraphics[width=\linewidth]{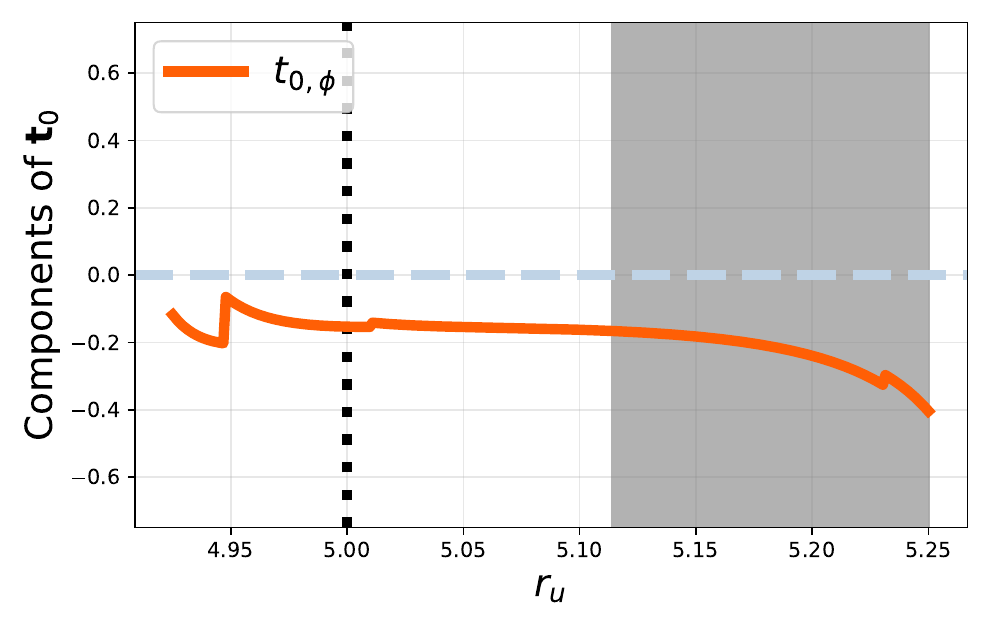}
        \caption{Biased model: $t_{0,\phi}$.}
    \end{subfigure}

    \caption{Calculation of the components of the vector $\mathbf{t}_0$ in the $\mathcal{O}(\varepsilon^2)$ component of the solutions of the weakly nonlinear analysis for both the unbiased and biased models, plotted as a function of the reaction parameters $r_u$. The remaining reaction parameters are taken from Table~\ref{table 1}, and the vertical dashed lines indicate the default values of $r_u$ and $\mu_u$. For the PDE parameters, we choose $D_u=D_\phi=0.01$ and use $D_v$ as the bifurcation parameter in the unbiased model. For the biased model, we choose $D_u=D_v=D_\phi=0.03$, set $w_v=1$, and use $w_u$ as the bifurcation parameter. The horizontal dashed lines indicate when the components $t_{0,u}$, $t_{0,v}$, and $t_{0,\phi}$ cross $0$, separating regimes in which pattern formation works to increase or decrease the total concentration of each species. The regions shaded in gray indicate values of $r_u$ for which we expect subcritical bifurcations.}
    \label{fig:t0-ru}
\end{figure}

\begin{figure}[htbp]
    \centering
    \begin{subfigure}{0.48\linewidth}
        \centering
        \includegraphics[width=\linewidth]{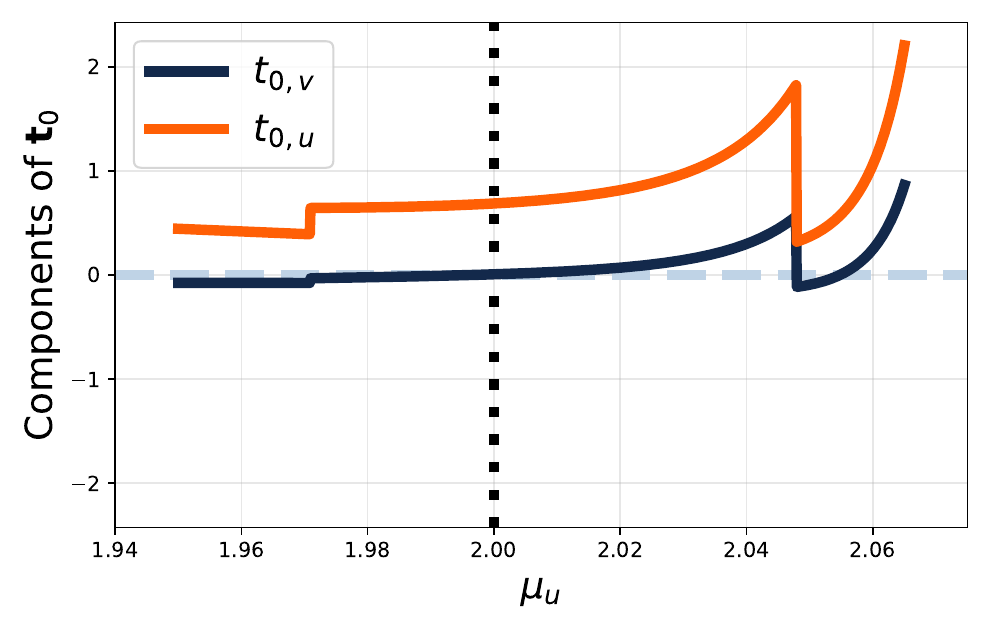}
        \caption{Unbiased model: $t_{0,u}$ and $t_{0,v}$.}
    \end{subfigure}
    \hfill
    \begin{subfigure}{0.48\linewidth}
        \centering
        \includegraphics[width=\linewidth]{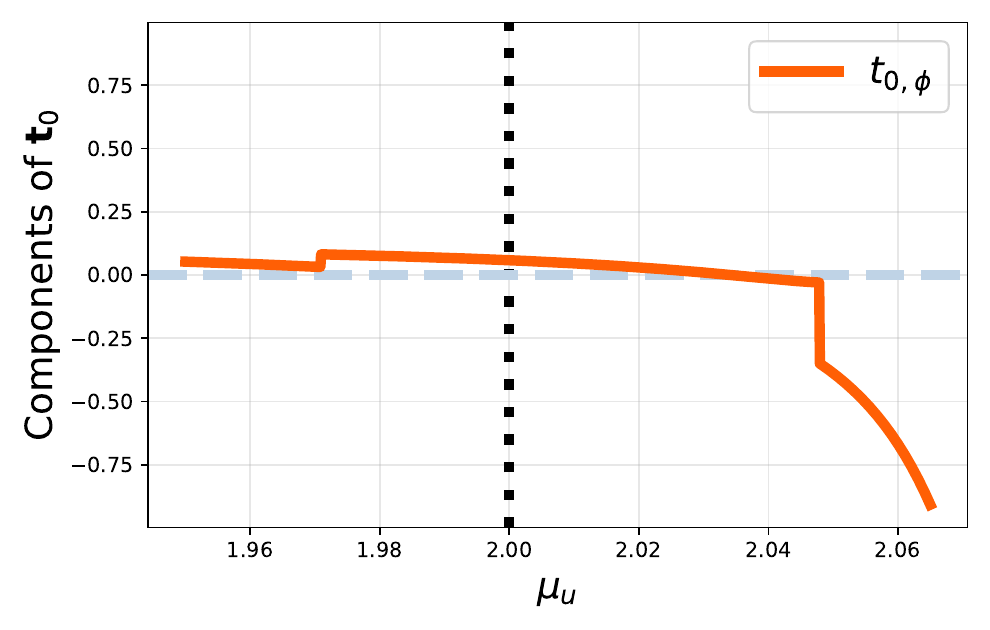}
        \caption{Unbiased model: $t_{0,\phi}$.}
    \end{subfigure}

    \vspace{0.3em}

    \begin{subfigure}{0.48\linewidth}
        \centering
        \includegraphics[width=\linewidth]{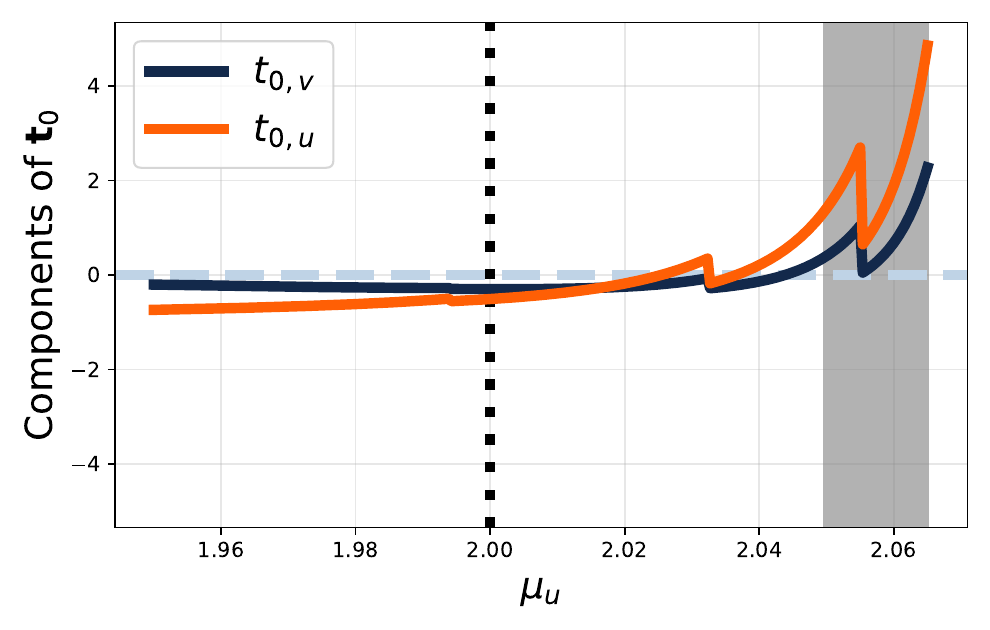}
        \caption{Biased model: $t_{0,u}$ and $t_{0,v}$.}
    \end{subfigure}
    \hfill
    \begin{subfigure}{0.48\linewidth}
        \centering
        \includegraphics[width=\linewidth]{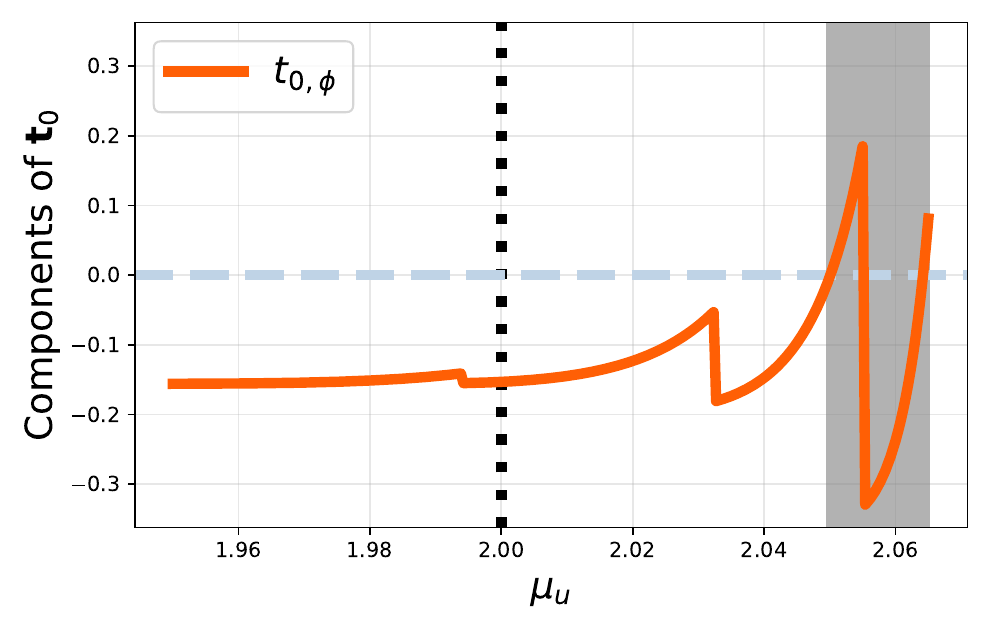}
        \caption{Biased model: $t_{0,\phi}$.}
    \end{subfigure}

    \caption{Calculation of the components of the vector $\mathbf{t}_0$ in the $\mathcal{O}(\varepsilon^2)$ component of the solutions of the weakly nonlinear analysis for both the unbiased and biased models, plotted as a function of the reaction parameters $\mu_u$. The remaining reaction parameters are taken from Table~\ref{table 1}, and the vertical dashed lines indicate the default values of $r_u$ and $\mu_u$. For the PDE parameters, we choose $D_u=D_\phi=0.01$ and use $D_v$ as the bifurcation parameter in the unbiased model. For the biased model, we choose $D_u=D_v=D_\phi=0.03$, set $w_v=1$, and use $w_u$ as the bifurcation parameter. The horizontal dashed lines indicate when the components $t_{0,u}$, $t_{0,v}$, and $t_{0,\phi}$ cross $0$, separating regimes in which pattern formation works to increase or decrease the total concentration of each species. The regions shaded in gray indicate values of $\mu_u$ for which we expect subcritical bifurcations.}
    \label{fig:t0-muu}
\end{figure}

So far, we have seen that, in the weakly nonlinear regime, the total concentration of all three species increase under diffusion-driven instability and decrease under chemotactic instability for the default parameters in Table \ref{table 1}. However, we can also explore how robust these qualitative behaviors are under changes in reaction parameters, and we can use an approach similar to our weakly nonlinear analysis in Section \ref{sec:robustness-WNA} to see how the components of $\mathbf{t}_0$ vary under changes in $r_u$ and $\mu_u$. Using the threshold parameters for the unbiased and biased models, we compute the the components of $\mathbf{t}_0$ based on \eqref{eq:unbiased-t} and \eqref{eq:biased-t} and display the resulting values in Figures~\ref{fig:t0-ru} and \ref{fig:t0-muu}. We see that there is a reasonable large range of $r_u$ and $\mu_u$ for which the components of $\mathbf{t}_0$ have the same sign as in the case of the default parameter values, although the range is larger for the components $t_{0,u}$ and $t_{0,\phi}$ associated with the change in total concentration for the cooperator and public good. We also compare the parameter regions featuring reversal of $\mathbf{t}_0$ from the default values in Table \ref{table 1} and the regions featuring subcritical pattern-forming bifurcations (shaded in gray in Figures \ref{fig:t0-ru} and \ref{fig:t0-muu}), and we see that supercritical bifurcation is neither a necessary nor sufficient condition for seeing a change in the signs of the total concentration levels for each species.

\begin{remark}
A particularly notable feature we observe in Figures~\ref{fig:t0-ru} and \ref{fig:t0-muu} is that while the components of $\mathbf{t}_0$ have the same sign for the default parameters, we can see that there are many parameter regimes in which the sign of $t_{0,u}$ disagrees with either $t_{0,v}$ or $t_{0,\phi}$. The disagreement between the sign for change in total concentration of cooperators and of defectors in the patterned state is consistent with our finding that the coefficients $q_u$ and $q_v$ for cooperators and defectors in our solution for the $\mathcal{O}(\varepsilon)$ problem, suggesting that the spatial anti-correlation of cooperators and defectors could correspond to a difference in how spatial patterns benefit the two types. However, the disagreement between $t_{0,u}$ and $t_{0,\phi}$ may be a bit more surprising, as we show in Section \ref{sec:patterncorrelation} that the spatial profiles of cooperators and defectors will be correlated due to the fact that the coefficients $q_u$ and $q_{\phi}$ for the solution to the linear problem are guaranteed to have the same sign. This means that, even though our mechanisms of diffusion-driven and chemotactic instable produce patterns that collocate the concentration of cooperators and public goods, the resulting patterns may still lead to a different prediction on the total concentration of cooperators and public goods across the spatial domain.
\end{remark}

\begin{figure}[!htbp]
    \centering
    \begin{subfigure}[b]{0.48\textwidth}
        \includegraphics[width=\linewidth]{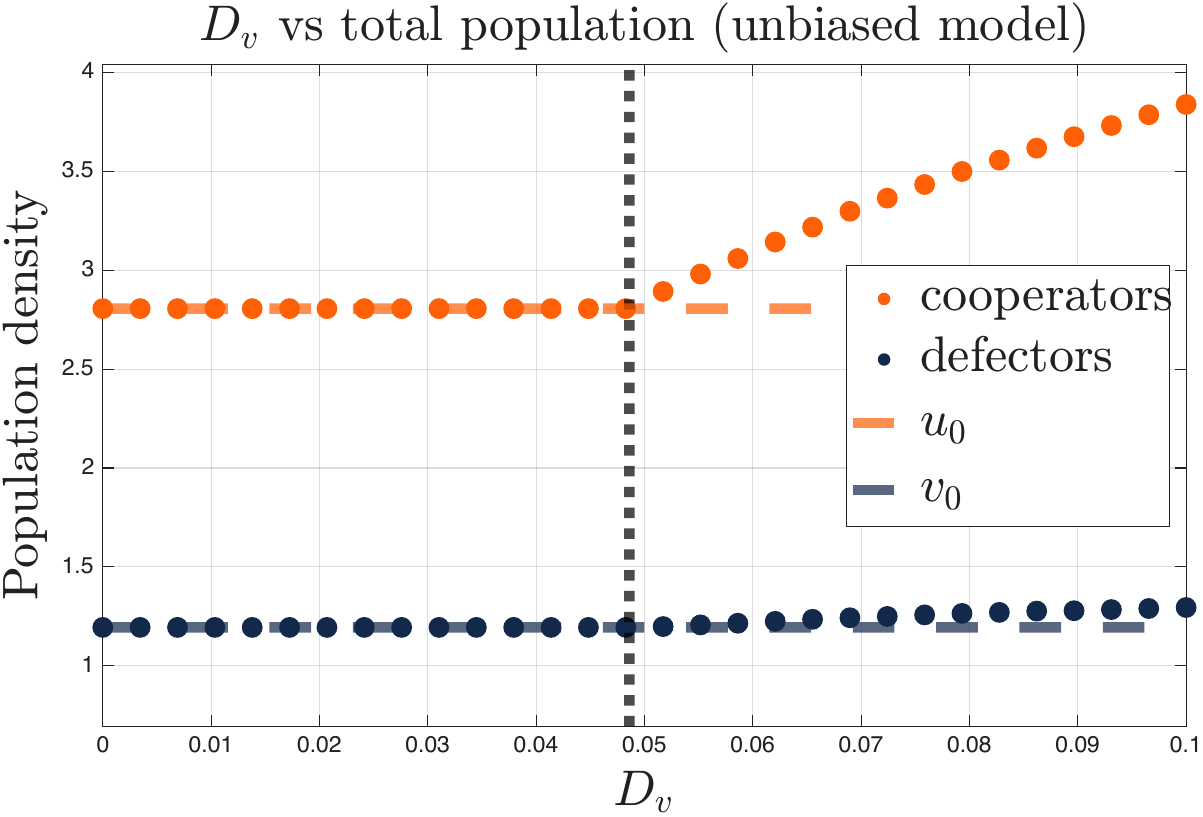}
        \caption*{(a1) Total population $\int_{0}^{L}u(t,x)\,dx$ and $\int_{0}^{L}v(t,x)\,dx$}
    \end{subfigure}
    \hfill
    \begin{subfigure}[b]{0.48\textwidth}
        \includegraphics[width=\linewidth]{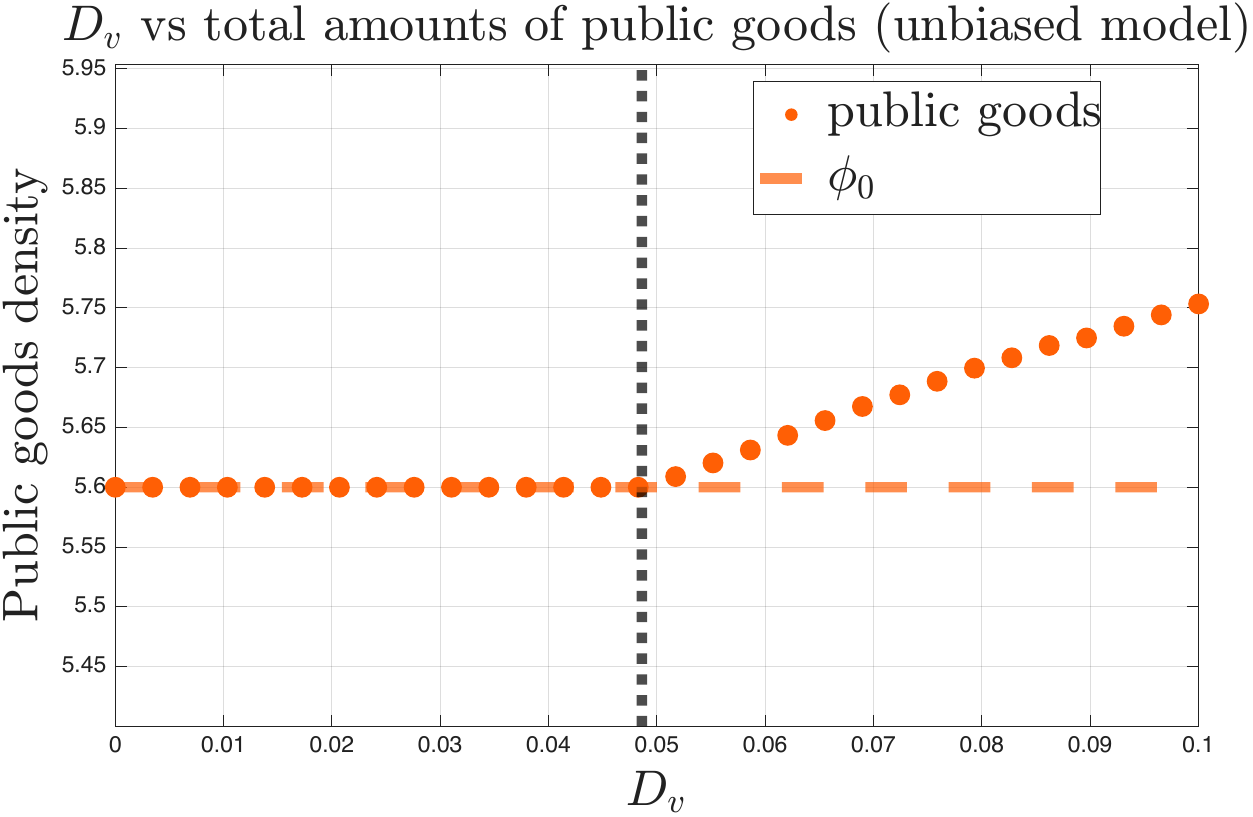}
        \caption*{(a2) Total amount of goods $\int_{0}^{L}\phi(t,x)\,dx$}
    \end{subfigure}\vspace{-12pt}
    \begin{subfigure}[b]{\textwidth}
        \centering
        \vspace{1.6em}
        \caption*{(a) The unbiased model at $t=50{,}000$}
    \end{subfigure}

    \vspace{1em}

    \begin{subfigure}[b]{0.48\textwidth}
        \includegraphics[width=\linewidth]{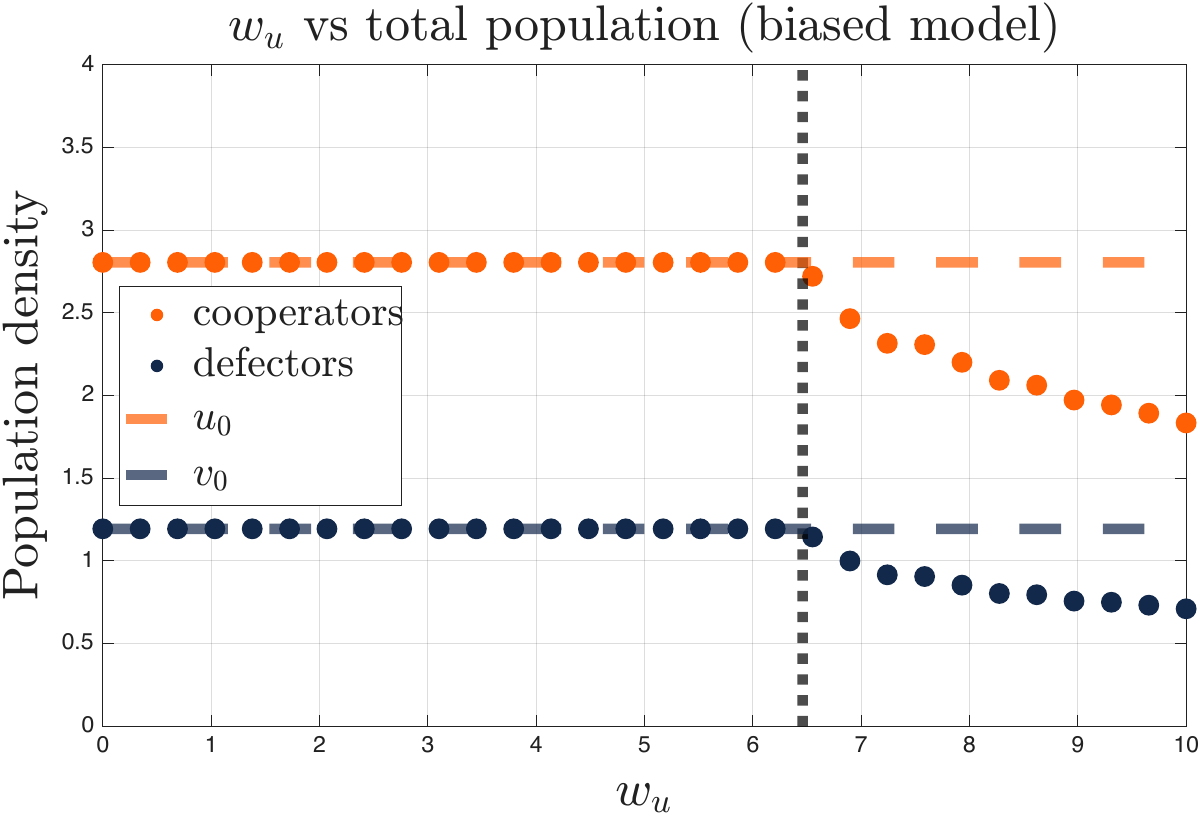}
        \caption*{(b1) Total population $\int_{0}^{L}u(t,x)\,dx$ and $\int_{0}^{L}v(t,x)\,dx$}
    \end{subfigure}
    \hfill
    \begin{subfigure}[b]{0.48\textwidth}
        \includegraphics[width=\linewidth]{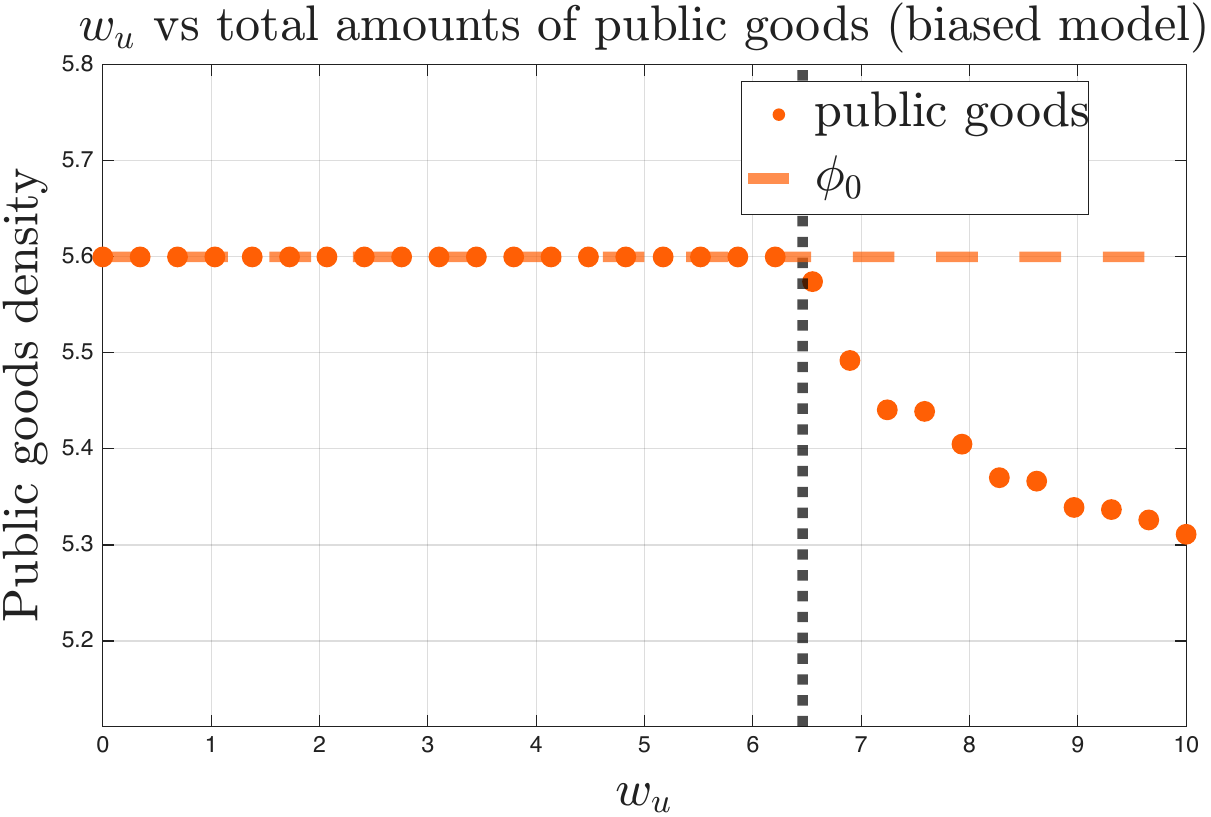}
        \caption*{(b2) Total amount of goods $\int_{0}^{L}\phi(t,x)\,dx$}
    \end{subfigure}\vspace{-12pt}
    \begin{subfigure}[b]{\textwidth}
        \centering
        \vspace{2em}
        \caption*{(b) The biased model at $t=80{,}000$}
    \end{subfigure}

    \caption{Calculation of the total long-time population sizes for cooperators and defectors and the total level of public good across the spatial domain for both the biased and unbiased model. We plot these quantities as a a function of our bifurcation parameter of interest ($D_v$ for the unbiased model and $w_u$ for the biased model).  The horizontal dashed lines represent density achieved in the uniform coexistence state, while the vertical dashed lines indicate the threshold values $D_v^\ast$ and $w_u^\ast$ for the onset of pattern formation. The horizontal axis shows the parameters $D_v$ in (a) and $w_u$ in (b), whereas the vertical axis represents the total population and the total amount of goods as the system evolves until certain time $t$. We use the same random initial conditions as in Figure~\ref{fig:simulation-unbiased} for (a) and Figure~\ref{fig:simulation-biased} for (b).}\label{fig:unbiased-biased-scatter}
\end{figure}

An interesting observation we see from the weakly nonlinear analysis is that, for spatial movement parameters close to the onset of pattern formation, it appears that diffusive and biased spatial motion produce different predictions on how the overall population benefits from the formation of spatial patterns. For the parameters considered, the case of purely random diffusion with increased defector diffusivity appears to lead to increased cooperation and a greater total amount of public goods for the parameters while directed motion results in decreased population sizes for cooperators and defectors and a reduced total amount of the public good. We further explore this behavior in Figure \ref{fig:unbiased-biased-scatter}, exploring whether the trend of increase or decrease in cooperation and public goods provisions continue as we increase the spatial movement parameters away from the threshold for pattern formation. For a fixed large time $t$, we vary the control parameters used in the weakly nonlinear analysis (which are $D_v$ for the unbiased model and $w_u$ for the biased model), then plot the total mass of cooperators and defectors across the domain as a function of the bifurcation parameter. For the baseline parameters we considered in the simulations for Figures \ref{fig:simulation-unbiased} and \ref{fig:simulation-biased}, we see that increasing defector diffusivity $D_v$ in the unbiased model appears to increase the size of both populations and the overall level of public goods, while increasing the cooperator movement sensitivity $w_u$ tends to result in the decrease in cooperators, defectors, and the public good. 

\begin{figure}[!htbp]
    \centering

    \begin{subfigure}[b]{0.48\textwidth}
        \includegraphics[width=\linewidth]{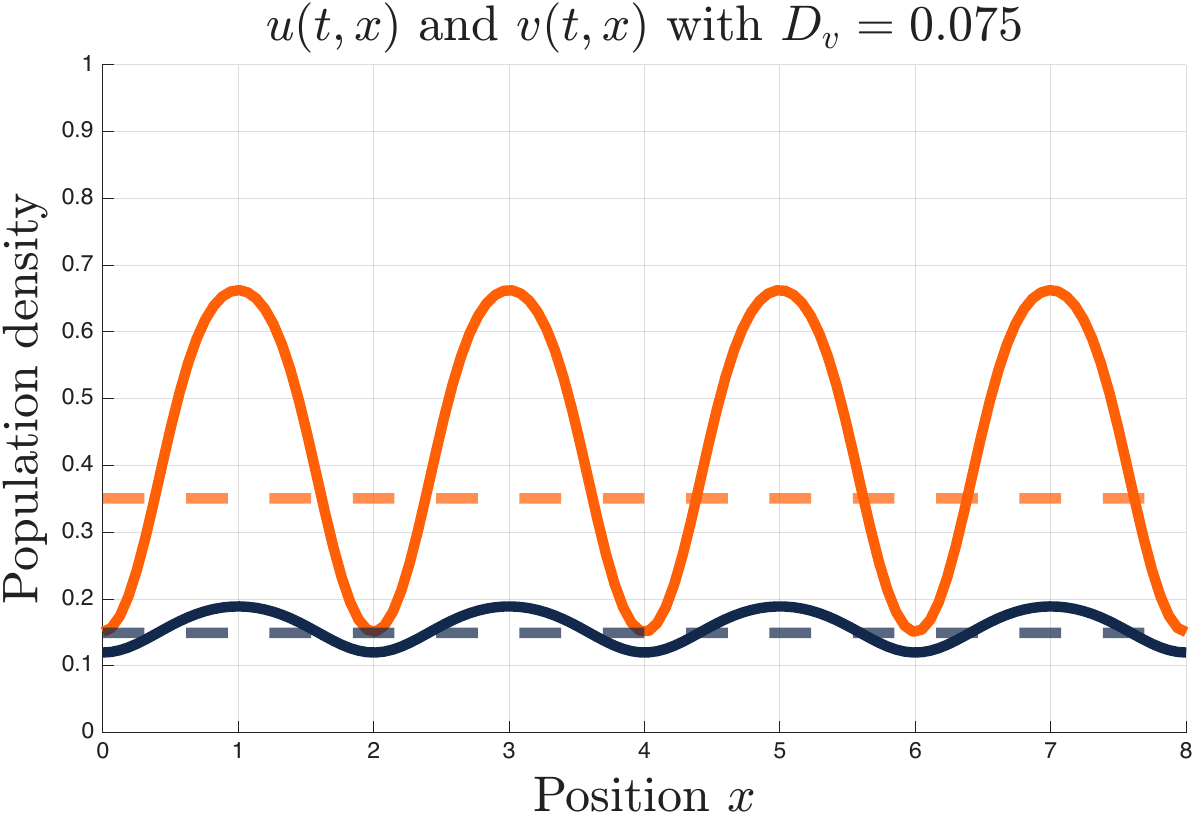}
    \end{subfigure}
    \hfill
    \begin{subfigure}[b]{0.48\textwidth}
        \includegraphics[width=\linewidth]{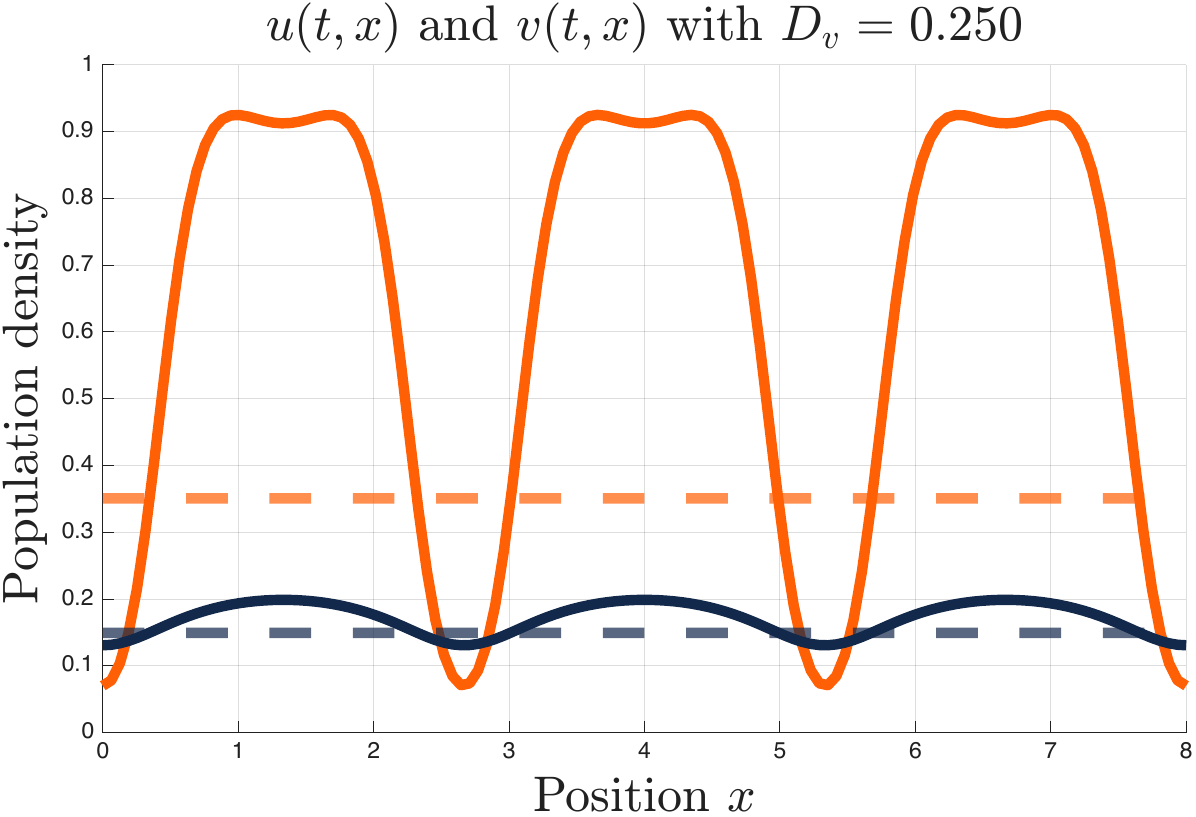}
    \end{subfigure}\vspace{-24pt}
    \begin{subfigure}[b]{\textwidth}
        \centering
        \vspace{2em}
        \caption*{(a) Snapshots of the unbiased model at $t=50{,}000$}
    \end{subfigure}

    \vspace{1em}

    \begin{subfigure}[b]{0.48\textwidth}
        \includegraphics[width=\linewidth]{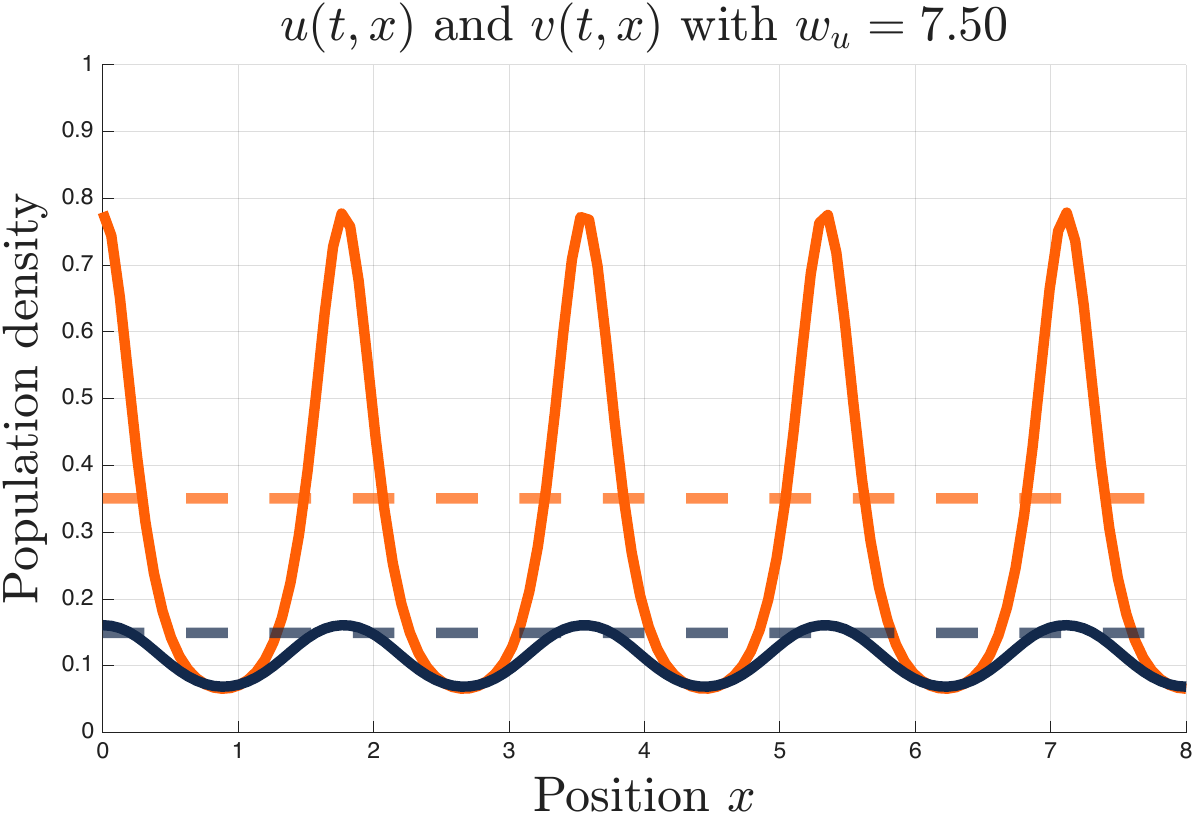}
    \end{subfigure}
    \hfill
    \begin{subfigure}[b]{0.48\textwidth}
        \includegraphics[width=\linewidth]{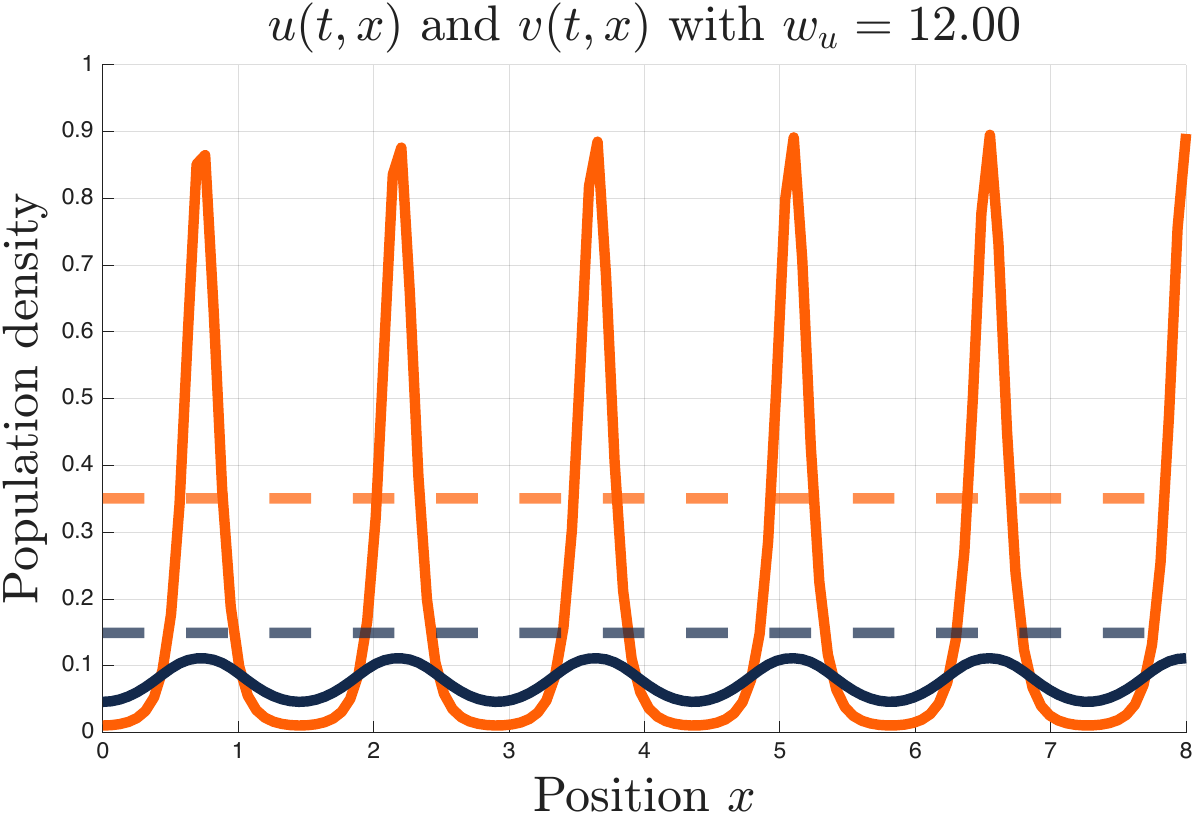}
    \end{subfigure}\vspace{-24pt}
    \begin{subfigure}[b]{\textwidth}
        \centering
        \vspace{2em}
        \caption*{(b) Snapshots of the biased model at $t=80{,}000$}
    \end{subfigure}

    \caption{Plots of numerical solutions for the spatial profiles of cooperators and defectors achieved after $t = 50,000$ for the unbiased model (in part (a)) and after $t = 80,000$ time-steps for unbiased model (in part (b)). For part (a), we simulate the unbiased model with defector diffusivity of $D_v = 0.075$ (left) and $D_v = 0.25$ (right), each greater than $D_v^\ast=0.04861$. For part (b), we simulate the unbiased model for cases of the cooperator's sensitivity to public good concentration given by $w_u = 7.5$ (left) and $w_u = 12$ (right), each greater than $w_u^\ast=6.4603$. All other parameters are the same as in Figure~\ref{fig:simulation-unbiased} and Figure~\ref{fig:simulation-biased}. The orange solid lines denote $u(t,x)$, the dark blue lines denote $v(t,x)$, and the dashed lines show the corresponding steady states $u_0$ and $v_0$.}
    \label{fig:spike-pattern}
\end{figure}

We can also explore the long-time spatial profiles achieved for spatial movement parameters further from the pattern-forming threshold. In Figure \ref{fig:spike-pattern}, we show the emergence of mesa-like patterns in the reaction-diffusion case for cheater diffusivities $D_u$ much larger than the threshold $D_u^*$, with the densities of cooperators and defectors both exceeding the values supported in the spatially uniform state across most of our spatial domain. For the case of directed motion, we see the spatial profiles of cooperators and defectors concentrate upon several spike-like concentrations as the chemotactic sensitivity $w_u$ of cooperators grows past the threshold level $w_u^*$ required for the onset of spatial patterns, with the population achieving densities much lower than present in the uniform state at most locations in the spatial domain. This formation of mesa patterns in the reaction-diffusion case and the spike patterns in the chemotaxis case are consistent with the qualitative behaviors observed in related models for far-from-equilibrium behavior of reaction-diffusion models \cite{kolokolnikov2007self,mckay2012stability} and in models of chemotaxis \cite{wang2015global,kong2022stability,kong2023existence,potapov2005metastability}.

\FloatBarrier

\section{Discussion}
\label{sec:Discussion}

In this paper, we have characterized how the interactions between cooperators, defectors, and an explicitly modeled public good can help to bring about patterns in the spatial profile of cooperative behavior and the provision of public goods, with emergent patterns shaped by the rules of spatial motion of individuals and public goods in continuous space. Starting from an individual-based description of birth and death events for cooperators and defectors, the production and decay of public goods, and the rules for spatial motion of individuals and goods due to random walks on a discrete spatial lattice, we derive and analyze systems of PDEs that can describe spatial pattern formation in the presence of purely diffusive motion for the individuals (our unbiased model) or in the case of a chemotaxis-like directed motion of individuals toward increasing concentrations of public goods (our model of biased directed motion). We see that either an increased diffusivity of defectors or an increased sensitivity of cooperators for directed motion can help to destabilize a spatially uniform coexistence equilibrium, and we use a mix of analytical and numerical exploration to further study the emergent patterns that arise through these rules for spatial motion. 

Through linear stability analysis, we were able to characterize distinct routes to pattern formation through either a Turing instability due to increased defector diffusivity or through a chemotactic instability through increased directed motion of cooperators. For the case of directed motion, we also saw that the threshold for instability had an increasing linear dependence on the strength $w_v$ for directed motion of defectors, indicating that increased movement of defectors towards regions of high public good concentration would help to suppress pattern formation through increased consumption of public goods at nascent peaks of cooperator and public good density. Although the the threshold on defector diffusivity $D_v$ required for Turing pattern formation has a more complicated dependence on the diffusivities of cooperators $D_u$ and public good $D_v$, we can still potentially understand the three-species system in terms of activator-inhibitor dynamics by interpreting defectors as an inhibitor and interpreting cooperators and public goods as an activator subnetwork. Through the incorporation of weakly nonlinear analysis, we were able to observe that, close to their respective thresholds for instability, patterns formed due to either mechanism feature the emergence of spatial correlation of cooperator and public good density, resulting in hubs of abundant public goods and high densities of cooperation. By contrast, the spatial profiles of defectors could either be positively or negatively correlated with the density of cooperators, depending on the parameters chosen for the reaction dynamics and spatial motion. As weakly nonlinear analysis is only meant to reflect behavior of patterns only in a neighborhood of the pattern-forming threshold, we perform further numerical simulations to show that similar spatial clusters also emerge further from onset and observe that a similar correlation between cooperator and public good densities holds across a wide range of movement parameters.

To further explore the qualitative properties of the pattern-forming instability, we combined a weakly nonlinear stability analysis and numerical simulations to analyze the stability of emergent patterns and to characterize how individuals of each strategy and the total level of public good either gain or lose from the spatial movement rules that promote spatially heterogeneous distributions of strategies and resources. An interesting feature in our results is that,  for a wide range of parameters including the default parameter set from Table \ref{table 1}, the Turing instability resulted in spatial patterns with an increased size of the cooperator population and an increased overall abundance of public goods across the spatial domain, even though the spatial motion corresponds to undirected motion via simple random walks. Although weakly nonlinear stability analysis only characterizes behavior of patterns close to the threshold for instability and only predicts solutions whose mass deviates from the uniform state by a term of size $\mathcal{O}(\varepsilon^2)$, the predictions made by weakly nonlinear analysis for the default parameters agreed with the results obtained by numerical simulation when our bifurcation parameters moved further beyond the threshold and achieved a much more substantial deviation from the uniform equilibrium values. A similar behavior was seen in a reaction-diffusion model with demographic dynamics arising from the payoff of a hawk-dove game and logistic density-dependent regulation \cite{yao2026pattern}, in which average payoff and average population size increased with diffusion-driven pattern formation. By contrast, we saw that directed motion of cooperators towards increasing levels of public goods actually results in spatial patterns with a worse collective outcome for both strategies than the population would experience in the absence of spatial motion. A similar behavior was seen numerically in a model of evolutionary games with environmental feedback and environment-driven motion \cite{yao2025spatial}, suggesting that this may be a fairly common phenomenon of a spatial social dilemma in the presence of directed motion towards public goods or shared resources.

The use of weakly nonlinear analysis and the derivation of the Stuart-Landau amplitude equation for emergent patterns was helpful in our analysis to characterize our pattern-forming bifurcation as a supercritical pitchfork bifurcation for both the models with unbiased and biased motion for a relatively wide range of parameters around the default parameters from Table \ref{table 1}. By considering a wider range of reaction parameters, we were also able to observe parameter regions in which the pattern-forming bifurcations could correspond to subcritical pattern-forming bifurcations, suggesting the possible bistability between the uniform coexistence state and a spatially patterned state when reaction parameters were close to a threshold for biologically instability of the coexistence equilibrium state. This region of subcriticality is reminiscent of related results highlighting the role played by spatial pattern formation as a resilience mechanism near collapse of uniform equilibria of the reaction dynamics \cite{deangelis2012self,martinez2013vegetation,rietkerk2021evasion,van2025vegetation,banerjee2026rethinking}, and suggests that there may be a rich variety of spatial behaviors exhibited for different parameter regimes in both the unbiased and biased model \cite{dawes2008localized,brena2014subcritical,paige2026towards}.

In addition to characterizing the type of bifurcation displayed by our pattern-forming instability, the resulting approximate expression we obtained from weakly nonlinear stability analysis close to the threshold for instability allowed us to obtain approximate expressions for the population sizes and public good abundance across the spatial domain in our patterned states. While the use of weakly nonlinear approximations has been used in the past to characterize qualitative features of species abundance in Turing patterns for predator-prey dynamics \cite{segel1976application}, there is substantial room to further incorporate this approach to achieve a greater understanding of the levels of cooperation and average payoff in models of spatial pattern formation and in evolutionary games \cite{helbing2009outbreak,helbing2009pattern,deforest2013spatial,deforest2018game,xu2017strong,olson2022community}. Such asymptotic expansions for key functionals of the spatial profiles achieved in patterned states may also relevant in a range of PDE models for other models of social systems, such as models for economic aggregation into cities by diffusive or capital-induced labor migration \cite{juchem2015capital,neto2018economic,neto2019returns,zincenko2018economic,zincenko2021turing,morsky2025urban} or other models of collective spatial dynamics in cities \cite{hasan2020transport,shaw2024dynamical,mauro2025dynamic,lindstrom2020qualitative,rogers2011unified,rogers2012jamming}.

As the bifurcation parameters moved beyond the threshold for the pattern-forming instability, we saw the emergence of spike patterns for the directed motion model with strong cooperator sensitivity $w_u$, as well as the formation of mesa or plateau patterns for the reaction-diffusion model in the case of fast defector diffusivity. Asymptotic analysis of chemotaxis and reaction-diffusion models has explored the existence and stability of spike and mesa patterns \cite{kolokolnikov2007self} in other PDE models displaying spatial pattern formation far from the pattern-formation threshold. In particular, both interior and boundary spikes have been seen as long-time behaviors for the Keller-Segel model with logistic population dynamics \cite{kong2022stability,kong2023existence}, and analogous behavior has been demonstrated in chemotaxis-like models for economic aggregation \cite{kong2025boundary} and urban crime \cite{tse2016hotspot,tse2018asynchronous,buttenschoen2020cops}. The behavior we observe in our numerical simulations suggests that it may be possible to explore mesa-like or spike-like patterns in the far-from-threshold regime, which may allow us to further characterize the qualitative behavior of emergent patterns in the case of strong defector diffusion or fast directed motion of cooperators towards regions of increasing public good concentration.

While we have focused on a particular choice of directed motion based on the spatial gradient of the concentration of public goods $\phi(t,x)$ in this paper, there are many ways we can formulate our models for directed motion based on the underlying rules for public goods production and the population dynamics of cooperators and defectors. In particular, one could consider directed motion of cooperators and defectors could based on the per-capita growth rate of each strategy, which would work in analogously to models of payoff-driven motion in evolutionary game models that describe how individuals move in the direction of increasing payoff for their strategy \cite{helbing2009pattern,deforest2013spatial,xu2017strong,deforest2018game,yao2026pattern}. Alternatively, one could consider other sensitivity functions like $f(w,A) = 1 + wA$, which would produce PDE models with spatial movement rules in the form of logarithmic chemotactic sensitivity that often arise in models arising as continuum limits of agent-based models of directed motion in biological and social systems \cite{sleeman2005existence,kang2007stability,short2010nonlinear,rodriguez2010local}. For further study, we provide detailed computations for the continuum limit in Appendix \ref{sec:PDEderivation}, systematic computations for the perturbations equations for the PDE model of Equation \eqref{eq:DM-model} in Appendix \ref{sec:weakly-nonlinear-expansions}, which we believe would benefit and facilitate future researchers.

One complicating factor in interpreting analytical results for our model is that studying the spatial dynamics of cooperators, defectors, and the public good produces a system of three PDEs, and the resulting stability analysis of these models requires studying the eigenvalues of relatively full $3\times3$ matrices. It may be possible to consider simplified versions of this model in future work to obtain more interpretable formulas for the pattern-forming instability or the weakly nonlinear approximation by writing the spatial profile of the public good as a function of the strategic distribution of the population. Young and Belmonte are able to make such a reduction by assuming that the public good does not diffuse between spatial locations and the reaction dynamics of the public good are sufficiently fast relative to the demographic and movement dynamics of individuals in the population. In our PDE model featuring spatial diffusion of the public good, we could potentially try to assume fast spatiotemporal dynamics of the public good and use a parabolic-elliptic system or a Green's function to represent the steady-state distribution of public goods in terms of the spatial densities of cooperators and defectors. Such a strategy has been frequently applied in establishing the connection between systems of local PDEs used to describe chemotaxis based on populations of individuals and their chemoattractants or chemorepellants \cite{topaz2006nonlocal,milewski2008simple,buttenschon2024cells}, but additional care may be required in performing similar reductions of our public goods model due to the nonlinearity in the reaction dynamics for the public good due to consumption of the good by cooperators and defectors. 

The production of public goods and the sustainable harvest of common-pool resources can result in closely related social dilemmas for the evolution of cooperation behavior with a self-organized change in environments. Our analysis of the dynamics of cooperation with an explicit public good is therefore complementary to models that have studied the evolution of cooperative extraction of renewable or exhaustible resources \cite{weitz2016oscillating,tilman2020evolutionary,mitchener2022symmetric,wang2020eco}, and we can look to compare the spatio-temporal dynamics achieved in our public goods models with related spatial models of evolutionary games with resource dynamics \cite{lin2019spatial,li2026phase,betz2024evolutionary,cheng2024evolution,cheng2026spatio}. Related spatial models for the short-range facilitation via public goods provision and long-range competition for resources have further shown a range of pattern-forming behaviors \cite{hermsen2022emergent,doekes2024multiscale} as associated evolution of cooperative behaviors, providing connections between models of the evolution of cooperation and related non-local PDE models for vegetation dynamics and ecological competition over natural resources \cite{aydogmus2015patterns,aydogmus2018phase,martinez2013vegetation,martinez2014minimal,kuehn2013nonlocal}. Between these different frameworks for exploring cooperation and competition over public goods and common-pool resources, there is substantial room for further exploration of PDE models for the role of spatial dynamics in the emergence and stability of cooperative behavior. 

While much past work on spatial patterns and cooperative behavior has focused on game-theoretic models and rules for reproduction and spatial motion that may depend on the payoff from interactions generating payoff from games, the model considered in this paper works more directly with the ecological dynamics associated with the production, consumption, and movement of an explicit public good. We have seen that spatial patterns can emerge due to the interaction between the reaction terms introduced by Young and Belmonte and either differences in the diffusivities or the chemotactic-like sensitivity to gradients in public good concentration between the two strategies. In particular, both the Turing-type and chemotaxis-type mechanisms have generated spatial patterns via finite-wavenumber instabilities, allowing us to perform weakly nonlinear analysis and characterize the effects of the spatial movement rules on the total population size and public goods concentration across the spatial domain. By contrast, prior work on game-theoretic models has shown the possibility of infinite-number instabilities in spatial models with payoff-driven directed motion, so the cases of Turing instability and instability due to directed motion are not as directly comparable in models using payoff directly to determine environmental quality. For this reason, models of ecological feedback through either public goods production or the depletion of common-pool resources can be a particularly salient model problem for understanding how rules for directed and undirected spatial motion can help to promote the evolution of cooperative behavior and the promotion of socially beneficial steady-state profiles for the strategic composition of populations. 

\begin{quote}
\begin{center}
    \textbf{Acknowledgments}
\end{center}
DBC’s research is partially supported by the Simons Foundation through the
Travel Support for Mathematicians Program (grant MPS-TSM-00007872).
\end{quote}

\begin{quote}
\begin{center}
    \textbf{Statement on Code and Data Availability}
\end{center}
    The numerical simulations in this paper were implemented in Python and MATLAB. The code used to generate the figures and results is publicly available in the GitHub repository: \href{https://github.com/yxuanzhao/BioMath-Public-Goods-Dilemma}{yxuanzhao/BioMath-Public-Goods-Dilemma}. There is no additional data required to reproduce the results in the paper. 
\end{quote}

\bibliographystyle{unsrt} 
\bibliography{main}

\appendix
\section*{Summary of Appendix Sections}

In the appendix, we provide additional derivations and calculations for formulating and analyzing our PDE models for the spatial public good with diffusive and directed motion. In Section \ref{sec:PDEderivation}, we provide the full details for the heuristic derivation of our PDE model from a stochastic version of the Young-Belmonte public goods dilemma in which cooperators and defectors perform biased random walks towards patches with greater concentrations of the public good. We also provide additional details about the linear stability analysis for the reaction diffusion case in Section \ref{sec:linear-Routh-Hurwitz}, highlighting when we can deduce that instability of the uniform state can be deduced from the determinant of the linearization matrix becoming positive. 

We also provide addition derivations and calculations for the weakly nonlinear stability analysis in the appendix. We provide the details for the derivation of the hierarchy of PDEs for our asymptotic approximation of the PDE model close to the onset of the pattern-forming instability, which we then apply in Section \ref{sec:weakly-nonlinear-analysis} to study approximate solutions to our PDE model in this parameter regime. We also provide additional calculations for the solutions in the weakly nonlinear regime in Section \ref{sec:eigenvectors}, highlighting our analytical characterization of the correlation between the concentration of cooperators and the public good in the patterned state and highlighting how the conclusions for the form of Stuart-Landau equation are invariant under certain choices for the form of the solutions to our perturbation expansion. 

\section{Derivation of PDE Model from Stochastic Spatial Model}
\label{sec:PDEderivation}

In this section we provide a full heuristic derivation for the PDE system of Equation \eqref{eq:general-continuum-PDE} starting from the expected change in the density of cooperators, defectors, and the public good presented in Equation \eqref{eq:discrete-main-equations} for the underlying stochastic spatial model. Our goal is to show that the continuum limit of Equation \eqref{eq:general-continuum-PDE} arises from performing a diffusive / parabolic scaling limit of biased random walks and local reaction events on a regular spatial lattice. Although the analysis of our PDE model focuses on the case of a one-dimensional spatial domain, we perform this derivation for a more general $d$-dimensional lattice with a von Neumann neighborhood (meaning that individuals have two neighbors in each of the $d$ axes on the lattice).  

We consider a stochastic spatial model occurring on a $d$-dimensional lattice with spatial distance $\ell$ between the nearest neighbors on the lattice, with the temporal dynamics occuring discrete time with time-steps of length $\Delta t$. During each time interval from time $t$ to time $t+\Delta t$, cooperators, defectors, and public goods undergo a single update containing two stages: first individuals and the public good undergo reaction dynamics based on the local state at each site and then individuals and the good can perform movement between neighboring sites on the lattice. In the reaction stage, the local population dynamics at each lattice node follows the reaction dynamics specified by the Young-Belmonte model, featuring reaction rates corresponding to the ODE dynamics formulated in Equation \eqref{eq:Young-Belmonte-ODE}. In particular, cooperators and defectors reproduce, public goods are produced and consumed, individuals die, and public goods decay. In the movement stage, cooperators, defectors, and public goods have probabilities $\chi_u$, $\chi_v$, and $\chi_{\phi}$ to consider moving to a neighboring lattice size during this time-step, and then individuals that consider moving will either move to a neighboring site or remain at their current site according to a biased random walk based on the strategic and public good composition within their local neighborhood. The jump probabilities in the biased random walk depend on the local environmental attractiveness, described by the functions $A_i$, $B_i$, and $C_i$, so that individuals preferentially move toward move favorable locations.

We show that, when the spatial step size $\ell\to 0$ and the time step $\Delta t\to 0$ with $\ell^2\sim \Delta t$, the terms describing the biased random walk will converge to a linear diffusion term and a nonlinear advection term in our continuum PDE limit. The diffusion term describes undirected components of the underlying random walk, while the drift term arises from the spatial bias in the jump probabilities due to differences in the attractiveness of neighboring lattice sites. This limiting procedure leads to a system of reaction-advection-diffusion PDEs.

We interpret $u_i(t), v_i(t)$, and $\phi_i(t)$ as expected population densities at lattice site $i$. Using the rates of reactions from the Young-Belmonte model and considering a biased random walk based on the attractiveness $A_i$, $B_i$, and $C_i$ of the lattice site $i$ for cooperators, defectors, and the public good allow us to derive the following deterministic lattice equations for the expected densities of the three quantities at site $i$:
\begin{subequations}
    \begin{align}
        u_i(t+\Delta t)=&\begin{aligned}[t]
            &\,\,\displaystyle \left\{ u_i(t) + u_i(t)\left[r_u\phi_i(t)-c-\gamma(u_i(t)+v_i(t))-\mu_u\right]\Delta t \right\} \left( 1 - \chi_u + \chi_u \mathds{P}_u(i \to i) \right) \\
            &\,\,\displaystyle+\chi_u \sum_{j\in N(i)}\mathds{P}_{u}(j\to i) \left\{ u_j(t) +  u_j(t)\left[r_u\phi_j(t)-c-\gamma(u_j(t)+v_j(t))-\mu_u\right]\Delta t\right\},
            \end{aligned}\label{eq:discrete-equation-u}\\
        v_i(t+\Delta t)=&\begin{aligned}[t]
            &\,\,\displaystyle \left\{ v_i(t) + v_i(t)\left[r_v\phi_i(t)-\gamma(u_i(t)+v_i(t))-\mu_v\right]\Delta t \right\}\left( 1 - \chi_v + \chi_v \chi_u \mathds{P}_v(i \to i) \right) \\
            &\,\,\displaystyle+\chi_v \sum_{j\in N(i)}\mathds{P}_{v}(j\to i) \left\{ v_j(t) + v_j(t)\left[r_v\phi_j(t)-\gamma(u_j(t)+v_j(t))-\mu_v\right]\Delta t \right\},
        \end{aligned}\label{eq:discrete-equation-v}\\
        \phi_i(t+\Delta t)=&\begin{aligned}[t]
            &\,\,\displaystyle  \left\{ \phi_i(t) + \left(cu_i(t)-\left[\kappa(u_i(t)+v_i(t))+\delta\right]\phi_i(t)\right)\Delta t \right\} \left( 1 - \chi_{\phi} + \chi_{\phi}  \mathds{P}_{\phi}(i \to i) \right) \\
            &\,\,\displaystyle+ \chi_{\phi} \sum_{j\in N(i)}\mathds{P}_{\phi}(j\to i) \left\{ \phi_j(t) + \left(cu_j(t)-\left[\kappa(u_j(t)+v_j(t))+\delta\right]\phi_j(t)\right)\Delta t \right\},
        \end{aligned}\label{eq:discrete-equation-phi}
    \end{align}
\end{subequations}
where $N(i)$ is the set of all neighboring nodes of $i$. In this section, we will explicitly carry out the derivation of the continuum PDE corresponding to the discrete dynamics of Equation \eqref{eq:discrete-equation-u} for the expected cooperator density, and we can carry out similar derivations for the defector and public good densities with an analogous approach. 

The effect of the biased movement rules can be written as the difference between incoming and outgoing fluxes at each lattice size. We use $P$ and $Q$ to respectively denote the the expected inflows into site $i$ from neighboring sites the expected outflow from site $i$ to its neighbors, and we see that $P$ and $Q$ are given by the following expressions:
\begin{equation}
    \begin{aligned}
        P&=\chi_u \sum_{j\in N(i)}\mathds{P}_{u}(j\to i) \left\{ u_j +  u_j\left[r_u\phi_j-c-\gamma(u_j+v_j)-\mu_u\right]\Delta t\right\},\\
        Q&=\left\{u_i+ u_i\left[r_u\phi_i-c-\gamma(u_i+v_i)-\mu_u\right]\Delta t \right\} \chi_u \left( 1 - \mathds{P}_u(i \to i)\right).
    \end{aligned}
    \nonumber
\end{equation}
To pass to the continuum limit, we assume a regular square lattice in $d$ dimensions with a von Neumann neighborhood and interpret lattice quantities as samples of a sufficiently smooth function $g(x)$ at $x_i=i\ell$. As $\ell\to0^+$, we can use the symmetry of the von Neumann neighborhood to see that
\begin{equation} \label{eq:gmovementcalculation}
\begin{aligned}
\sum_{j \in N(i)} g(x_j) &= \sum_{\alpha=1}^d  \left[g(x_i) +  \ell g_{\alpha}(x_i) + \frac{\ell^2}{2} g_{\alpha \alpha}(x_i) + \frac{\ell^2}{6} g_{\alpha \alpha \alpha}(x_i) +  \mathcal{O}\left(\ell^4\right)  \right] \\
&+ \sum_{\alpha = 1}^{d} \left[g(x_i) - \ell g_{\alpha}(x_i) + \frac{\ell^2}{2} g_{\alpha \alpha}(x_i) - \frac{\ell^3}{6} g_{\alpha \alpha \alpha}(x_i) +  \mathcal{O}\left(\ell^4\right)  \right]  \\  
&= \sum_{\alpha = 1}^{d} \left[ 2 g(x_i)  + \ell^2 g_{\alpha \alpha}(x_i) + \mathcal{O}(\ell^4) \right] \\
&=  zg(x_i)+\ell^2\Delta g(x_i)+\mathcal{O}(\ell^4)m
\end{aligned}
\end{equation}
where  $z = 2d$ denotes the number of neighbors of lattice site $i$ and  $g_{\alpha}(x_i)$ denotes the partial derivative of $g(x_i)$ with respect to the variable in each of the dimensions in $\{1,2,\cdots,d\}$ on our $d$-dimensional lattice.

This expansion tells us that movement between nearest neighboring sites on the lattice produces a second-order spatial operator that tends to the Laplacian in the continuum limit under appropriate scaling of the spatial and temporal step-sizes. Here we allow an individual not only to move to one of its neighboring sites, but also to remain at its current site during a time step with a probability depending on the attractiveness of the current site. This means that the movement rule will depend on the attractiveness of the $z$ neighboring sites and the original site $i$ itself, so the total weight required for normalization of the transition probabilities involves calculated the attractiveness of $1+z$ possible destination sites. Using the expression from the nearest-neighbor expansion from Equation \eqref{eq:gmovementcalculation} to approximate the normalization term for the jumped probabilities in the biased random walk, we can expand $P$ and $Q$ in powers of $\ell$. For convenience, we write $S_j:=u_j +  u_j\left[r_u\phi_j-c-\gamma(u_j+v_j)-\mu_u\right]\Delta t$, then 
\begin{equation}
    \begin{aligned}
        P=&\,\, \sum_{j\in N(i)}\frac{\chi_uf(w_u, A_i)}{f(w_u, A_j)+\left[\ell^2\Delta f(w_u, A_j)+zf(w_u, A_j)\right]}S_j+\mathcal{O}(\ell^4)\\
        =&\,\,\sum_{j\in N(i)}\frac{\chi_u f(w_u, A_i)}{(1+z)f(w_u, A_j)+\ell^2\Delta f(w_u, A_j)}S_j+\mathcal{O}(\ell^4)\\
        =&\,\,\frac{\chi_u f(w_u, A_i)}{1+z}\sum_{j\in N(i)}\left[\frac{1}{f(w_u, A_j)}-\frac{\ell^2\Delta f(w_u, A_j)}{(1+z)f^2(w_u, A_j)}\right]S_j+\mathcal{O}(\ell^4)\\
        =&\,\,\frac{\chi_u f(w_u, A_i)}{1+z}\left[\frac{zS_i}{f(w_u, A_i)}+\ell^2\Delta\left(\frac{S_i}{f(w_u, A_i)}\right)-\frac{\ell^2z}{1+z}\cdot\frac{S_i}{f^2(w_u, A_i)}\Delta f(w_u, A_i)\right]+\mathcal{O}(\ell^4)\\
        =&\,\,\chi_u\left[\frac{zS_i}{1+z}+\frac{\ell^2f(w_u, A_i)}{1+z}\Delta\left(\frac{S_i}{f(w_u, A_i)}\right)-\frac{\ell^2z}{(1+z)^2}\cdot\frac{S_i}{f(w_u, A_i)}\Delta f(w_u, A_i)\right]+\mathcal{O}(\ell^4).
    \end{aligned}
    \nonumber
\end{equation}
Since we assume $\ell^2/\Delta t\to D$, then $\mathcal{O}(\ell\Delta t)=\mathcal{O}(\ell^4)$, so we can put all $\ell^2\Delta t$ terms into $\mathcal{O}(\ell^4)$, then
\begin{equation}
    \begin{aligned}
        P=&\,\,\chi_u\left[\frac{zu_i}{1+z}+\frac{\ell^2f(w_u, A_i)}{1+z}\Delta\left(\frac{u_i}{f(w_u, A_i)}\right)-\frac{\ell^2z}{(1+z)^2}\cdot\frac{u_i}{f(w_u, A_i)}\Delta f(w_u, A_i)\right]\\
        &\,\,+\frac{\chi_u z}{1+z}u_i\left[r_u\phi_i-c-\gamma(u_i+v_i)-\mu_u\right]\Delta t+\mathcal{O}(\ell^4).
    \end{aligned}
    \nonumber
\end{equation}
Similarly, we can compute $Q$ as 
\begin{equation}
    \begin{aligned}
        Q=&\,\,-\frac{\chi_u f(w_u,A_i)}{f(w_u,A_i)+\sum_{k\in N(i)}f(w_u,A_k)}S_i+\chi_u S_i\\
        =&\,\,-\frac{\chi_uf(w_u,A_i)}{(1+z)f(w_u,A_i)+\ell^2\Delta f(w_u,A_i)}S_i+\chi_uS_i+\mathcal{O}(\ell^4)\\
        =&\,\,-\chi_uf(w_u,A_i)\left[\frac{1}{(1+z)f(w_u,A_i)}-\frac{\ell^2\Delta f(w_u,A_i)}{(1+z)^2f^2(w_u,A_i)}\right]S_i+\chi_uS_i+\mathcal{O}(\ell^4)\\
        =&\,\,-\left[\frac{\chi_uS_i}{1+z}-\frac{\chi_u\ell^2\Delta f(w_u,A_i)}{(1+z)^2f(w_u,A_i)}S_i\right]+\chi_uS_i+\mathcal{O}(\ell^4)\\
        =&\,\,-\frac{\chi_uu_i}{1+z}-\frac{\ell^2}{(1+z)^2}\cdot\frac{\chi_uu_i}{f(w_u,A_i)}\Delta f(w_u,A_i)-\frac{\chi_u}{1+z}u_i\left[r_u\phi_i-c-\gamma(u_i+v_i)+\mu_u\right]\\
        &\,\,+\chi_u u_i-\chi_uu_i\left[r_u\phi_i-c-\gamma(u_i+v_i)+\mu_u\right]+\mathcal{O}(\ell^4).
    \end{aligned}
    \nonumber
\end{equation}
The expression $P-Q$ represents the net movement flux at lattice site $i$, which can be written in the following form:
\begin{equation}
    \begin{aligned}
        P-Q=&\,\,\frac{\chi_u\ell^2}{1+z}f(w_u, A_i)\Delta\left(\frac{u_i}{f(w_u, A_i)}\right)-\frac{\chi_u\ell^2}{1+z}\cdot\frac{u_i}{f(w_u, A_i)}\Delta f(w_u, A_i)+\mathcal{O}(\ell^4)\\
        =&\,\,\frac{\chi_u\ell^2}{1+z}\left[f(w_u, A_i)\Delta\left(\frac{u_i}{f(w_u, A_i)}\right)-\frac{u_i}{f(w_u, A_i)}\Delta f(w_u, A_i)\right]+\mathcal{O}(\ell^4)\\
        =&\,\,\frac{\chi_u\ell^2}{1+z}\left[\Delta u_i-2\nabla\left(\frac{u_i}{f(w_u, A_i)}\right)\cdot\nabla f(w_u, A_i)-\frac{2u_i}{f(w, A_i)}\Delta f(w_u, A_i)\right]+\mathcal{O}(\ell^4)\\
        =&\,\, \frac{\chi_u\ell^2}{1+z}\left[\Delta u_i-2\nabla\cdot\left(\frac{u_i}{f(w_u, A_i)}\nabla f(w_u, A_i)\right)\right]+\mathcal{O}(\ell^4)\\
        =&\,\, \frac{\chi_u\ell^2}{1+z}\nabla\cdot\left[\nabla u_i-\frac{2u_i}{f(w_u, A_i)}\nabla f(w_u, A_i)\right]+\mathcal{O}(\ell^4).
    \end{aligned}
    \nonumber
\end{equation}
Writing the net movement term in divergence form identifies the corresponding flux, which allows us to understand the rules for directed and diffusive motion for our continuum limit. In particular, we see that the flux term features two terms reminiscent of the flux in  models for chemotaxis \cite{keller1970initiation,hillen2009user,painter2019mathematical,SHORT200808}, where the first component corresponds to ordinary Fickian diffusion and the second component describes directed movement up the gradient of $\ln f(w,A_i)$ toward regions where the attractiveness $f(w,A_i)$ is larger. Now, we can apply our expression for net spatial flux $P-Q$ to Equation \eqref{eq:discrete-equation-u} to see that the expected change in the density of cooperators $u_i(t)$ satisfies 
\begin{equation}
    \begin{aligned}
        \frac{u_i(t+\Delta t)-u_i(t)}{\Delta t}=&\,\,u_i(t)\left[r_u\phi_i(t)-c-\gamma(u_i(t)+v_i(t))-\mu_u\right]\\
        &\,\,+\frac{\chi_u\ell^2}{(1+z)\Delta t}\nabla\cdot\left[\nabla u_i-\frac{2u_i}{f(w_u, A_i)}\nabla f(w_u, A_i)\right]+\frac{\mathcal{O}(\ell^4)}{\Delta t}.
    \end{aligned}
    \nonumber
\end{equation}
We can take this finite different approximation for the time derivative of $u_i(t)$ and apply the limit as $\Delta t \to 0$, $\ell \to 0$, and the the parabolic scaling assumption that $\ell^2\propto\Delta t$ to describe the continuum limit of our model. In particular, we assume that there is a positive constant $D$ such that $\ell^2/\Delta t\to D$ as $\Delta t, \ell \to 0$, which allows us to balance the spatial and time-steps our model to maintain a nonzero, finite diffusivity of the individuals and public good as our spatial and temporal steps tend to $0$. Taking this scaling limit on both sides of the equation and denoting the densities $u(t,\mathbf{x})$, $v(t,\mathbf{x})$, and $\phi(t,\mathbf{x})$ as the limiting values of the lattice densities, we are able to deduce that the mean-field limit of our stochastic model as the following system of PDEs:
\begin{equation}\label{equation a2}
    \begin{aligned}
        \frac{\partial u}{\partial t}&=\frac{\chi_u D}{1+z}\nabla\cdot\left[\nabla u-\frac{2u}{f(w_u, A)}\nabla f(w_u, A)\right]+u\left[r_u\phi-c-\gamma(u+v)-\mu_u\right]\\
        &=\frac{\chi_uD}{1+z}\nabla\cdot\left[\nabla u-2u\nabla\left(\ln f(w_u,A)\right)\right]+u\left[r_u\phi-c-\gamma(u+v)-\mu_u\right],\\
        \frac{\partial v}{\partial t}&=\frac{\chi_v D}{1+z}\nabla\cdot\left[\nabla v-\frac{2v}{f(w_v, B)}\nabla f(w_v, B)\right]+v\left[r_v\phi-\gamma(u+v)-\mu_v\right]\\
        &=\frac{\chi_vD}{1+z}\nabla\cdot\left[\nabla v-2v\nabla\left(\ln f(w_v,B)\right)\right]+v\left[r_v\phi-\gamma(u+v)-\mu_v\right]\\
        \frac{\partial \phi}{\partial t}&=\frac{\chi_\phi D}{1+z}\nabla\cdot\left[\nabla \phi-\frac{2\phi}{f(w_\phi, C)}\nabla f(w_\phi, C)\right]+cu-\left[\kappa(u+v)+\delta\right]\phi\\
        &=\frac{\chi_\phi D}{1+z}\nabla\cdot\left[\nabla \phi-2\phi\nabla\left(\ln f(w_\phi,C)\right)\right]+cu-\left[\kappa(u+v)+\delta\right]\phi,
    \end{aligned}
\end{equation}
where $A(t,\mathbf{x})$, $B(t,\mathbf{x})$, and $C(t,\mathbf{x})$ describe the attractiveness of the location $x$ in our domain at time $t$ to cooperators, defectors, and the public good. If we denote the diffusivities of cooperators, defectors, and the public good by 
\begin{equation}
    D_u:=\frac{\chi_u D}{1+z},\quad D_v:=\frac{\chi_vD}{1+z},\quad D_\phi:=\frac{\chi_\phi D}{1+z},
    \nonumber
\end{equation}
then we can obtain the following system of partial differential equations that describes a general form of the continuum limit of the Young-Belmonte model for public good production and arbitrary rules for directed motion for cooperators, defectors, and the public good:
\begin{equation}\label{equation a3}
    \begin{aligned}
        \frac{\partial u}{\partial t}&=D_u\nabla\cdot\left[\nabla u-2u\nabla\ln f(w,A)\right]+u\left[r_u\phi-c-\gamma(u+v)-\mu_u\right],\\
        \frac{\partial v}{\partial t}&=D_v\nabla\cdot\left[\nabla v-2v\nabla\ln f(w,B)\right]+v\left[r_v\phi-\gamma(u+v)-\mu_v\right],\\
        \frac{\partial \phi}{\partial t}&=D_\phi\nabla\cdot\left[\nabla \phi-2\phi\nabla\ln f(w,C)\right]+cu-\left[\kappa(u+v)+\delta\right]\phi.
    \end{aligned}
\end{equation}
In the resulting PDE system, the diffusion terms arise from unbiased movement, while the advection terms originate from the spatial bias in the jump probabilities. The reaction terms correspond directly to the local birth-death processes in the stochastic model.

\section{Further Calculations for the Onset of Turing Instability via the Routh-Hurwitz Criteria}
\label{sec:linear-Routh-Hurwitz}

In this section, we will provide further justification for relying on the condition of a positive determinant $Q_3(k) = - \mathrm{tr}(M(k)) > 0$ in the linearization matrix to identify the onset for pattern-forming bifurcation for the reaction-diffusion dynamics in the unbiased model (as discussed in Section \ref{RD:Linearstability}. In particular, we will show that if the quantity $Q_{3,v}(k) > 0$, then it is not possible to achieve the condition $Q_3(k) > Q_1(k) Q_2(k)$ when the diffusivities $D_u$ and $D_{\phi}$ of cooperators and the public good are sufficiently close to $0$. This will allow us to show that there is a parameter regime in which we can use the condition $Q_3(k) = -\det(M(k)) < 0$ to identify the threshold for the onset of Turing pattern formation in the unbiased model.

By writing our expressions for the quantities $Q_1(k) = -\mathrm{tr}(M(k))$, $Q_2(k) = S_2(M(k))$, and $Q_3(k) = - \det(M(k))$ as
\begin{subequations}
\begin{align}
Q_1(k) &= Q_{1,v}(k) D_v + Q_{1,c}(k) \\ 
Q_2(k) &= Q_{2,v}(k) D_v + Q_{2,c}(k) \\ 
Q_3(k) &= -Q_{3,v}(k) D_v + Q_{3,c}(k),
\end{align}
\end{subequations}
we can express the condition $Q_3(k) > Q_1(k) Q_2(k)$ in the form 
\begin{equation}
 -Q_{3,v} D_v + Q_{3,c} > \left( Q_{1,v}(k) D_v + Q_{1,c}(k) \right) \left(Q_{2,v}(k) D_v + Q_{2,c}(k) \right).
\end{equation}
This condition is a quadratic inequality in the defector diffusivity $D_v$ that can be rewritten in the form
\begin{equation} \label{eq:quadraticinequalityTuring}
Q_{1,v}(k) Q_{2,v}(k) D_v^2 + \left[ Q_{1,c}(k) Q_{2,v}(k) + Q_{1,v}(k)  Q_{2,c}(k) + Q_{3,v}  \right] D_v + \left[Q_{1,c}(k) Q_{2,c}(k) - Q_{3,c} \right] < 0.
\end{equation}
We showed in Section \ref{RD:Linearstability} that $Q_{1,v}(k), Q_{1,c}(k), Q_{2,v}, Q_{2,c}, Q_{3,c} > 0$ for all parameters, so the quadratic and linear terms in our inequality will be positive for the case in which $Q_{3,v} > 0$ (which is required to allow the condition $Q_3(k) = -\det(M(k)) > 0$ for sufficiently large $D_v$). 

To understand the sign of the constant term in our quadratic inequality, we can use the expressions from Equations \eqref{eq:Q1RD}, \eqref{eq:Q2RD}, and \eqref{eq:Q3RD} to see that, in the limit of small $D_u$ and $D_{\phi}$, we have that
\begin{subequations}
\begin{align}
\lim_{D_u, D_{\phi} \to 0} Q_{1,c}(k) &= - \det(J) \\
\lim_{D_u, D_{\phi} \to 0} Q_{2,c}(k) &=  S_2(J) \\
\lim_{D_u, D_{\phi} \to 0} Q_{1,c}(k) &= - \mathrm{tr}(J).
\end{align}
\end{subequations}
Because the coexistence equilibrium $E_3$ is stable under the reaction dynamics, we know that $\det(J) > \mathrm{tr}(J) S_2(J)$, and we can therefore deduce that, for a given wavenumber $k$, $Q_{1,c}(k) < Q_{2,c}(k) Q_{2,c}(k)$ for sufficiently small $D_u$ and $D_v$. 

This allows us to see that, if $Q_{3,v} > 0$, all three coefficients of the powers of $D_v$ on the lefthand side of the inequality from Equation \eqref{eq:quadraticinequalityTuring} will be positive for sufficiently small $D_u$ and $D_v$, and therefore the lefthand side cannot be negative for those parameter values. As Turing instability can occur for sufficiently large $D_v$ when $Q_{3,v} > 0$, we can therefore deduce that the minimal value of $D_v$ for which $Q_3(k) = -\det(M(k)) > 0$ will be our threshold for the onset of pattern formation in the parameter regime for which $D_u$ and $D_{\phi}$ are sufficiently small. We will therefore use the determinant condition for instability as our key condition for pattern formation in the reaction-diffusion model, and we will use this threshold value of $D_v$ as our critical parameter for the weakly nonlinear stability in Section \ref{sec:weakly-nonlinear-analysis}.

\section{Calculation of the Weakly Nonlinear Approximation}\label{sec:weakly-nonlinear-expansions}

In this section, we  provide the details of the derivation of the order-by-order system of PDEs we derive from our perturbation equations as part of the weakly nonlinear stability analysis. In particular, we consider the dynamics of our PDE model when our bifurcation of parameters $D_v$ or $w_u$ are close to the threshold for the onset of pattern formation for the cases of the unbiased and biased models, in which case we expect the exponential growth rate to be approximately zero starting from a small perturbation near the coexistence equilibrium. To perform this analysis, we introduce a small parameter $\varepsilon$ to measure the distance from the bifurcation threshold, assuming that the bifurcation parameters take the form $D_v = D_v^* + \varepsilon^2 \tilde{D}$ and $w_u = w_u^* + \varepsilon^2 \tilde{w}$ respectively for the unbiased and biased model, which corresponds to exponential growth rates $\lambda$ proportional to $\epsilon^2 t$ for the linearization performed in Section \ref{sec:PDElinear}.

With this slow growth of the amplitude of our sinusoidal perturbation arising in the linear analysis, we can look to perform a perturbation expansion in terms of our small parameter $\varepsilon$ to analyze the behavior of our solution beyond the linear expansion. With this perturbation expansion, we can obtain a hierarchy of PDEs for our expanded solution that starts with the linearized PDE derived previously and that includes the effects of nonlinear interactions and nonlinear rules for directed motion to determine how the growth of patterns can saturate to produce a spatially heterogeneous steady state. By solving the hierarchy of PDEs in order, we encounter the possibility of having a resonant term at order $\mathcal{O}(\varepsilon^3)$, so we need to apply the solvability condition from the Fredholm alternative to obtain an equation satisfied by the amplitude of our emergent patterns. We then use the resulting amplitude equation to characterize the form of the pattern-forming bifurcation and to describe an approximate expression for a patterned steady state for our PDE model.  

When the bifurcation parameter for the biased or unbiased model differs from the threshold value by a quantity on order $\varepsilon^2$, we expect the temporal evolution of the pattern to occur on a similar timescale, This motivates the introduction of a slow time variable $T=\varepsilon^2 t$, which we will use to describe the temporal evolution of the amplitude of our spatial parameters. We then assume that, close to the onset of pattern formation, the solution of our PDE models can be written in terms of the following perturbation expansion in terms of the small parameter $\varepsilon$:
\begin{equation}
    \begin{aligned}
        u(t,x)&=u_0+\varepsilon u_1(T,x)+\varepsilon^2u_2(T,x)+\varepsilon^3u_3(T,x)+\cdots,\\
        v(t,x)&=v_0+\varepsilon v_1(T,x)+\varepsilon^2v_2(T,x)+\varepsilon^3v_3(T,x)+\cdots,\\
        \phi(t,x)&=\phi_0+\varepsilon\phi_1(T,x)+\varepsilon^2\phi_2(T,x)+\varepsilon^3\phi_3(T,x)+\cdots,
    \end{aligned}
    \nonumber
\end{equation}
where each term in expansion is either constant in time or depends on time only through the slow time variable $T = \varepsilon^2 t$. 

We will plug these perturbation expansions into both sides of our PDE model from Equation \eqref{eq:general-continuum-PDE}. To represent the expansion of the righthand sides of each PDE, we introduce the notation $F_u^{(j)},F_v^{(j)},F_\phi^{(j)}$ to represent the terms obtained at order $\varepsilon^j$ in the $u$-, $v$-, and $\phi$-equations, respectively. This allows us to write the expansion of the righthand sides of Equation \eqref{eq:general-continuum-PDE} as 
\begin{equation}
    \begin{aligned}
        F_u^{(1)}=&\,\,D_u\nabla^2\left(u_1-2w_uu_0\phi_1\right)+u_0\left[r_u\phi_1-\gamma(u_1+v_1)\right],\\
        F_v^{(1)}=&\,\,D_v\nabla^2\left(v_1-2w_vv_0\phi_1\right)+v_0\left[r_v\phi_1-\gamma(u_1+v_1)\right],\\
        F_\phi^{(1)}=&\,\,D_\phi\nabla^2\phi_1+cu_1-\left(\kappa(u_0+v_0)+\delta\right)\phi_1-\kappa\phi_0(u_1+v_1),\\
        F_u^{(2)}=&\,\,D_u\nabla^2\left(u_2-2w_uu_0\phi_2\right)-2w_uD_u\nabla\cdot\left(u_1\nabla \phi_1\right)\\
        &\,\,+u_0\left(r_u\phi_2-\gamma(u_2+v_2)\right)+u_1\left(r_u\phi_1-\gamma(u_1+v_1)\right),\\
        F_v^{(2)}=&\,\,D_v\nabla^2\left(v_2-2w_vv_0\phi_2\right)-2w_vD_v\nabla\cdot\left(v_1\nabla \phi_1\right)\\
        &\,\,+v_0\left(r_v\phi_2-\gamma(u_2+v_2)\right)+v_1\left(r_v\phi_1-\gamma(u_1+v_1)\right),\\
        F_\phi^{(2)}=&\,\,D_\phi\nabla^2 \phi_2+cu_2-\left(\kappa(u_0+v_0)+\delta\right)\phi_2-\kappa\phi_0(u_2+v_2)-\kappa\phi_1(u_1+v_1),\\
        F_u^{(3)}=&\,\,D_u\nabla^2\left(u_3-2w_uu_0\phi_3\right)-2w_uD_u\nabla\cdot\left(u_1\nabla\phi_2+u_2\nabla\phi_1\right)\\
        &\,\,+u_2[r_u\phi_1-\gamma(u_1+v_1)]+u_1[r_u\phi_2-\gamma(u_2+v_2)]+u_0[r_u\phi_3-\gamma(u_3+v_3)],\\
        F_v^{(3)}=&\,\,D_v\nabla^2\left(v_3-2w_vv_0\phi_3\right)-2w_vD_v\nabla\cdot\left(v_1\nabla\phi_2+v_2\nabla\phi_1\right)\\
        &\,\,+v_2[r_v\phi_1-\gamma(u_1+v_1)]+v_1[r_v\phi_2-\gamma(u_2+v_2)]+v_0[r_v\phi_3-\gamma(u_3+v_3)],\\
        F_\phi^{(3)}=&\,\,D_\phi\nabla^2\phi_3+cu_3-(k(u_0+v_0)+\delta)\phi_3-\kappa\phi_0(u_3+v_3)-\kappa\phi_1(u_2+v_2)-\kappa\phi_2(u_1+v_1).
    \end{aligned}
    \nonumber
\end{equation}
We can also apply our expansion of the bifurcation parameter $D_v = D_v^* + \varepsilon^2 \tilde{D}$ or $w_u = w_u^* + \varepsilon^2 \tilde{w}$ to our key parameter for the unbiased and biased model and use the fact that the right-hand side only depends on the slow timescale $T = \epsilon^2 t$ to see that we can collect all of the terms in our PDE at the order $\mathcal{O}(\varepsilon)$ to obtain the following system of equations
\begin{equation}
    \begin{aligned}
        0&=D_u\nabla^2\left(u_1-2w_uu_0\phi_1\right)+u_0\left[r_u\phi_1-\gamma(u_1+v_1)\right],\\
        0&=D_v\nabla^2\left(v_1-2w_vv_0\phi_1\right)+v_0\left[r_v\phi_1-\gamma(u_1+v_1)\right],\\
        0&=D_\phi\nabla^2\phi_1+cu_1-\left(\kappa(u_0+v_0)+\delta\right)\phi_1-\kappa\phi_0(u_1+v_1),
    \end{aligned}
    \nonumber
\end{equation}
which corresponds to the case of the linearization of our PDE model right at the critical bifurcation parameter (as this critical parameter value results in neither exponential decay nor exponential growth of the perturbation over time). We can then rewrite this linearized solution in the form 
\begin{equation}
    \mathbf{0}=\left(\underbrace{\begin{bmatrix}
        D_u & 0 & -2w_uu_0\\
        0 & D_v & -2w_vv_0\\
        0 & 0 & D_\phi
    \end{bmatrix}}_{-R}\nabla^2+\underbrace{\begin{bmatrix}
        -\gamma u_0 & -\gamma u_0 & r_uu_0\\
        -\gamma v_0 & -\gamma v_0 & r_vv_0\\
        c-\kappa\phi_0 & -\kappa\phi_0 & -\kappa(u_0+v_0)-\delta
    \end{bmatrix}}_{J}\right)\begin{bmatrix}
        u_1\\
        v_1\\
        \phi_1
    \end{bmatrix},
    \nonumber
\end{equation}
which highlights the dependence of this linearized system on both the linearization $J$ of the reaction dynamics at the coexistence equilibrium and a movement matrix $R$ describing the effects of diffusive and directed motion in our PDE model. We can then use these two matrices to write our linearized operator for the system as $\mathcal{L}=-R\nabla^2 + J$, and we can use our perturbation expansion to see that a copy of this operator will appear at each differential equation at order $\mathcal{O}(\varepsilon^j)$ due to its action on the terms $u_j$, $v_j$, and $\phi_j$ in our expansion for the cooperator, defector, and public goods densities. Notably, this linear operator will be evaluated in terms of either the critical parameter $D_v^*$ or $w_u^*$ depending on whether we are analyzing the unbiased or biased model, but we retain the single notation for the linear operator $\mathcal{L}$ for simplicity in this section.

We can similarly collect terms in the expanded form of our equation at order $\mathcal{O}(\varepsilon^2)$ to obtain our next system of equations
\begin{equation}
    \begin{aligned}
        0=&\,\,D_u\nabla^2\left(u_2-2w_uu_0\phi_2\right)-2w_uD_u\nabla\cdot\left(u_1\nabla \phi_1\right)\\
        &\,\,+u_0\left(r_u\phi_2-\gamma(u_2+v_2)\right)+u_1\left(r_u\phi_1-\gamma(u_1+v_1)\right),\\
        0=&\,\,D_v\nabla^2\left(v_2-2w_vv_0\phi_2\right)-2w_vD_v\nabla\cdot\left(v_1\nabla \phi_1\right)\\
        &\,\,+v_0\left(r_v\phi_2-\gamma(u_2+v_2)\right)+v_1\left(r_v\phi_1-\gamma(u_1+v_1)\right),\\
        0=&\,\,D_\phi\nabla^2 \phi_2+cu_2-\left(\kappa(u_0+v_0)+\delta\right)\phi_2-\kappa\phi_0(u_2+v_2)-\kappa\phi_1(u_1+v_1),
    \end{aligned}
    \nonumber
\end{equation}
which includes how our linearized operator acts upon the terms $u_2$, $v_2$, and $\phi_2$ and a nonlinear term describing how the reaction dynamics and rules for directed spatial motion depend on nonlinear terms arising from products of the $\mathcal{O}(\epsilon)$ solutions $u_1$, $v_1$, and $\phi_1$ and their derivatives. We can then rewrite this system in terms of the linearized operator $\mathcal{L}$
\begin{equation}
    -\mathcal{L}\begin{bmatrix}
        u_2\\
        v_2\\
        \phi_2
    \end{bmatrix}=\begin{bmatrix}
        -2w_uD_u\nabla\cdot\left(u_1\nabla \phi_1\right)\\
        -2w_vD_v\nabla\cdot\left(v_1\nabla \phi_1\right)\\
        0
    \end{bmatrix}+\begin{bmatrix}
        r_uu_1\phi_1 - \gamma u_1^2-\gamma u_1v_1\\
        r_vv_1\phi_1 - \gamma u_1v_1 -\gamma v_1^2\\
        -\kappa\phi_1(u_1+v_1)
    \end{bmatrix}.
    \nonumber
\end{equation}
Notably, the right-hand side features nonlinear terms highlighting interactions between the densities of cooperators, defectors, and public goods, with the nonlinearities depending only on terms $u_1$, $v_1$, and $\phi_1$ that can be obtained by solving the $\mathcal{O}(\varepsilon)$-problem. 

To describe the behavior of our model at order $\mathcal{O}(\varepsilon^3)$, we will further have to include how the linearized operator acts on the $\mathcal{O}(\varepsilon^3)$ solution $u_3$, $v_3$, and $\phi_3$, as well as nonlineriaties arising from the products of solutions at orders $\mathcal{O}(\varepsilon)$ and $\mathcal{O}(\varepsilon^3)$ and their derivatives. In addition, we see that additional terms arise at this order due to our assumptions on the form of the bifurcation parameters ($D_v = D_v^* + \varepsilon^2 \tilde{D}$ for the unbiased model and $w_u = w_u^* + \varepsilon^2 \tilde{w}$ for the biased model), which will result in $\mathcal{O}(\varepsilon^3)$ terms due to the fact that $u_0$ and $v_0$ are constant densities and that we can write out the relevant movement terms in the linearized operator for the biased model as
\begin{equation}
\begin{aligned}
D_v \doubledelsame{v(t,x)}{x} &= \left( D_v^* + \varepsilon^2 \tilde{D} \right) \left( v_0 + \varepsilon v_1 + \varepsilon v_2 + \cdots \right) \\
&= \varepsilon \left[ D_v^* \doubledelsame{v_1}{x} \right] + \varepsilon^2 \left[ D_v^* \doubledelsame{v_2}{x}\right] + \varepsilon^3 \left[ D_v^* \doubledelsame{v_3}{x} + \tilde{D} \doubledelsame{v_1}{x} \right] + \mathcal{O}(\varepsilon^4)
\end{aligned}
\end{equation}
and we can write the relevant term in the biased model as
\begin{equation}
\begin{aligned}
w_u \dsdel{}{x} \left( u_0 \dsdel{v(t,x)}{x} \right) &= w_u u_0 \doubledelsame{v(t,x)}{x} \\
&= \left( w_u^* + \varepsilon^2 \tilde{w} \right) u_0 \doubledelsame{}{x} \left[v_0 + \varepsilon v_1 + \varepsilon^2 v_2 + \varepsilon^3 v_3 + \cdots \right] \\
&= \varepsilon \left[ w_u^* u_0 \doubledelsame{v_1}{x} \right] + \varepsilon^2 \left[ w_u^* u_0 \doubledelsame{v_2}{x} \right] + \varepsilon^3 \left[ w_u^* u_0 \doubledelsame{v_3}{x} + \tilde{w} u_0 \doubledelsame{v_1}{x} \right] + \mathcal{O}\left( \varepsilon^4 \right).
\end{aligned}
\end{equation}

We can similarly use the assumption of the presence of the slow timescale with the new time variable $T = \epsilon^2 t$ to see that time-derivatives of the quantities $u_1$, $v_1$, and $\phi_1$ will first appear at this order. In particular, we can see use the perturbation expansion for our solution and the fact that $u_0$ is a constant density to see that the time derivatives of our solution for the density of cooperators is given by
\begin{equation}
\begin{aligned}
\dsdel{u(t,x)}{t} &= \dsdel{}{t} \left[u_0 + \varepsilon u_1(T,x) + \varepsilon^2 u_2(T,x) + \cdots \right] \\ &= \varepsilon \dsdel{u_1(T,x)}{T} \dsdel{T}{t} + \varepsilon^2 \dsdel{u_2(T,x)}{T} \dsdel{T}{t} + \cdots \\ &= \varepsilon^3 \dsdel{u_1(T,x)}{T} + \mathcal{O}\left( \varepsilon^4\right),
\end{aligned}
\end{equation}
and therefore we can deduce that the term $\dsdel{u_1}{T}$ will appear to represent the time-derivative of the cooperator density in the $\mathcal{O}(\varepsilon^3)$ equation. We can similarly see that the terms $\dsdel{v_1}{T}$ and $\dsdel{\phi_1}{T}$ will appear in their respective equations at this order.

Putting together all of the terms we have discussed above, we see that the problem for the unbiased model at order $\mathcal{O}(\varepsilon^3)$ is given by
\begin{equation}
    \begin{aligned}
        \frac{\partial u_1}{\partial T}=&\,\,D_u\nabla^2\left(u_3-2w_uu_0\phi_3\right)-2w_uD_u\nabla\cdot\left(u_1\nabla\phi_2+u_2\nabla\phi_1\right)\\
        &\,\,+u_2[r_u\phi_1-\gamma(u_1+v_1)]+u_1[r_u\phi_2-\gamma(u_2+v_2)]+u_0[r_u\phi_3-\gamma(u_3+v_3)],\\
        \frac{\partial v_1}{\partial T}=&\,\,D_v\nabla^2\left(v_3-2w_vv_0\phi_3\right)-2w_vD_v\nabla\cdot\left(v_1\nabla\phi_2+v_2\nabla\phi_1\right)+\tilde{D}\nabla^2v_1\\
        &\,\,+v_2[r_v\phi_1-\gamma(u_1+v_1)]+v_1[r_v\phi_2-\gamma(u_2+v_2)]+v_0[r_v\phi_3-\gamma(u_3+v_3)],\\
        \frac{\partial \phi_1}{\partial T}=&\,\,D_\phi\nabla^2\phi_3+cu_3-(k(u_0+v_0)+\delta)\phi_3-\kappa\phi_0(u_3+v_3)-\kappa\phi_1(u_2+v_2)-\kappa\phi_2(u_1+v_1),
    \end{aligned}
    \nonumber
\end{equation}
while the $\mathcal{O}(\varepsilon^3)$ problem for the biased model is given by
\begin{equation}
    \begin{aligned}
        \frac{\partial u_1}{\partial T}=&\,\,D_u\nabla^2\left(u_3-2w_uu_0\phi_3\right)-2w_uD_u\nabla\cdot\left(u_1\nabla\phi_2+u_2\nabla\phi_1\right)\\
        &\,\,+u_2[r_u\phi_1-\gamma(u_1+v_1)]+u_1[r_u\phi_2-\gamma(u_2+v_2)]+u_0[r_u\phi_3-\gamma(u_3+v_3)]-2\tilde{w}D_uu_0\nabla^2\phi_1,\\
        \frac{\partial v_1}{\partial T}=&\,\,D_v\nabla^2\left(v_3-2w_vv_0\phi_3\right)-2w_vD_v\nabla\cdot\left(v_1\nabla\phi_2+v_2\nabla\phi_1\right)\\
        &\,\,+v_2[r_v\phi_1-\gamma(u_1+v_1)]+v_1[r_v\phi_2-\gamma(u_2+v_2)]+v_0[r_v\phi_3-\gamma(u_3+v_3)],\\
        \frac{\partial \phi_1}{\partial T}=&\,\,D_\phi\nabla^2\phi_3+cu_3-(k(u_0+v_0)+\delta)\phi_3-\kappa\phi_0(u_3+v_3)-\kappa\phi_1(u_2+v_2)-\kappa\phi_2(u_1+v_1).
    \end{aligned}
    \nonumber
\end{equation}
We can further rewrite the problems at order $\varepsilon^3$ respectively for the unbiased and biased models in terms of the linearized operator $\mathcal{L}$ as
\begin{equation}
    \begin{aligned}
        -\mathcal{L}\begin{bmatrix}
            u_3\\
            v_3\\
            \phi_3
        \end{bmatrix}=&\,\,-\frac{\partial}{\partial T}\begin{bmatrix}
            u_1\\
            v_1\\
            \phi_1
        \end{bmatrix}+\begin{bmatrix}
            -2w_uD_u\nabla\cdot\left(u_1\nabla\phi_2+u_2\nabla\phi_1\right)\\
            -2w_vD_v\nabla\cdot\left(v_1\nabla\phi_2+v_2\nabla\phi_1\right)\\
            0
        \end{bmatrix}+\begin{bmatrix}
            0\\
            \tilde{D}\nabla^2v_1\\
            0
        \end{bmatrix}\\
        &\,\,+\begin{bmatrix}
            u_2[r_u\phi_1-\gamma(u_1+v_1)]+u_1[r_u\phi_2-\gamma(u_2+v_2)]\\
            v_2[r_v\phi_1-\gamma(u_1+v_1)]+v_1[r_v\phi_2-\gamma(u_2+v_2)]\\
            -\kappa\phi_1(u_2+v_2)-\kappa\phi_2(u_1+v_1)
        \end{bmatrix}
    \end{aligned}
    \nonumber
\end{equation}
and
\begin{equation}
    \begin{aligned}
        -\mathcal{L}\begin{bmatrix}
            u_3\\
            v_3\\
            \phi_3
        \end{bmatrix}=&\,\,-\frac{\partial}{\partial T}\begin{bmatrix}
            u_1\\
            v_1\\
            \phi_1
        \end{bmatrix}+\begin{bmatrix}
            -2w_uD_u\nabla\cdot\left(u_1\nabla\phi_2+u_2\nabla\phi_1\right)\\
            -2w_vD_v\nabla\cdot\left(v_1\nabla\phi_2+v_2\nabla\phi_1\right)\\
            0
        \end{bmatrix}+\begin{bmatrix}
            -2\tilde{w}D_uu_0\nabla^2\phi_1\\
            0\\
            0
        \end{bmatrix}\\
        &\,\,+\begin{bmatrix}
            u_2[r_u\phi_1-\gamma(u_1+v_1)]+u_1[r_u\phi_2-\gamma(u_2+v_2)]\\
            v_2[r_v\phi_1-\gamma(u_1+v_1)]+v_1[r_v\phi_2-\gamma(u_2+v_2)]\\
            -\kappa\phi_1(u_2+v_2)-\kappa\phi_2(u_1+v_1)
        \end{bmatrix}.
    \end{aligned}
    \nonumber
\end{equation}
Using this form of the problem will help us apply the solvability condition from the Fredholm alternative, which will allow us to obtain the amplitude equation for the pattern-forming bifurcation and assess qualitative properties of emergent spatial patterns in both models. 

\section{Additional Insights from Weakly Nonlinear Analysis: Correlation of Cooperators and Public Good in Spatial Patterns }\label{sec:eigenvectors}

In this section, we note several relevant properties that can be learned about spatial patterns from the solution to the $\mathcal{O}(\varepsilon)$ problem from weakly nonlinear stability analysis. We first show that the qualitative behavior of the amplitude equation and resulting patterns is independent of the choice of a representative element $\mathbf{q}$ of the  kernel of the linearization matrix $M(k)$, which we present in Section \ref{sec:scalingamplitude}. We then  apply this in Section \ref{sec:patterncorrelation} to show that we can pick a particular representation of the kernel that will help us determine that the spatial patterns of cooperators and public goods will feature similar peaks and troughs in their profiles across the domain. 

\subsection{Independence of Qualitative Behavior of Amplitude Equation on Scaling of Kernel Vector} \label{sec:scalingamplitude}

In this section, we show that the choice of vector $\mathbf{q}$ in the kernel of $M_{D_u^\ast}(k^\ast)$ in \eqref{eq:linearization-wavenumber-reaction-diffusion} does not affect the results of the weakly nonlinear analysis. In particular, we can use the expression we found in Section \ref{sec:weakly-nonlinear-analysis} to see that the weakly nonlinear approximation for our solution to both the biased and unbiased model can be expressed in the following form:
\begin{equation}\label{eq:weakly-nonlinear-expansions-appendix}
    \begin{bmatrix}
        u(T,x)\\
        v(T,x)\\
        \phi(T,x)
    \end{bmatrix}=\begin{bmatrix}
        u_0\\
        v_0\\
        \phi_0
    \end{bmatrix}+\varepsilon\mathbf{q}A(T)\cos\left(\frac{\pi k^\ast}{L}x\right)+\varepsilon^2A^2(T)\left[\mathbf{t}_0+\mathbf{t}_2\cos\left(\frac{2\pi k^\ast}{L}x\right)\right]+\mathcal{O}(\varepsilon^3). 
\end{equation}
where $A(T)$ denotes the amplitude, which satisfies the cubic Stuart-Landau equation \eqref{eq:unbiased-SL-equation} with initial condition $A(0)$. 

Because we would like to consider the rescaling of the vector $\mathbf{q}$, we can try to divide $\mathbf{q}$ by a given constant and see if we can multiply the same constant by the amplitude $A(T)$ to leave the qualitative behavior of the pattern unchanged. In particular, we need to show that multiplying the initial amplitude by a constant will produce a solution to the amplitude equation that is multiplied by the same constant. This motivates the following lemma, which will allow us to transfer a constant between the kernel vector $\mathbf{q}$ and the amplitude $A(T)$ of our spatial pattern. 

\begin{lemma}\label{lemma:S-L-ODE-rescaling}
Assume that $\lambda\neq 0$. Let $A_1$ and $A_2$ satisfy
\begin{equation}
    \begin{aligned}
        \frac{dA_1}{dT}&=\eta A_1 + \beta A_1^3,\\
        \frac{dA_2}{dT}&=\eta A_2 + \lambda^2\beta A_2^3.
    \end{aligned}
    \nonumber
\end{equation}
If $A_1(0)=\lambda A_2(0)$, then $A_1(T)=\lambda A_2(T)$ for all $T$. Therefore we see that the nonzero equilibria of the second amplitude equation are given by
\begin{equation}
A_2^* = \pm \lambda \sqrt{\frac{-\beta}{\eta}},
\end{equation}
so the resulting equilibrium amplitude satisfies $A_2^* = \lambda A_1^*$ for each equilibrium of the original amplitude equation for $A_1(T)$. 
\end{lemma}

Rescaling $\mathbf{q}$ by a scalar factor $\lambda\neq 0$ rescales $\mathbf{p}$ by $1/\lambda$ since $\mathbf{p}\cdot\mathbf{q}=1$, while $\mathbf{t}_0$ and $\mathbf{t}_{2}$ in \eqref{eq:unbiased-t} are multiplied by $\lambda^2$. The coefficient $\eta$ in the Stuart-Landau equation \eqref{eq:unbiased-SL-equation} remains unchanged, whereas $\beta$ is multiplied by $\lambda^2$, and the initial condition $A(0)$ is scaled by $1/\lambda$. Applying Lemma~\ref{lemma:S-L-ODE-rescaling}, these changes compensate each other, and all dependence on $\lambda$ cancels in \eqref{eq:weakly-nonlinear-expansions-appendix}.

Since the choice of eigenvector $\mathbf{q}$ does not affect the weakly nonlinear expansion, we may fix a $q_\phi=1$ for $\mathbf{q}=(q_u,q_v,q_\phi)$. This will allow us to simplify our analysis of qualitative properties of emergent patterns, as the signs of $q_v$ and $q_{\phi}$ can then be used to determine whether the peaks and troughs of our spatial patterns will be synchronized across the cooperator, defector, and public good densities.

\subsection{Correlation of Cooperators and Public Goods in Spatial Patterns}
\label{sec:patterncorrelation}

In this section, we will look to calculate the kernel element $\mathbf{q}$ for the linearization matrix $M(k)$ for the unbiased model. This will allow us use the weakly nonlinear expansion from Equation \eqref{eq:weakly-nonlinear-expansions-appendix} to determine whether we expect the cooperator, defector, and public good densities to be correlated in the patterned state. We will consider the expression for the linearization matrix $M(k)$ for general values of the diffusivities $D_u$, $D_v$, and $D_{\phi}$ and the directed motion sensitivities $w_u$ and $w_v$ so that our analysis can be applied directly to understand properties of both the unbiased and biased models.

We start with assuming a kernel vector of the form $\mathbf{q} = \bpm q_u & q_v & 1\epm^T$, and our goal is to compute analytic expressions for $q_u$ and $q_v$ that allow us to see that $M(k) \mathbf{q} = \mathbf{0}$. The matrix $M(k)$ is given by
\begin{equation}
    M(k)=\begin{bmatrix}
        -\left(\frac{\pi k}{L}\right)^2 D_u - \gamma u_0 & -\gamma u_0 & 2\left(\frac{\pi k}{L}\right)^2w_uD_u u_0 + r_u u_0\\
        -\gamma v_0 & -\left(\frac{\pi k}{L}\right)^2D_v - \gamma v_0 & 2\left(\frac{\pi k}{L}\right)^2w_v D_v v_0 + r_v v_0\\
        c-\kappa\phi_0 & -\kappa \phi_0 & -\left(\frac{\pi k}{L}\right)^2D_\phi -\kappa(u_0 + v_0)-\delta
    \end{bmatrix}.
    \nonumber
\end{equation}
Assuming the critical eigenvalue is simple at the bifurcation point, the remaining two equations are linearly independent and uniquely determine $q_u$ and $q_v$. The relation $M(k) \mathbf{q}=0$ yields the linear system
\begin{equation}
    \begin{aligned}
        -\gamma v_0q_u-\left[\left(\frac{\pi k}{L}\right)^2D_v+\gamma v_0\right]q_v+2\left(\frac{\pi k}{L}\right)^2w_v D_v v_0 + r_v v_0&=0,\\
        \big(c-\kappa\phi_0\big)q_u-\kappa \phi_0q_v-\left(\frac{\pi k}{L}\right)^2D_\phi -\kappa(u_0 + v_0)-\delta&=0.
    \end{aligned}
    \nonumber
\end{equation}
Then we can solve for $q_u$ and $q_v$ to see that
\begin{subequations}
    \begin{align}
        q_u&=\frac{\kappa\phi_0 v_0\Big[2\left(\frac{\pi k}{L}\right)^2w_vD_v+r_v\Big]+\Big[\left(\frac{\pi k}{L}\right)^2D_v+\gamma v_0\Big]\Big[\left(\frac{\pi k}{L}\right)^2D_\phi+\kappa\big(u_0+v_0\big)+\delta\Big]}{\gamma\kappa v_0\phi_0+\Big[\left(\frac{\pi k}{L}\right)^2D_v+\gamma v_0\Big]\big(c-\kappa \phi_0\big)}, \label{eq:quappendix}\\
        q_v&=\frac{\big(c-\kappa\phi_0\big)\Big[2\left(\frac{\pi k}{L}\right)^2w_vD_vv_0+r_vv_0\Big]-\gamma v_0\Big[\left(\frac{\pi k}{L}\right)^2D_\phi+\kappa\big(u_0+v_0\big)+\delta\Big]}{\gamma\kappa v_0\phi_0+\Big[\left(\frac{\pi k}{L}\right)^2D_v+\gamma v_0\Big]\big(c-\kappa \phi_0\big)} \label{eq:qvappendix}
    \end{align}
\end{subequations}
Furthermore, we can use the scaling behavior demonstrated in Section \ref{sec:scalingamplitude} to see that, for a more general element of the kernel $\mathbf{q} = \bpm q_u & q_v & q_{\phi} \epm^T$, the ratios of the coefficients of $q_u$ and $q_v$ with $q_{\phi}$ are given by
   \begin{subequations}
    \begin{align}
    \label{eq:uphiratioappendix}
    \frac{q_u}{q_{\phi}} &=  \frac{\kappa\phi_0 v_0\Big[2\left(\frac{\pi k}{L}\right)^2w_vD_v+r_v\Big]+\Big[\left(\frac{\pi k}{L}\right)^2D_v+\gamma v_0\Big]\Big[\left(\frac{\pi k}{L}\right)^2D_\phi+\kappa\big(u_0+v_0\big)+\delta\Big]}{\gamma\kappa v_0\phi_0+\Big[\left(\frac{\pi k}{L}\right)^2D_v+\gamma v_0\Big]\big(c-\kappa \phi_0\big)} \\
    \label{eq:vphiratioappendix} \frac{q_v}{q_{\phi}} &= \frac{\big(c-\kappa\phi_0\big)\Big[2\left(\frac{\pi k}{L}\right)^2w_vD_vv_0+r_vv_0\Big]-\gamma v_0\Big[\left(\frac{\pi k}{L}\right)^2D_\phi+\kappa\big(u_0+v_0\big)+\delta\Big]}{\gamma\kappa v_0\phi_0+\Big[\left(\frac{\pi k}{L}\right)^2D_v+\gamma v_0\Big]\big(c-\kappa \phi_0\big)}.
    \end{align}
    \end{subequations}
Near the bifurcation point, the spatial pattern is well approximated by the critical linear mode,
\begin{equation}
    \begin{bmatrix}
        u(T,x)\\
        v(T,x)\\
        \phi(T,x)
    \end{bmatrix}=\begin{bmatrix}
        u_0\\
        v_0\\
        \phi_0
    \end{bmatrix} + \varepsilon A(T)\,\mathbf{q}\cos\!\left(\frac{\pi k^\ast}{L}x\right)+\mathcal{O}(\varepsilon^2),
\end{equation}
where $\mathbf{q}$ is the a representative vector in the kernel of $M(k^*)$. Consequently, the signs of the components of $\mathbf{q}$ determine whether the variables attain maxima and minima at the same spatial locations.

Because we assume that $c>\kappa\phi_0$  to satisfy the equilibrium conditions for the reaction dynamics, we can deduce from Equation \eqref{eq:quappendix} that $\frac{q_u}{q_{\phi}}>0$. Hence the $u$- and $\phi$-components of the critical mode have the same sign and therefore attain spatial maxima and minima at the same locations in the patterned state. This calculation matches with our observation from numerical simulations  that the cooperator density and the public good concentration exhibit similar peak and trough locations in both the unbiased and biased models. In contrast, it does not appear to be possible to determine the sign of $q_v$ analytically from Equation \eqref{eq:qvappendix}, and therefore it appears that we have to rely on numerically evaluating the expression for $q_v$ for given parameter values.

\end{document}